\documentclass{statsoc}

\usepackage{lineno,hyperref}
\usepackage[a4paper]{geometry}
\usepackage{graphicx}
\usepackage[textwidth=8em,textsize=small]{todonotes}
\usepackage{amsmath}
\usepackage{natbib}
\usepackage{graphics,amssymb,epsfig,color}

\usepackage{enumerate}
\usepackage{algorithm,algorithmicx,algpseudocode}
\usepackage{bm}
\usepackage{booktabs}
\usepackage{multirow} 
\usepackage{float}
\usepackage{amsfonts}
\usepackage{algcompatible,amsmath}
\usepackage{rotating}[figuresright]
\usepackage[shortlabels]{enumitem}
\usepackage{dashrule}
\usepackage{diagbox}
\usepackage{amssymb}

\DeclareMathOperator{\Cov}{Cov}
\DeclareMathOperator{\Ex}{E}

\DeclareMathOperator*{\argmin}{argmin}

\DeclareMathOperator{\Med}{Med}
\DeclareMathOperator{\FuNMAD}{FuNMAD}
\usepackage{array}
\usepackage{booktabs,amsfonts,dcolumn}
\usepackage{soul}
\newcolumntype{M}[1]{>{\centering\arraybackslash}m{#1}}

\title{ROFANOVA}
\title[Robust Functional ANOVA]{\textbf{Robust Functional ANOVA with Application to Additive Manufacturing}}
\author[Centofanti {\it et al.}]{Fabio Centofanti}\coaddress{fabio.centofanti@unina.it}
\address{Department of Industrial Engineering, University of Naples Federico II, Naples, Italy}
\author[Centofanti {\it et al.}]{Bianca Maria Colosimo }
\address{Department of Mechanical Engineering, Politecnico di Milano, Milan,  Italy}
\author[Centofanti {\it et al.}]{ Marco Luigi Grasso }
\address{Department of Mechanical Engineering, Politecnico di Milano, Milan,  Italy}
\author[Centofanti {\it et al.}]{ Alessandra Menafoglio  }
\address{ Department of Mathematics, Politecnico di Milano,  Milan,  Italy}
\author[Centofanti {\it et al.}]{Biagio Palumbo}
\address{Department of Industrial Engineering, University of Naples Federico II,  Naples, Italy}
\author[Centofanti {\it et al.}]{Simone Vantini  }
\address{ Department of Mathematics, Politecnico di Milano,  Milan,  Italy}

\author[Centofanti {\it et al.}]{}

\begin{document}
\begin{abstract}
	The development of data acquisition systems is facilitating the collection of data that are apt to be modelled as functional data. In some applications, the interest lies in the identification of significant differences in group functional means defined by varying experimental conditions, which is known as functional analysis of variance (FANOVA). With real data, it is common that the sample under study is contaminated by some outliers, which can strongly bias the analysis. In this paper, we propose a new robust nonparametric functional ANOVA method (RoFANOVA) that reduces the weights of outlying functional data on the results of the analysis. It is implemented through a permutation test based on a test statistic obtained via a functional extension of the classical robust $ M $-estimator. By means of an extensive Monte Carlo simulation study, the proposed test is compared with some alternatives already presented in the literature, in both one-way and two-way designs. The performance of the RoFANOVA is demonstrated in the framework of a motivating real-case study in the field of additive manufacturing that deals with the analysis of spatter ejections.
	The RoFANOVA method is implemented in the  \textnormal{\sffamily R} package \textnormal{\sffamily rofanova}, available online at \url{https://github.com/unina-sfere/rofanova}.
	
\end{abstract}

\section{Introduction}
\label{sec_int}
The development of data acquisition methods allow the analysis of complex systems in several operating conditions as never before. Several examples may be found in the current Industry 4.0 framework, which is  reshaping the variety of signals and measurements that can be gathered during manufacturing processes. Experimental data are more and more characterized by complex and novel formats, like images, videos, dense point clouds. These data may be acquired not only off line, during post-process inspections on the product, but also in line,  during the production process, by exploiting a variety of sensors installed and embedded into the system. The rich information enclosed in such big data streams allows one to monitor and optimize industrial processes, as well as to improve the productivity and efficiency of production plants and enable  several benefits of the ongoing digital transition. As a consequence, the focus of many applications in industrial statistics is  moving from product quality characteristics to in-line process measurements, thanks to enhanced sensing and monitoring capabilities. Moreover, novel production paradigms are characterized by several controllable factors and complex process dynamics that impose the need for effective and efficient experimental approaches to determine optimal process conditions and gather  deeper comprehension of underlying physical phenomena.

In this framework, a number of novel challenges shall be faced, with respect to how the quality of products is monitored, modelled and continuously improved. In many cases, statistical methods require a transformation of input data that are characterized by complex and/or high dimensional formats (e.g., multi-channel signals, images, videos, point clouds) into a format that is  easier to handle and, at the same time, able to capture the  information content and in order to draw  reliable and robust decisions. A family of statistical methods suitable to tackle this problem is known as functional data analysis (FDA). For a comprehensive overview of FDA methods and applications we refer the reader to \cite{ramsay2005functional,horvath2012inference,kokoszka2017introduction} and, for further theoretical insights, to \cite{hsing2015theoretical,bosq2012linear}. FDA allows the representation of observation units in terms of functions in a 1D, 2D or higher dimensional domain with a  general validity this is not limited to manufacturing applications. Such functional representation makes statistical inference methods applicable also in cases where the complexity of the input data goes far beyond traditional univariate or multivariate domains.
A large variety of industrial applications where sensor signals and metrology data can be represented and analyzed as functional data have been presented so far \citep{noorossana2011statistical}. Examples include signals with  cyclic patterns, calibration curves and coordinate measurements of profiles that can be treated as 1D functions \citep{Paynabar20131235, Guo2019, Qiu2010265, Colosimo20101575, Grasso20146110}. Other examples include spatial measurements and surface data that can be treated as 2D functions \citep{Zang2018379, Colosimo201495}. FDA resulted to be effective in modelling complex spatial or spatio-temporal patterns of image and video-image data as well, with various applications. Examples from this research line were reviewed by \cite{Megahed201183}. Other examples include process monitoring and quality modelling applications where data are modelled as functional data and lead to effective  anomaly detection  \citep{Wang2005677, Menafoglio2018497, Wells20131267,capezza2021functional1, centofanti2021functional,  capezza2021clus, Colosimo2021}.

\begin{figure}
  	\centering
  	\begin{tabular}{cc}
  \vspace{0.32cm} \includegraphics[width=0.3\textwidth]{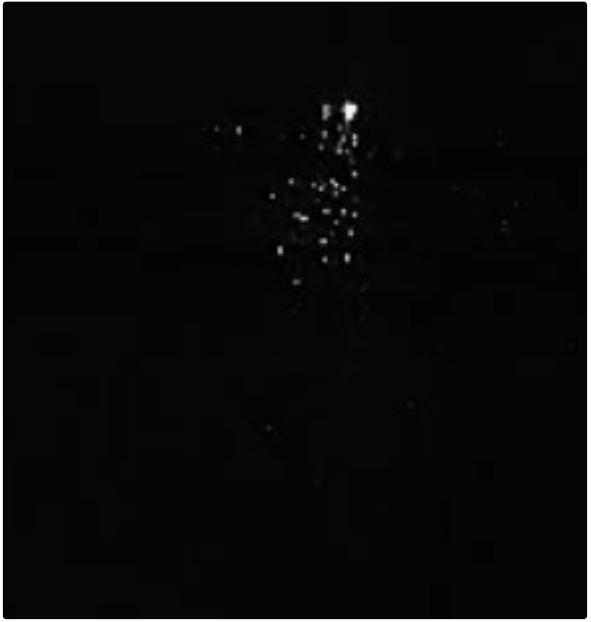} &
  \vspace{0pt} \includegraphics[width=0.335\textwidth]{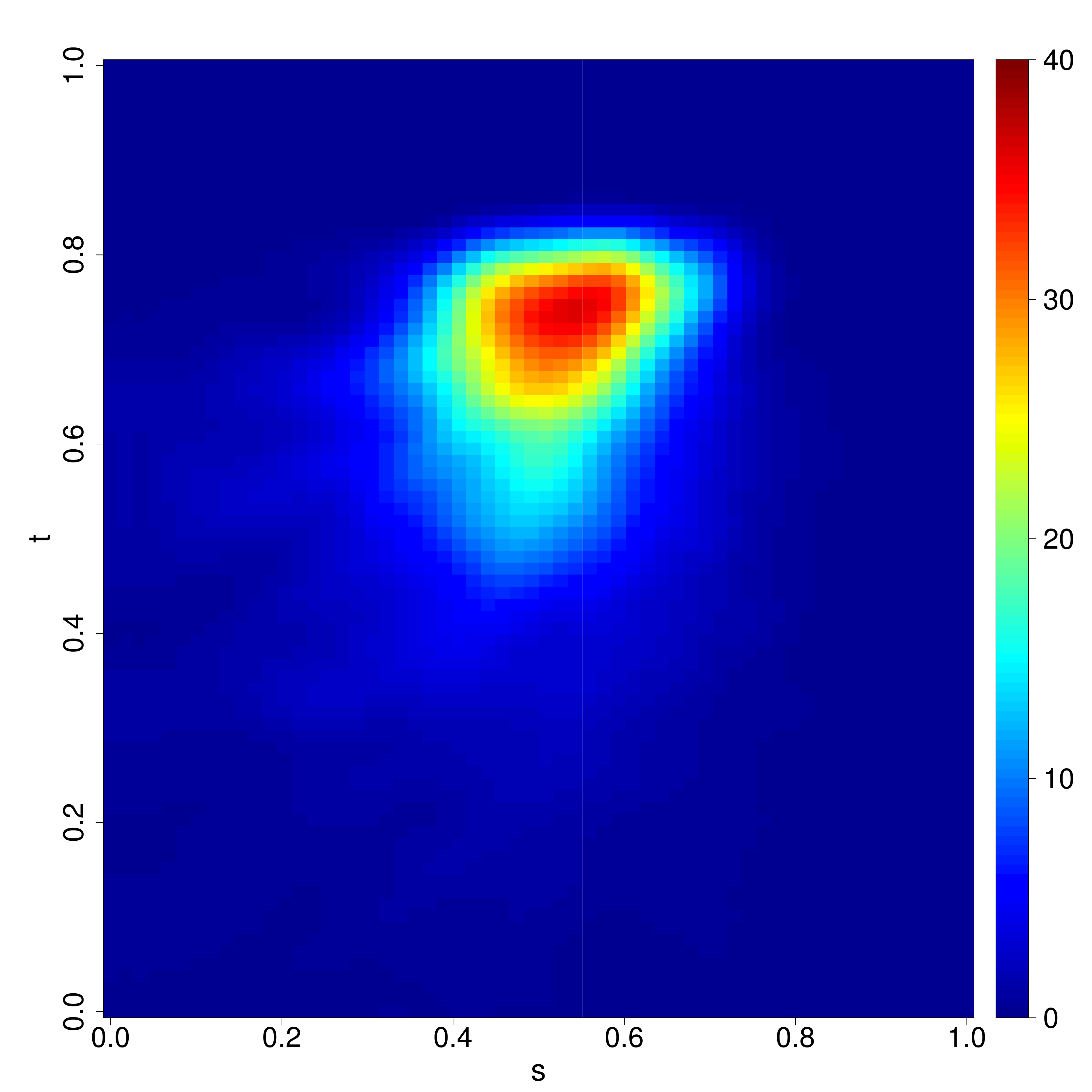}\\
  (a) & (b)\\
\end{tabular}

  	\caption{\label{fig_introduction_1}Example of a video frame acquired during an L-PBF process showing ejected spatters as bright spots (a) and corresponding spatter intensity function (b). }
  \end{figure}

One example of the use of FDA to translate video-image data into a functional form is presented here below and motivated the present study. It regards the analysis of process stability in a metal additive manufacturing process known as laser powder bed fusion (L-PBF) by means of high speed videos acquired during the process \citep{colosimo2018opportunities,colosimo2020machine,Grasso2021}. L-PBF is an additive process suitable to produce metal parts by means of a laser beam that selectively melts  thin layers of metal powder. The process is repeated layer by layer, with the material solidified in one layer being welded to the material in underneath layers, enabling the fabrication of products with complex geometries and innovative properties \citep{gibson2014additive}. Fig. \ref{fig_introduction_1}, left panel, shows an example of video frame acquired during this process. The small white particles, which are visible in the image, are spatters produced by the laser-material interaction, whereas the bigger white spot is the heat affected region where the laser is melting the material. This is just one frame of a high-speed (1000 frames per second) video, where spatters exhibit a complex time-variant dynamic pattern that is representative of the process stability. It is evident that the application of statistical inference methods to video-image data like these can be applied only if the information content is transformed, modelled or synthesised into a different format. One possible way consists of estimating synthetic quantities (like the number of spatters, their size, etc.) and translating the original video frame into a multivariate vector of descriptors \citep{yang2020monitoring,andani2017spatter,repossini2017use}. This approach entails an intrinsic information loss and an arbitrary and problem dependent choice of descriptors. Another approach consists in transforming the image into a functional format. An example of this transformation is shown in Fig. \ref{fig_introduction_1}, right panel, where a 2D function depicts spatter  spread in space over the video frame. This function, which will be referred to as \textit{spatter intensity} function in this study, maps the amount of spatters observed in any region of the bi-dimensional video-frame space, $(s,t)$. The term \textit{intensity} here refers to the occurrence of spatters in a given location. A high spatter intensity at given spatial coordinates $(s,t)$ means that a large amount of spatters was captured in the video image stream in that specific location. Such  representation allows one to capture  spatial information on spatter spread in space and to make inference in a FDA fashion. 
The example shown in Fig. \ref{fig_introduction_1} can be regarded as just one of many real applications where a functional data representation is suitable to deal with complex patterns and data types. A  functional representation similar to that of Fig. \ref{fig_introduction_1} can be suitable  in all processes where spatters and hot ejections are generated, like welding or laser cutting.

A classical statistical problem consists in the identification of significant differences in group functional means belonging to  a sample with varying experimental conditions. In the literature, this problem is known as functional analysis of variance (FANOVA) that is the FDA extension  of the classical (non-functional) ANOVA problem. Referring to the example in Fig. \ref{fig_introduction_1}, the FANOVA approach may be used to study the effect of different process conditions on the spatter behaviour, which is a problem that attracted  great interest in the additive manufacturing community, because the spatter behaviour can be regarded as  a proxy of  process stability and quality \citep{yang2020monitoring,andani2017spatter,repossini2017use,ly2017metal,bidare2018fluid}.
\cite{ramsay2005functional} proposed a functional ANOVA test, based on a pointwise $ F $-test statistic, that relies on the normality assumption of the error function. If the observed statistics is larger than the critical value, calculated as a percentile of the Fisher distribution, for each domain value, then  the hypothesis of no differences among the groups can be safely rejected. 
\cite{cuevas2004anova} proposed a FANOVA test based on the  integrated squared difference among group functional means, for both the homoscedastic and heteroscedastic cases.
The $ L^2 $-norm-based test proposed by \cite{faraway1997regression,zhang2007statistical} uses a statistic based on the integrated squared differences between the group mean and the global mean, whose distribution is approximately proportional to a chi-squared random variable.
\cite{shen2004f,zhang2011statistical} proposed an $ F $-type test based on the fraction of the sum of the integrated squared differences between the group means and the global mean, and, the sum of the integrated squared differences between the functional observations and the group means. Under certain conditions, this statistic has a Fisher distribution. Bootstrap versions of both  $ L^2 $-norm-based and $ F $-type tests were proposed by \cite{zhang2013analysis}. Finally, \cite{zhang2014one} introduced a globalized version of the pointwise $ F $-test. Note that all the aforementioned works deal with the  one-way FANOVA design. 

\begin{figure}
  		\centering
  	\begin{tabular}{ccc}
  	\includegraphics[width=0.3\textwidth]{Figures/Case-study_introduction_heat-eps-converted-to.pdf} &
  	\includegraphics[width=0.3\textwidth]{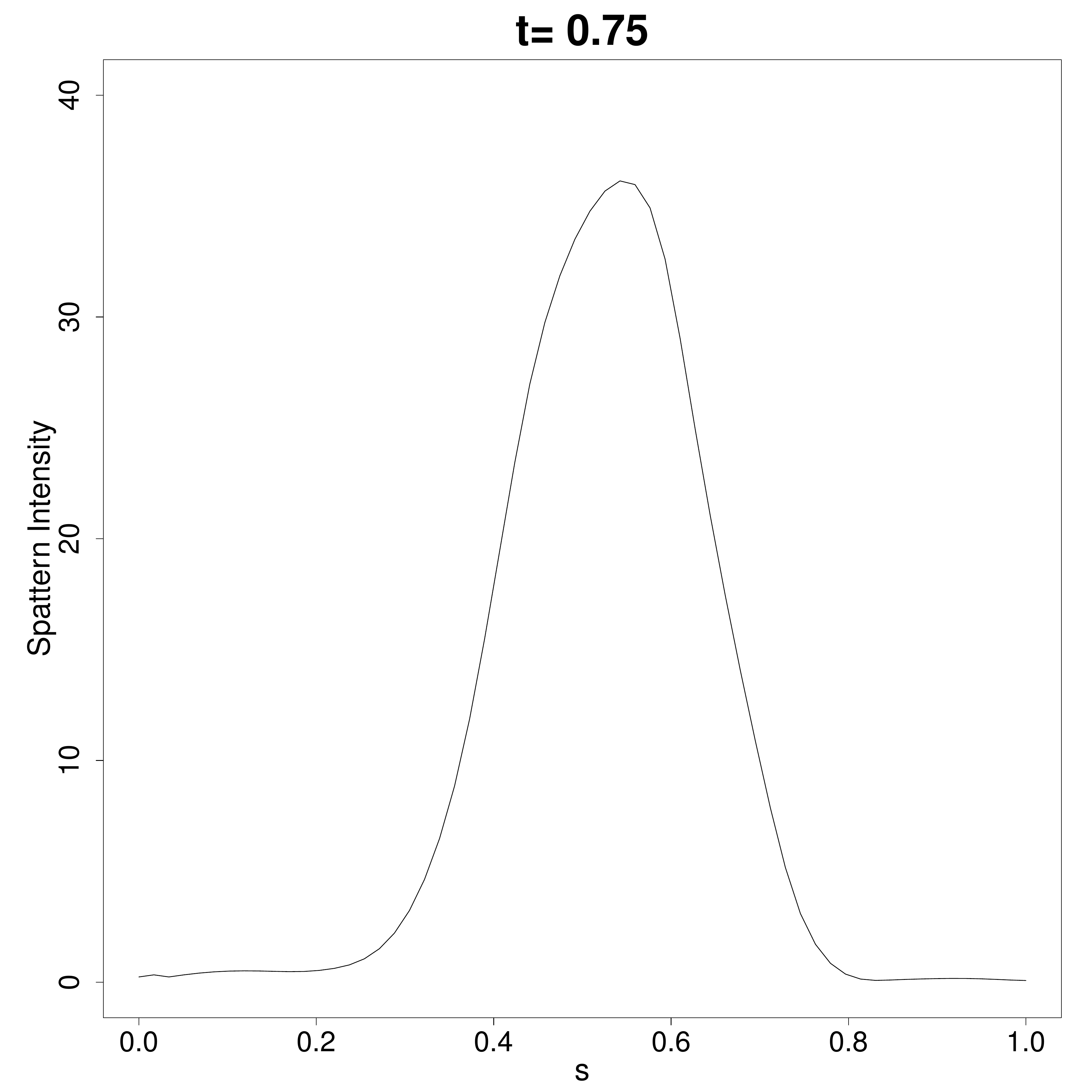} &
  	\includegraphics[width=0.3\textwidth]{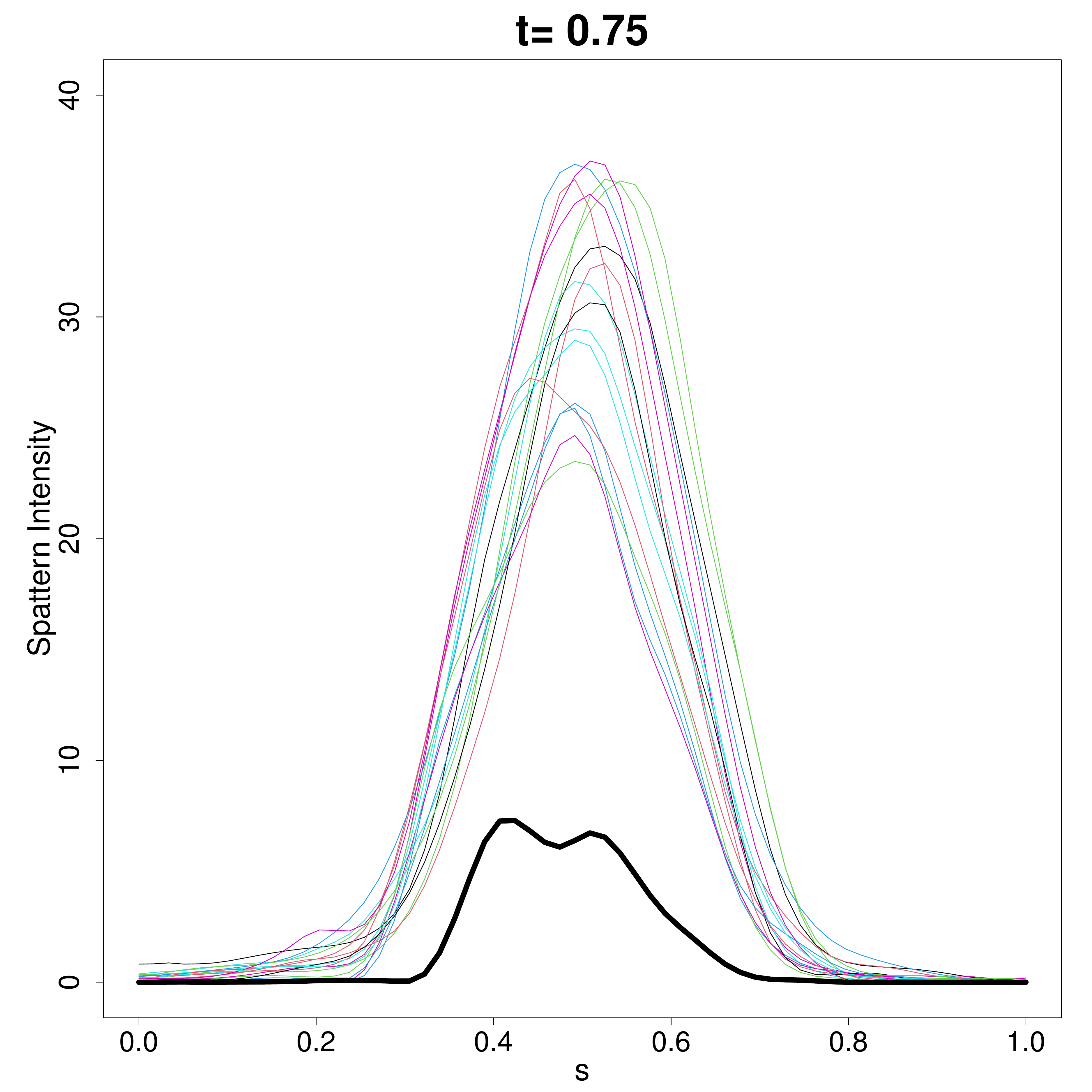} \\
  (a) & (b)& (c)\\
  	\end{tabular}
  	\caption{\label{fig_introduction} Example of a spatter intensity function (a), one cross-section of the spatter intensity function at t = 0.75 (b) and a superimposition of cross-sections corresponding to different experimental realizations of the spatter intensity function, where an outlying pattern is highlighted with a thick black line (c).}
  \end{figure}

The  multi-way functional ANOVA design has been much less studied than the one-way counterpart. In particular, \cite{brumback1998smoothing,guo2002inference,gu2013smoothing} proposed tests that are able to deal with more complicated designs that rely on the use of  smoothing splines (SS-ANOVA). A simple technique  was proposed by \cite{cuesta2010simple} who  transform functional data into univariate data by means of random projections. \cite{pini2018domain} proposed a non‐parametric domain‐selective multi-way functional ANOVA able to identify the specific subdomains where group functional means differ.
In this study, we address the functional analysis of variance in the presence of nuisance effects associated to outlying patterns in the experimental dataset. The proposed real-case study in additive manufacturing highlights the need for novel and effective methods in this framework.
% Indeed, additive manufacturing, like many other manufacturing processes and thermal treatments, is characterized by complex dynamics and many transient and local phenomena that not only affect the natural variability of the measured quantities, but could also lead to outlying patterns.
In the motivating case study considered by this paper, an outlying spatter ejection behaviour may be observed as a consequence of a variety of possible root causes. 
Fig. \ref{fig_introduction} shows an example of an outlying pattern in the spatter intensity function. For the sake of graphical clarity, functions corresponding to different realizations under the same experimental treatment are compared by looking at their cross-sections at a fixed coordinate $t$. The cross-section shown with a solid thick line in Fig. \ref{fig_introduction} represents an outlying spatter behaviour, consisting of a lower amount of spatters spread in space, possibly caused by a transient laser beam attenuation that occurred at a given point in time. Additional details about the real-case study can be found in Section 4.

From a design-of-experiments perspective, outlying patterns like the one in Fig. \ref{fig_introduction} represent a nuisance, as they may inflate the variability and mask effects of potential interest. From a statistical process monitoring perspective, instead, outliers  commonly drive  relevant information, being potential indicators of anomalies and flaws. In this study, we refer to the former perspective, aiming at proposing an effective approach for the analysis of variance in the presence of outliers that contaminate the experimental functional data. 
Due to the many different dynamics involved in the process, determining whether an experimental point is an outlier and identifying its root cause can be a difficult task, but similar challenges can be faced in many different manufacturing applications, due to the complex nature of the response variables and the complex underlying physical phenomena.

All the one-way and multi-way FANOVA design cited above  combine in a different quadratic fashion the functional mean to obtain the test statistic.
However, as in the case of finite dimensional data,  the functional mean, as well as quadratic forms, are shown to be  highly sensitive to the presence of outliers.
\cite{hubert2015multivariate} set up a taxonomy of functional outliers.
To deal with outliers,  the \textit{diagnostic} and the \textit{robust} approaches  are the two common alternatives.
The diagnostic approach  is based on standard estimates after the removal of sample units identified as outliers. Even though it is  criticized as it is subject to  the analyst's personal decision, it can often be safely applied, such as in the case depicted in Fig. \ref{fig_introduction}, where the marked curve can be safely deleted. However, as we will see below, it is not always easy to label an observation as outlier, especially when complex process dynamics and lack of measurable covariates make the search for root causes a difficult task.
On the contrary, the robust approach produces  parameter estimators as well as associated tests and confidence intervals that limit the influence of outliers on final results and decisions without the need for searching and explicitly removing them before the estimation. For a general perspective on this topic in the classical setting see \cite{huber2004robust,hampel2011robust,maronna2019robust}.

In the very last years, several works have explored robust estimation for functional data.
\cite{fraiman2001trimmed} defined  trimmed means for functional data based on a functional depth defined as an integral of the univariate depths for each domain value.
To obtain  robust estimates of the center of a functional distribution, \cite{cuesta2006impartial} extended the notion of impartial trimming to a functional data framework.
Other location estimators based on depth functions for functional data were proposed by \cite{cuesta2008random,cuevas2009depth,lopez2009concept,lopez2011half}.
The above methods are all extensions of the classical linear combination type estimators (i.e., $ L $-estimator) \citep{maronna2019robust} to the functional setting.
More recently, \cite{sinova2018m} extended the notion of maximum likelihood type estimators (i.e., $  M $-estimators) to the functional data setting. $ M $-estimators \citep{huber1964robust} 
are less influenced by outliers than the standard least-squares or maximum likelihood estimators, because they are based on  loss functions that  increase less rapidly than the usual square loss.
These estimators have been applied by \cite{kalogridis2019robust} to the functional linear model.

The  FANOVA methods  are not  necessarily robust against outliers, as they  rely on both  the functional mean and   quadratic forms, which are known to be highly sensitive to outlying observations. In the classical setting,  robust ANOVA methods have been proposed by \cite{schrader1977robust,schrader1980robust}, who adapted  Huber's $ M $-estimates to be used in   a modified $ F $-statistic and a likelihood  ratio type test.
However, to the best of our knowledge, no robust  ANOVA has been introduced so far in the functional setting.

In this paper, we propose a  robust  functional ANOVA method (RoFANOVA)  that is able to test, in a nonparametric fashion, the differences among group functional means.  The RoFANOVA method is based  on a functional generalization of  the test statistic proposed by \cite{schrader1977robust} included in a permutational framework \citep{good2013permutation,pesarin2010permutation}. Applications of nonparametric methods in FDA can be found in \cite{ramsay2005functional,corain2014new,pini2017interval,pini2018domain}.
 Moreover, to obtain the test statistic, we introduce a functional extension of the normalized median absolute deviation (NMAD) estimator, referred to as functional normalized median absolute deviation (FuNMAD) estimator, as well as an equivariant version of the functional $ M $-estimator proposed by \cite{sinova2018m}.
An extensive Monte Carlo simulation study is presented to quantify the performance of the RoFANOVA with respect to  FANOVA tests already appeared in the literature before,  both in one-way and two-way designs. The application of the proposed approach to the real-case study in the additive manufacturing field also highlights its effectiveness over competing  methods in identifying interaction effects that are relevant to get deeper insights about the functional response variable of interest.

%\textbf{The proposed method has been motivated by a real case-study, which consists of an  additive manufacturing (AM) process, where the influence of the level of fluency and acquisition layer on the spatter distribution is studied.
%Aggiungere qualche info riguardante l'additive il problema in esame ed eventuali references.}

The paper is organized as follows. In Section \ref{sec_met}, the robust functional analysis of variance is introduced together with the  functional normalized median absolute deviation and the scale equivariant functional $ M $-estimator. Section \ref{sec_sim} presents a Monte Carlo simulation study that compares the RoFANOVA with  competing methods both in one-way and two-way designs. Then, in Section \ref{sec_real} the RoANOVA is applied to the real-case study devoted to the study of the spatter behaviour in the L-PBF process.
Conclusion is provided in Section \ref{sec_con}.
All computations and plots have been created by using R software \citep{Rcoreteam2020}. The RoFANOVA method is implemented in the \textnormal{\sffamily R}  package \textnormal{\sffamily rofanova},
openly available online at \url{https://github.com/unina-sfere/rofanova}.

\section{The robust functional analysis of variance }
\label{sec_met}

\subsection{The scale equivariant functional $ M $-estimator and the functional normalized median absolute deviation estimator}
\label{subsec_mestimator}
This section introduces the equivariant functional $ M $-estimator and the functional normalized median absolute deviation estimators.
Let us consider the random element  $ X $ with value in $ L^2\left(\mathcal{T}\right) $, the Hilbert space of square integrable functions defined on the compact set $ \mathcal{T}\subset \mathbb{R}^p $, with the usual norm $ ||f||=\left(\int_{\mathcal{T}}f^2\left(t\right) dt\right)^{1/2} $, for $f\in L^2\left(\mathcal{T}\right)$, having mean function $ \mu\left(t\right)=\Ex\left[X\left(t\right)\right] $ and covariance function $ \gamma\left(s,t\right)=\Cov\left[X\left(s\right),X\left(t\right)\right] $, for $ s,t\in\mathcal{T}$. Moreover, let $ \bm{X}=\left(X_1,\dots,X_n\right)^T $ be a vector whose elements  $X_i$ are independent realizations  of $ X $.
Recently, \cite{sinova2018m} proposed a functional $ M $-estimator of location defined as
\begin{equation}\label{eq_sin}
\hat{\mu}_{s}=\argmin_{y\in L^2\left(\mathcal{T}\right) }\sum_{i=1}^{n}\rho\left(||X_i-y||\right),
\end{equation}
where  $ \rho:\mathbb{R}^+\rightarrow \mathbb{R}$ is the \textit{loss function}, which is continuous, non-decreasing and satisfies $ \rho\left(0\right)=0 $. As shown by  \cite{sinova2018m}, each version of $ \hat{\mu}_{s} $ is well-defined and enjoys good theoretical properties, e.g., it has maximal breakdown value and is strong consistent under suitable model assumptions.
Unfortunately, these estimators are not scale equivariant. This means that, if all $ X_i $ are equally scaled,  the resulting robust estimator is not necessarily  equally scaled, in analogy with the multivariate case \citep{maronna2019robust}.
Following \cite{maronna2019robust}, we propose a scale equivariant $ M $-estimator of location defined as
\begin{equation}\label{eq_scaleq}
\hat{\mu}=\argmin_{y\in L^2\left(\mathcal{T}\right) }\sum_{i=1}^{n}\rho\left(\Big\lVert\frac{X_i-y}{\sigma}\Big\lVert\right),
\end{equation}
where $ \sigma\left(t\right)=\sqrt{\gamma\left(t,t\right) }$, for $ t\in\mathcal{T} $.  If $ \sigma $ is known, the problem can be reduced to the case of a $ L^2 $ random  element with $ \sigma=1 $. However, $ \sigma $ can be rarely assumed as known, and thus it should be substituted by a robust scale estimator.
In this regard, we define the FuNMAD estimator of $\sigma$  as follows
\begin{equation}\label{eq_funmad}
\FuNMAD\left(\bm{X}\right)=\frac{1}{c}\Med\left(|\bm{X}-\hat{\mu}_{s,med}|\right),
\end{equation}
with $c=0.6745$ and where $ \hat{\mu}_{s,med}$ denotes the functional generalization of the median obtained as the solution of the optimization problem in equation \eqref{eq_sin} with $ \rho^{med}\left(\cdot\right)= |\cdot| $; $|\bm{X}-\hat{\mu}_{s,med}|=\left(|X_1-\hat{\mu}_{s,med}|,\dots,|X_n-\hat{\mu}_{s,med}|\right)^T$ and $ \Med\left(\cdot \right) $ transforming a vector of functions to a function of pointwise medians.  The constant $ c $ makes $ \FuNMAD $  an asymptotically pointwise consistent estimator of $\sigma$ as shown in the Supplementary Material.

Because the minimization problem in equation \eqref{eq_sin} has not a closed-form solution, \cite{sinova2018m} proposed a standard iteratively re-weighted least-squares algorithm to  approximate  $ \hat{\mu}_{s} $. The algorithm is specifically modified to approximate   $ \hat{\mu} $  in equation \eqref{eq_scaleq} with $ \sigma $ estimated through $\FuNMAD\left(\bm{X}\right)$, and  can be summarized in the following steps.
\begin{enumerate}[Step 1.]
	\item Select initial weight vector $\bm{w}^{\left(0\right)}=\left(w_1^{\left(0\right)},\dots,w_n^{\left(0\right)}\right)\in \mathbb{R}^n$ such that $ w_i^{\left(0\right)}\geq 0 $ and $ \sum_{i=1}^{n}w_i^{\left(0\right)}=1$.
	\item Generate a sequence $ \lbrace \hat{\mu}^{\left(k\right)}\rbrace_{k\in\mathbb{N}} $ iterating the following procedure:
	\begin{equation*}
	\hat{\mu}^{\left(k\right)}=\sum_{i=1}^{n}w_i^{\left(k-1\right)}X_i,\quad \quad w_i^{\left(k\right)}=\frac{\psi\left(\Big\lVert\frac{X_i-\hat{\mu}^{\left(k\right)}}{\sigma}\Big\lVert\right)}{\sum_{i=1}^{n}\psi\left(\Big\lVert\frac{X_i-\hat{\mu}^{\left(k\right)}}{\sigma}\Big\lVert\right)},
	\end{equation*}
	where $ \psi $ is the first derivative of the loss function $ \rho $.
	\item Terminate the algorithm when, for a tolerance $ \varepsilon>0 $, the following condition is met
	\begin{equation*}
	\frac{|J\left(\hat{\mu}^{\left(k\right)}\right)-J\left(\hat{\mu}^{\left(k-1\right)}\right)|}{J\left(\hat{\mu}^{\left(k-1\right)}\right)}<\varepsilon,
	\end{equation*}
	where $ J\left(h\right) =\sum_{i=1}^{n}\rho\left(\Big\lVert\frac{X_i-h}{\hat{\sigma}}\Big\lVert\right)$.
\end{enumerate}
The initial weight vector can be chosen with $ w_i^{\left(0\right)}=\frac{\psi\left(\Big\lVert\frac{X_i-\hat{\mu}^{\left(0\right)}}{\sigma}\Big\lVert\right)}{\sum_{i=1}^{n}\psi\left(\Big\lVert\frac{X_i-\hat{\mu}^{\left(0\right)}}{\sigma}\Big\lVert\right)} $ where $ \hat{\mu}^{\left(0\right)} $ is a robust  initial estimate  of $ \mu $.

The loss function $ \rho $ in equation \eqref{eq_scaleq} defines the properties of the resulting estimator  $ \hat{\mu} $. 
For instance, the \emph{Huber's family} of loss functions \citep{huber1964robust}, which generates  monotone functional
$ M $-estimators of location, is given by
\begin{equation*}\label{eq_huber}
\rho_{a}^{HU}\left(x\right)=\begin{cases} 
x^2/2 & \text{if } 0\leq x\leq a \\
a\left(x-a/2\right) & \text{if }a< x ,
\end{cases}
\end{equation*}
with tuning parameter $ a>0 $. It gives less importance to large errors compared to the standard least-squares loss function $ \rho^{sqr}\left(x\right)=x^2 $.
Functional $ M $-estimators arise from the \emph{bisquare} or \emph{Tukey's biweight family} of  loss functions \citep{beaton1974fitting} defined as
\begin{equation*}\label{eq_bisquare}
\rho_{a}^{BI}\left(x\right)=\begin{cases} 
a^2/6\left[1-\left(1-\left(x/a\right)^2\right)^3\right] & \text{if } 0\leq x\leq a \\
a^2/6 & \text{if }a< x ,
\end{cases}
\end{equation*}
with tuning parameter $ a>0 $. $ M $-estimators obtained by using $ \rho_{a}^{BI} $ are redescending, that is values of $ x>a $  give the same contribution to the loss, regardless of their distance from $ a $.

Another very used family of loss functions is the \emph{Hampel's one} \citep{hampel1974influence}, which is  defined as
\begin{equation*}\label{eq_hampel}
\rho_{a,b,c}^{HA}\left(x\right)=\begin{cases} 
x^2/2 & \text{if } 0\leq x<a \\
a\left(x-a/2\right) & \text{if } a\leq x<b\\
\frac{a\left(x-c\right)^2}{2\left(b-c\right)}+a\left(b+c-a\right)/2 & \text{if } b\leq x<c \\
a\left(b+c-a\right)/2& \text{if }c\leq x,
\end{cases}
\end{equation*}
with tuning parameter $ a,b,c>0 $. $ M $-estimators obtained by using $ \rho_{a,b,c}^{HA} $ are redescending as well.
Finally, the \emph{optimal family} of loss functions \citep{maronna2019robust} is defined as
\begin{equation*}\label{eq_opt}
\rho_{a}^{OP}\left(x\right)=\int_{0}^{x}\left(-\frac{\Phi'\left(|x|\right)+a}{\Phi\left(|x|\right)}\right)_{+}dx,
\end{equation*}
where $ \Phi $ is the standard normal density, $ a>0 $ is a tuning parameter and $ \left(t\right)_{+} $ denotes the positive part of $ t $.
The tuning parameters used in $ \rho_{a}^{HU},\rho_{a}^{BI},\rho_{a,b,c}^{HA} $ and $ \rho_{a}^{OP} $ are chosen in order to ensure given asymptotic efficiency with respect to the normal distribution \citep{maronna2019robust}.
% The loss functions $ \rho_{a}^{HU},\rho_{a}^{BI},\rho_{a,b,c}^{HA} $ and $ \rho_{a}^{OP} $  with tuning constants chosen to achieve $ 95\%$ asymptotic efficiency, along with $ \rho^{sqr} $ and $ \rho^{med} $, are displayed in Fig. \ref{fig_rho}.

% \begin{figure}
% 	\centering
% 	\makebox{\includegraphics[width=0.4\textwidth]{Figures/different_rho-eps-converted-to.pdf}}
% 	\caption{\label{fig_rho}The loss functions $ \rho_{a}^{HU}$ (HUB), $ \rho_{a}^{BI} $ (BIS), $ \rho_{a,b,c}^{HA} $ (HAM), $ \rho_{a}^{OP} $ (OPT),  with tuning constants chosen to achieve $ 95\%$ asymptotic efficiency, and, $ \rho^{sqr} $ (SQR) and $ \rho^{med} $ (MED).}
% \end{figure}

\subsection{The proposed robust method for the functional analysis of variance}
\label{subsec_prop}
The aim of this section is to describe the proposed RoFANOVA for the  multiway functional ANOVA design. Without loss of generality, and for ease of notation, we will focus on the  two-way functional ANOVA design with  interaction, but the extension to more complex designs is straightforward.
To introduce the  two-way functional ANOVA design with  interaction, let us consider a functional response $ X $, which is a   random element  with values in $ L^2\left(\mathcal{T}\right) $,  $ \mathcal{T}\subset \mathbb{R}^p $,  and is possibly affected by two factors, say A and B (with $ I $ and $ J $ levels, respectively). In this model, $ X $ will be expressed as the sum of two main effects and an interaction between them, plus a random error. Our aim is to test the statistical significance of the main effects and  interaction term.
 For $ k=1,\dots,n_{ij} $, let $ X_{ijk} $, denote the  realizations of $ X $ at level  $ i $ of the factor A,  $ i=1,\dots,I $, and  level  $ j $ of the factor B,  $ j=1,\dots,J $. Then, the two-way functional ANOVA model with interaction to be tested is
\begin{equation}\label{eq_modanova}
X_{ijk}\left(t\right)=m\left(t\right)+f_i\left(t\right)+g_j\left(t\right)+h_{ij}\left(t\right)+\varepsilon_{ijk}\left(t\right) \quad t\in \mathcal{T},
\end{equation}
 where $ m $ is the functional grand mean, which describes the overall shape of the process, $ f_i $ and $ g_j $ are the functional main effects and $ h_{ij} $ is the interaction term. All these terms have values in $ L^2\left(\mathcal{T}\right) $.
 The functional errors  $ \varepsilon_{ijk} $ are assumed to be independent and identically distributed  random functions with zero-mean and covariance function $ \gamma$. They are not required to be Gaussian.
 In order to make the model identifiable, we will assume that $ \sum_{i=1}^{I}\sum_{j=1}^{J}n_{ij}f_i\left(t\right)=\sum_{j=1}^{J}\sum_{i=1}^{I}n_{ij}g_j\left(t\right)=\sum_{i=1}^{I}\sum_{j=1}^{J}n_{ij}h_{ij}\left(t\right)=0 $.
To test the significance of the coefficients in the model \eqref{eq_modanova}, (that is, to extend the classical ANOVA test to the functional data setting), we consider the following null and alternative hypotheses  
 \begin{align}
 \label{eq_H0A}& H_{0,A}:f_1=\dots=f_I=\bm{0}, \quad  H_{1,A}: \left(H_{0,A}\right)^{C},\\
   \label{eq_H0B}&H_{0,B}:g_1=\dots=g_J=\bm{0}, \quad  H_{1,B}: \left(H_{0,B}\right)^{C},\\
   \label{eq_H0AB} &H_{0,AB}:h_{11}=\dots=h_{IJ}=\bm{0}, \quad\quad  H_{1,AB}: \left(H_{0,AB}\right)^{C},
 \end{align}
 where $\bm{0}$ is a function almost everywhere equal to zero.
 The hypotheses $ H_{0,A} $ against $ H_{1,A} $ and $ H_{0,B} $ against $ H_{1,B} $ involve the effects of the main factors A and  B, respectively, whereas, the hypothesis $ H_{0,AB} $ against $ H_{1,AB} $ involves the interaction term between them.
 
Each test is carried out through a nonparametric permutational approach.
In this regard, we introduce a test statistic that is a functional extension of the robust F-statistic proposed by  \cite{schrader1977robust}. The authors considered a robust version of the classical $ F $-test statistic, defined as the fraction of  the drop in residual sum of squares between  the full model (i.e., the  model when $ H_0 $ is false) and the reduced model (i.e., the  model when $ H_0 $ is true), and  the  standard deviation  of the error distribution, where all  quantities are estimated by using the least-squares approach. The $ F $-test statistic was modified by  a specific residual sum of dispersions corresponding to a  loss function as those described in Section \ref{subsec_mestimator} in place  of the  residual sum of squares, and  a robust estimate, instead of the  least-squares estimate, of the  standard deviation  of the error distribution.

Specifically, to test the hypotheses \eqref{eq_H0A}, we propose  the following test statistic
\small
\begin{multline*}
F_A=\left(I-1\right)^{-1}\left[\sum_{i=1}^{I}\sum_{j=1}^{J}\sum_{k=1}^{n_{ij}}\rho\left(\Big\lVert\frac{X_{ijk}-\bar{X}_{r}-\bar{X}_{r,ij}+\bar{X}_{r,i\cdot}}{\hat{\sigma}_{r,e}}\Big\lVert\right)\right.\\-\sum_{i=1}^{I}\sum_{j=1}^{J}\sum_{k=1}^{n_{ij}}\rho\left(\Big\lVert\frac{X_{ijk}-\bar{X}_{r,ij}}{\hat{\sigma}_{r,e}}\Big\lVert\right)\Bigg],
\end{multline*}\normalsize
where $ \rho $ is a given loss function,  $ \hat{\sigma}_{r,e} $ is a robust estimate of the functional  standard deviation  of the error distribution, and $ \bar{X}_{r} $, $ \bar{X}_{r,i\cdot} $, and $ \bar{X}_{r,ij} $ are, respectively, scale equivariant functional $ M $-estimators (Section \ref{subsec_mestimator}) of the functional grand mean $ m $, group means  of $ \lbrace X_{ijk} \rbrace_{k=1,\dots n_{ij},i=1,\dots I } $ and $ \lbrace X_{ijk} \rbrace_{k=1,\dots n_{ij} } $.
 In detail, $ \bar{X}_{r} $, $ \bar{X}_{r,i\cdot} $, $ \bar{X}_{r,ij} $ and $ \hat{\sigma}_{r,e} $ are defined as
 \begin{align*} 
 \bar{X}_{r}&=\argmin_{y\in L^2\left(\mathcal{T}\right) }\sum_{i=1}^{I}\sum_{j=1}^{J}\sum_{k=1}^{n_{ij}}\rho\left(\Big\lVert\frac{X_{ijk}-y}{\hat{\sigma}_{r}}\Big\lVert\right), \quad &  \hat{\sigma}_{r}&=\FuNMAD\left(\lbrace X_{ijk} \rbrace_{\begin{aligned}
 \scriptscriptstyle i= & \scriptscriptstyle 1,\dots, I\\[-1em]
\scriptscriptstyle j= & \scriptscriptstyle 1,\dots, J\\[-1em]
\scriptscriptstyle k= & \scriptscriptstyle 1,\dots, n_{ij}\\[-1em]
\end{aligned}} \right),\\
 \bar{X}_{r,i\cdot}&=\argmin_{y\in L^2\left(\mathcal{T}\right) }\sum_{j=1}^{J}\sum_{k=1}^{n_{ij}}\rho\left(\Big\lVert\frac{X_{ijk}-y}{\hat{\sigma}_{r,i\cdot}}\Big\lVert\right),\quad &  \hat{\sigma}_{r,i\cdot}&=\FuNMAD\left(\lbrace X_{ijk} \rbrace_{\begin{aligned}
 \scriptscriptstyle j= & \scriptscriptstyle 1,\dots, J\\[-1em]
\scriptscriptstyle k= & \scriptscriptstyle 1,\dots, n_{ij}\\[-1em]
\end{aligned} } \right),\\
  \bar{X}_{r,ij}&=\argmin_{y\in L^2\left(\mathcal{T}\right) }\sum_{k=1}^{n_{ij}}\rho\left(\Big\lVert\frac{X_{ijk}-y}{\hat{\sigma}_{r,ij}}\Big\lVert\right),\quad  &\hat{\sigma}_{r,ij}&=\FuNMAD\left(\lbrace X_{ijk} \rbrace_{\begin{aligned}
\scriptscriptstyle k= & \scriptscriptstyle 1,\dots, n_{ij}\\[-1em]
\end{aligned}} \right),
 \end{align*}
 \vspace{-0.2cm}
 \begin{equation*}
 \hat{\sigma}_{r,e}=\frac{1}{0.6745}\Med\left(|\lbrace X_{ijk}-\bar{X}_{r,ij}\rbrace_{\begin{aligned}
\scriptscriptstyle i= & \scriptscriptstyle 1,\dots, I\\[-1em]
\scriptscriptstyle j= & \scriptscriptstyle 1,\dots, J\\[-1em]
\scriptscriptstyle k= & \scriptscriptstyle 1,\dots, n_{ij}\\[-1em]
\end{aligned}} |\right).
 \end{equation*}
 The test statistic $F_A$ represents the mean  difference between the standardized residual sum of dispersions under the reduced model  and  the full model, and is analogous to that used by \cite{schrader1977robust} in the classical setting.
 Intuitively, it is a measure of the discrepancy between  residuals of the model under  $H_{0,A} $ and under $ H_{1,A} $, obtained through robust statistics. 
%  In this latter case, $F_A$  assumes high values.
 Analogously, to test the hypotheses \eqref{eq_H0B} and \eqref{eq_H0AB}, we define
\small{
\begin{multline*}
F_B=\left(J-1\right)^{-1}\left[\sum_{i=1}^{I}\sum_{j=1}^{J}\sum_{k=1}^{n_{ij}}\rho\left(\Big\lVert\frac{X_{ijk}-\bar{X}_{r}-\bar{X}_{r,ij}+\bar{X}_{r,\cdot j}}{\hat{\sigma}_{r,e}}\Big\lVert\right)\right.\\-\sum_{i=1}^{I}\sum_{j=1}^{J}\sum_{k=1}^{n_{ij}}\rho\left(\Big\lVert\frac{X_{ijk}-\bar{X}_{r,ij}}{\hat{\sigma}_{r,e}}\Big\lVert\right)\Bigg],
\end{multline*}
\begin{multline*}
F_{AB}=\left(\left(I-1\right)\left(J-1\right)\right)^{-1}\left[\sum_{i=1}^{I}\sum_{j=1}^{J}\sum_{k=1}^{n_{ij}}\rho\left(\Big\lVert\frac{X_{ijk}-\bar{X}_{r,i\cdot}-\bar{X}_{r,\cdot j}+\bar{X}_{r}}{\hat{\sigma}_{r,e}}\Big\lVert\right)\right.\\-\sum_{i=1}^{I}\sum_{j=1}^{J}\sum_{k=1}^{n_{ij}}\rho\left(\Big\lVert\frac{X_{ijk}-\bar{X}_{r,ij}}{\hat{\sigma}_{r,e}}\Big\lVert\right)\Bigg],
\end{multline*}}
\normalsize
where 
\small{
\begin{align*} 
\bar{X}_{r,\cdot j}&=\argmin_{y\in L^2\left(\mathcal{T}\right) }\sum_{i=1}^{I}\sum_{k=1}^{n_{ij}}\rho\left(\Big\lVert\frac{X_{ijk}-y}{\hat{\sigma}_{r,\cdot j}}\Big\lVert\right),\quad &  \hat{\sigma}_{r,\cdot j}&=\FuNMAD\left(\lbrace X_{ijk} \rbrace_{\begin{aligned}
 \scriptscriptstyle i= & \scriptscriptstyle 1,\dots, I\\[-1em]
\scriptscriptstyle k= & \scriptscriptstyle 1,\dots, n_{ij}\\[-1em]
\end{aligned} } \right).
\end{align*}
}
\normalsize
Different versions of the proposed test statistics may emerge by the choice of the loss function $ \rho $ as defined in Section \ref{subsec_mestimator}, and by the use of $ \hat{\sigma}_{r,ij}=\hat{\sigma}_{r,e} $ to estimate $ \bar{X}_{r,ij} $.

 Another element to choose in a permutation test is the approximation method for  the distribution of the considered statistic under the null hypothesis. In our case, we selected the Manly's  scheme \citep{gonzalez1998analysis,manly2006randomization} that consists of  simply permuting the raw data without restrictions.
 Although other schemes could be used, the Manly's  one has demonstrated  good performance and simplicity, especially when the sample size,  at  given factor levels, is small. See \cite{gonzalez1998analysis} and \cite{anderson2001permutation} for further details.
 
 Lt $ F $ generically denotes the   statistic (resp., $ F_A $ or  $ F_B $ or $ F_{AB} $) to test, at level $ \alpha $, $ H_0 $ against $ H_1 $ (resp.,  $H_{0,A} $ against $ H_{1,A} $; or $ H_{0,B} $ against  $ H_{1,B} $; or $ H_{0,AB} $ against $ H_{1,AB} $). Then, the proposed permutation test can be outlined by the following steps. 
\begin{enumerate}[Step 1.]
	\item Compute the observed value of the test statistic $ F_{obs} $, by considering the original sample $ \lbrace X_{ijk} \rbrace_{k=1,\dots n_{ij},i=1,\dots I,j=1,\dots J } $.
	\item Randomly permute the data, among the Factor A and Factor B combinations, $ B $ times, and for each permuted sample compute the value $ F^*_1,\dots, F^*_B $ of the statistic $ F $.
	\item Compute the approximated p-value as
	\begin{equation*}
	p=\frac{1}{B}\sum_{i=1}^{B}I\left(F^*\geq  F_{obs}\right),
	\end{equation*}
	where $ I\left(E\right) $ takes values 1 or 0 depending on whether E is true or
	false.
	\item Accept $ H_0 $ if $ p>\alpha $, otherwise reject $ H_0 $.
\end{enumerate}
This  is an approximate (asymptotically exact) level-$ \alpha  $ test for $ H_0 $ against $ H_1 $ \citep{anderson2001permutation}.  
The larger the number of permutations $B$,  the lower the approximation error. We suggest to select the number of permutations $B$ equal to or larger than 1000 \citep{good2013permutation}.

\section{Simulation study}
\label{sec_sim}
In this section, by means of an extensive Monte Carlo simulation study, the performance of the proposed method is assessed  in terms of empirical size and power  of the test. In particular, the following two scenarios are investigated: 
\begin{enumerate}[label=Scenario \arabic*]
	\item A one-way FANOVA model (i.e., model \eqref{eq_modanova} with $ m=0 $, $ g_1=\dots=g_J=0$ and $ h_{11}=\dots=h_{IJ}=0$) is considered  (Section \ref{sec_oneway}).
	\item A two-way FANOVA model (i.e., model \eqref{eq_modanova}) is considered   (Section \ref{sec_twoway}).
\end{enumerate}
In each scenario, the FANOVA model is contaminated by different type of outlying curves. To do so, we use the same contamination models as in previous works on robust FDA \citep{fraiman2001trimmed,lopez2009concept,sinova2018m}.
All the details about the data generation process are provided in the Supplementary Materials.

\subsection{One-way functional analysis of variance}
\label{sec_oneway}	
The proposed simulation study framework for one-way FANOVA  has been inspired by \cite{cuevas2004anova,gorecki2015comparison}.
Three different model M1, M2 and M3, with 3 level main effect $ f_i$, $i=1,2,3$, are considered, and without loss of generality, we assume the curve domain  $\mathcal{T}=\left[0,1\right]$.
Model M1 corresponds to  $ H_0 $:  $ f_1=f_2=f_3$   true, whereas, M2 and M3 provide examples, with $ H_0 $ false, of monotone functions with different increasing patterns. In particular, M2 simulates $f_i$ differences that are smaller than M3, where $ f_i $ are quite separated.
In model M1, we use as performance measure the empirical size, whereas in M2 and M3 we use the empirical power. 
Moreover, to simulate different types of outlying curves, seven contamination models denoted by C0-6 are considered.
The  model C0 is representative of  no contamination. C1-4 represents magnitude contaminations, i.e., curves are generated  far from the center, with, in particular, C1-2 (resp., C3-4) representing  symmetric and partial trajectory contamination models, that are independent (resp.,
dependent) from the level of the main effect.
Models C5-6 are shape contamination models \citep{lopez2009concept,sinova2018m}.

In all the cases considered, the response curves are independent realizations of a  Gaussian process with covariance function $ \gamma\left(s,t\right)=\sigma^2 e^{\left(-|s-t|10^{-5}\right)}$ and are observed  through $ 25 $ evenly spread discrete points with $ \sigma $  equal to  $ \sigma_{1}=1/25 $, $ \sigma_{2}=1.8/25 $, $ \sigma_{3}=2.6/25 $, $ \sigma_{4}=3.4/25 $, $ \sigma_{5}=4.2/25 $, $ \sigma_{6}=5/25 $ \citep{cuevas2004anova}.
We expect that the higher  $\sigma$, the worse the performance in terms of both empirical size and power.
Five versions of the RoFANOVA method are considered, which are defined by different choices of the loss function, viz., the RoFANOVA with  median loss $ \rho^{med} $, referred to as RoFANOVA-MED,  Huber loss $ \rho_{a}^{HU}$, referred to as RoFANOVA-HUB,   bisquare loss $ \rho_{a}^{BI} $, referred to as RoFANOVA-BIS,  Hampel loss $ \rho_{a,b,c}^{HA} $, referred to as RoFANOVA-HAM, and,  optimal loss $ \rho_{a}^{OP} $, referred to as RoFANOVA-OPT.  The tuning constants are chosen to achieve $ 95\%$ asymptotic efficiency, the number of permutations $ B $ are set equal to $ 1000 $ and the functional $0.8\%$ deepest curve following the FM criteria \citep{febrero2012fdausc} is chosen as starting value to compute the robust equivariant functional $ M $-estimators (Section \ref{subsec_mestimator}).
The proposed tests are compared with some non-robust methods already appeared in the literature before. In particular, we consider  the method proposed by \cite{gorecki2015comparison}, referred to as FP, which is a permutation test that relies on a basis function representation of the response function; the method proposed by \cite{zhang2014one}, referred to as GPF, based on a globalized version of the pointwise $ F $-test;  the method proposed by \cite{zhang2007statistical}, referred to as L\textsuperscript{2}B, a $ L^2 $-norm-based test with the bias-reduced method to estimate the unknown parameters; and the method proposed by \cite{zhang2011statistical}, referred to as FB,  which is an $ F $-type test based on the bias-reduced estimation method. All these methods are implemented with the default settings  of the R package \texttt{fdANOVA} \citep{gorecki2018fdanova}. In addition, the method proposed by \cite{cuesta2010simple}, based on randomly chosen one-dimensional projections, with both the Bonferroni  (referred to as TRPbon)	and the  false discovery rate (referred to as TRPfdr) corrections, is considered. The TRPbon and TRPfdr are run with 30 random projections through the R package \texttt{fda.usc} \citep{febrero2012fdausc}.

For each triplet (M$l$,C$m$,$ \sigma_n $), $ l=1,\dots,3 $, $ m=0,\dots,6 $, $ n=1,\dots,6 $, the five proposed  and the seven competing methods are applied $ N=500 $ times to the generated functional sample to test  $ H_0 $:  $ f_1=f_2=f_3$  against  $ H_1 $:   $ (H_0)^C $ at level $ \alpha=0.05 $.
Then, for each case, the empirical sizes (for model M1) and powers (for models M2 and M3) of the tests were  computed as the proportion  of rejections out of the $ N $ replications whose standard deviation is equal at most to 0.0224, which corresponds to the case of probability of rejection equal to 0.5.

\begin{figure}
		\centering
	\resizebox{1\textwidth}{!}{
		\begin{tabular}{M{0.5\textwidth}M{0.5\textwidth}M{0.5\textwidth}M{0.5\textwidth}}
			\multirow{2}{*}{\includegraphics[width=0.5\textwidth]{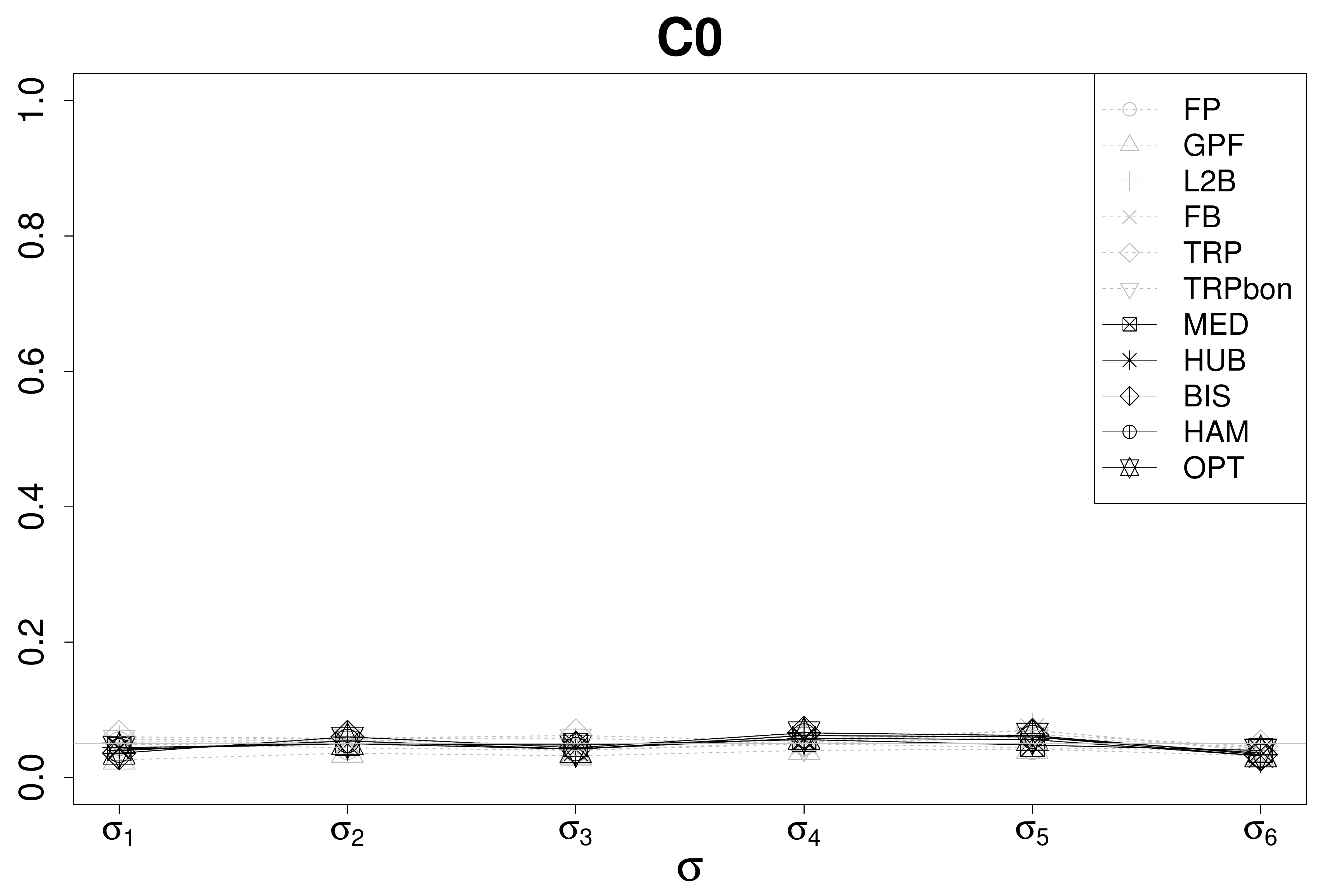}}&\includegraphics[width=0.5\textwidth]{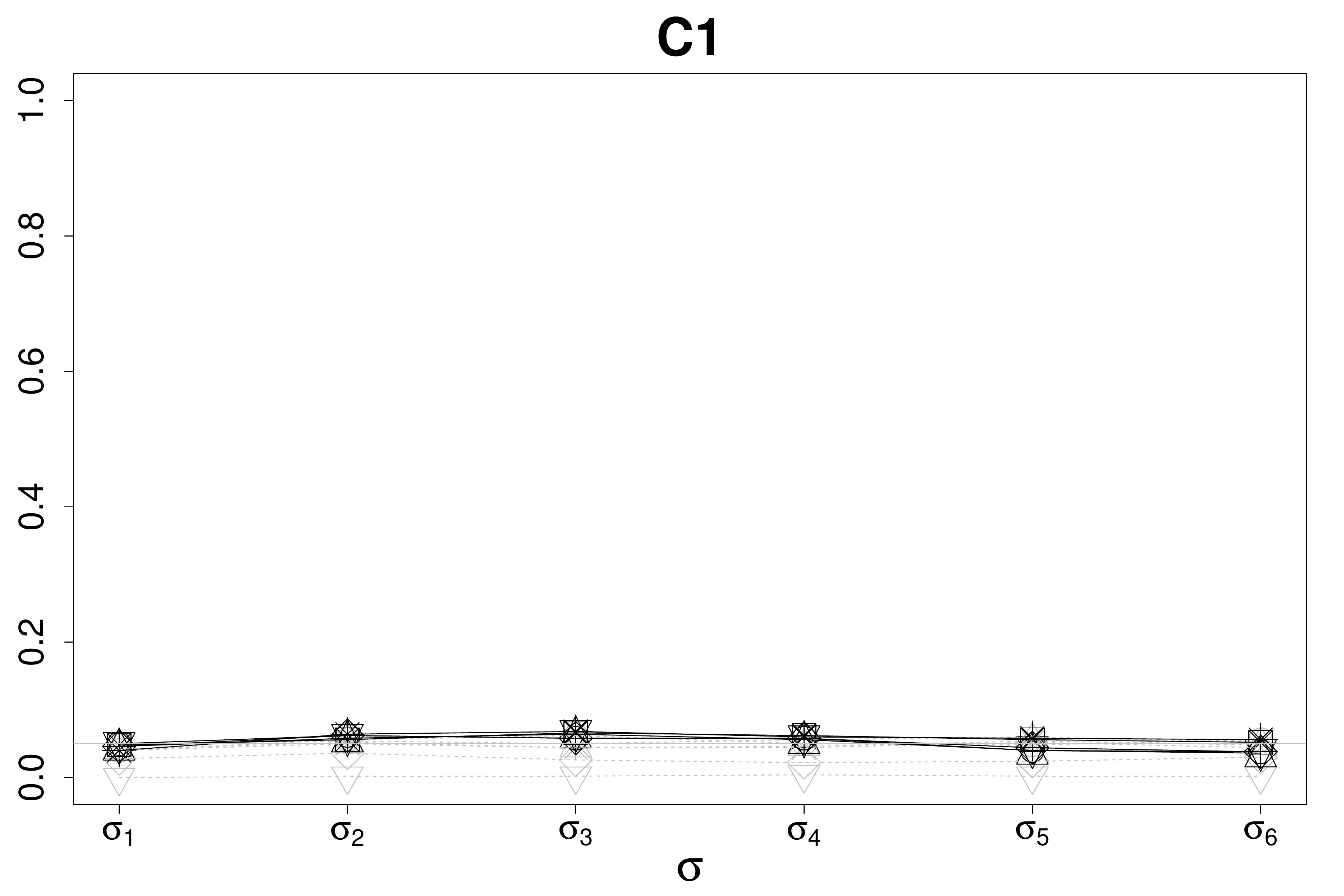}&\includegraphics[width=0.5\textwidth]{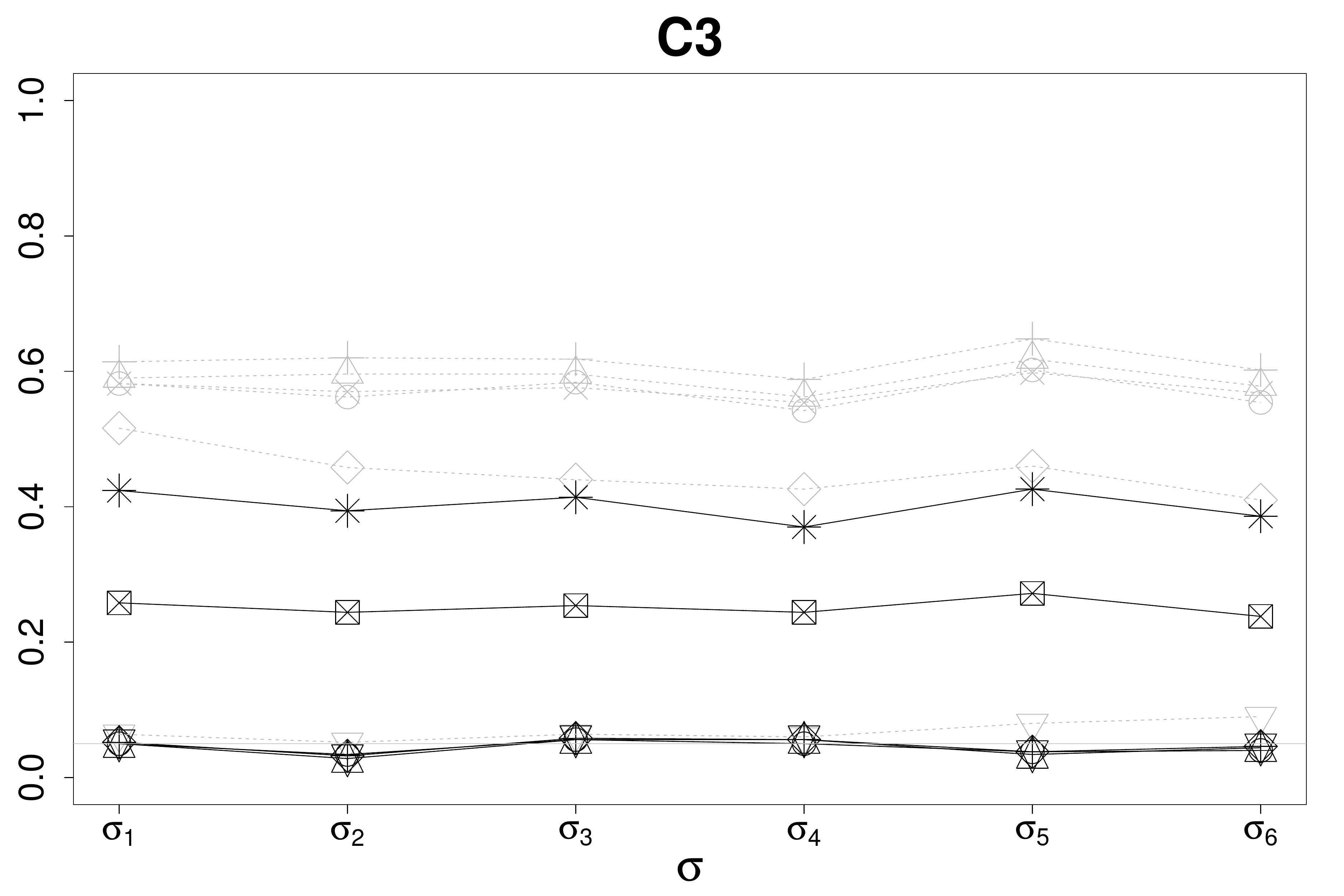}&\includegraphics[width=0.5\textwidth]{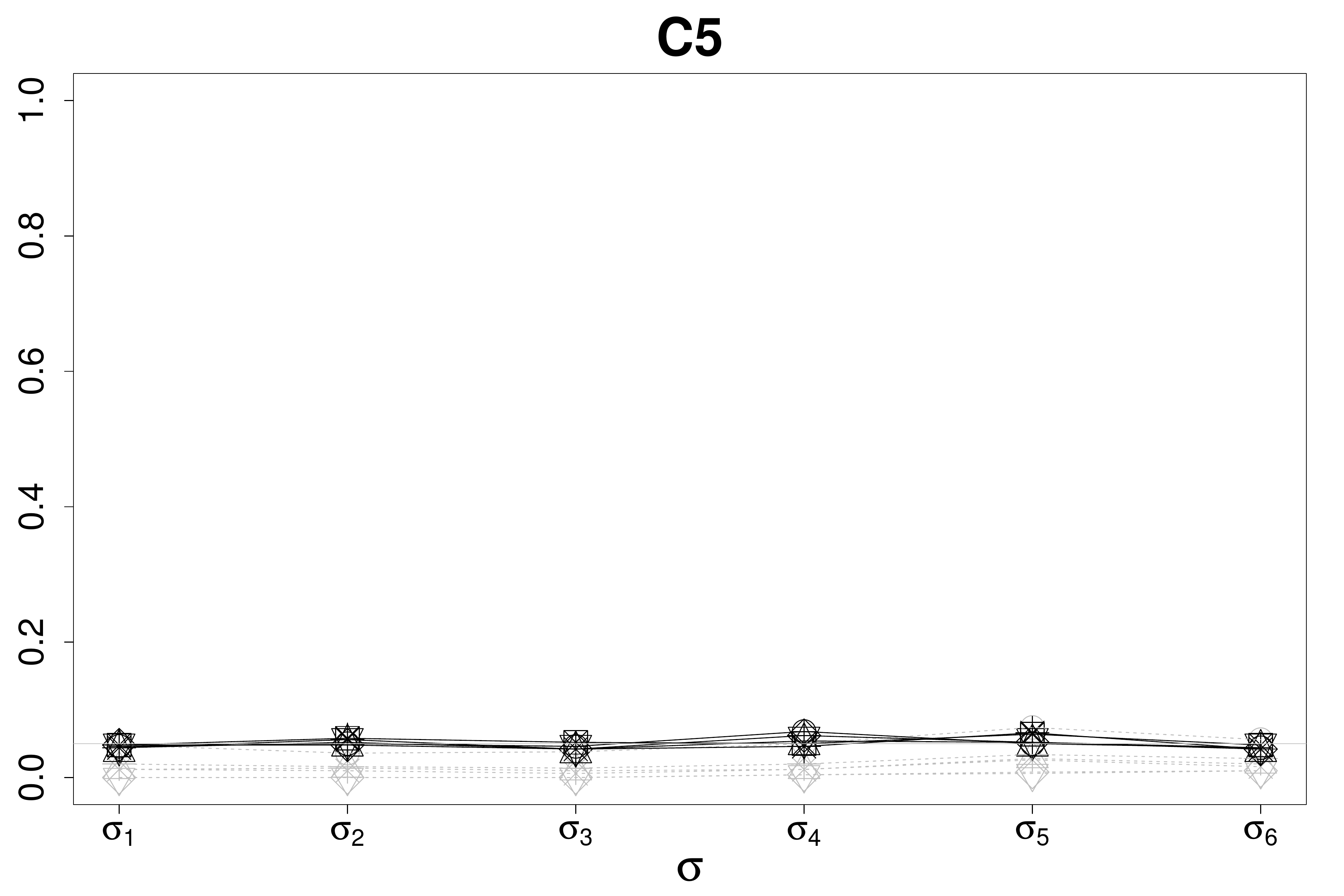}\\
			&\includegraphics[width=0.5\textwidth]{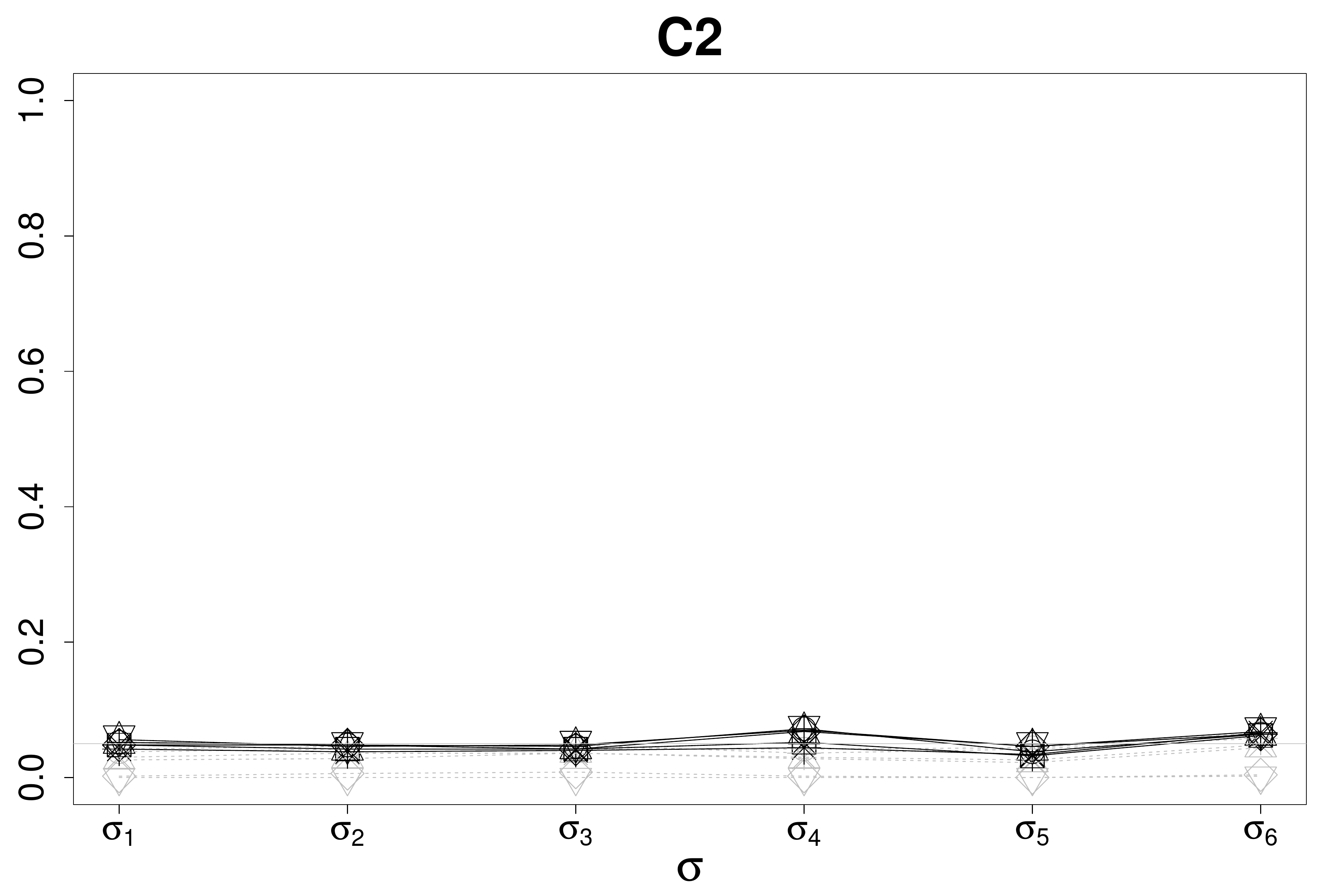}&\includegraphics[width=0.5\textwidth]{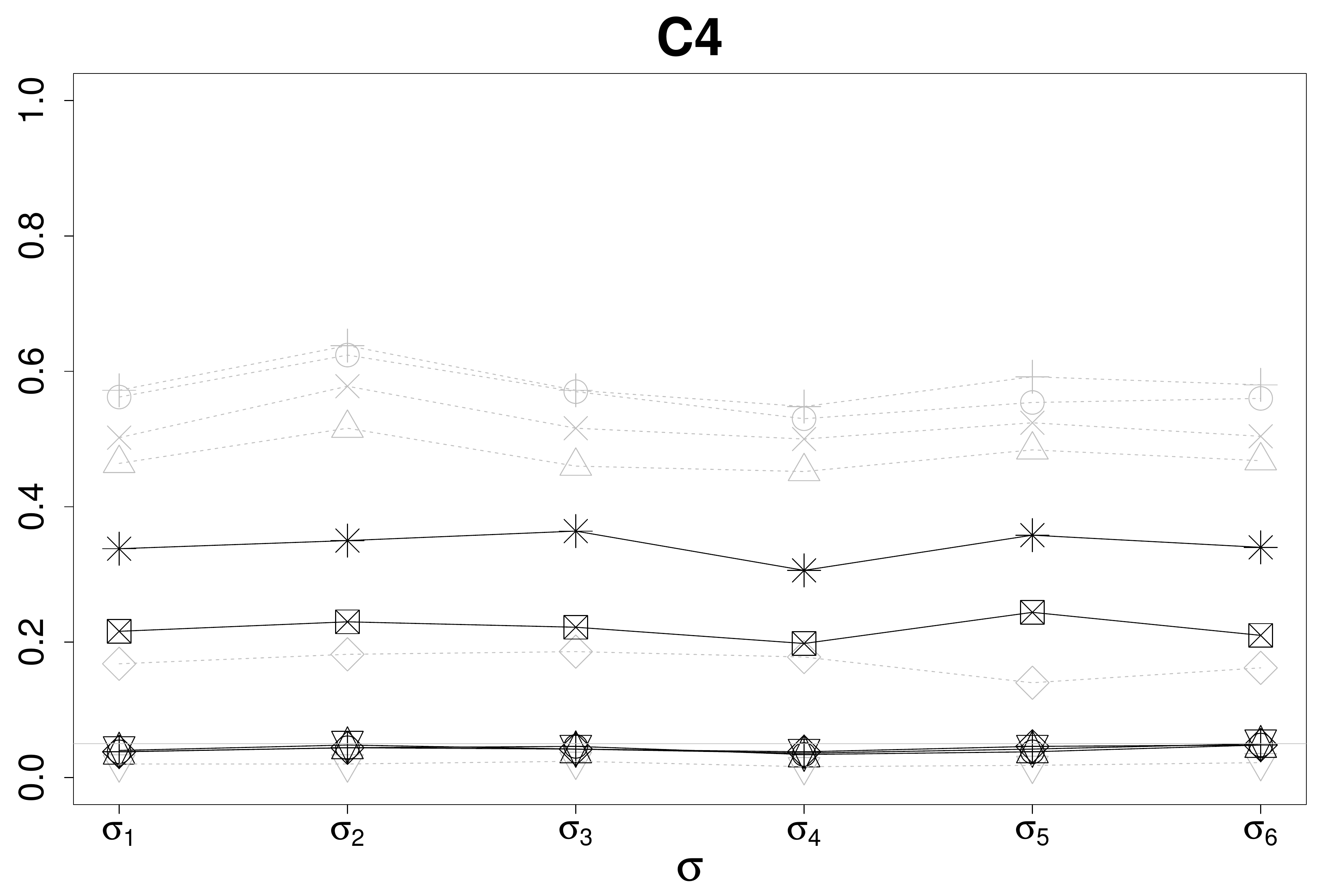}&\includegraphics[width=0.5\textwidth]{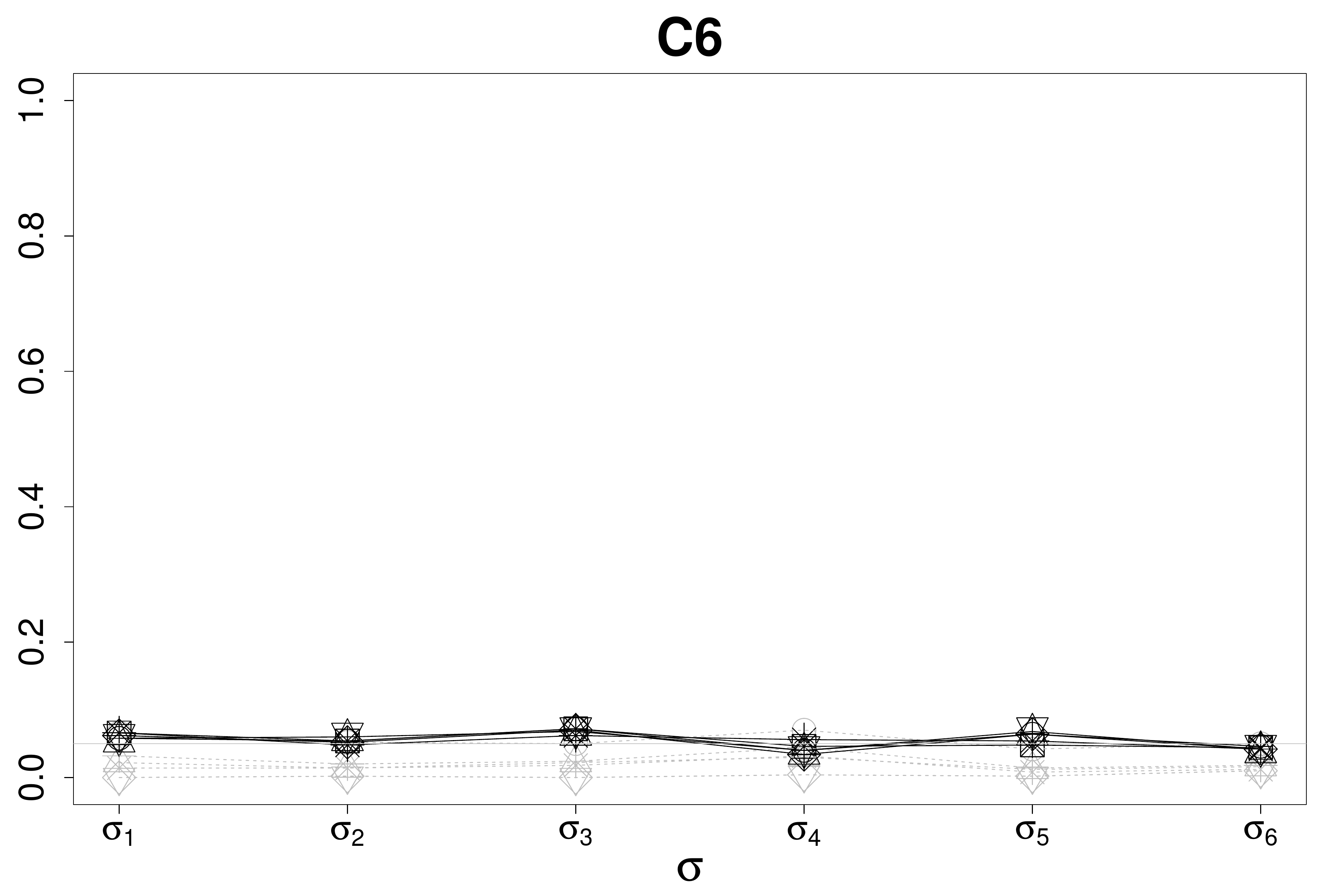}\\
	\end{tabular}}
	
	\caption{\label{fig_M1}Empirical size of all tests for $ H_{0} $ against  $ H_{1} $ (at level $ \alpha=0.05 $) as a function of $ \sigma_n $, $ n=1,\dots,6 $, for  contamination models C0-6 in  model M1 of Scenario 1. The proposed and competing tests are displayed as  black and grey lines, respectively.  }
\end{figure}

Fig. \ref{fig_M1}  displays the results for model M1, that is the empirical size of the eleven tests as a function of $ \sigma_n $, $ n=1,\dots,6 $, for  contamination models C0-6. 
In this case, the tests provide satisfactory results in  controlling the level $ \alpha $, i.e., the empirical size is approximately less than or equal to $ 0.05 $, in case of no contamination (C0), symmetric magnitude contamination (C1-2) and shape contamination, both symmetric (C5) and asymmetric (C6).
On the contrary, for asymmetric magnitude contamination (C3-4), only the RoFANOVA tests based on redescending loss functions, i.e.,  RoFANOVA-BIS, RoFANOVA-HAM and RoFANOVA-OPT, are able to control the level $ \alpha $ by ensuring an empirical size  approximately less or equal than $ 0.05 $. This was somehow expected, as it is known that rededescending estimators give no weight to observations that are far from the center \citep{maronna2019robust}. 
The estimators used in  RoFANOVA-MED and RoFANOVA-HUB tests do not have this property and, thus, they suffer from the presence of contaminations depending on the level of the main factor. Note that, among the competitors, the TRPbon approximately controls the level for contamination model C4, while it is slightly affected by outliers in  model C5. This comes from the Bonferroni correction property of  being  conservative for high-dimensional multiple comparisons \citep{lehmann2006testing}.

\begin{figure}
		\centering
	\resizebox{1\textwidth}{!}{
		\begin{tabular}{M{0.5\textwidth}M{0.5\textwidth}M{0.5\textwidth}M{0.5\textwidth}}
			\multirow{2}{*}{\includegraphics[width=0.5\textwidth]{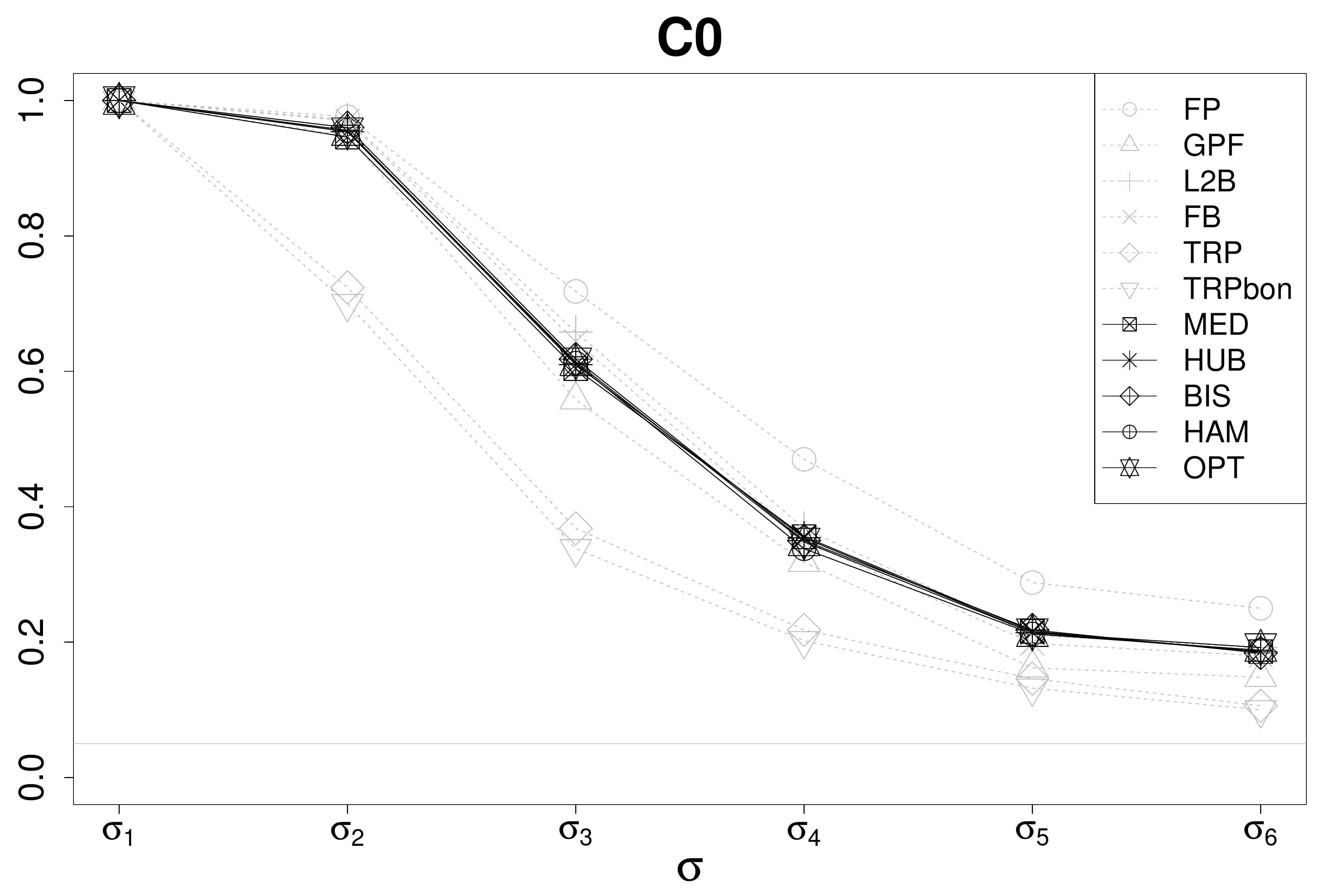}}&\includegraphics[width=0.5\textwidth]{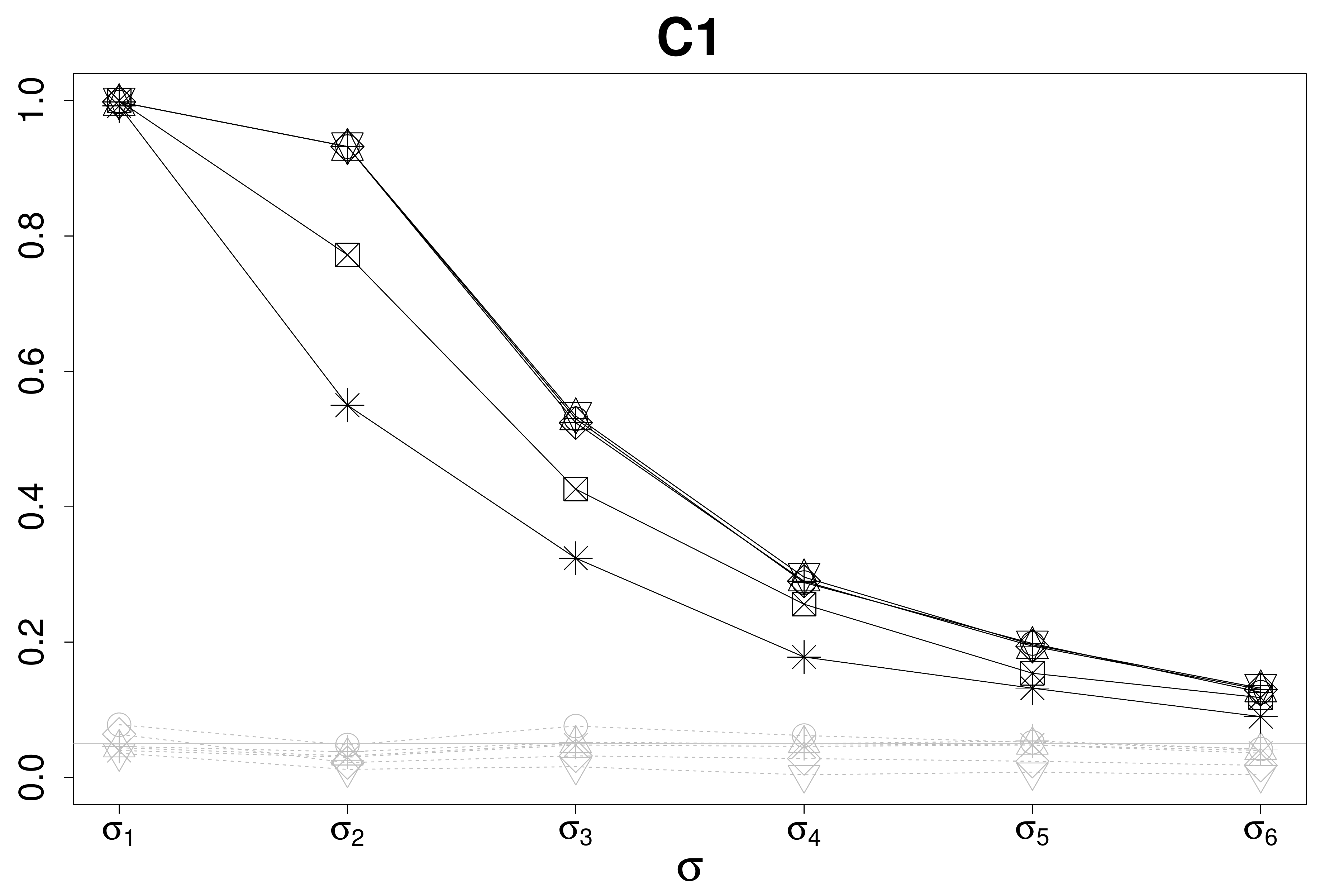}&\includegraphics[width=0.5\textwidth]{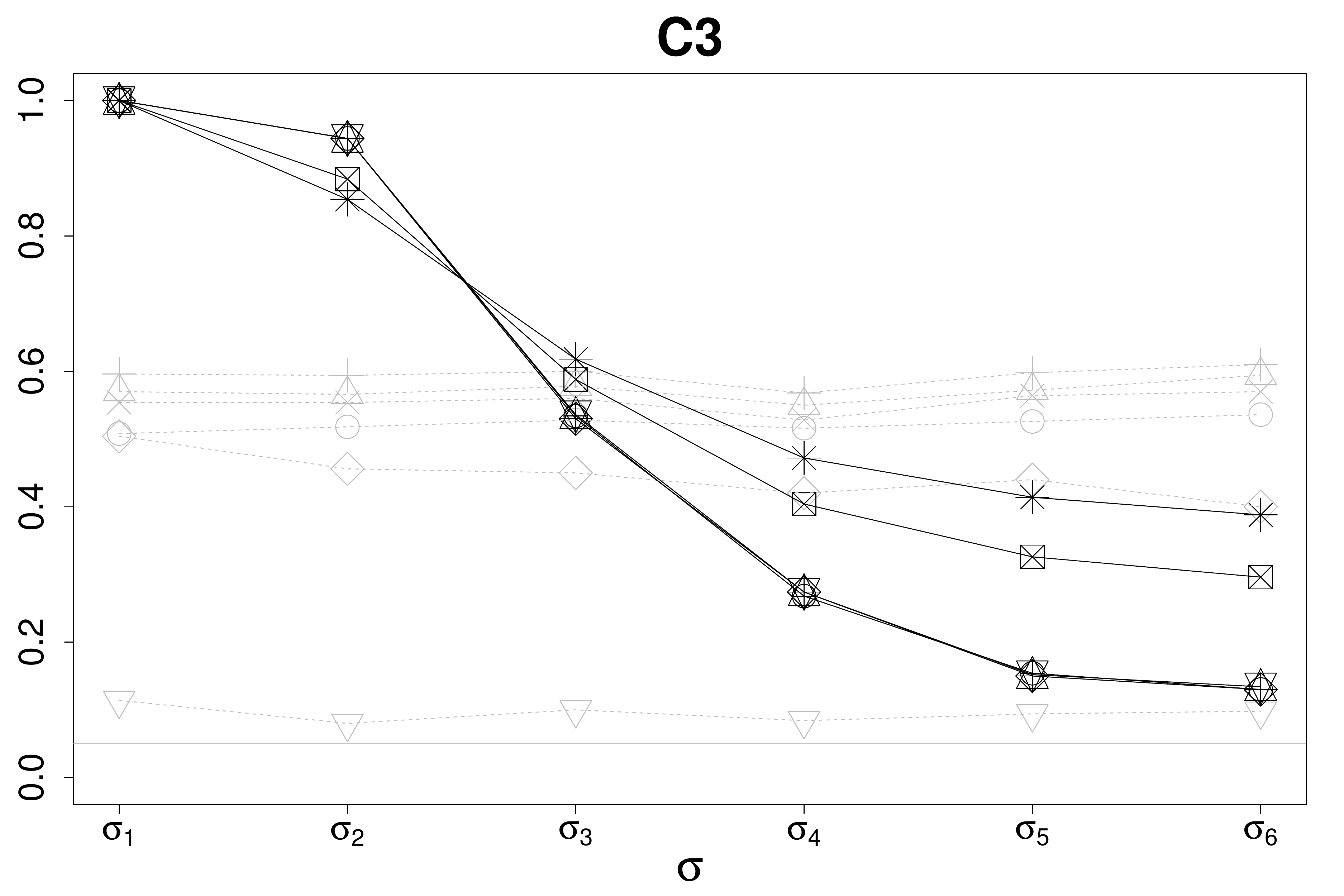}&\includegraphics[width=0.5\textwidth]{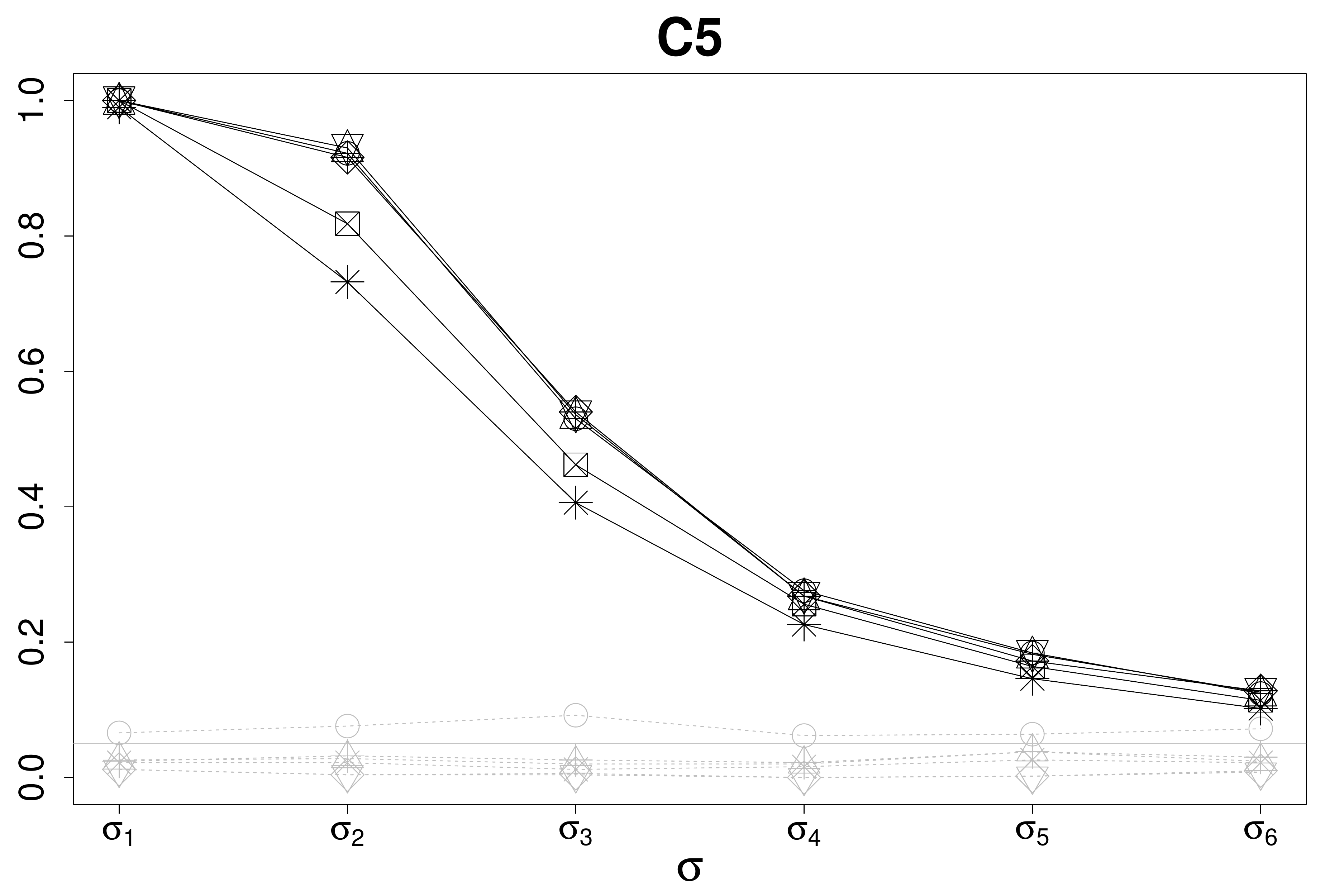}\\
			&\includegraphics[width=0.5\textwidth]{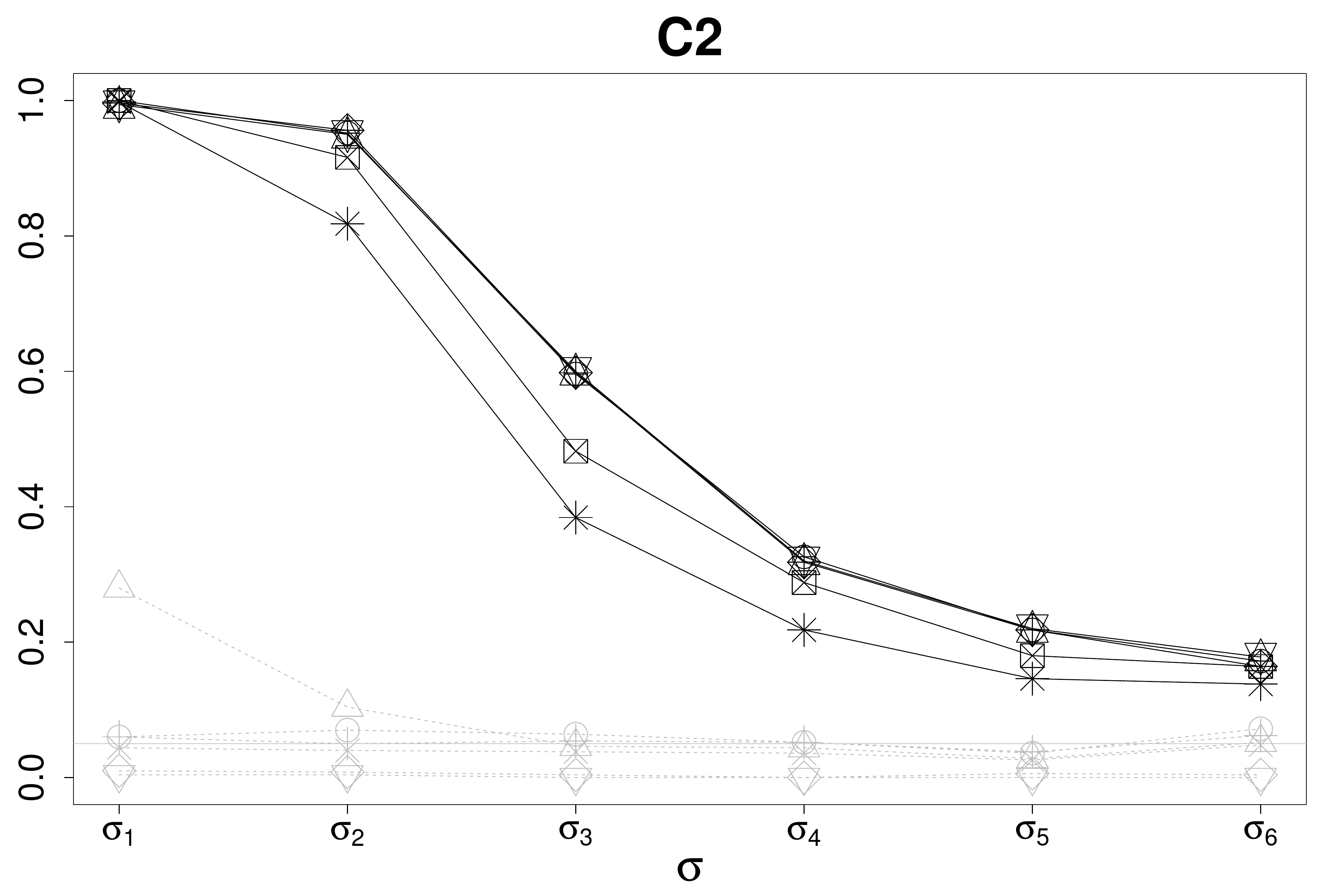}&\includegraphics[width=0.5\textwidth]{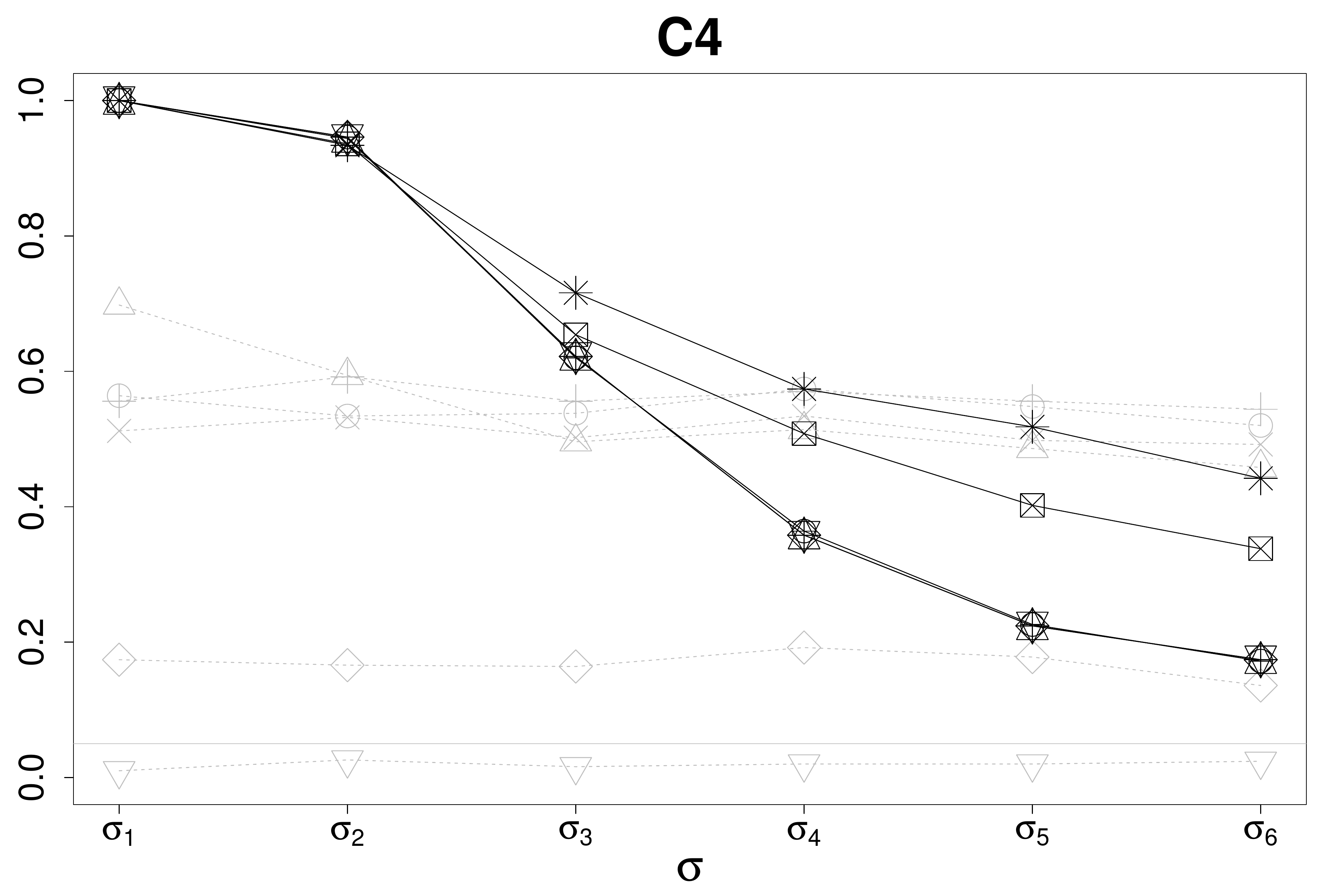}&\includegraphics[width=0.5\textwidth]{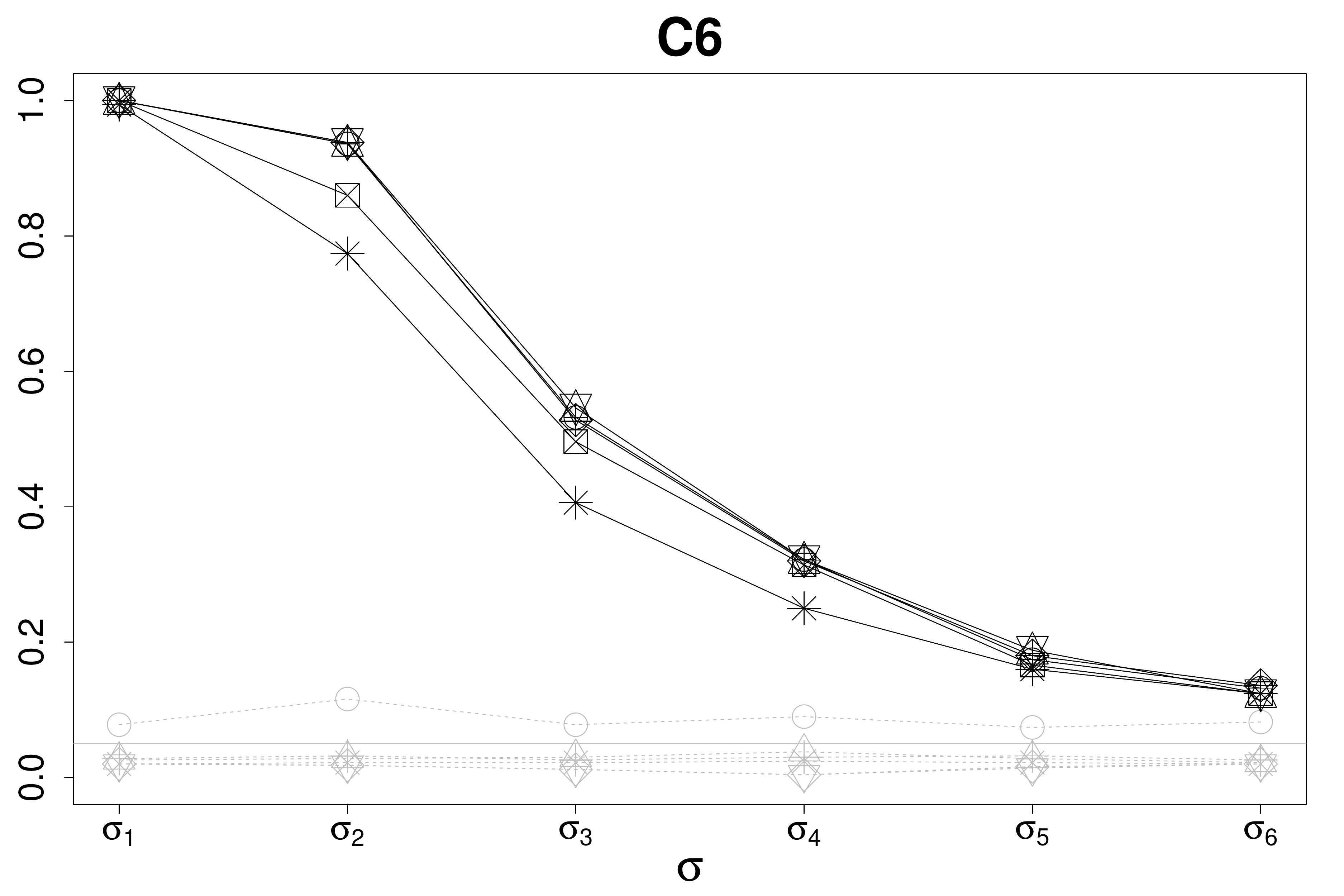}\\
	\end{tabular}}
	
	\caption{\label{fig_M2}Empirical power of all tests  for $ H_{0} $ against  $ H_{1} $ (at level $ \alpha=0.05 $) as a function of $ \sigma_n $, $ n=1,\dots,6 $,  for  contamination models C0-6 in  model M2 of Scenario 1. The proposed and competing tests are displayed as  black and grey lines, respectively. }
\end{figure}

Fig. \ref{fig_M2} shows the results for model M2 in terms of empirical power. These tend to get worse as $ \sigma_n$ increases. In case of no contamination (C0), the FP test achieves the largest empirical power, even though all  RoFANOVA tests have  comparable results. For contamination model C1-6, it is extremely  clear  the proposed RoFANOVA tests outperform all  competitors. In particular, among  RoFANOVA tests, those based on redescending  functional $ M $-estimators (viz., RoFANOVA-BIS, RoFANOVA-HAM and RoFANOVA-OPT) are the best ones.  Note that, for contaminations C3-4, only the RoFANOVA-BIS, RoFANOVA-HAM and RoFANOVA-OPT tests and the TRPbon test (only for C4) should be considered, because all the other methods are not able to successfully control the level $ \alpha $ (see Fig. \ref{fig_M1}).

\begin{figure}
	\centering
	\resizebox{1\textwidth}{!}{
		\begin{tabular}{M{0.5\textwidth}M{0.5\textwidth}M{0.5\textwidth}M{0.5\textwidth}}
			\multirow{2}{*}{\includegraphics[width=0.5\textwidth]{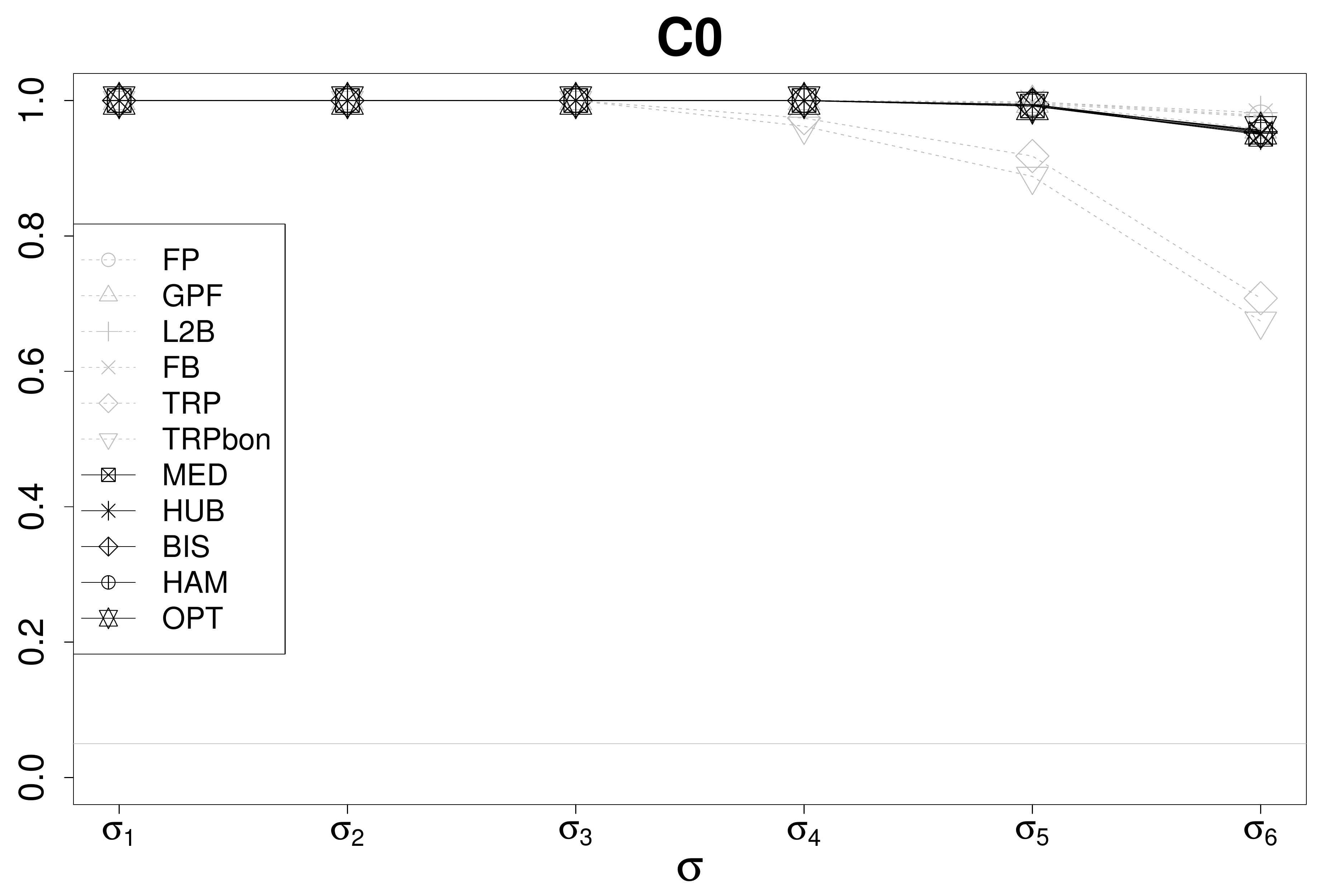}}&\includegraphics[width=0.5\textwidth]{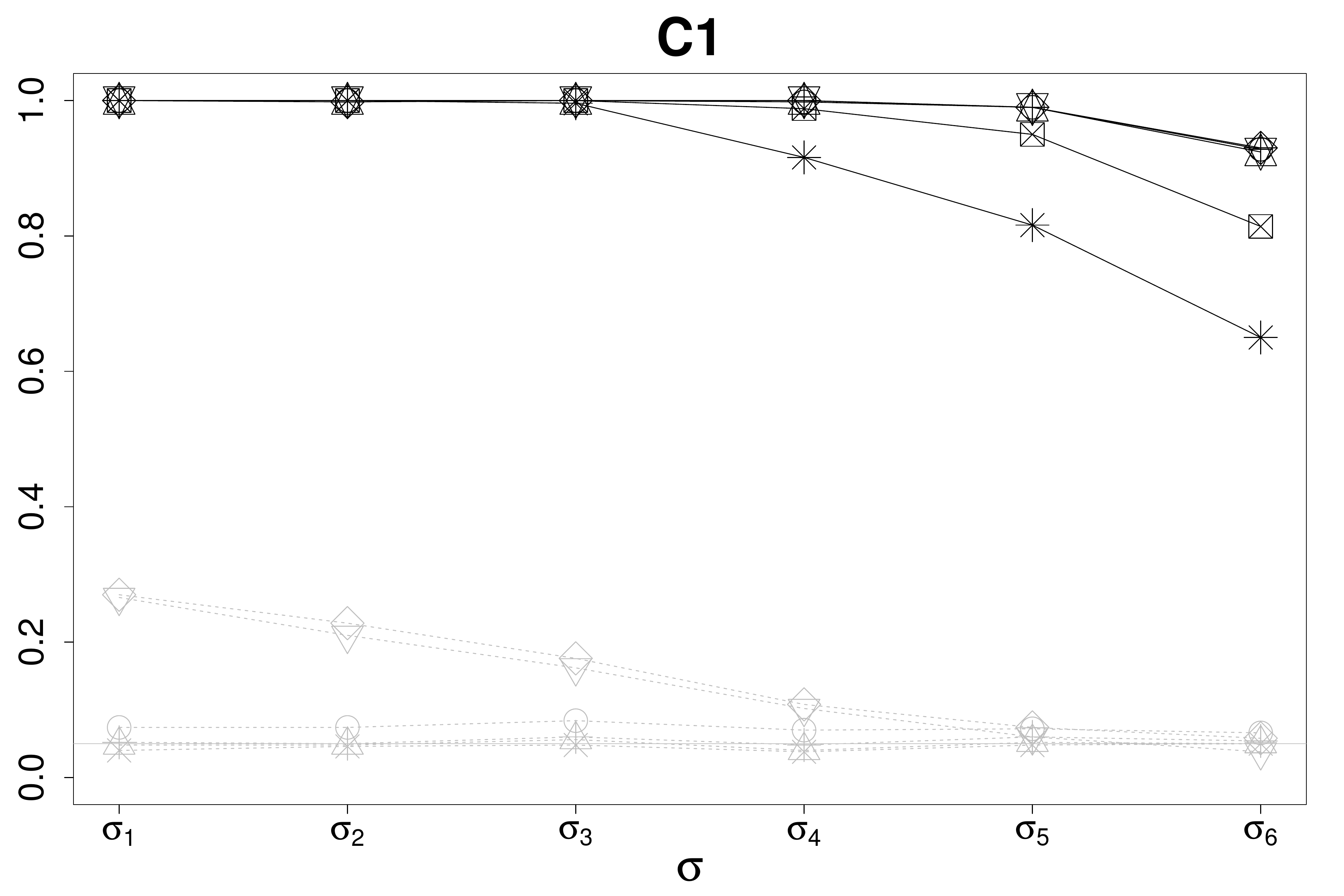}&\includegraphics[width=0.5\textwidth]{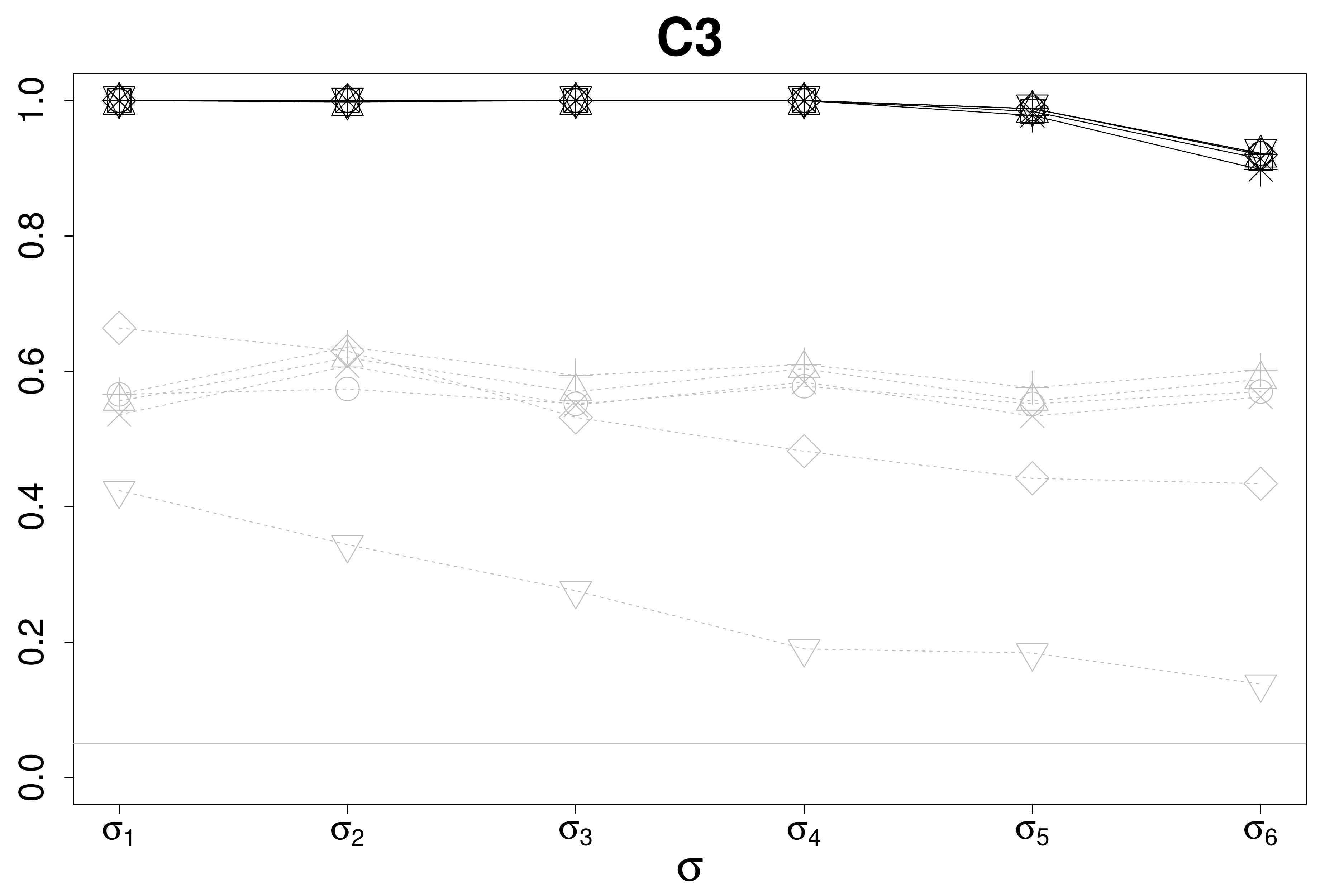}&\includegraphics[width=0.5\textwidth]{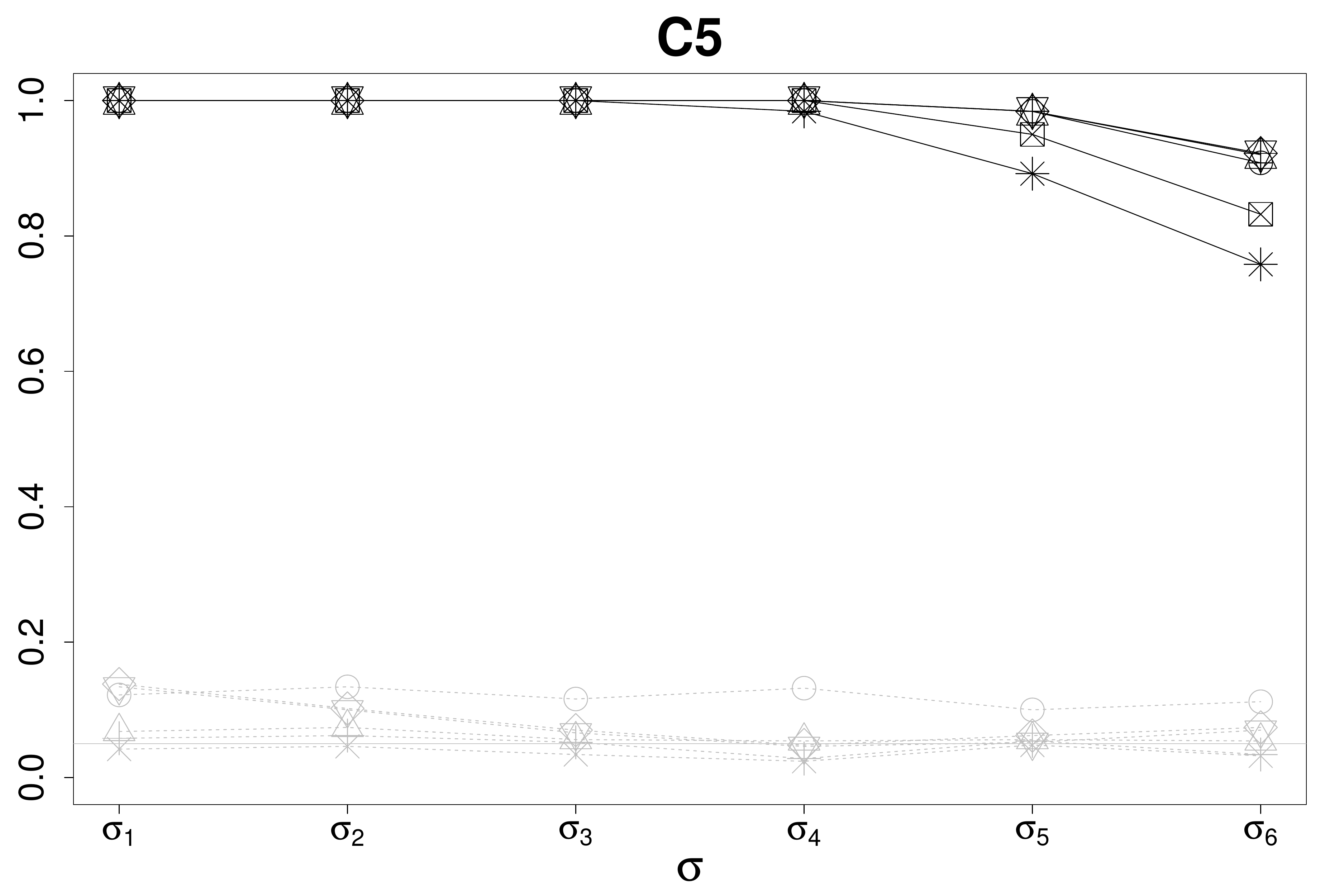}\\
			&\includegraphics[width=0.5\textwidth]{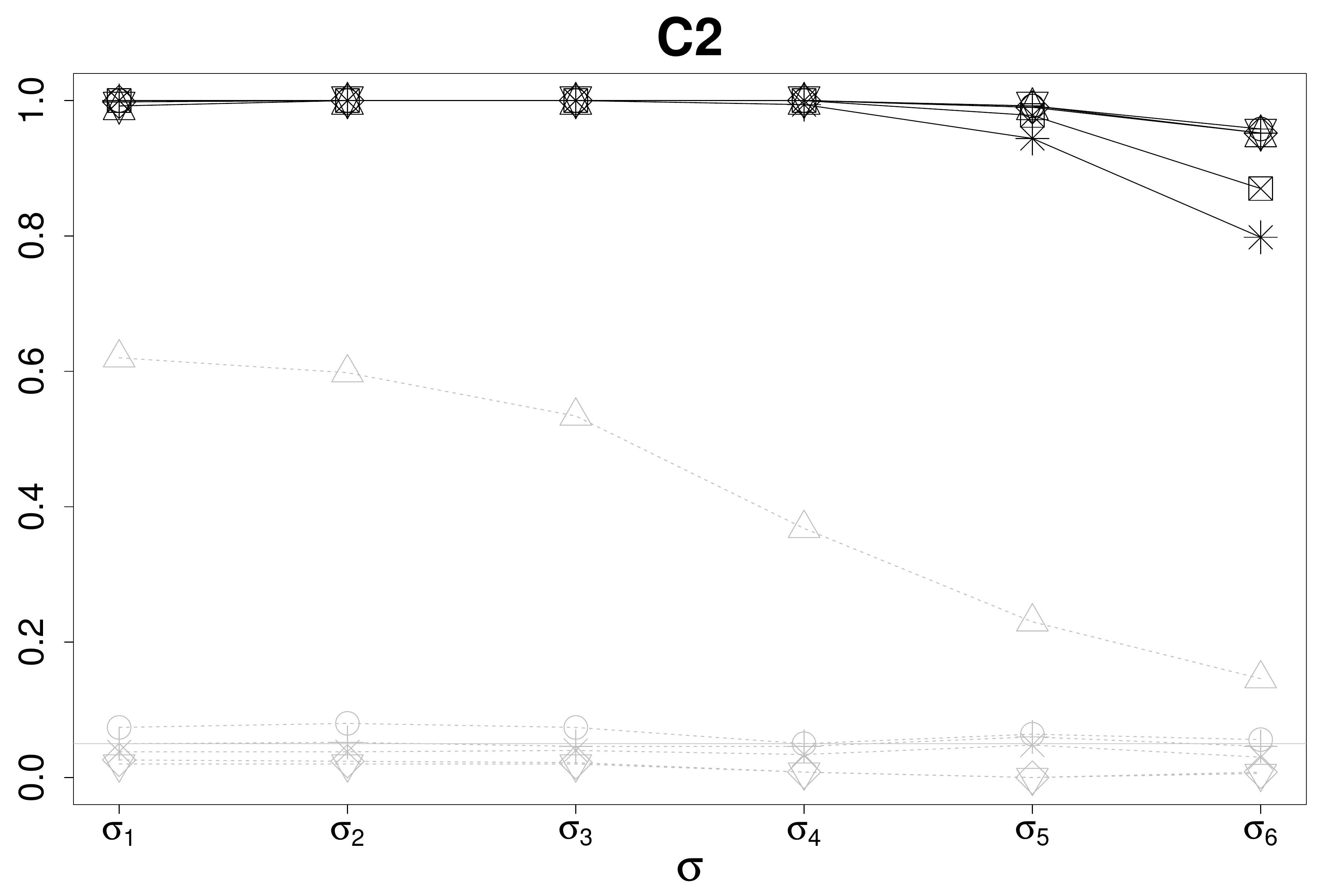}&\includegraphics[width=0.5\textwidth]{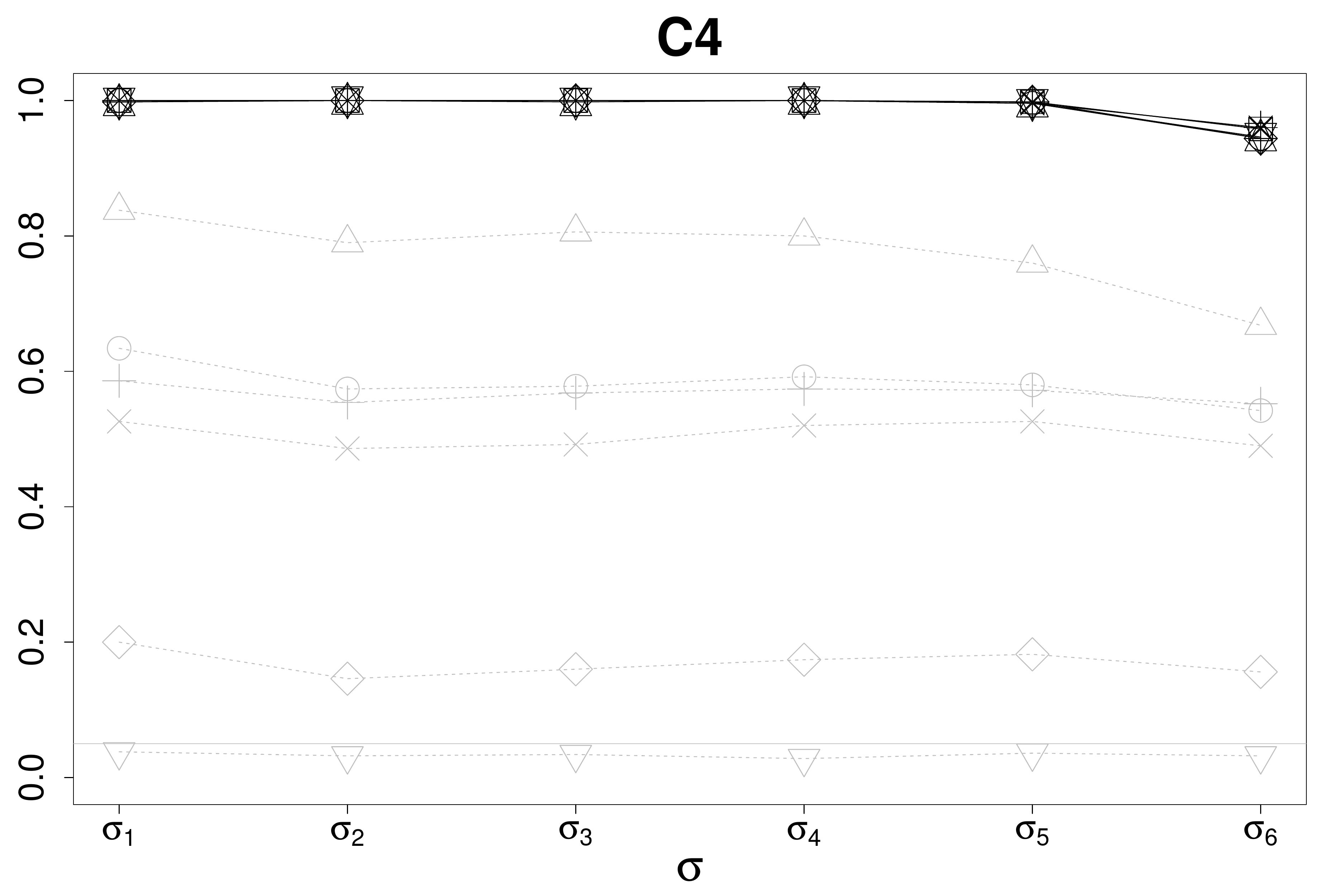}&\includegraphics[width=0.5\textwidth]{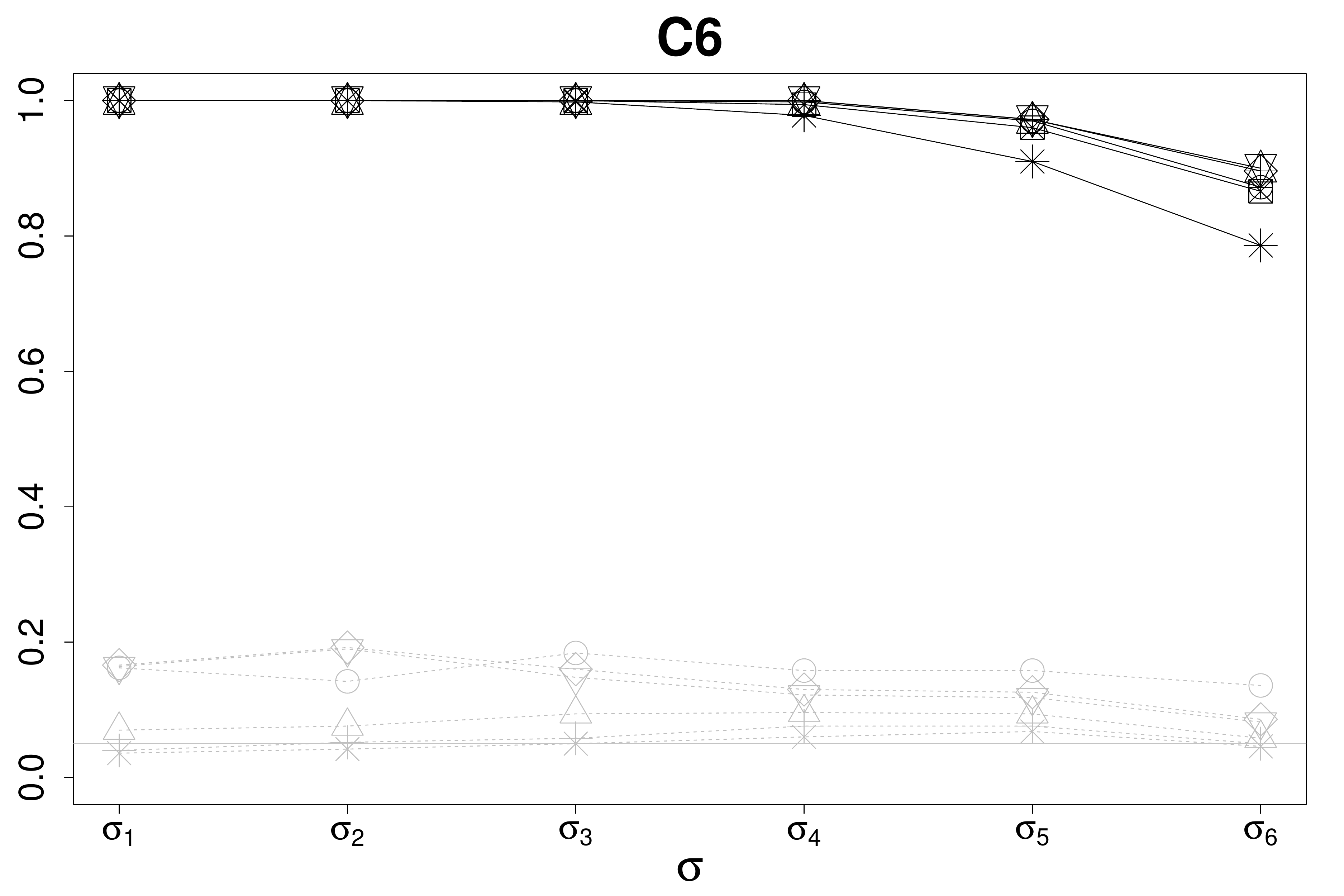}\\
	\end{tabular}}
	
	\caption{\label{fig_M3}Empirical power of all tests for $ H_{0} $ against  $ H_{1} $ (at level $ \alpha=0.05 $) as a function of $ \sigma_n $, $ n=1,\dots,6 $, for  contamination models C0-6 in  model M3 of Scenario 1. The proposed and competing tests are displayed as  black and grey lines, respectively.}
\end{figure}

Fig. \ref{fig_M3} shows the empirical power for model M2  generally become smaller when $ \sigma_n $ increases.
The results are similar to those for model M2, even though the empirical power tends to be larger, due to  the more apparent separation of the main effects.
Again, the proposed RoFANOVA tests outperform the competitors in case of contamination (C1-6), and have satisfactory power in case of no contamination (C0). The best results are achieved by the RoFANOVA-BIS, RoFANOVA-HAM and RoFANOVA-OPT tests.

\subsection{Two-way functional analysis of variance}
\label{sec_twoway}	
In this section, we consider the two-way  FANOVA model introduced in equation \eqref{eq_modanova}. The simulation design is inspired by that of  \cite{cuesta2010simple}.

As for Scenario 1, we assume  $ \mathcal{T}=\left[0,1\right] $ and the functional response depending on a grand mean $ m $, 2 level main effects $ f_i$ and $ g_i $, and on an interaction term $ h_{ij} $ through two parameters $ a $ and $ b $ with values in $ \lbrace 0,0.05,0.10,0.25,0.50\rbrace  $. Here, the larger the values of $a$ and $b$ the more  $ f_i$  and $g_i$ deviate, respectively, from the grand mean $ m $. Thus, the empirical power should be an increasing function of $a$  and $b$. The empirical size is studied for $a=0$ or $b=0$.
% The grand mean $ m $, as well as the  2 level main effects $ f_i$ and $ g_i $ are shown in Figure \ref{fig_two_way_ex}.

As in Scenario 1, seven contamination models C0-6 are considered, and the response curves are assumed as independent realizations of a  Gaussian process with covariance function $ \gamma\left(s,t\right)=\sigma^2 e^{\left(-|s-t|10^{-5}\right)}$. Data are observed  through $ 25 $ evenly spread discrete points with $ \sigma=0.3 $.
Also in this scenario, we consider the five versions of the proposed method, viz., RoFANOVA-MED, RoFANOVA-HUB, RoFANOVA-BIS, RoFANOVA-HAM, and RoFANOVA-OPT, with tuning parameters chosen as in Scenario 1.
As competitors we consider: (i) the permutation version of the method proposed by \cite{zhang2011statistical}, referred to as FNDP,  which is permutation test based on a $ F $-type statistic; and (ii) the global version of the method proposed by \cite{pini2017interval}, which is the two-way extension of the method of \cite{zhang2014one}, referred to as TGPF.
Both for the FNDP and TGPF methods, the distribution of the test statistic is approximated by using the Manly's scheme \citep{manly2006randomization} with 1000 random permutations.
Moreover, also the TRPbon and TRPfdr method (Section \ref{sec_oneway}) are considered with 30 random projections.

For each triplet (C$m$,$ a$,$ b $),  with $ m=0,\dots,6 $, and $ a,b\in \lbrace 0,0.05,0.10,0.25,0.50\rbrace  $, the five proposes  and the four competing methods are applied $ N=500 $ times to the generated functional sample to test, at level $ \alpha=0.05 $,   $ H_{0,A} $, $ H_{0,B} $ and $ H_{0,AB} $ against $ H_{1,A} $, $ H_{1,B} $ and $ H_{1,AB} $, respectively.
Then, for each triplet and  test, the empirical size and  empirical power of the test were  computed as the fraction of rejections out of $ N $ replications (also in this case, with  maximum standard deviation equal to 0.0224). The former is considered when $ a=b=0 $, for $ H_{0,A} $, $ H_{0,B} $  against $ H_{1,A} $, $ H_{1,B} $, and when $ a<0.25 $ for  $H_{0,AB} $ against $ H_{1,AB} $; whereas the latter is considered when $ a\neq 0$ or $b\neq0 $, for $ H_{0,A} $, $ H_{0,B} $  against $ H_{1,A} $, $ H_{1,B} $, and $ a\geq0.25 $ for  $H_{0,AB} $ against $ H_{1,AB} $. 

For the sake of conciseness, we  summarize the results for cases that are statistically equivalent. For instance, when  analyzing the null hypothesis $ H_{0,A} $ (resp., $ H_{0,B} $),  for each value of $ a $ (resp., $ b $),  the five values corresponding to  $ b=\lbrace 0,0.05,0.10,0.25,0.50\rbrace $ (resp., $ a=\lbrace 0,0.05,0.10,0.25,0.50\rbrace $) are summarized through their median. Similarly, when  analyzing $ H_{0,AB} $, the values corresponding to  $ a<0.25 $ are substituted by their median for each value of $ b $.

\begin{figure}
	\centering
	\resizebox{1\textwidth}{!}{
		\begin{tabular}{M{0.5\textwidth}M{0.5\textwidth}M{0.5\textwidth}M{0.5\textwidth}}
			\multirow{2}{*}{\includegraphics[width=0.5\textwidth]{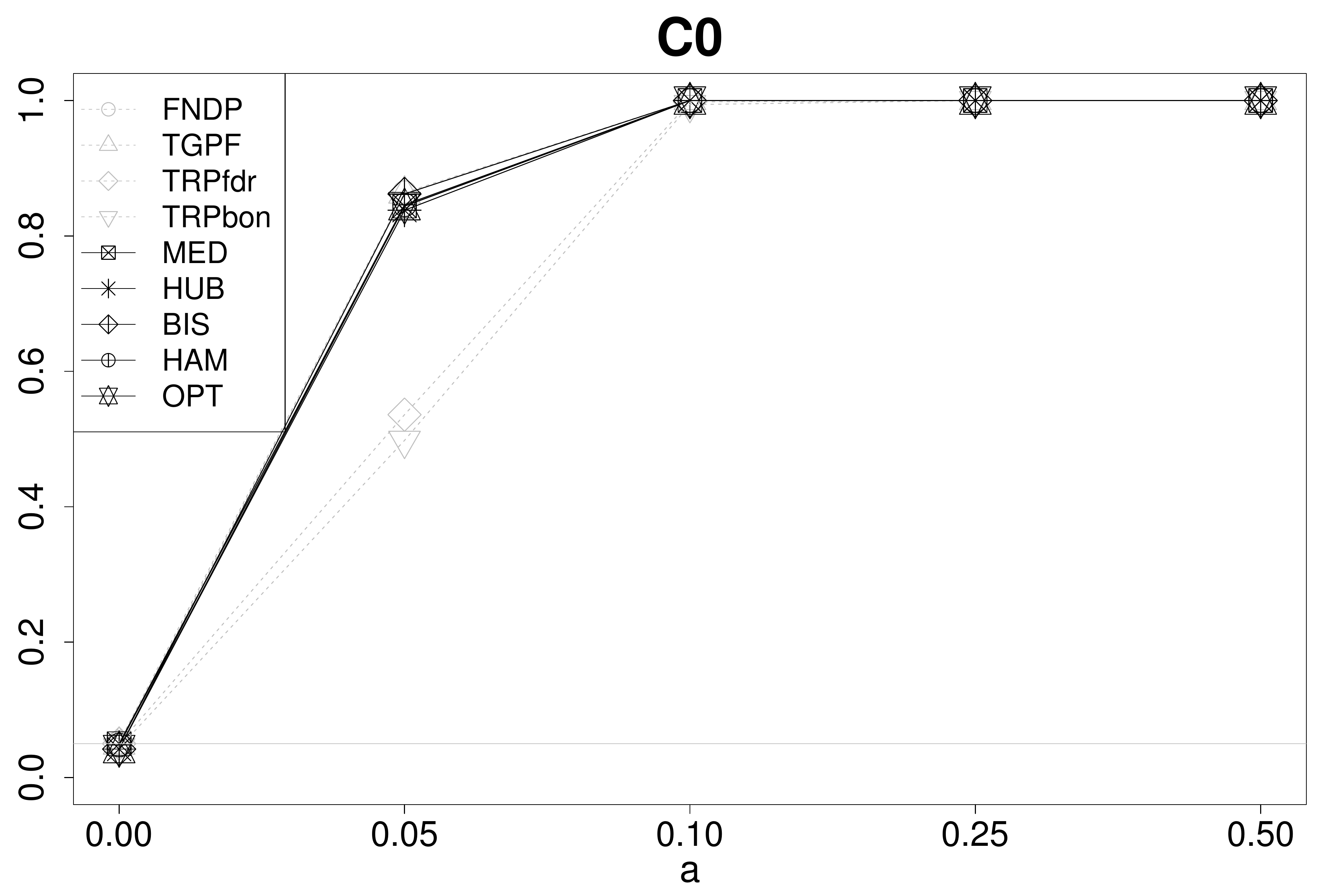}}&\includegraphics[width=0.5\textwidth]{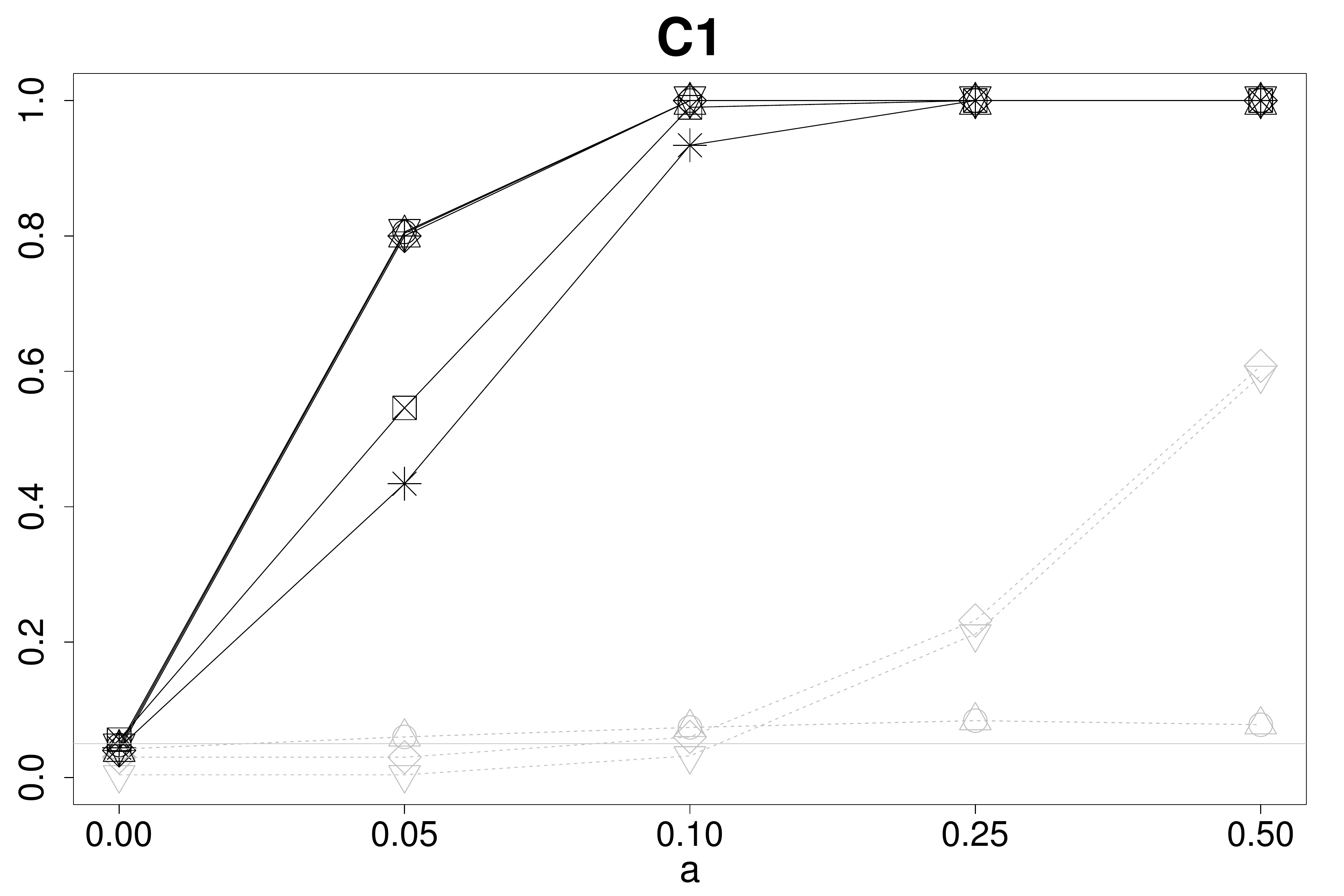}&\includegraphics[width=0.5\textwidth]{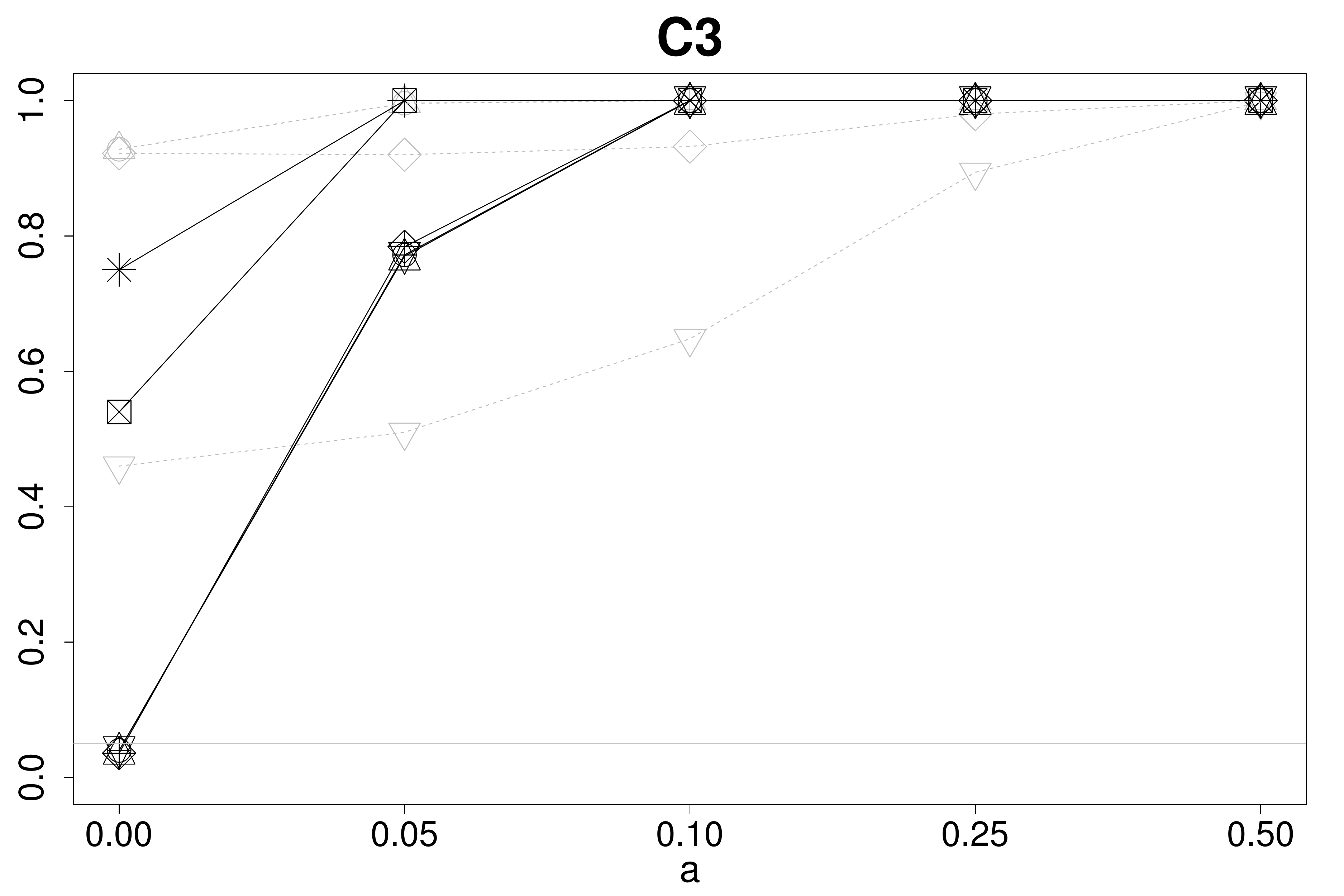}&\includegraphics[width=0.5\textwidth]{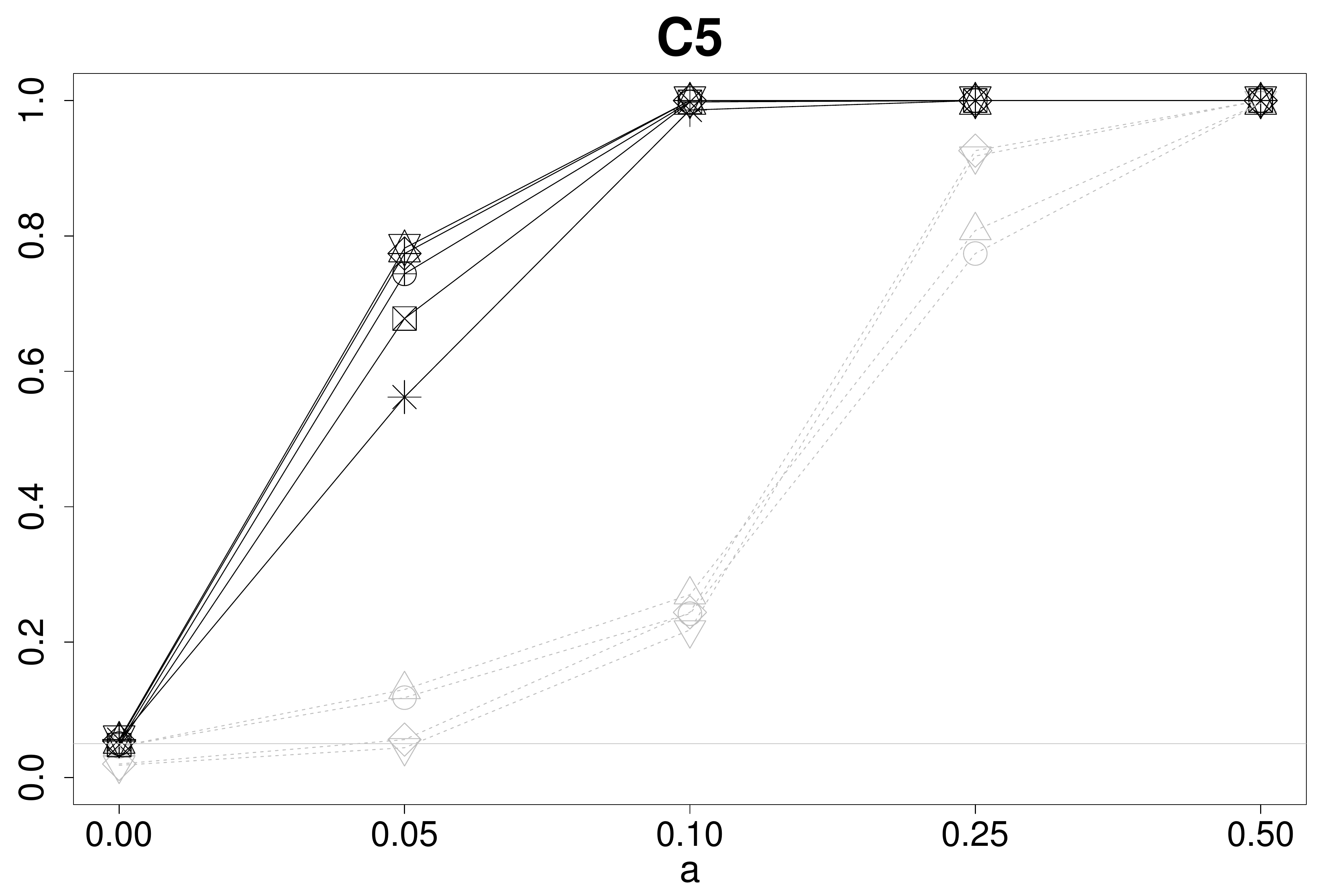}\\
			&\includegraphics[width=0.5\textwidth]{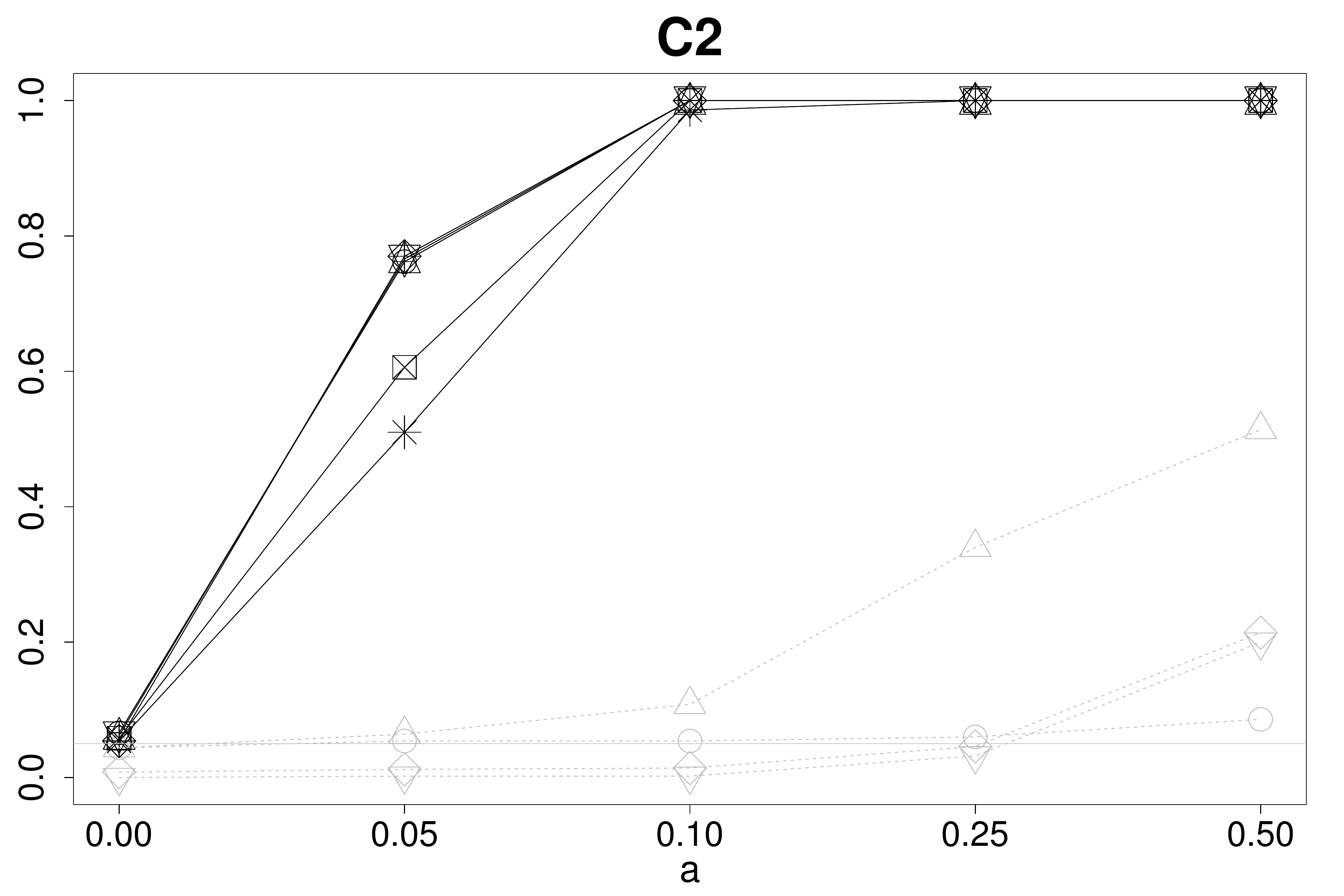}&\includegraphics[width=0.5\textwidth]{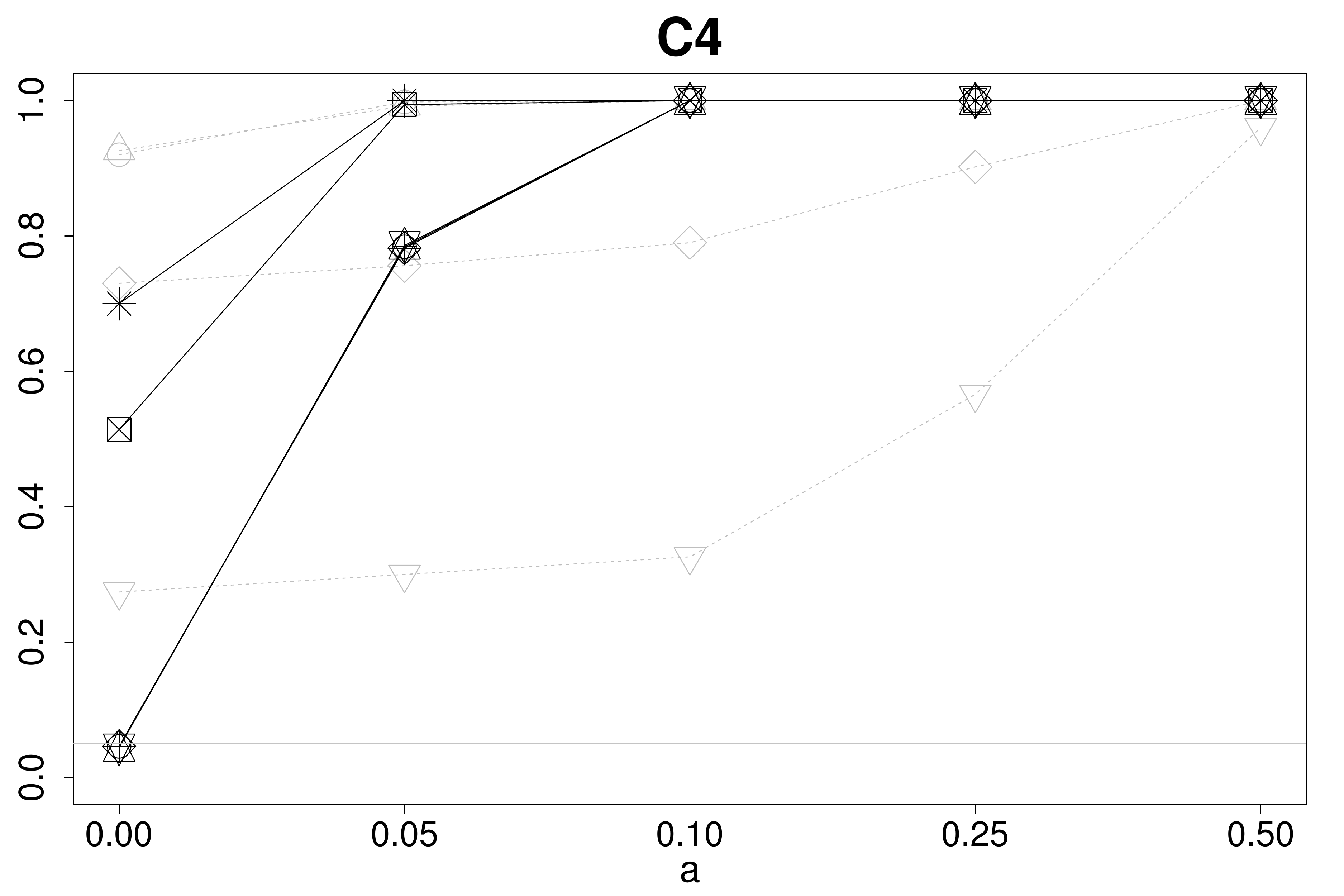}&\includegraphics[width=0.5\textwidth]{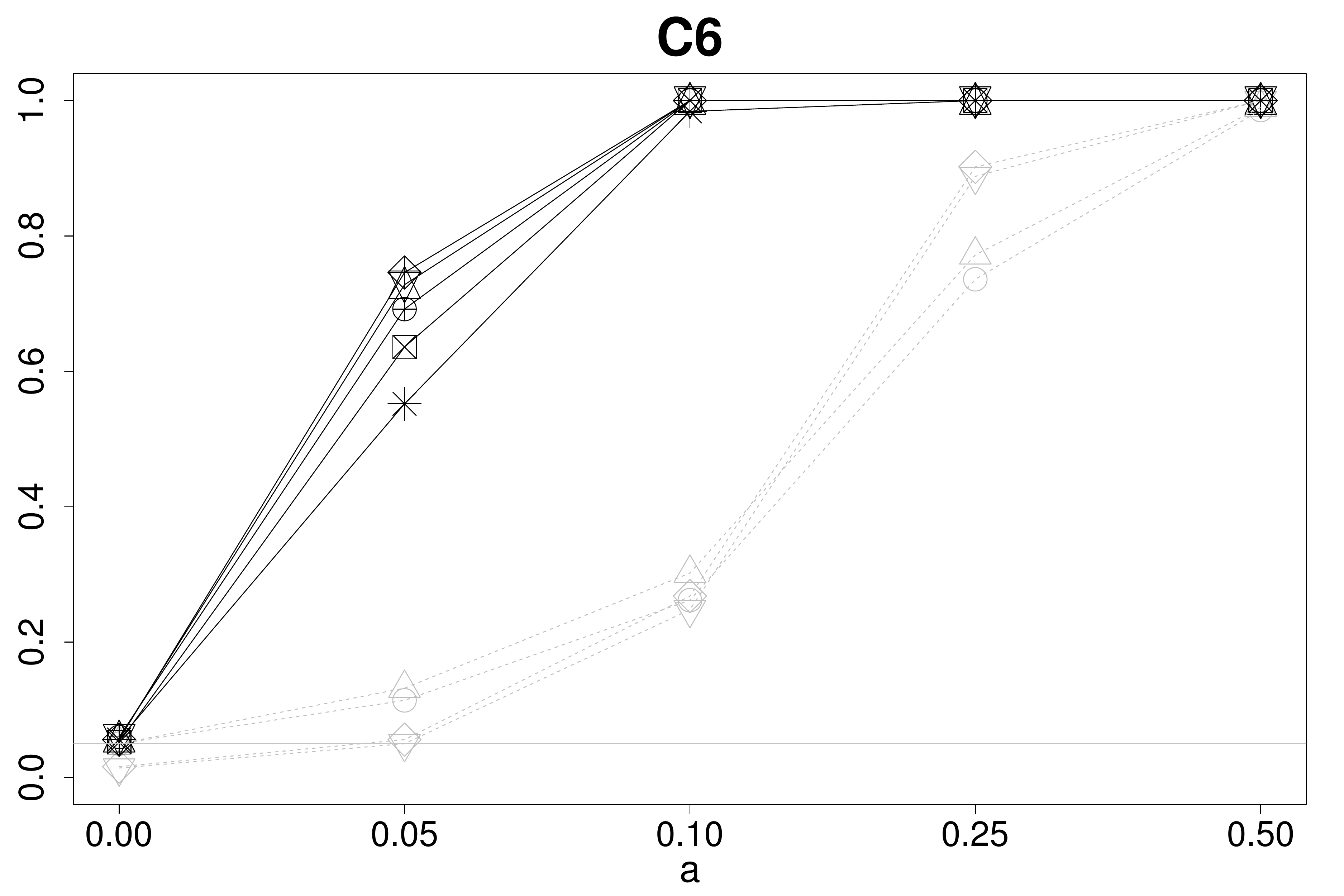}\\
		\end{tabular}}
	
	\caption{\label{fig_twoa}Empirical size ($ a=0 $) and  power  ($ a\neq0 $) of all tests for $ H_{0,A} $ against  $ H_{1,A} $ (at level $ \alpha=0.05 $) as a function of $ a $, for different contamination models (C0-6) in Scenario 2. The proposed and competing tests are displayed as  black and grey lines, respectively.}
\end{figure}

Fig. \ref{fig_twoa} shows the empirical size ($ a=0 $) and  power  ($ a\neq0 $) of all tests for $ H_{0,A} $ against  $ H_{1,A} $ as a function of $ a $. When $ a $ increases, the performance of all the methods in rejecting $ H_{0,A} $ enhances. In terms of empirical size (i.e., when $ a=0 $), the results are quite satisfactory for all the methods in case of no contamination (C0), symmetric magnitude contamination (C1-2) and  both symmetric (C5) and asymmetric (C6) shape contamination.
However, in case of asymmetric magnitude contamination (C3-4), only the RoFANOVA-BIS, RoFANOVA-HAM and RoFANOVA-OPT tests are able to control the level $ \alpha $, being approximately less than or equal to 0.05.
This behavior is analogous to that achieved in Scenario 1  of Section \ref{sec_oneway}.
In terms of empirical power ($ a\neq0 $), the proposed RoFANOVA test has comparable performance  when there are no outliers (C0); whereas it is far better than the competitors for the contamination models C1-6. Note that, for asymmetric magnitude contamination (C3-4), only the RoFANOVA-BIS, RoFANOVA-HAM and RoFANOVA-OPT tests should be considered, being the only ones able to control the level $ \alpha $.

\begin{figure}
		\centering
	\resizebox{1\textwidth}{!}{
		\begin{tabular}{M{0.5\textwidth}M{0.5\textwidth}M{0.5\textwidth}M{0.5\textwidth}}
			\multirow{2}{*}{\includegraphics[width=0.5\textwidth]{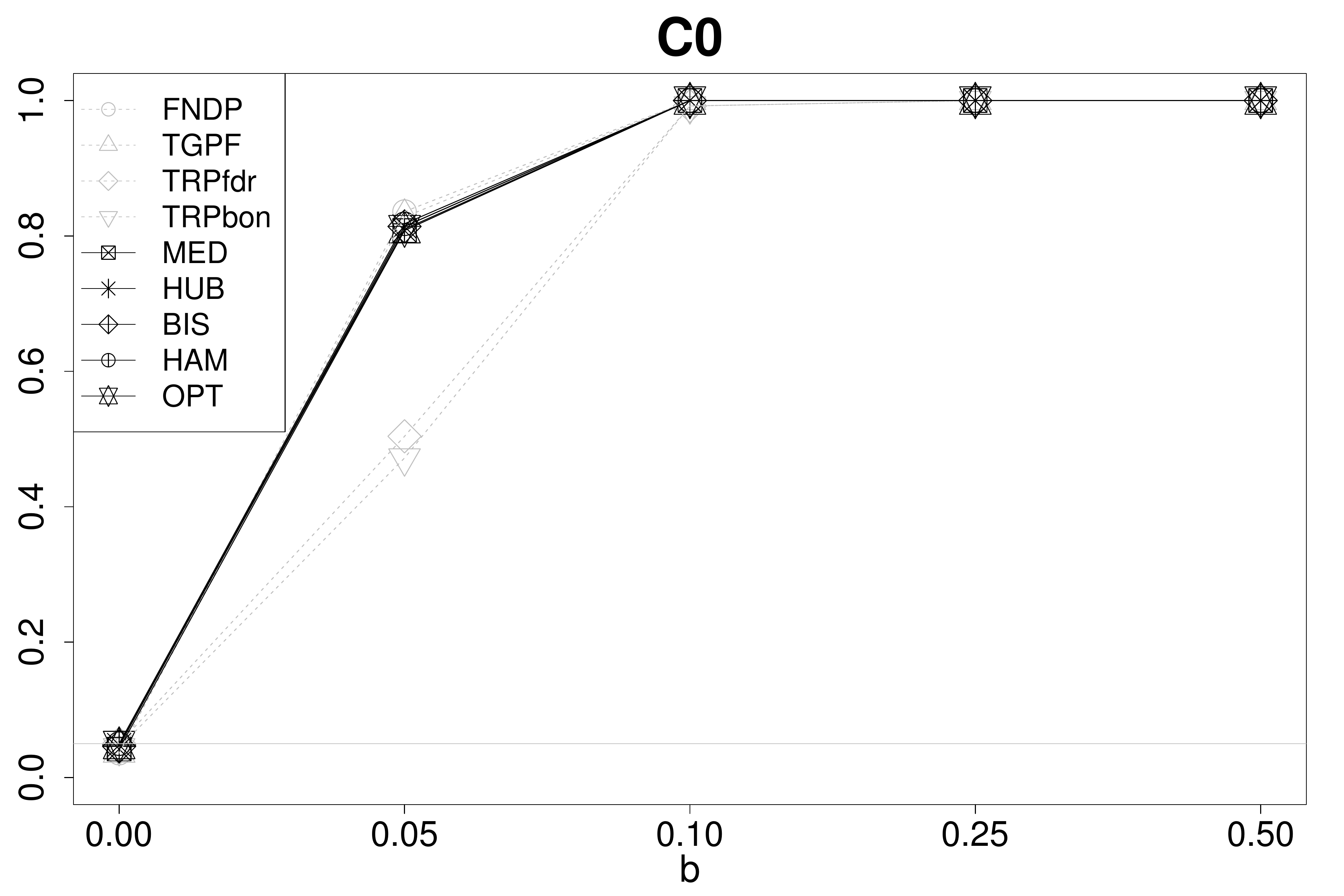}}&\includegraphics[width=0.5\textwidth]{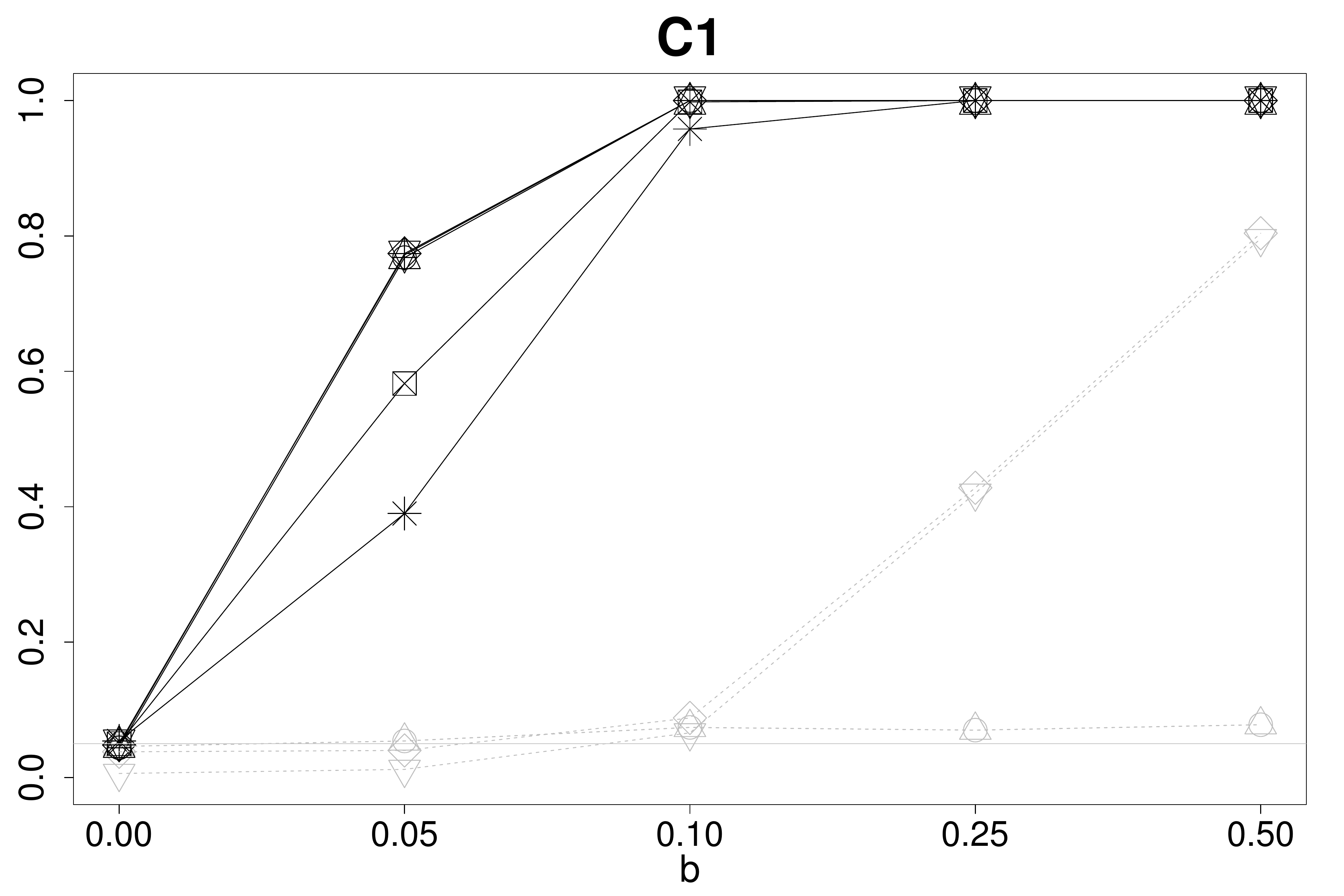}&\includegraphics[width=0.5\textwidth]{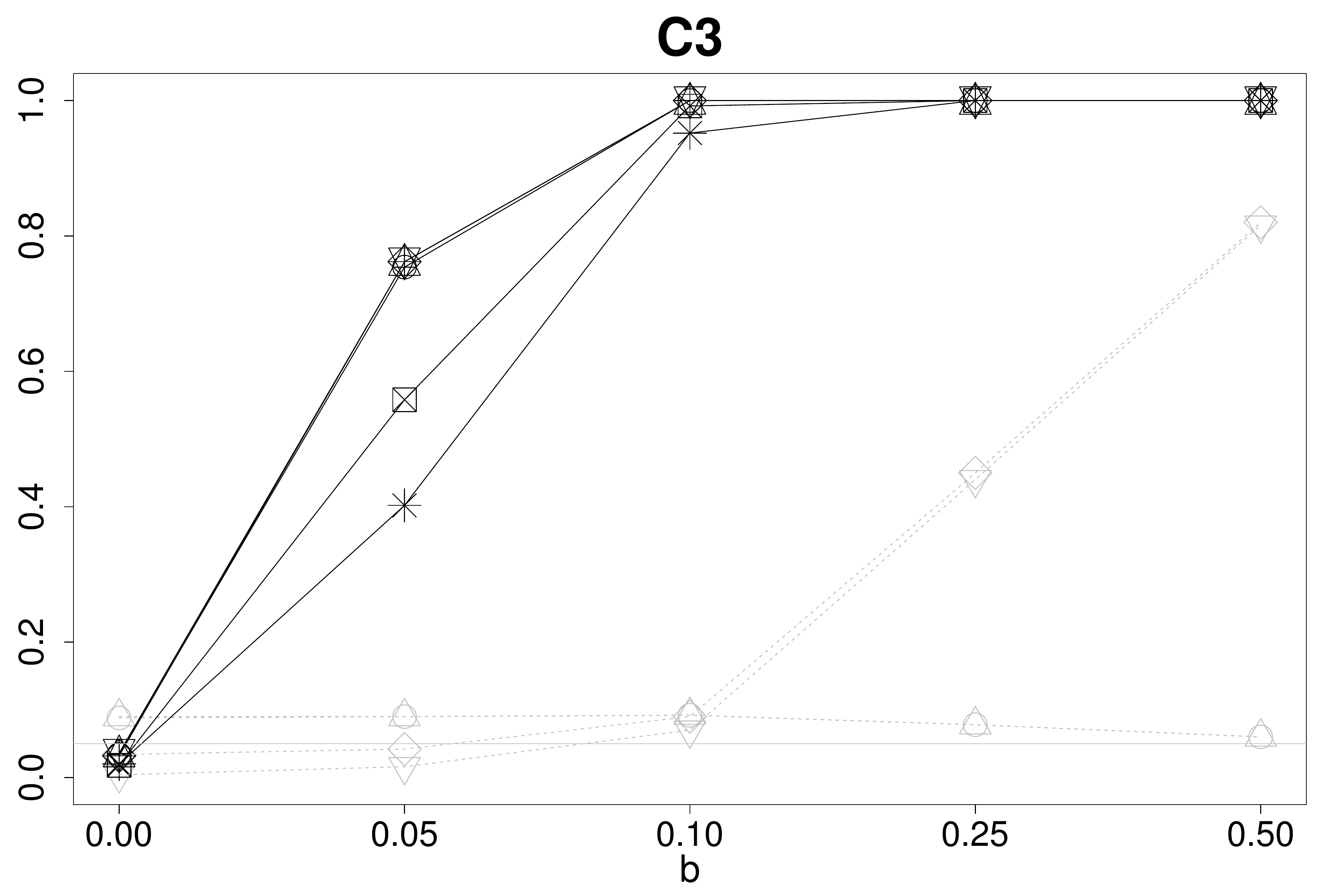}&\includegraphics[width=0.5\textwidth]{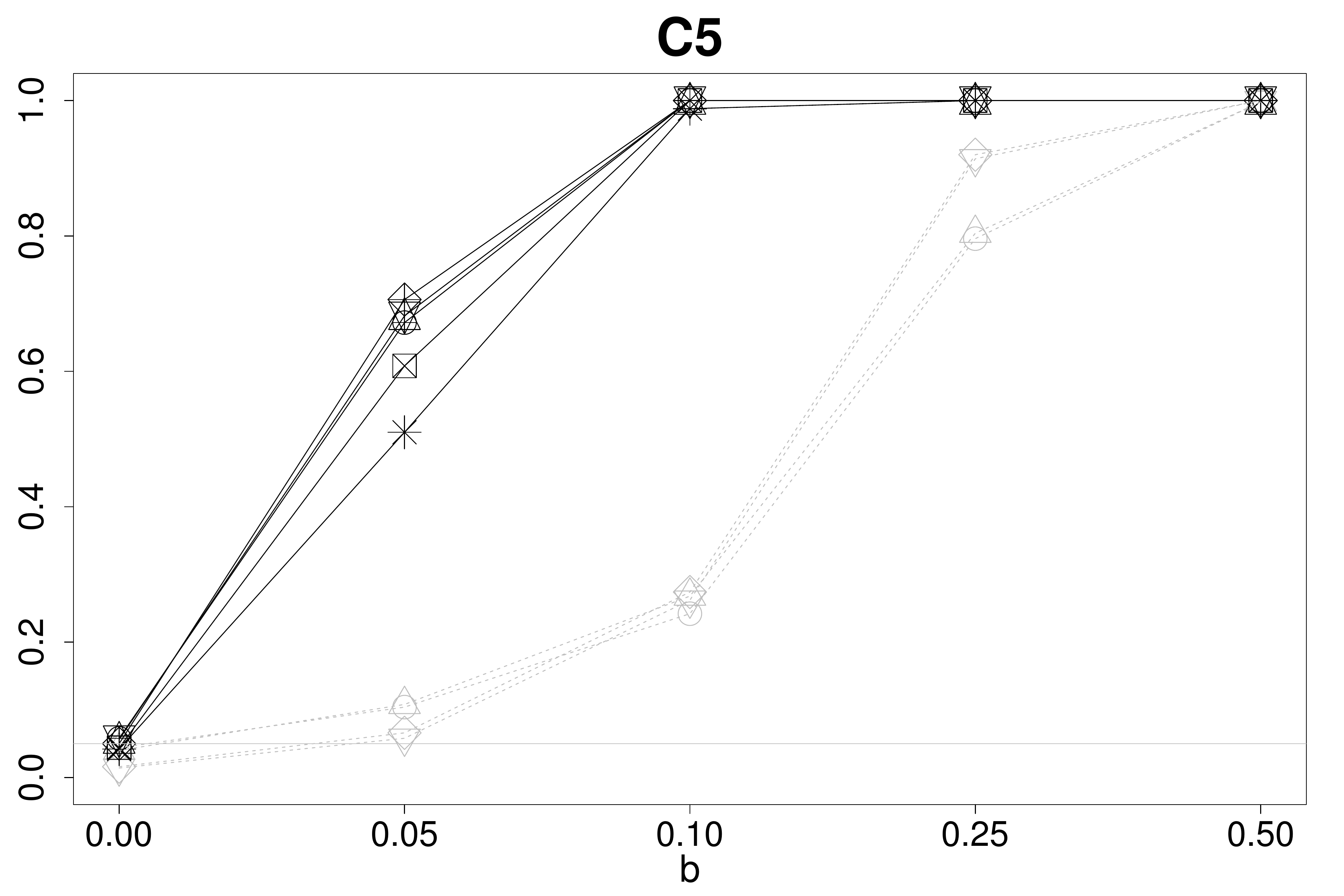}\\
			&\includegraphics[width=0.5\textwidth]{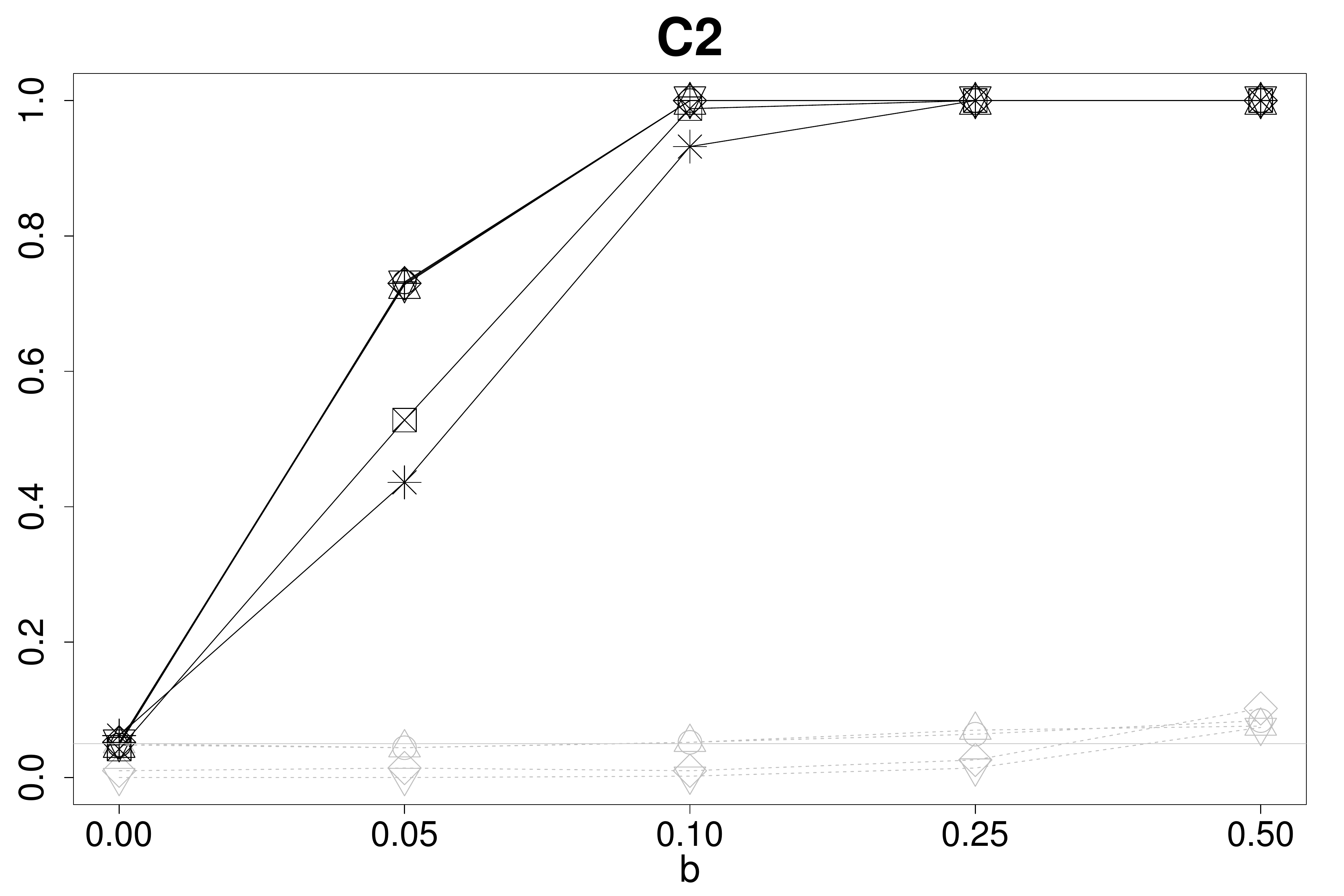}&\includegraphics[width=0.5\textwidth]{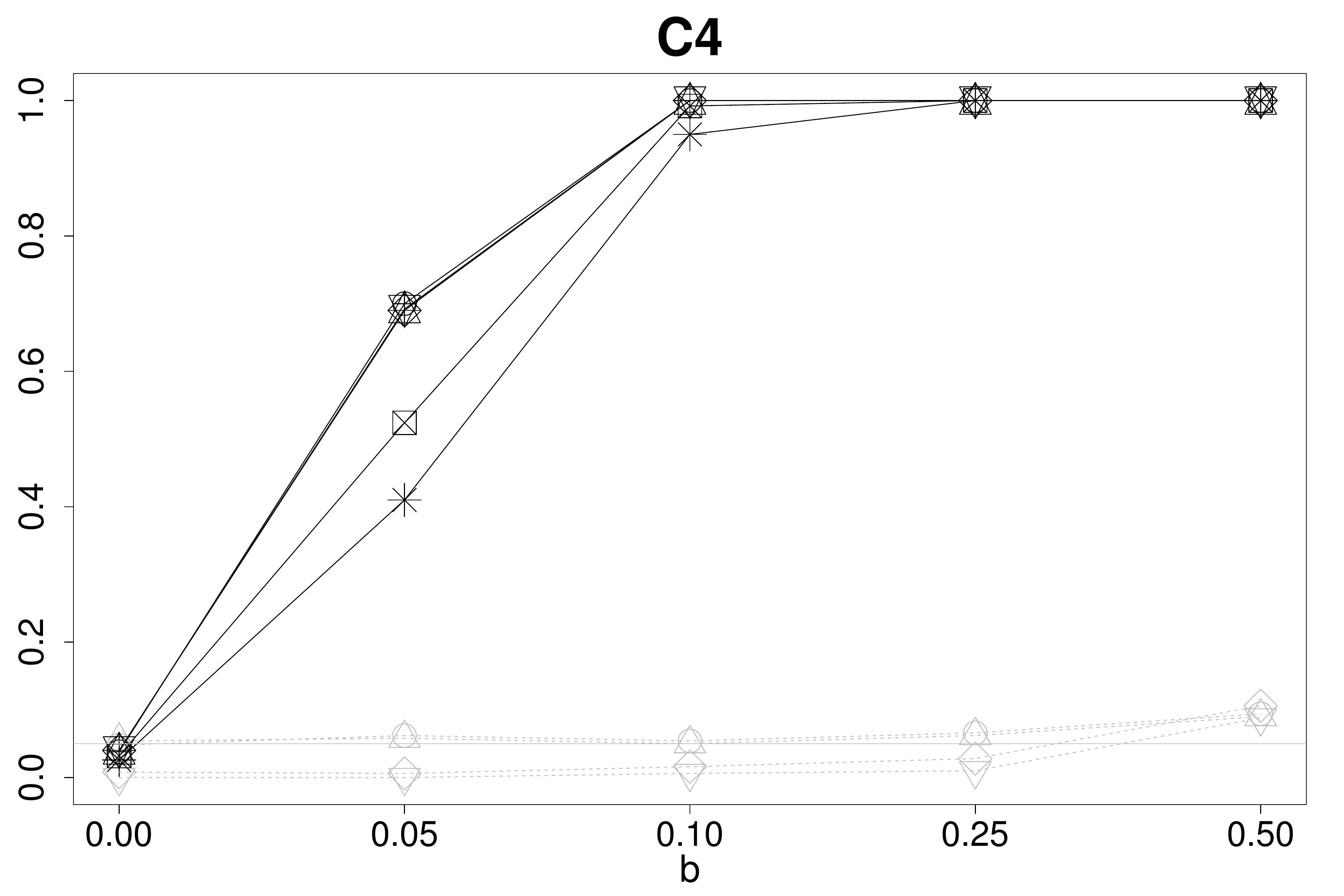}&\includegraphics[width=0.5\textwidth]{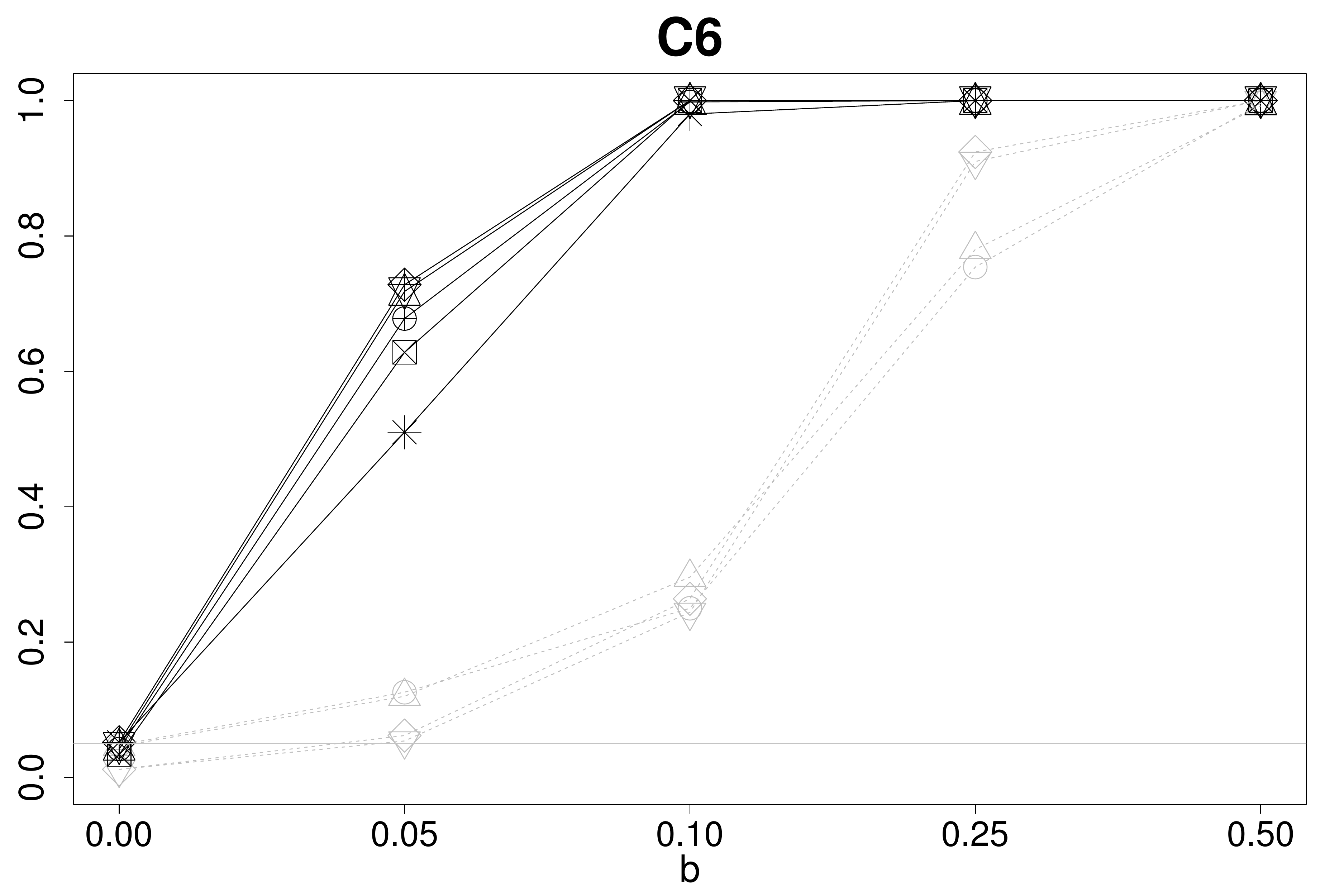}\\
	\end{tabular}}
	
	\caption{\label{fig_twob}Empirical size ($ b=0 $) and  power  ($ b\neq0 $) of all tests for $ H_{0,B} $ against  $ H_{1,B} $ (at level $ \alpha=0.05 $) as a function of $ b $, for different contamination models (C0-6) in Scenario 2. The proposed and competing tests are displayed as  black and grey lines, respectively.}
\end{figure}

In Fig. \ref{fig_twob}, the empirical size ($ b=0 $) and the empirical power  ($ b\neq0 $) of all tests for $ H_{0,B} $ against  $ H_{1,B} $ (at level $ \alpha=0.05 $) are displayed as a function of $ b $. Also in this case, the proposed tests outperform the competitors, in terms of power, for contamination models C1-6. They  simultaneously have, in fact, comparable performance in absence of  contamination (C0). Moreover, differently from Scenario 1, all the tests are able to approximately control the level $ \alpha $, even for the contamination models C3-4. This is expected in this case, because the asymmetry in the contamination affects  the main effect $ f_i $, only,  and not  $ g_i $.
Among the proposed tests, the  RoFANOVA-BIS, RoFANOVA-HAM and RoFANOVA-OPT  tend to perform better than the ones based on monotonic functional $ M $-estimator, viz., the RoFANOVA-MED and RoFANOVA-HUB tests.
To test  $ H_{0,AB} $ against  $ H_{1,AB} $, results are presented in the Supplementary Material where it is shown that the empirical power of the proposed tests is much larger than that of the competitors, for all the contamination model C1-6. Moreover, and that among the RoFANOVA tests, the RoFANOVA-BIS, RoFANOVA-HAM and RoFANOVA-OPT  achieve  the best performance.

\section{Real-case study: analysis of variance of applied to the analysis of spatter behaviour in laser powder bed fusion}
\label{sec_real}
To demonstrate the potential of the proposed approach, this Section presents the real-case study in additive manufacturing. In L-PBF, spatters are process by-products that can be ejected either by the melt pool, i.e., the region when the thin layer of powder is locally melted by the laser, in the form of hot and liquid droplets or by the powder bed regions surrounding the melt pool \citep{young2020types,ly2017metal,bidare2018fluid}. In the latter case, spatters consist of powder particles entrained by convective motions above and around the melt pool. For more details about the spatter generation mechanism, the reader is referred to \cite{young2020types,ly2017metal,bidare2018fluid}, and the literature cited therein.
The analysis of spatters in the L-PBF process has gathered an increasing interest in the last years because they can  drive  relevant information about the process state and the final quality of the manufactured part.
% \citep{yang2020monitoring,tan2020neural,yin2020correlation,andani2017spatter,repossini2017use,ly2017metal,bidare2018fluid}.
Studying the effect of controllable process factors and other operating conditions on the spatter behaviour allows getting a deeper comprehension of underlying physical phenomena. Such knowledge may be used to tune the process condition and enhance the quality and mechanical performances of the products, or to design in-line and real-time process monitoring methodologies \citep{colosimo2020machine}.

Hot spatters ejected as a consequence of the laser-material interaction can be observed by means of high-speed cameras installed into the L-PBF machine or placed outside its viewports. The mainstream literature devoted to spatter analysis and monitoring in L-PBF relies on video image processing methods to compute synthetic indices that capture salient aspects of the spatter behaviour, e.g., the number of ejected spatters in each video frame, their size, speed, etc. \citep{grasso2017phase,everton2016review}. In the real-case study presented in this section, instead of treating synthetic descriptors of the spatter ejections as univariate or multivariate variables, the spatter behaviour is translated into a functional format by means of the so-called spatter intensity function introduced in Section 1. Such function captures the spatial spread of ejected spatters and can be estimated for each manufactured layer and for each test treatment. 
Section 4.1 presents the main experimental settings, whereas the results of the analysis and the comparison against benchmark methods are reported in Section 4.2.

%REFERENCE DA AGGIUNGERE
%Tan, Z., Fang, Q., Li, H., Liu, S., Zhu, W., & Yang, D. (2020). Neural network based image segmentation for spatter extraction during laser-based powder bed fusion processing. Optics & Laser Technology, 130, 106347.
%Yin, J., Wang, D., Yang, L., Wei, H., Dong, P., Ke, L., ... & Zeng, X. (2020). Correlation between forming quality and spatter dynamics in laser powder bed fusion. Additive Manufacturing, 31, 100958.

\subsection{Experimental setting and data preprocessing}
\begin{figure}
  	\centering
  	\includegraphics[width=0.8\textwidth]{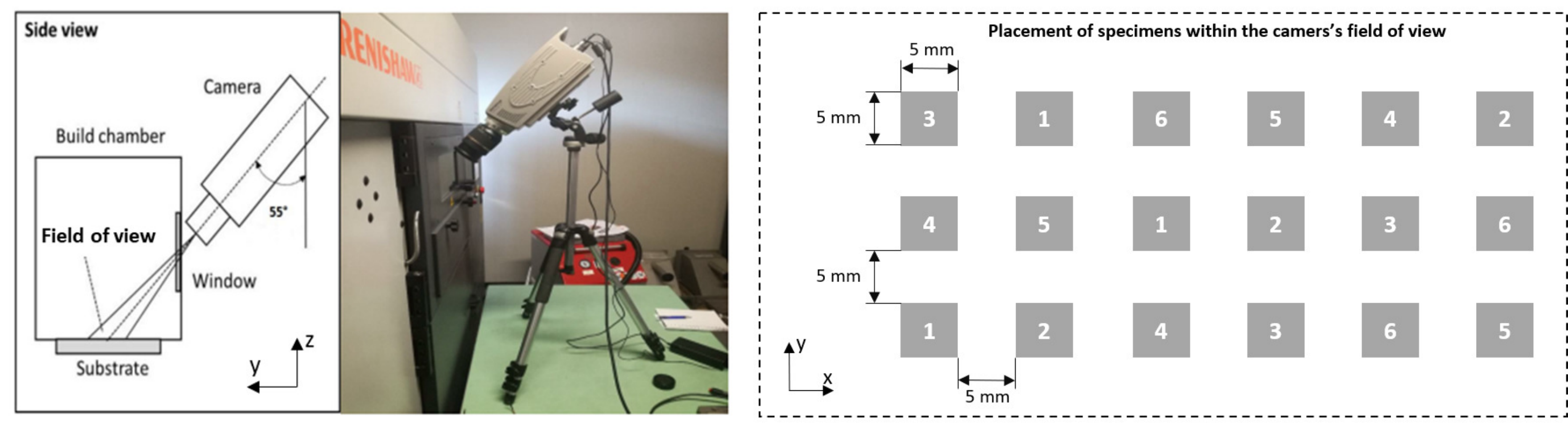} 
  	
  	\caption{\label{fig_setup} Setup of the high-speed camera in front of the Renishaw AM250 machine's viewport (left panel) and placement of manufactured specimens in the build area, within the camera's field of view (right panel): numbers shown in the specimens correspond to the energy density level, from 1 to 6, applied during the process.}
  \end{figure}

The case study involves the production of specimens of size 5 x 5 x 12 mm via L-PBF of 18Ni(300) maraging steel powder, a steel alloy commonly used for tooling applications, with average particle size between 25 and $35 \mu$m. An industrial L-PBF system, namely a Renishaw AM250, was used, with a high-speed camera in the visible range placed outside the front viewport of the machine as shown in Fig. \ref{fig_setup}, left panel. Videos were recorded during the production of six layers with a sampling rate of 1000 fps (frames per second) and a spatial resolution of about $200 \mu$m/pixel. Specimens are placed as shown in Fig. \ref{fig_setup}, right panel, and produced by varying the energy density provided by the laser to the material. Process parameters corresponding to the energy density levels are reported in the Supplementary Material.

The laser was displaced by a scanner along a predefined path consisting of parallel scan lines, whose orientation changed layer by layer, with a default rotation of about $67^{\circ}$ every layer.
Details about  locations and orientations of the six analysed layers  are provided in the Supplementary Material.
Along each scan line, the laser melts the material with a pulsed mode, i.e., by exposing points equispaced apart of a quantity $d$ along each scan line with a point exposure duration $t$. The energy density was varied by varying $t$ and $d$. 
% The energy density is known to be a factor of primary importance to determine the quality of the process, as insufficient or excessive energy densities may produce defects, like internal pores, and other deviations from the expected quality \citep{grasso2017phase,everton2016review,mani2015measurement}.
Within the build chamber, where the L-PBF process takes place, a laminar flow of inert gas, called shielding gas, is used to prevent ejected spatters from falling on the build area, with consequent potential contamination effects, and vaporized material from depositing on the laser window leading to possible attenuation of the laser beam \citep{anwar2018study}.
 
\begin{figure}
	\centering
		\resizebox{1\textwidth}{!}{
	\begin{tabular}{cccc}
		\includegraphics[width=0.25\textwidth]{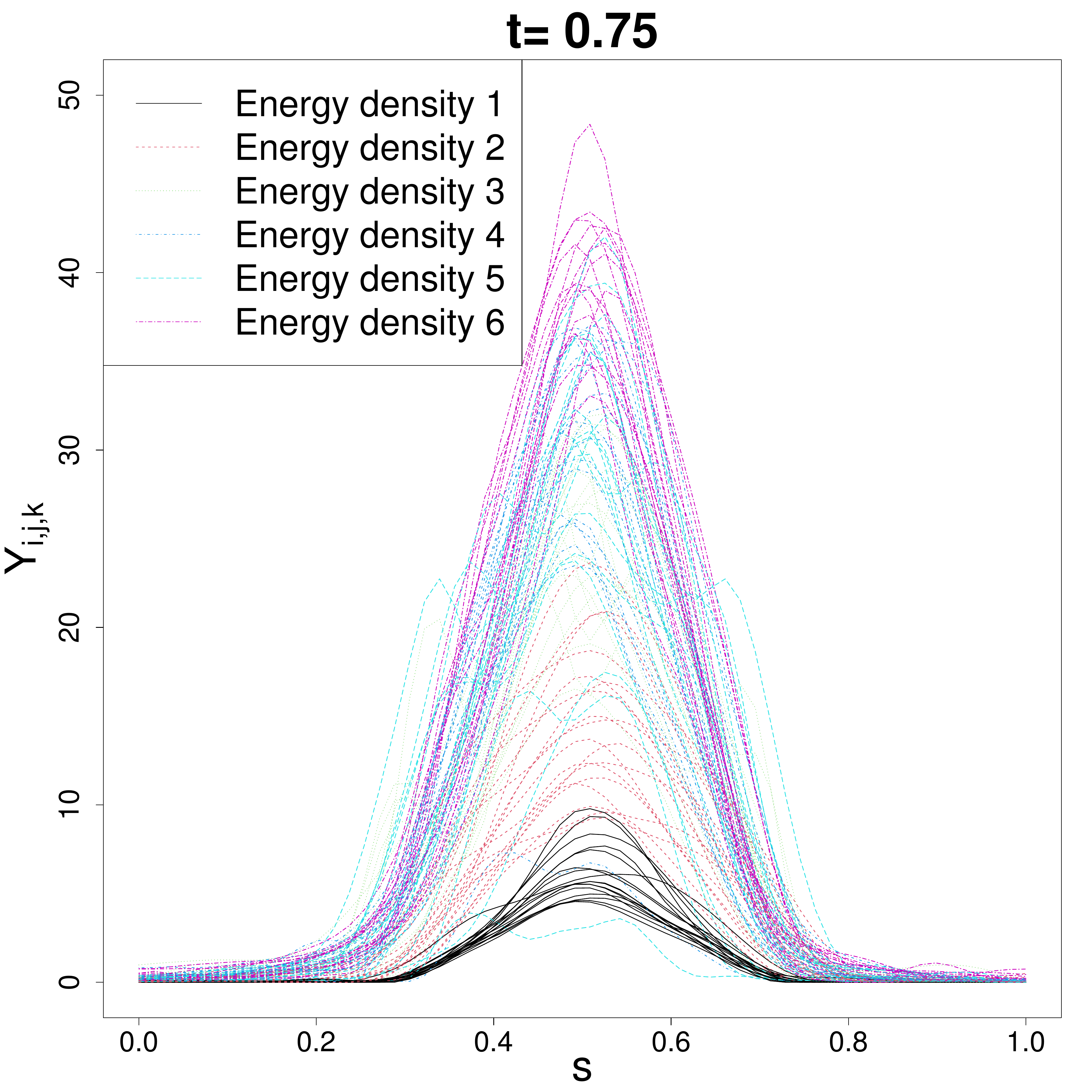} & \includegraphics[width=.25\textwidth]{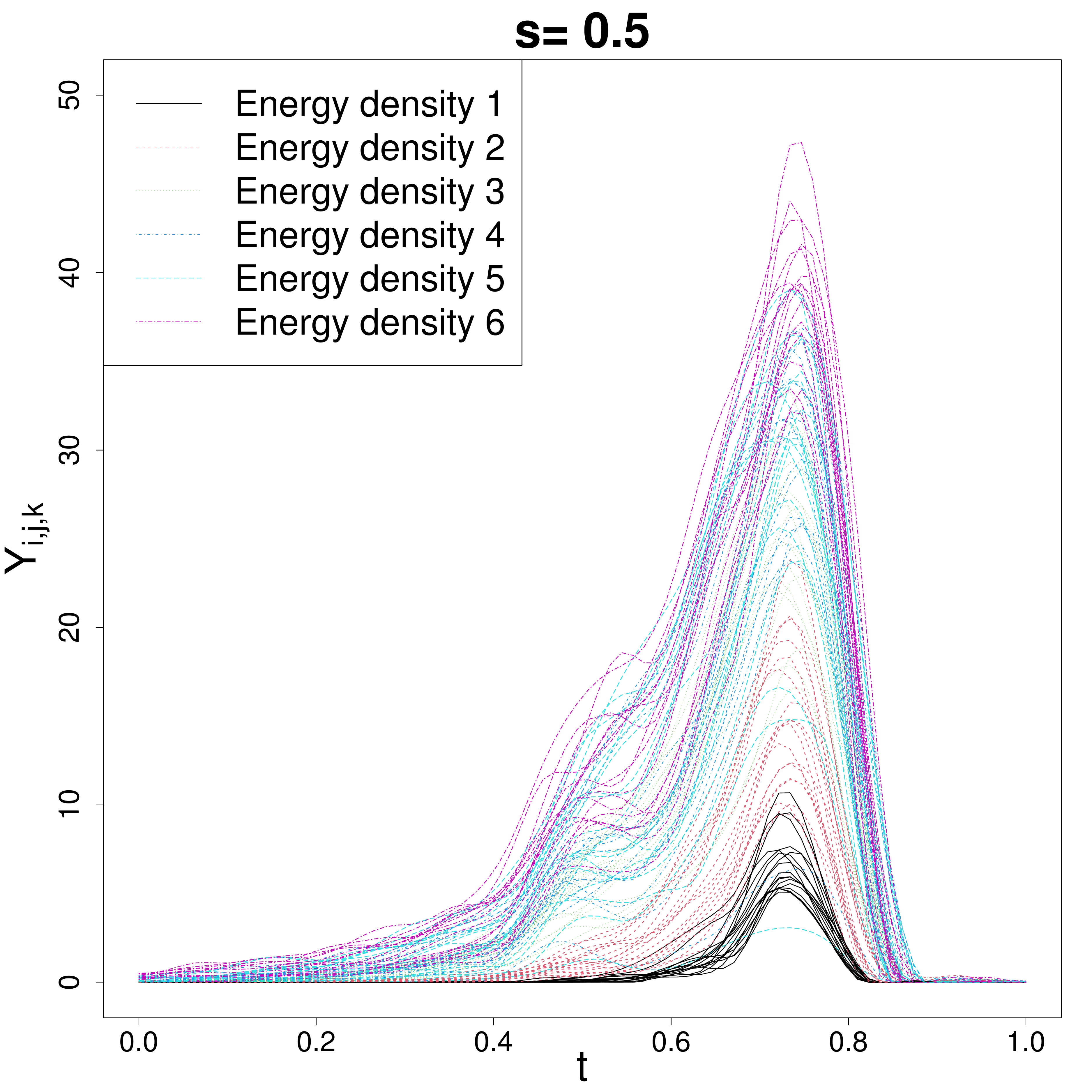}& \includegraphics[width=.25\textwidth]{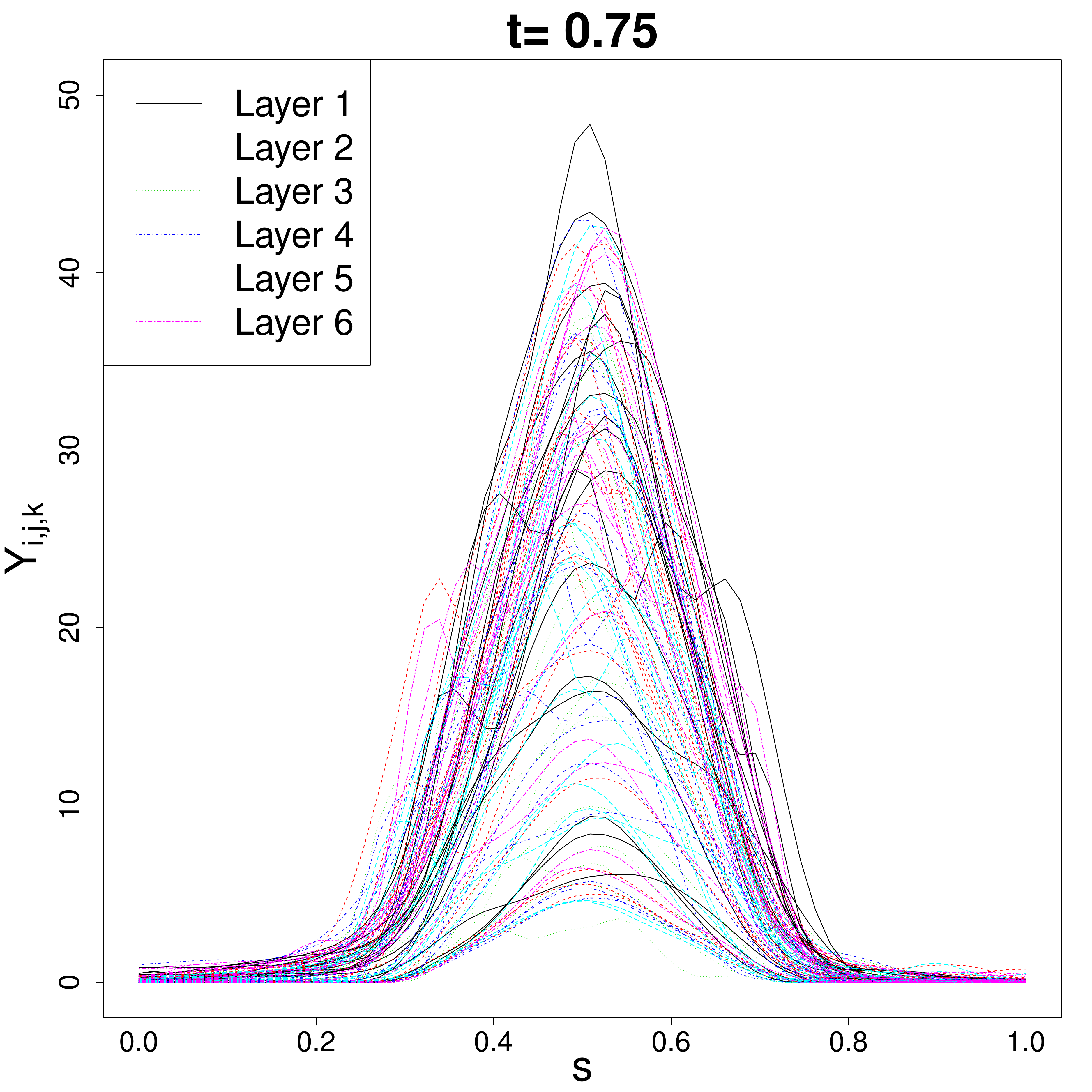} & \includegraphics[width=.25\textwidth]{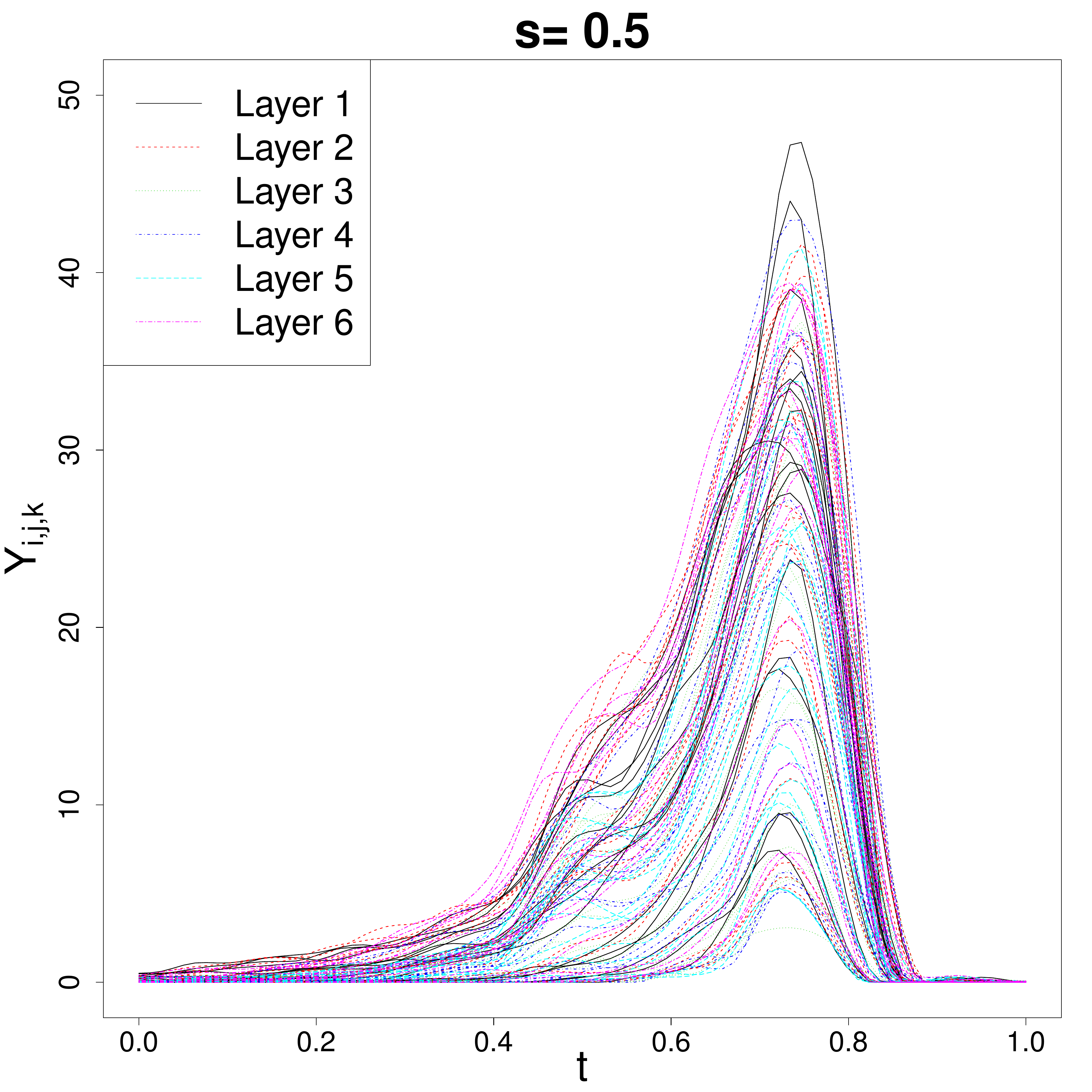}\\
			(a) & (b) &(c) & (d)
	\end{tabular}
}
	\caption{\label{fig_X}The functional observations $ Y_{i,j,k}$ for $ t=0.75 $, (a) and (c), and  $ s=0.5 $, (b) and (d), in the real-case study, for different energy density  levels ((a) and (b)) and different  scan strategies ((c) and (d)).}
\end{figure}

The functional response variable is the spatter intensity and was estimated by applying the video image pre-processing method presented in \cite{repossini2017use}. Thanks to this approach, the centroid of each spatter in the video frame was computed and used to determine the spatial coordinates $(s,t)$ of each detected spatter. All details about the video image pre-processing steps can be found in \cite{repossini2017use}.
In order to spatially map the amount of spatters ejected during the production of each specimen in each layer, three additional pre-processing operations were performed. First, the location of spatters was referred to a spatial domain centered in the center of the scanned area of each specimen, to allow comparing the functional response variables for specimens produced in different locations. Second, the spatial domain was discretized into 60 by 80 adjacent squared cells, in order to count the number of spatters ejected in each layer within each cell.
Based on these pre-processing steps, the spatial spread of the spatters, in each layer and for each specimen, could be summarized into the function $ Y_{i,j,k}\left(s,t\right) $ defined on the bi-dimensional domain $ \mathcal{T}=\left[0,1\right]\times \left[0,1\right] $, where indices $ i=1\dots,6 $, $ j=1,\dots 6 $ and $ k=1,\dots,n_{ij} $ indicate the energy density level, the layer, and the number of replicates (specimens) for each treatment, respectively.
The spatter intensity function $ Y_{i,j,k}\left(s,t\right) $ is a smoothed version of the actual amount of spatters counted in every location of the spatial domain. 
The number of replicates $ n_{ij} $ is fixed and equal to 3, as three specimens were produced for each energy density level, except for $ i=6 $ and $ j= 1$ where $ n_{ij}=2 $, due to a delamination occurred in initial layers prevented from producing one of the three specimens with the lowest energy density level.
The $ Y_{i,j,k} $ are obtained by means of a smoothing phase based on tensor product bases of cubic splines, with second derivative penalty as marginal smooths. The marginal basis dimensions, set equal to 30, and the smoothing parameters were chosen by using restricted maximum likelihood (REML) \citep{wood2017generalized}. The smoothing phase was performed by using the R package \texttt{mgcv} \citep{wood2017generalized}. 
Then, in order to reduce phase variability, a registration phase was performed \citep{ramsay2005functional}. It consists in the shifting of each $ Y_{i,j,k}$ along the $ s $ and $ t $ axes to minimize the $ L^2 $ distance with respect to the reference curve, which was chosen such that the mean of pairwise distances among the aligned curves is minimum.
The functional observations $ Y_{i,j,k}$, $ i=1\dots,6 $, $ j=1,\dots 6 $ and $ k=1,\dots,n_{ij} $,  for $ t=0.75 $ and  $ s=0.5 $ are represented in Fig. \ref{fig_X} at different energy density levels and in different layers.
The graphical representation of cross-sections of the spatter intensity function in Fig. \ref{fig_X} was adopted to aid the superimposition and direct comparison of functional patterns corresponding to different experimental treatments. 
% A global view of the original spatter intensity functions in the bi-dimensional domain is provided in Fig. \ref{fig_mat}.

%REFERENCE DA AGGIUNGERE
%Mani, M., Feng, S., Lane, B., Donmez, A., Moylan, S., & Fesperman, R. (2015). Measurement science needs for real-time control of additive manufacturing powder bed fusion processes. 
%Anwar, A. B., & Pham, Q. C. (2018). Study of the spatter distribution on the powder bed during selective laser melting. Additive Manufacturing, 22, 86-97. 

\subsection{Results}
\label{sec_cas_res}
The spatter intensity functions $ Y_{i,j,k}$ ($ i=1\dots,6 $, $ j=1,\dots 6 $ and $ k=1,\dots,n_{ij} $) are modeled according to  equation \eqref{eq_modanova}, where
 $ f_i $ is the energy density functional effect, $ g_i $ is the layer functional effect, $ h_{ij} $ is the interaction term between the energy density and the layer. 
The equivariant functional M-estimators (Section \ref{subsec_mestimator})  are shown in the Supplementary Material.

\begin{figure}
	\centering
	\begin{tabular}{cc}
		\includegraphics[width=0.3\textwidth]{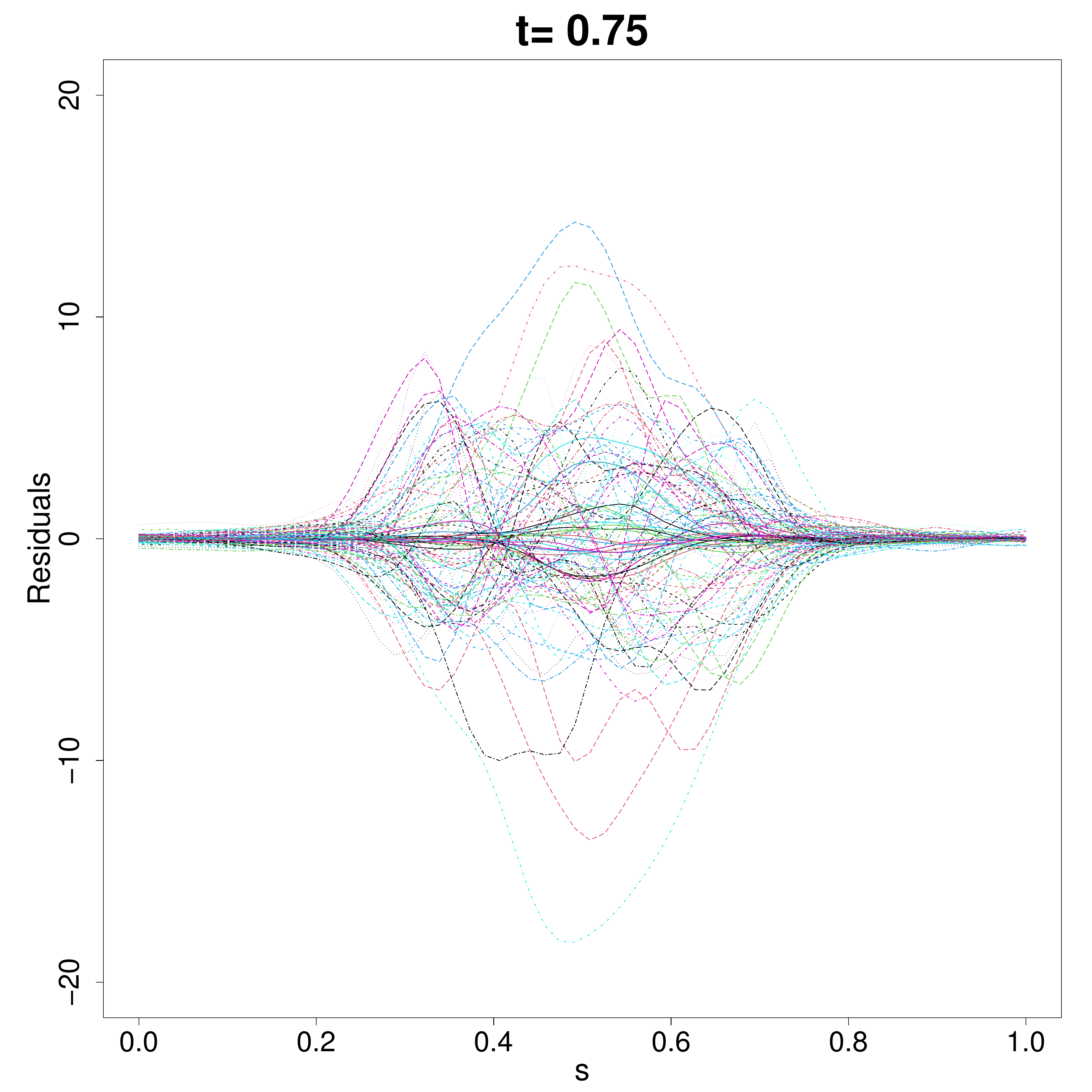} & \includegraphics[width=.3\textwidth]{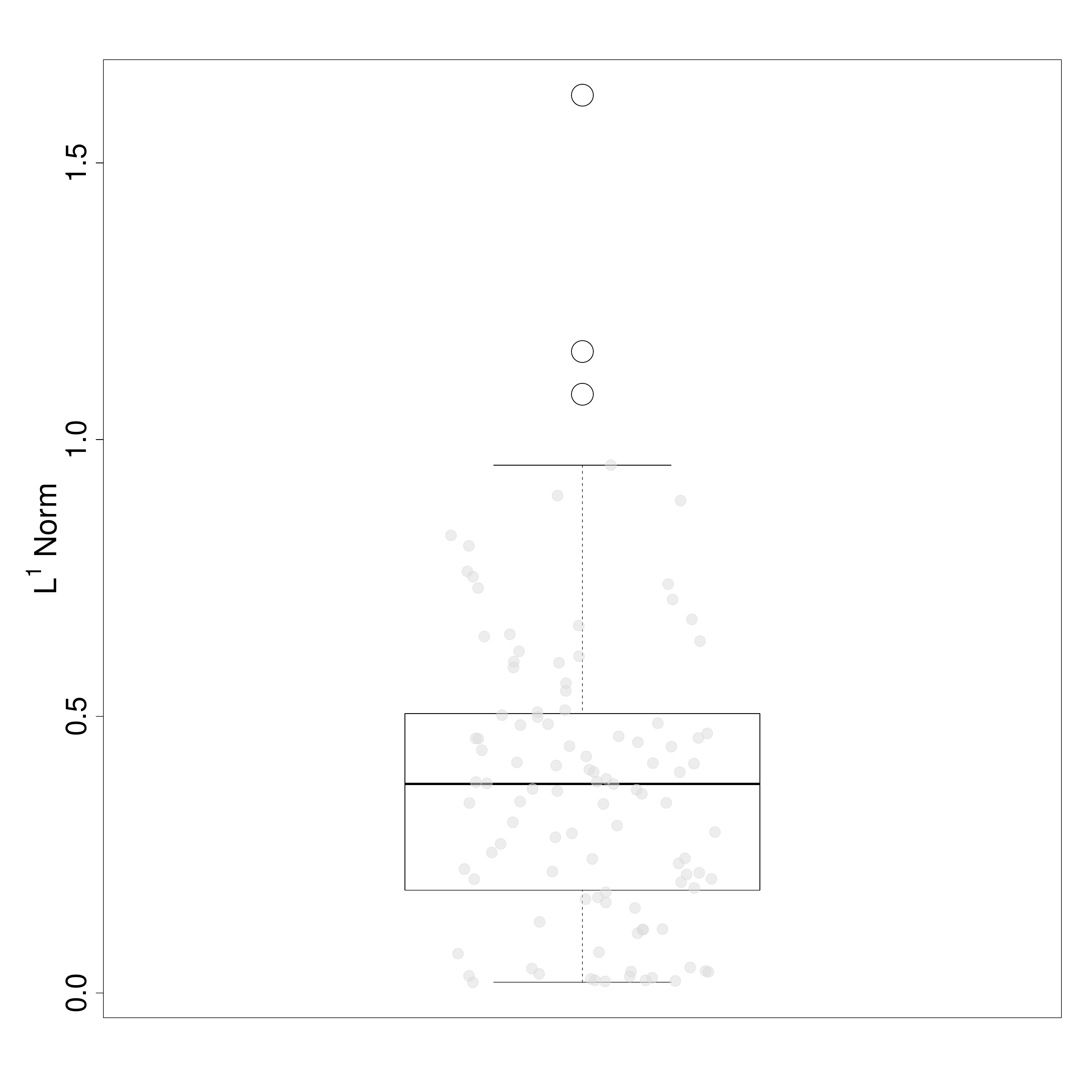} \\
		(a) & (b)\\
	\end{tabular}
	\caption{\label{fig_res}(a) Residuals of the fitted model for $t=0.75$, obtained by using the RoFANOVA-BIS test as implemented in Section \ref{sec_sim}, and (b)  boxplot of their $L^1$ norms.}
\end{figure}

The aim of the analysis is therefore to test the energy density effect $ H_{0,Flu}=H_{0,A} $ \eqref{eq_H0A}, the layer effect $ H_{0,Lay}=H_{0,B} $ \eqref{eq_H0B} (mainly related to the layer by layer variation of the laser scan direction) and their interaction effect $ H_{0,FluLay}=H_{0,AB} $ against the alternatives $ H_{1,Flu}=H_{1,A} $ \eqref{eq_H0A},  $ H_{1,Lay}=H_{1,B} $ \eqref{eq_H0B} and  $ H_{1,FluLay}=H_{1,AB} $.
In particular, Fig. \ref{fig_res}  shows (a) the residuals of the fitted model for $t=0.75$ (the approximate $t$ value of the spatter intensity peak), obtained by using the RoFANOVA-BIS test as implemented in Section \ref{sec_sim}, and (b) the boxplot of their $L^1$ norms, defined as $ ||f||_1=\int_{\mathcal{T}}|f\left(t\right)| dt $, for $f\in L^2\left(\mathcal{T}\right)$. Because $\mathcal{T}=\left[0,1\right]\times \left[0,1\right] $, the $L^1$ norm can be interpreted  as the average value of the function over its domain.
It is clear from Fig. \ref{fig_res} that some outliers are present in this real-case study. However, except from a few residuals that plot far from the bulk of the data, there are some points that could not be easily labeled as outliers. As mentioned in the introduction, the L-PBF process is characterized by complex dynamics with many transient and local phenomena that not only affect the natural variability of the measured quantities, but could lead also to outlying patterns. Determining whether an experimental point is an outlier or not,  and identifying its root causes can be a difficult task, which makes the diagnostic approach hardly applicable in the absence of additional data and information.

 Therefore, we applied the RoFANOVA test described in Section \ref{sec_sim}, viz, RoFANOVA-MED, RoFANOVA-HUB,  RoFANOVA-BIS,  RoFANOVA-HAM, and  RoFANOVA-OPT, specifically adapted for bi-dimensional functional data. As in the Monte Carlo simulation study (Section \ref{sec_sim}), the tuning constants are chosen such that the $ 95\%$ asymptotic efficiency is achieved, the number of permutations $ B $ are set equal to $ 1000 $. The functional sample mean is used as starting value to compute the robust equivariant functional $ M $-estimators (Section \ref{subsec_mestimator}).
The results are shown in Table \ref{tab_pvalue}. All the tests agree in considering significant the interaction between the energy density and the layer.
\begin{table}
	\caption{\label{tab_pvalue}p-values of the RoFANOVA tests for $ H_{0,Flu} $, $ H_{0,Lay} $ and $ H_{0,FluLay}$ against $ H_{1,Flu} $, $ H_{1,Lay} $ and $ H_{1,FluLay}$.}
	
	\centering
	\scriptsize
		\resizebox{1\textwidth}{!}{
	\begin{tabular}{cccccccc}\hline
&RoFANOVA-MED & RoFANOVA-HUB & RoFANOVA-BIS&RoFANOVA-HAM&RoFANOVA-OPT\\\hline
$ H_{0,FluLay}$&0.00&0.01 &0.00&0.00&0.00\\
$ H_{0,Flu}$&0.00&0.00&0.00&0.00&0.00\\
$ H_{0,Lay}$&0.00&0.00&0.00&0.00&0.00\\	\hline
	\end{tabular}
}
\end{table}

\begin{figure}
	\centering
	\includegraphics[width=0.3\textwidth]{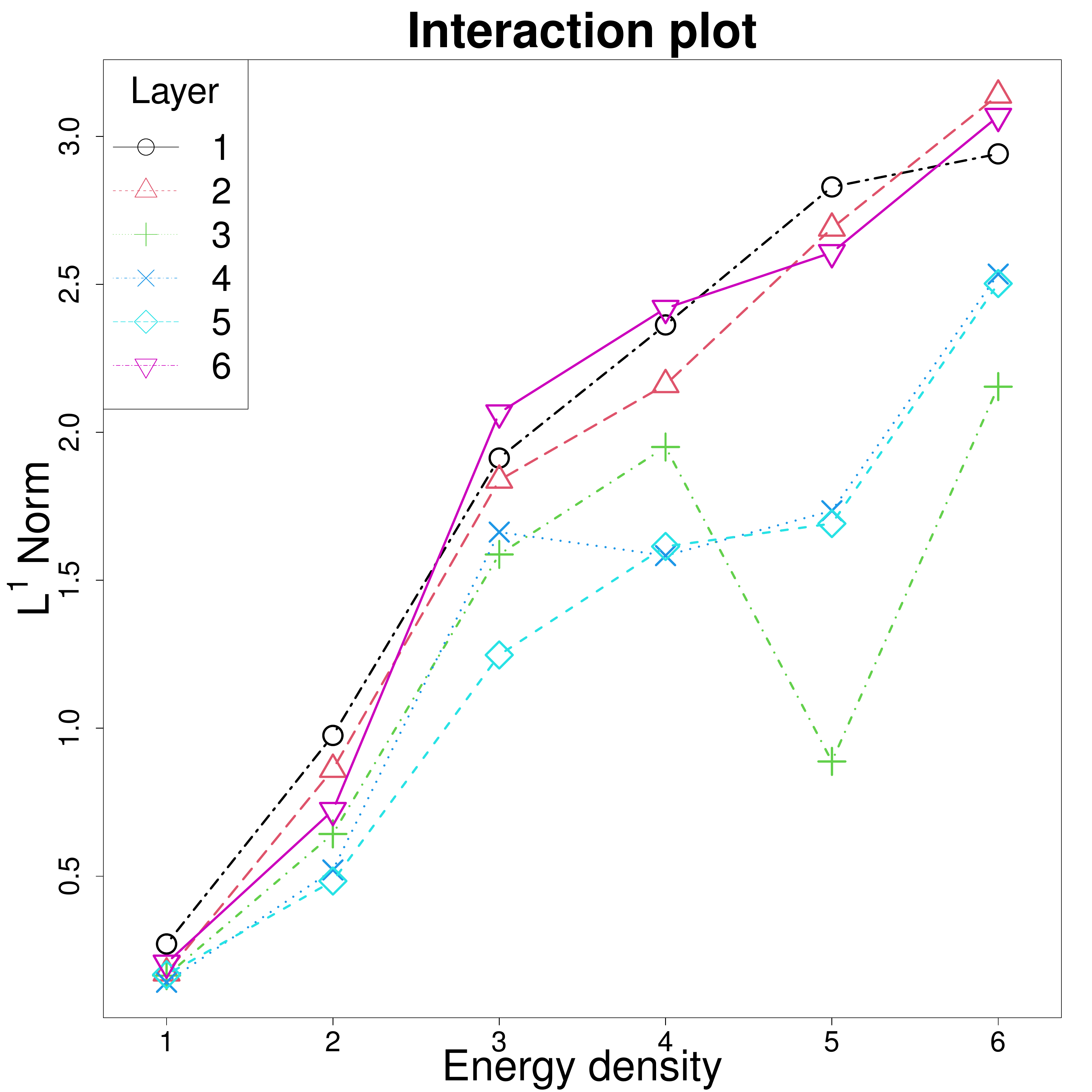} 
	\caption{\label{fig_int}Interaction plot as a function of  the energy density level and the layer in the real-case study. }
\end{figure}

When an the interaction effect is present, it is well-known that an  interpretation of the main effects becomes less straightforward than if the interaction is not significant \citep{miller1997beyond}, because the layer effect upon the spatter intensity will differ depending on the energy density level.
In this case, the best way to interpret the results is through the interaction plot \citep{montgomery2017design}, which graphically represents the response means at different factor levels.
Fig. \ref{fig_int} shows an interaction plot adapted to deal with bi-dimensional data. In particular, the $ L^1 $ norms of the group means, corresponding to the RoFANOVA-BIS test, are plotted as a function of the energy density level and the layer. In this case, if an interaction is present,  the trace of the average response across the levels of one factor, which is plotted separately for each level of the other factor, will not be parallel \citep{montgomery2017design}. 
Fig. \ref{fig_int} shows that, as the energy density increases, the spatter intensity tends to increase as well. This is in agreement with the fact that a higher energy density generates a larger and hotter melt pool with more intense convective and recoil motions, which translates into a more intense spatter ejection \citep{yang2020monitoring,repossini2017use,bidare2018fluid}. More interestingly, Fig. \ref{fig_int} shows different patterns corresponding to different layers. Indeed, in layers 1, 2, and 6,  the spatter intensity is increasing with respect to the energy density. These three levels were characterized by very similar laser scan directions, with a low angle relative to the shielding gas flow (between $10^{\circ}$ and $40^{\circ}$). When the scan direction is parallel (or little angled) to the gas flow, more powder bed particles are pushed along the laser path and  increase the occurrence of particles  heated up by the hot metal vapour emission and  ejected as hot spatters. Under these conditions, increasing the energy density increases the intensity of convective motions that entrap the powder particles into the hot vapour emission and hence the spatter intensity \citep{bidare2018fluid}.

A different influence of the energy density on the spatter intensity was observed in layers 3, 4 and 5. In these layers, the laser scan direction was almost perpendicular to the shielding gas flow direction, i.e., with angles in the range $80^{\circ}$ to $90^{\circ}$. Under these conditions, particles are dragged away from the scan path, and reduce the amount of powder particles ejected as hot spatters, and hence,  the overall spatter intensity  \citep{bidare2018fluid}. In addition, the analysis reveals that, when the laser scan direction was about perpendicular to the gas flow, there was a range of intermediate energy densities (from level 3 to level 5) at which the influence of the energy density itself on the spatter intensity  reduced or even inverted. This can be interpreted as follows. Conversely, when the laser scan direction is parallel to the gas flow, an increase of the energy density causes an increase of convective motions and metal vapour emissions that result also in higher spatter intensity. When the laser scan direction is perpendicular to the gas flow, an increase of the energy density still causes an increase of convective motions and metal vapour emissions, but such vapour emission has little effect on the spatter intensity, which makes the influence of the energy density mainly evident at very low or very high energy density levels only. Such interaction between the energy density and  laser scan direction on the spatter intensity was explored in a very few studies in the literature. Nevertheless, it is particularly relevant to understand the underlying spatter behaviour and to design either process optimization or process monitoring tools that rely on the in-line observation of such ejected particles.  
Finally,  we cannot confidently affirm  that  spatter intensity is affected by  layer (i.e., by  laser scan direction that changes layer by layer), because we cannot distinguish  if  differences among layers are due to  interactions, only, or to a systematic laser scan direction effect too.

Even if the use of  RoFANOVA  is recommended in light of the results shown by the Monte Carlo simulation study (Section \ref{sec_sim}), for the sake of completeness,  the bi-dimensional version of the FNDP and TGPF test have been applied. For the latter, the Manly's scheme \citep{manly2006randomization} with 1000 random permutations was used to approximate the test statistic distribution.
By comparing the additional results, which are shown in Table \ref{tab_pvalue2}, with  the proposed tests (Table \ref{tab_pvalue}) we note that they disagree in considering as significant the interaction between  energy density and  layer. In particular, the FNDP and the TGPF tests do not reject the null hypothesis of no interaction (i.e., large p-values).
This may suggest, in accordance with the Monte Carlo simulation results achieved in the two-way  FANOVA design case (Section \ref{sec_twoway}),  that FNDP and  TGPF tests may have not enough statistical power  to detect a technologically relevant interaction among the main factors. 
\begin{table}
	\caption{\label{tab_pvalue2}p-values of the FNDP and TGPF  tests for $ H_{0,Flu} $, $ H_{0,Lay} $ and $ H_{0,FluLay}$ against $ H_{1,Flu} $, $ H_{1,Lay} $ and $ H_{1,FluLay}$.}
	\centering
		\scriptsize
	\begin{tabular}{ccc}\hline
		&FNDP&TGPF\\\hline
		$ H_{0,FluLay}$&0.72&0.23\\
		$ H_{0,Flu}$&0.00&0.00\\
		$ H_{0,Lay}$&0.00&0.00\\	\hline
	\end{tabular}
\end{table}

\section{Conclusion}
\label{sec_con}
In this paper, we have proposed the  RoFANOVA test  for the functional analysis of variance problem. In particular, the proposed method has been designed to be robust against  functional outliers, which are increasingly common in complex  problems and, as it is well known, can severely bias the analyses. Robustness comes from the use of robust test statistics based on the functional equivariant $ M $-estimator and the functional normalized median absolute deviation, which are the  extensions of the classical $ M $-estimator and normalized median absolute deviation to functional data.
The test statistic is, then, incorporated in a  permutation test, in order to solve the FANOVA problem in a nonparametric fashion.
The proposed approach is demonstrated to be flexible to  different choices of  the loss function, and, to be applicable to both one-dimensional and bi-dimensional functional data.
To the best of the authors' knowledge, this is the first example  of a robust method for the FANOVA problem that is specifically designed to reduce the abnormal observation weights in the computation of the test statistic in comparisons with the standard least-squares loss function appeared in the literature, where attention has been mainly focused on non-robust methods.

The performance of the proposed method has been investigated by means of an extensive Monte Carlo simulation study, where the proposed RoFANOVA  have been compared with other methods already present in the literature. The results have shown that the proposed tests clearly outperform  the competitors in terms of both empirical size and empirical power when outlier contamination is present. Moreover, even in case of  no outlier contamination the loss of   power of the RoFANOVA tests with respect to  competitors is negligible.

The proposed method was applied to a motivating real-case study in the field of additive manufacturing.
% The aim of the analysis was to study the effects of the energy density and laser scan direction on the bi-dimensional distribution of the spatter intensity, i.e., a function mapping the amount of spatters ejected within the bi-dimensional field of view of the machine vision system used for in-line measurements.
Apart from the known influence of the energy density on the spatter intensity, in agreement with previous studies, the RoFANOVA test revealed a statistically significant interaction between the energy density and the laser scan direction relative to the shielding gas flow.
% The knowledge of such interaction is of particular importance to interpret and understand the underlying process dynamics and to design methods that use spatters as proxies of the process quality and stability. 
The statistical significance of the interaction between these two factors was not identified by the other non-robust tests, which confirms the effectiveness of the proposed approach to applications where complex process dynamics may lead to outlying patterns that contaminate the experimental dataset. The validity of the proposed approach is naturally not limited to the case study here presented and, in general, to manufacturing applications.

In future research, the effects of heteroscedasticity on the  RoFANOVA test should be investigated in order to be able to deal with a wider variety of settings. In addition, some efforts should be made to extend the proposed  method to more complex FANOVA designs.

\bibliographystyle{chicago}
\bibliography{References}
\end{document}

%%%%%%%%%% Merge with supplemental materials %%%%%%%%%%
%%%%%%%%%% Prefix a "S" to all equations, figures, tables and reset the counter %%%%%%%%%%
\setcounter{equation}{0}
\setcounter{figure}{0}
\setcounter{table}{0}
\setcounter{page}{1}
\makeatletter
\renewcommand{\theequation}{S\arabic{equation}}
\renewcommand{\thefigure}{S\arabic{figure}}

%%%%%%%%%% Prefix a "S" to all equations, figures, tables and reset the counter %%%%%%%%%%

\section{Derivation of the constant $c$ in the $ \FuNMAD $ expression}
\label{app_1}
Following Theorem 3.4 of \cite{sinova2018m}, $\hat{\mu}_{s,med}$ is a strongly consistent estimator of $\tilde{\mu}_{s,med}=\argmin_{y\in L^2\left(\mathcal{T}\right) }\Ex\left[||X_i-y||\right]$.
If we assume  $X$ is a Gaussian random process, then, by Proposition 3.2 of \cite{sinova2018m}, it follows that $\tilde{\mu}_{s,med}=\mu$, where $\mu$ is the mean function of the random element $X$.
So stated, from the population version of Equation \eqref{eq_funmad} and  the definition of univariate population median, for each $t\in\mathcal{T}$ we have asymptotically
\begin{equation*}
0.5=Pr\left[|X\left(t\right)-\mu\left(t\right)|<c\FuNMAD\left(t\right)\right]=Pr\left[|Z|<c\frac{\FuNMAD\left(t\right)}{\sigma(t)}\right],
\end{equation*}
where $Z$ is a standard normal random variable.
Therefore, we  have
\begin{equation*}
\Phi\left(c\frac{\FuNMAD\left(t\right)}{\sigma(t)}\right)-\Phi\left(-c\frac{\FuNMAD\left(t\right)}{\sigma(t)}\right)=0.5,
\end{equation*}
where $\Phi$ is the cumulative distribution function of the standard normal distribution.
Noticing that
\begin{equation*}
\Phi\left(-c\frac{\FuNMAD\left(t\right)}{\sigma(t)}\right)=1-\Phi\left(c\frac{\FuNMAD\left(t\right)}{\sigma(t)}\right),
\end{equation*}
then
\begin{equation}
c\frac{\FuNMAD\left(t\right)}{\sigma(t)}=\Phi^{-1}\left(3/4\right)=0.6745,
\end{equation}
% where $\Phi^{-1}$ is the quantile function for the standard normal distribution.
and thus $c=0.6745$ makes  $\FuNMAD$  an asymptotically pointwise consistent estimator of $\sigma$.

\section{Details on Data Generation}
\begin{figure}
	\centering

		 \includegraphics[width=.3\textwidth]{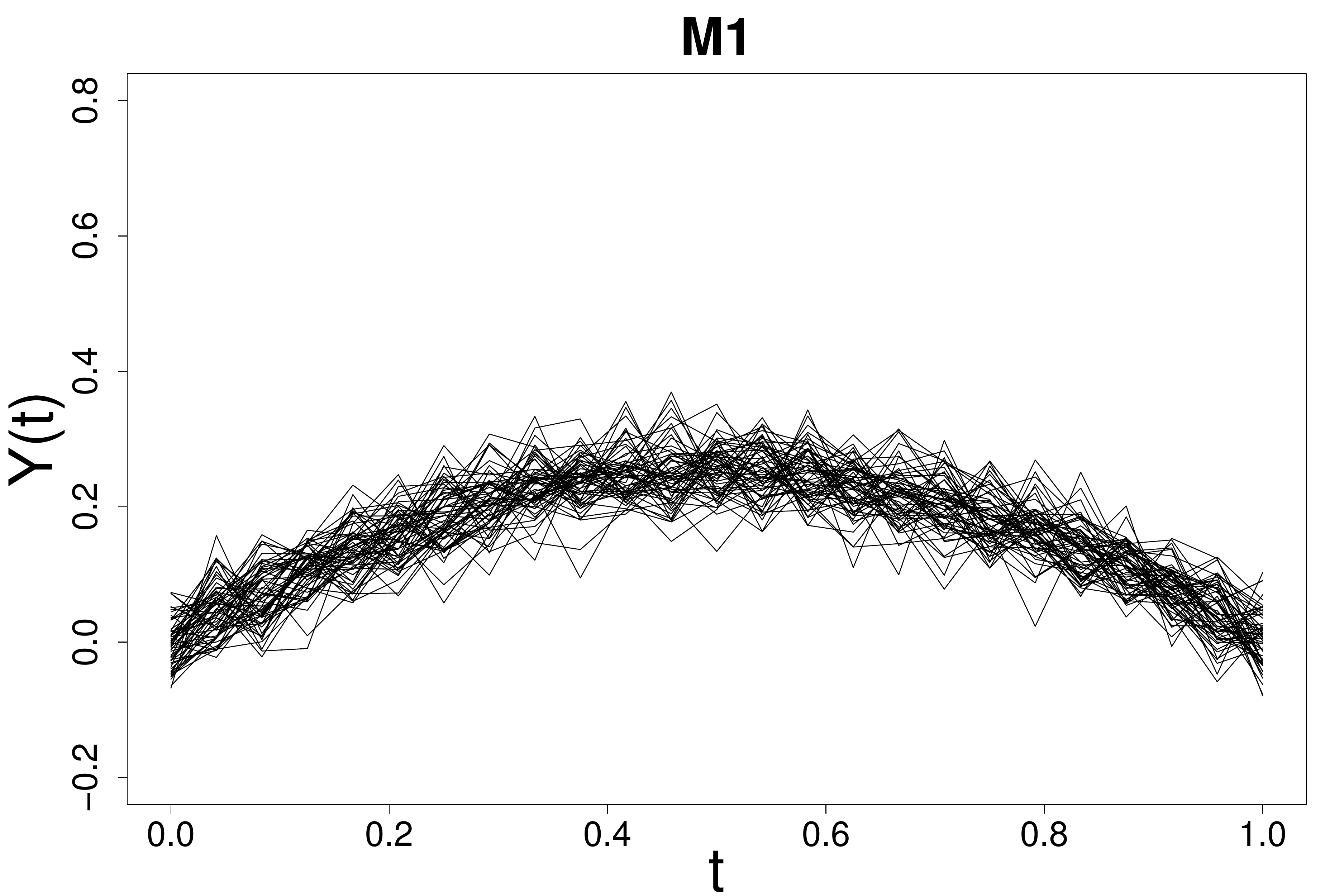} 
		  \includegraphics[width=.3\textwidth]{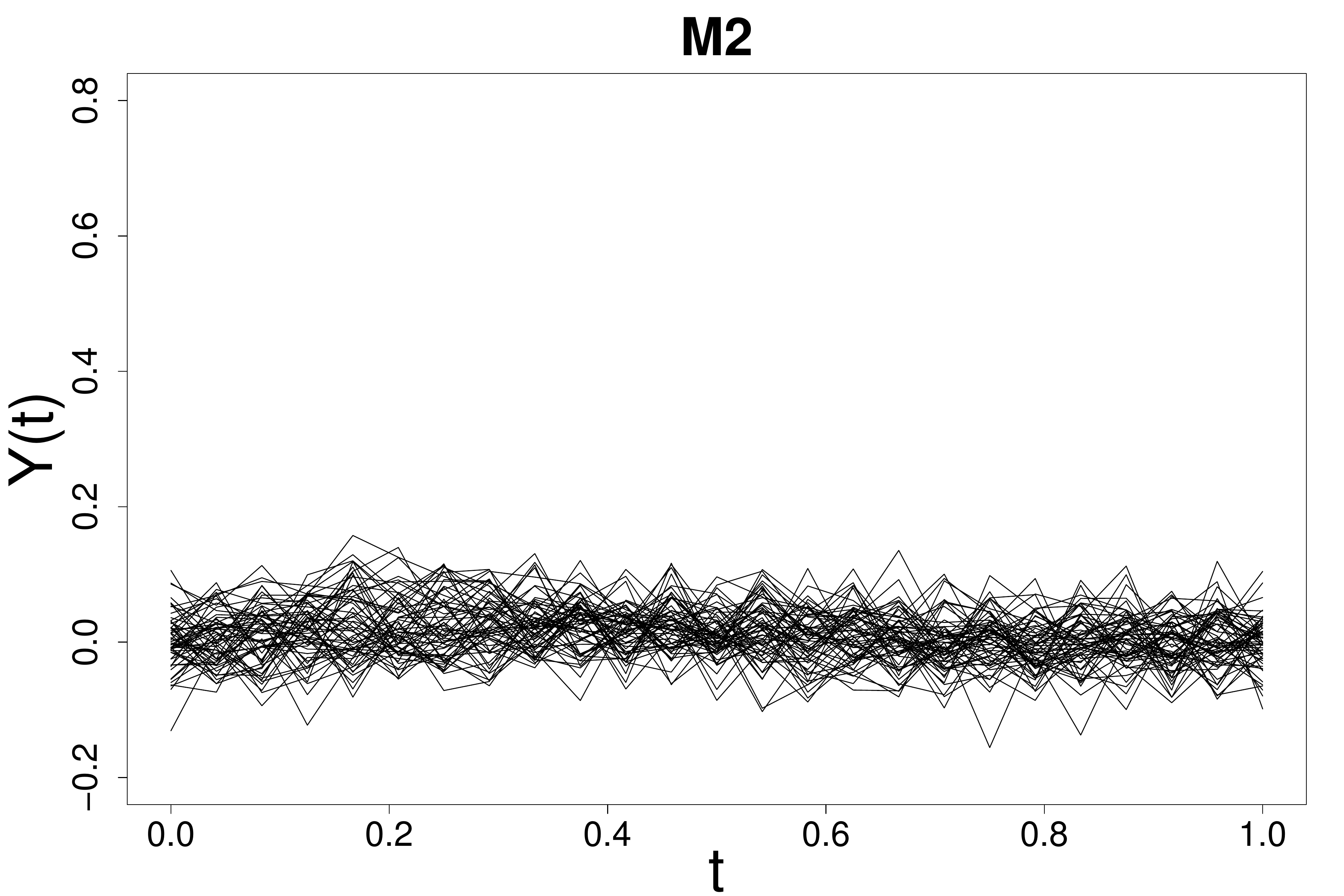} 
		   \includegraphics[width=.3\textwidth]{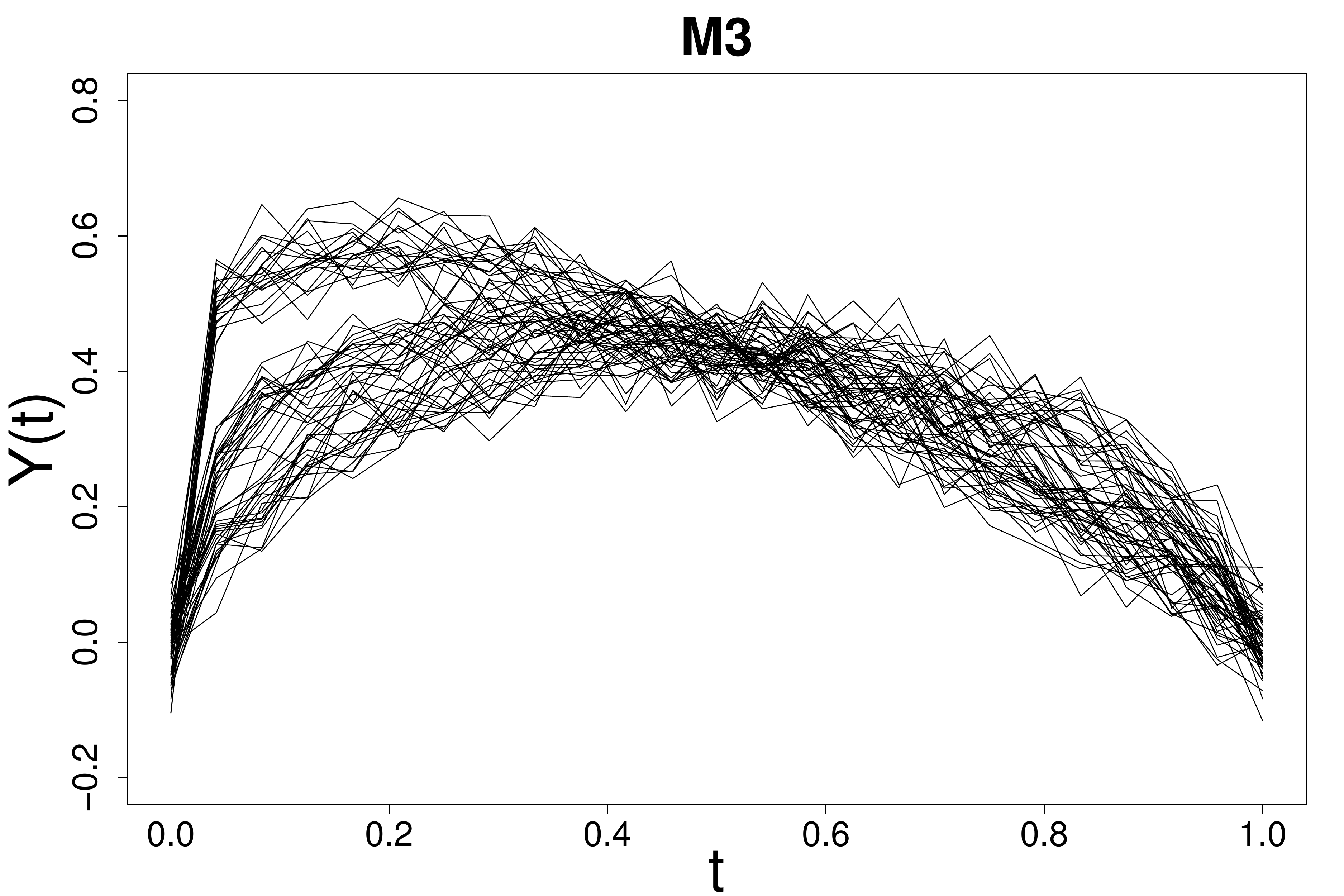}
	
	\caption{\label{fig_one_way ex}The  response curve   realizations for the three models M1, M2, and M3 in absence of  contamination (C0) for $\sigma=\sigma_{1}$ in Scenario 1.}
\end{figure}

In this section, the data generation process for Scenario 1 and Scenario 2 of the simulation study is described.
For Scenario 1, let $ \mathcal{T}=\left[0,1\right] $, then the three following different model with  3 level main effects $ f_i$ are considered
\begin{enumerate}[label=M\arabic*]
	\item $ f_i\left(t\right)= t\left(1-t\right) $ for $ t\in \left[0,1\right] $ and $ i=1,2,3 $,
	\item $ f_i\left(t\right)= t^i\left(1-t\right)^{6-i} $ for $ t\in \left[0,1\right] $ and $ i=1,2,3 $,
	\item $ f_i\left(t\right)= t^{i/5}\left(1-t\right)^{6-i/5} $ for $ t\in \left[0,1\right] $ and $ i=1,2,3 $.
\end{enumerate}
Model M1 corresponds to  $ H_0 $:  $ f_1=f_2=f_3$   true, whereas, M2 and M3 provide examples, with $ H_0 $ false, of monotone functions with different increasing patterns. In particular, M2 simulates $f_i$ differences that are smaller than M3, where $ f_i $ are quite separated.
To simulate different type of outlying curves, let $ B $ and $ U $ be two independent random variables following a Bernoulli (with parameter $ p $) and a discrete uniform (on $ \lbrace -1,1\rbrace $) distributions, respectively, and let $ T $ be  a random number generated from a uniform distribution on $ \left(0,0.75\right) $.
Then, the following four contamination models $ C_i $ are considered
\begin{enumerate}[label=C\arabic*]
	\setcounter{enumi}{-1} 
	\item $ C_i\left(t\right)= 0 $ for $ t\in \left[0,1\right] $ and $ i=1,2,3 $,
	\item $ C_i\left(t\right)=  BUM$ for $ t\in \left[0,1\right] $ and $ i=1,2,3 $,
	\item $ C_i\left(t\right)=  \begin{cases} 
	BUM & \text{if } t\geq T \\
	0 & \text{if }t< T ,
	\end{cases}$ for $ t\in \left[0,1\right] $ and $ i=1,2,3 $,
	\item $ C_i\left(t\right)=  \left(-1\right)^{i}BM$ for $ t\in \left[0,1\right] $ and $ i=1,2,3 $,
	\item $ C_i\left(t\right)=  \begin{cases} 
	\left(-1\right)^{i}BM & \text{if } t\geq T \\
	0 & \text{if }t< T ,
	\end{cases}$ for $ t\in \left[0,1\right] $ and $ i=1,2,3 $,
\end{enumerate}
with contamination size constant $ M =25 $ and $ p = 0.1 $.
The  model C0 is representative of  no contamination. C1-4 represents magnitude contaminations, i.e., curves are generated  far from the center, with, in particular, C1-2 (resp., C3-4) representing  symmetric and partial trajectory contamination models, that are independent (resp.,
dependent) from the level $i$ of the main effect.
Then, the curves $X_{ik}$ are generated, for $ i=1,2,3 $ and $ k=1,\dots,20 $, as
\begin{equation*}
X_{ik}\left(t\right)=  f_i\left(t\right)+ C_i\left(t\right)+\varepsilon_{ik}\left(t\right) \quad t\in \left[0,1\right],
\end{equation*}
where the errors  $ \varepsilon_{ik} $ are independent Gaussian processes with zero mean    and covariance function $ \gamma\left(s,t\right)=\sigma^2 e^{\left(-|s-t|10^{-5}\right)}$.
In what follows, we consider two shape contamination models \citep{lopez2009concept,sinova2018m}  that are both independent and dependent from the level $i$ of the main effect.
In this setting, the curves $X_{ik}$ are generated, for $ i=1,2,3 $ and $ k=1,\dots,20 $, as
\begin{equation*}
X_{ik}\left(t\right)=  (1-B)Y_{ik}\left(t\right)+ BZ_{ik}\left(t\right) \quad t\in \left[0,1\right],
\end{equation*}
with 
\begin{equation*}
Y_{ik}\left(t\right)=  f_i\left(t\right)+\varepsilon_{ik}\left(t\right),
\quad \quad
Z_{ik}\left(t\right)=  f_i\left(t\right)+\varepsilon_{ik,c}\left(t\right) \quad t\in \left[0,1\right],
\end{equation*}
where $ \varepsilon_{ik,c} $ are independent Gaussian processes with    zero mean and covariance function $ \gamma_{i,c}\left(s,t\right)=\sigma^2 e^{\left(-|s-t|k_{\gamma_{c},i}10^{-5}\right)}$.
The following choices for $ k_{\gamma_{c},i} $ are considered
\begin{enumerate}[label=C\arabic*]
	\setcounter{enumi}{4} 
	\item $ k_{\gamma_{c},i}=10^2 $ for $ i=1,2,3 $,
	\item $ k_{\gamma_{c},i}=10^{2+i} $ for $ i=1,2,3 $.
\end{enumerate}
In all the cases considered, the curves $ X_{ik} $ are observed  through $ 25 $ evenly spread discrete points and $ \sigma $ is equal to  $ \sigma_{1}=1/25 $, $ \sigma_{2}=1.8/25 $, $ \sigma_{3}=2.6/25 $, $ \sigma_{4}=3.4/25 $, $ \sigma_{5}=4.2/25 $, $ \sigma_{6}=5/25 $.
Fig. \ref{fig_one_way ex} shows the realizations of the response curve for the three models M1, M2, and M3 in presence of no contamination (C0) for $\sigma=\sigma_{1}$.

For Scenario 2, we assume  $ \mathcal{T}=\left[0,1\right] $ and the functional response depending on a grand mean $ m $, 2 level main effects $ f_i$ and $ g_i $, and on an interaction term $ h_{ij} $ through two parameters $ a $ and $ b $ as follows
\begin{itemize}
	\item $ m\left(t\right)=t(1-t) $ for $ t\in \left[0,1\right] $,
	\item $ f_i\left(t\right)=a(-1)^i |\sin(4\pi t)| $ for $ t\in \left[0,1\right] $ and $ i=1,2 $,
	\item $ g_j\left(t\right)=b(-1)^j I(t>0.5)$ for $ t\in \left[0,1\right] $ and $ j=1,2 $,
	\item $ h_{ij}\left(t\right)=-f_i\left(t\right)g_j\left(t\right) I(a\geq0.25)$ for $ t\in \left[0,1\right] $ and $ i=1,2 $, $ j=1,2 $,
\end{itemize}
with $ a,b\in \lbrace 0,0.05,0.10,0.25,0.50\rbrace  $.
For the contamination models C0-4 (Section \ref{sec_oneway}),  the curves $X_{ijk}$ are generated, for $ i=1,2 $, $ j=1,2 $ and $ k=1,\dots,20 $, as
\begin{equation*}
X_{ijk}\left(t\right)=  m\left(t\right)+f_i\left(t\right)+g_j\left(t\right)+h_{ij}\left(t\right)+C_i\left(t\right)+\varepsilon_{ijk}\left(t\right) \quad t\in \left[0,1\right],
\end{equation*}
where the errors  $ \varepsilon_{ijk} $ are independent Gaussian processes with mean   zero and covariance function $ \gamma\left(s,t\right)=\sigma^2 e^{\left(-|s-t|10^{-5}\right)}$.
The curves $ X_{ijk} $ are observed  through $ 25 $ evenly spread discrete points.

\begin{figure}
	\centering
	
	\includegraphics[width=.3\textwidth]{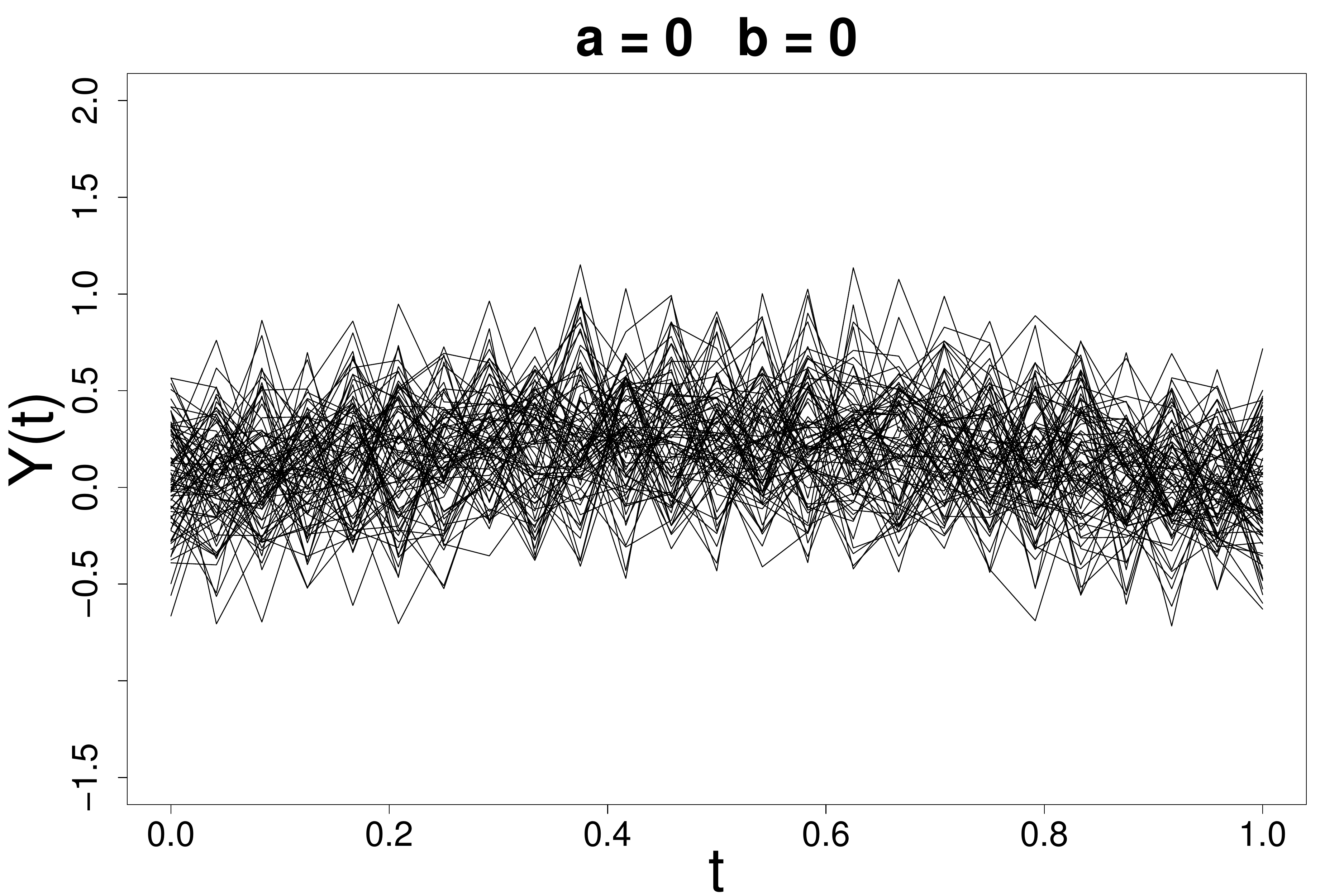} 
	\includegraphics[width=.3\textwidth]{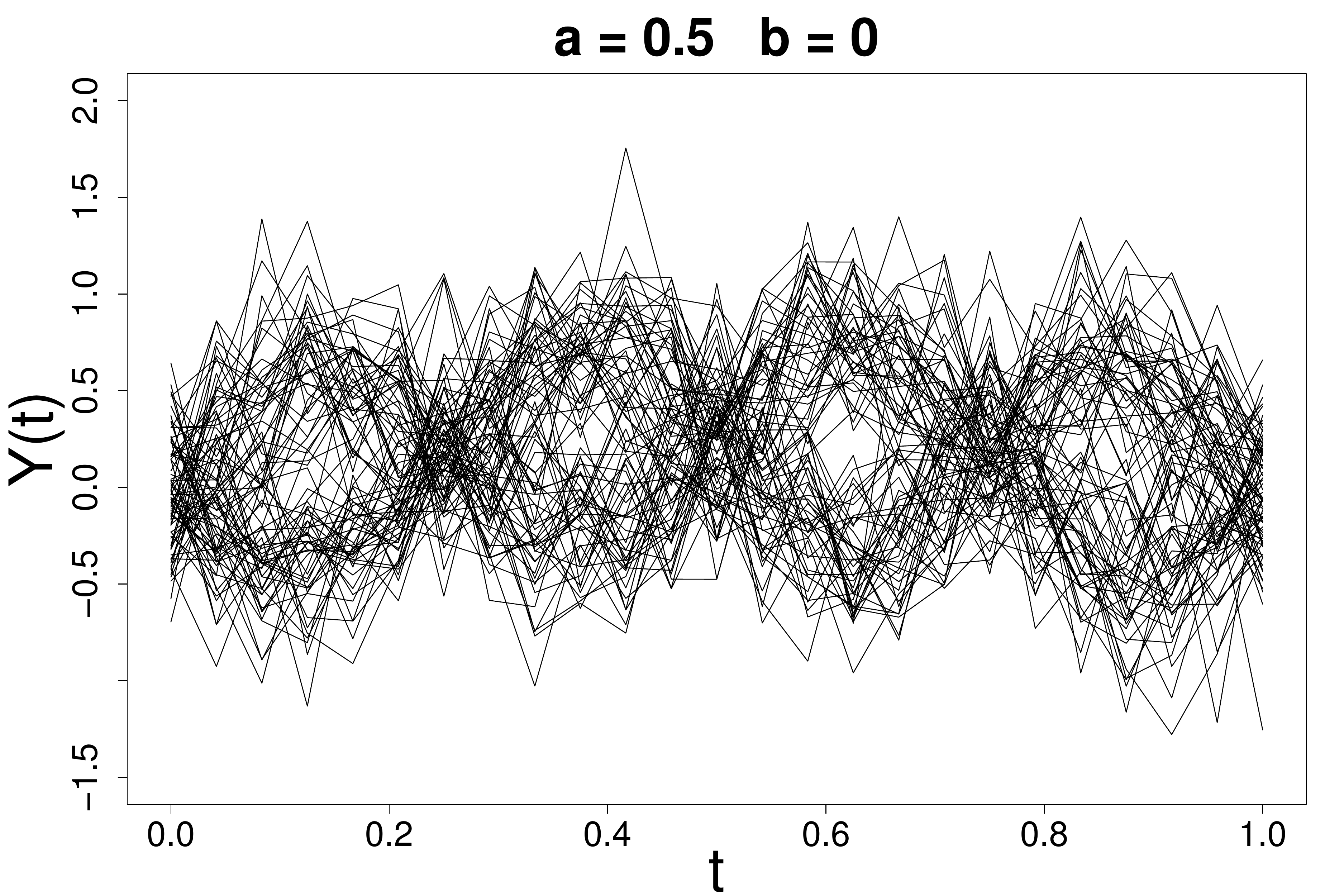} 
	\includegraphics[width=.3\textwidth]{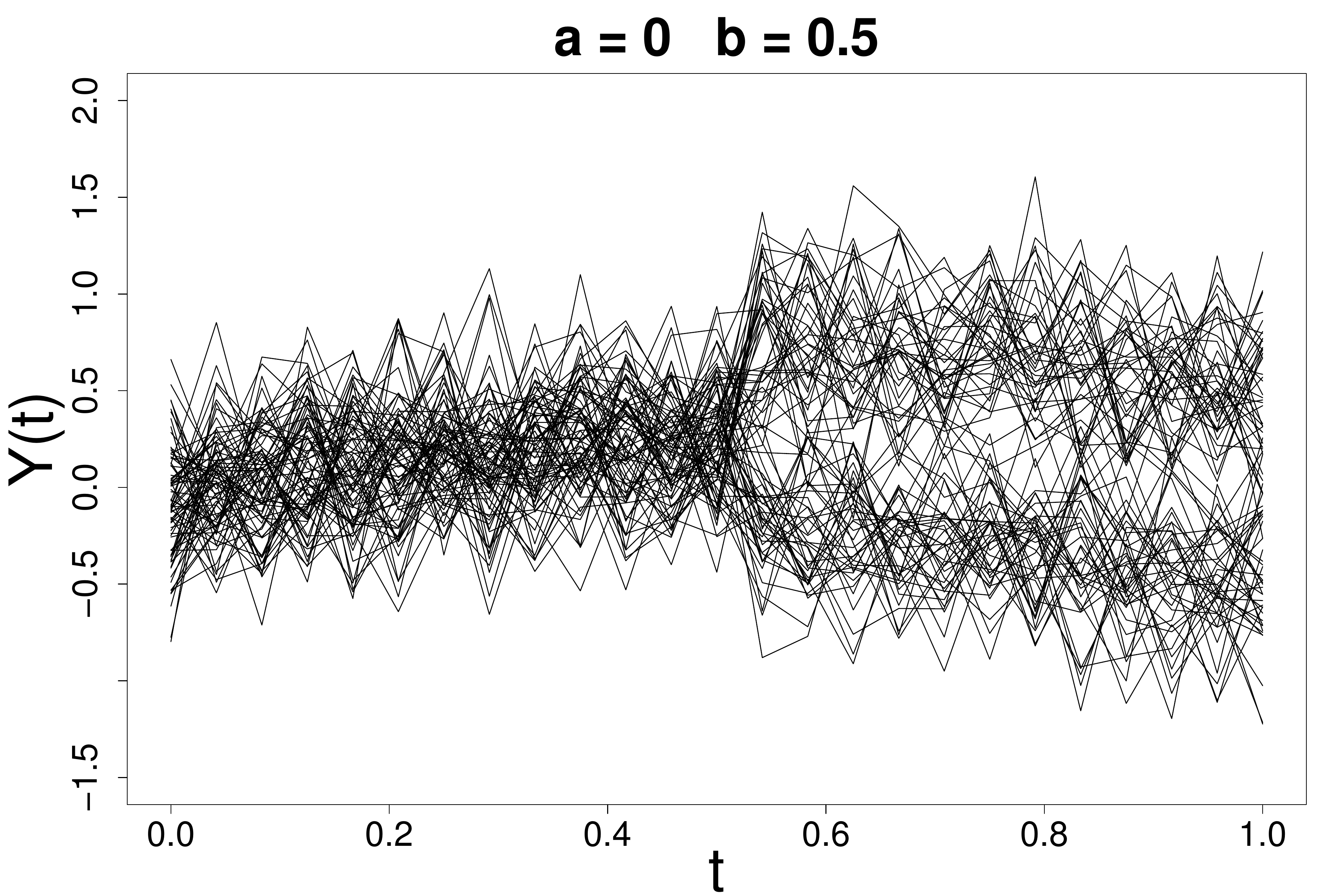}
	
	\caption{\label{fig_two_way_ex}The response curve  realizations for $a=b=0$, $a=0.5$ and $b=0$, and, $a=0$ and $b=0.5$ in presence of no contamination (C0)  in Scenario 2. }
\end{figure}

Whereas, for the contamination models C5-6 (Section \ref{sec_oneway}),  the curves $X_{ijk}$ are generated, for $ i=1,2 $, $ j=1,2 $ and $ k=1,\dots,20 $, as
\begin{equation*}
X_{ijk}\left(t\right)=  (1-B)Y_{ijk}\left(t\right)+ BZ_{ijk}\left(t\right) \quad t\in \left[0,1\right],
\end{equation*}
with 
\begin{equation*}
Y_{ijk}\left(t\right)=  m\left(t\right)+f_i\left(t\right)+g_j\left(t\right)+h_{ij}\left(t\right)+\varepsilon_{ijk}\left(t\right),
\end{equation*}
\begin{equation*}
Z_{ijk}\left(t\right)=  m\left(t\right)+f_i\left(t\right)+g_j\left(t\right)+h_{ij}\left(t\right)+\varepsilon_{ijk,c}\left(t\right),
\end{equation*}
for $ t\in \left[0,1\right] $, where $ \varepsilon_{ijk,c} $ are independent Gaussian process with mean   zero and covariance function $ \gamma_{ij,c}\left(s,t\right)=\sigma^2 e^{\left(-|s-t|k_{\gamma_{c},i}10^{-5}\right)}$ with, as for Scenario 1, $ k_{\gamma_{c},i}=10^2 $ for C5 and $ k_{\gamma_{c},i}=10^{2+i} $ for C6. The random variable $ B $  follows a Bernoulli (with parameter $ p=0.1 $) distribution.
In this case, the curves $ X_{ik} $ are observed  through $ 25 $ evenly spread discrete points with $ \sigma=0.3 $.
Fig. \ref{fig_two_way_ex} shows the realizations of the response curve for $a=b=0$, $a=0.5$ and $b=0$, and, $a=0$ and $b=0.5$, in absence of  contamination (C0).

\section{Additional results in the simulation study}
\begin{figure}
	\centering
	\resizebox{1\textwidth}{!}{
		\begin{tabular}{M{0.5\textwidth}M{0.5\textwidth}M{0.5\textwidth}M{0.5\textwidth}}
			\multicolumn{4}{c}{\huge $ \bm{a\leq 0.10} $}\\[0.2cm]
			\multirow{2}{*}{\includegraphics[width=0.5\textwidth]{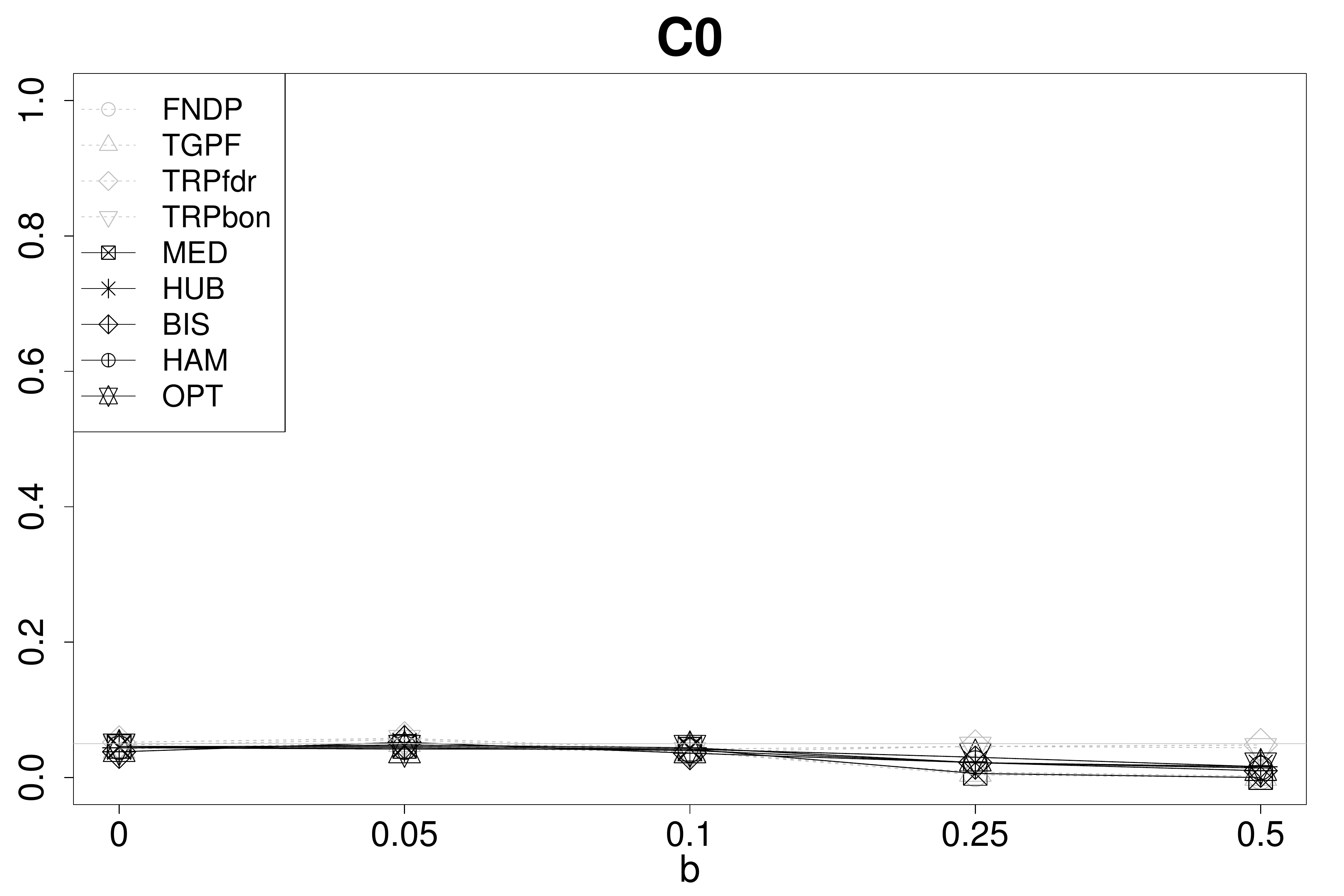}}&\includegraphics[width=0.5\textwidth]{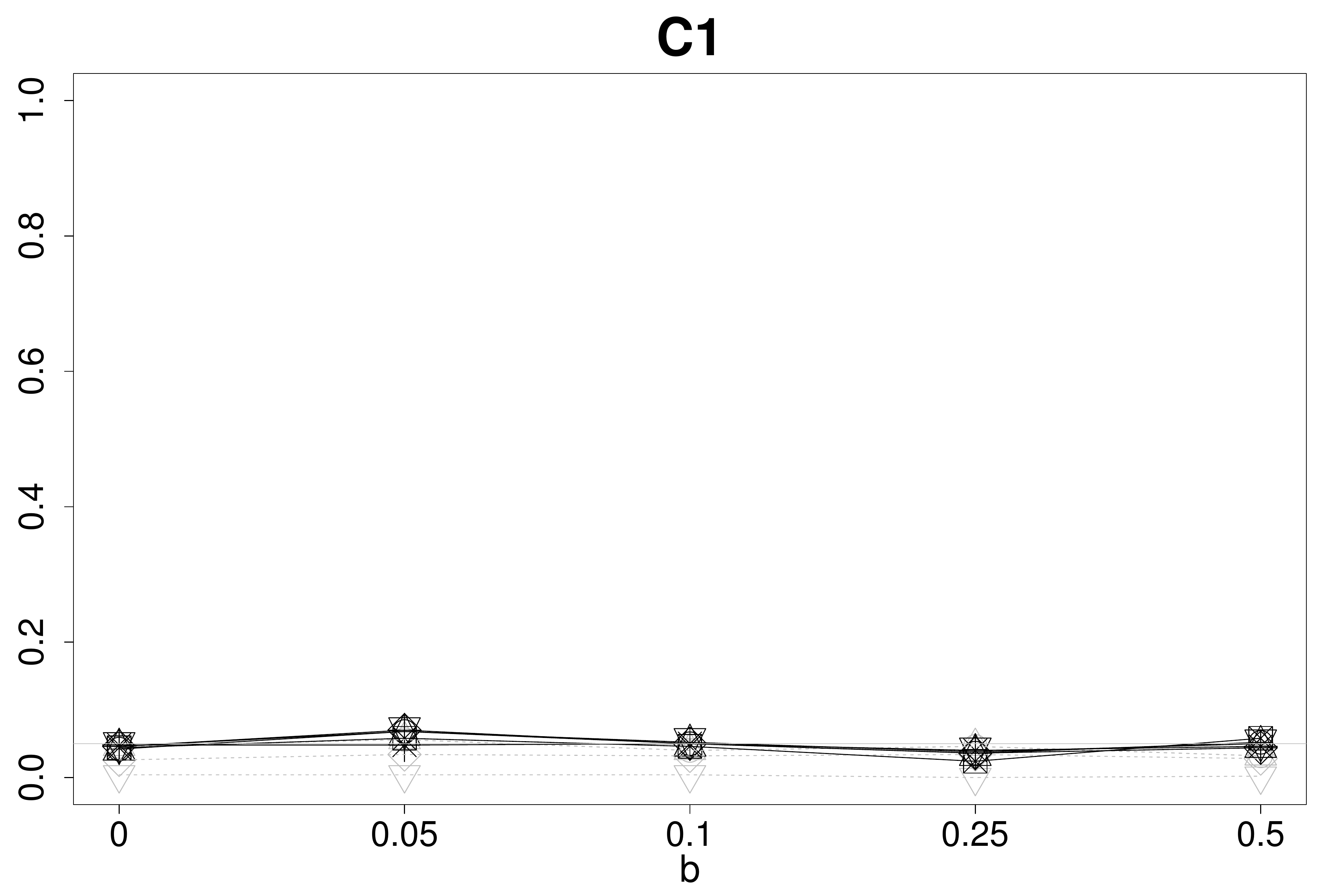}&\includegraphics[width=0.5\textwidth]{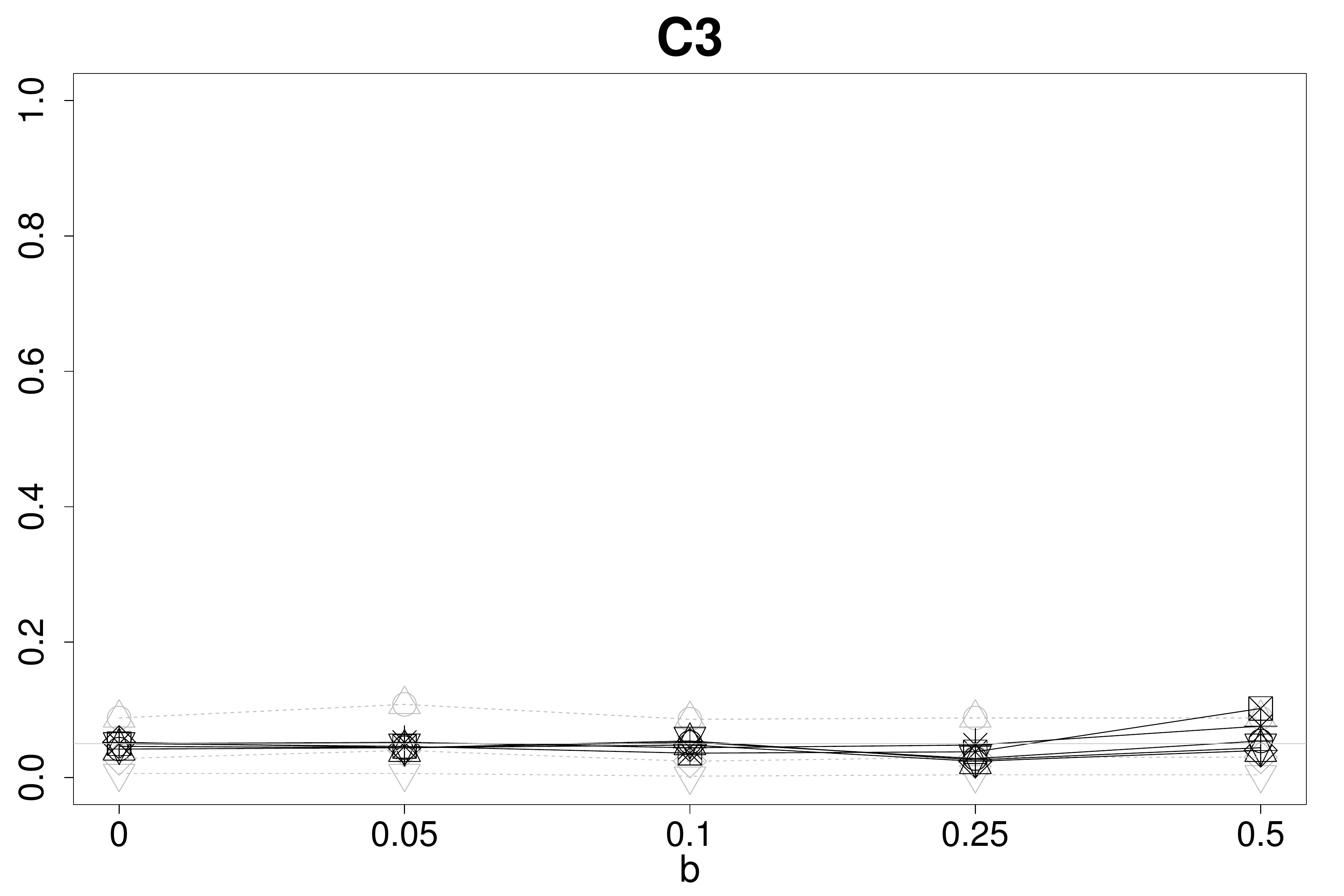}&\includegraphics[width=0.5\textwidth]{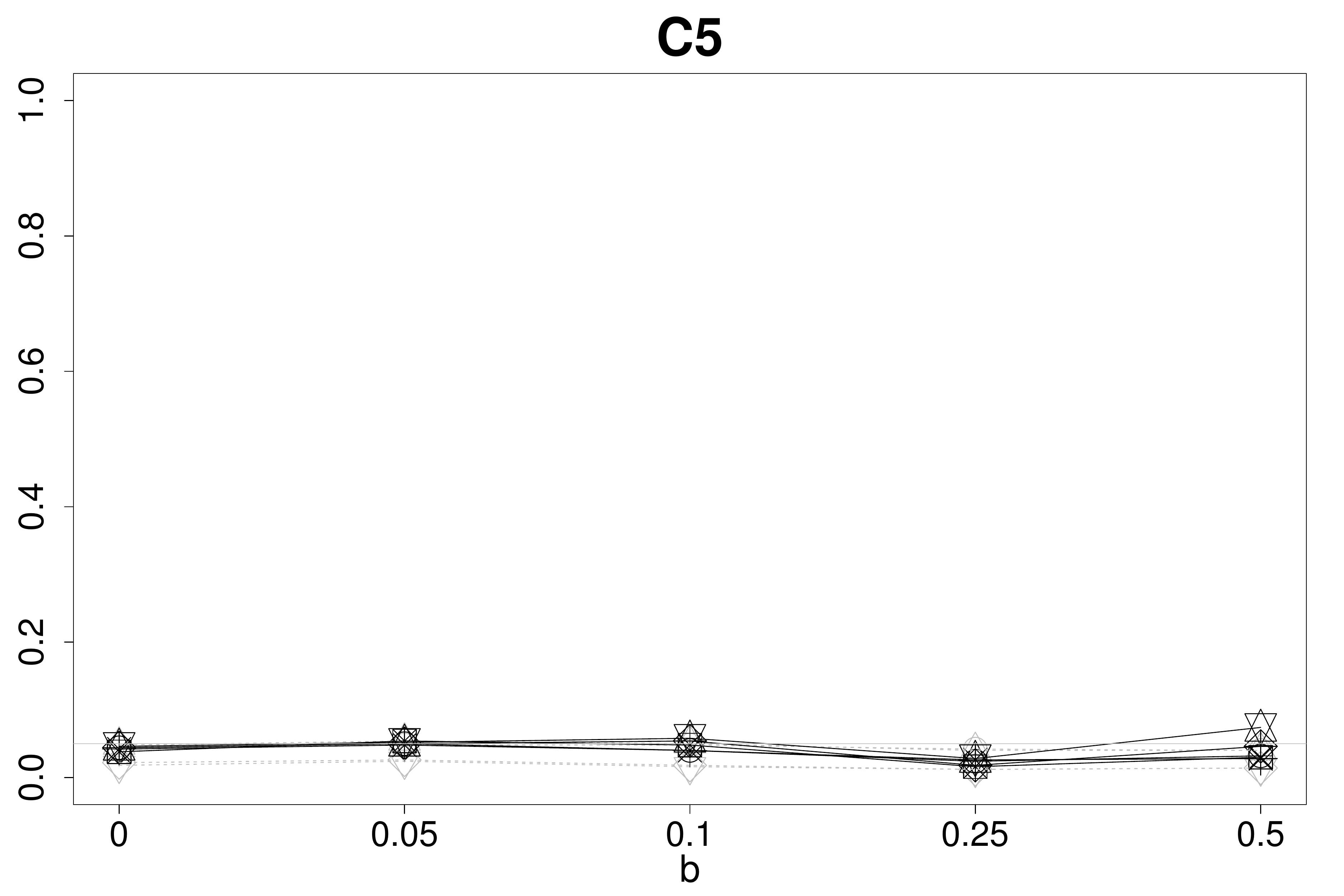}\\
			&\includegraphics[width=0.5\textwidth]{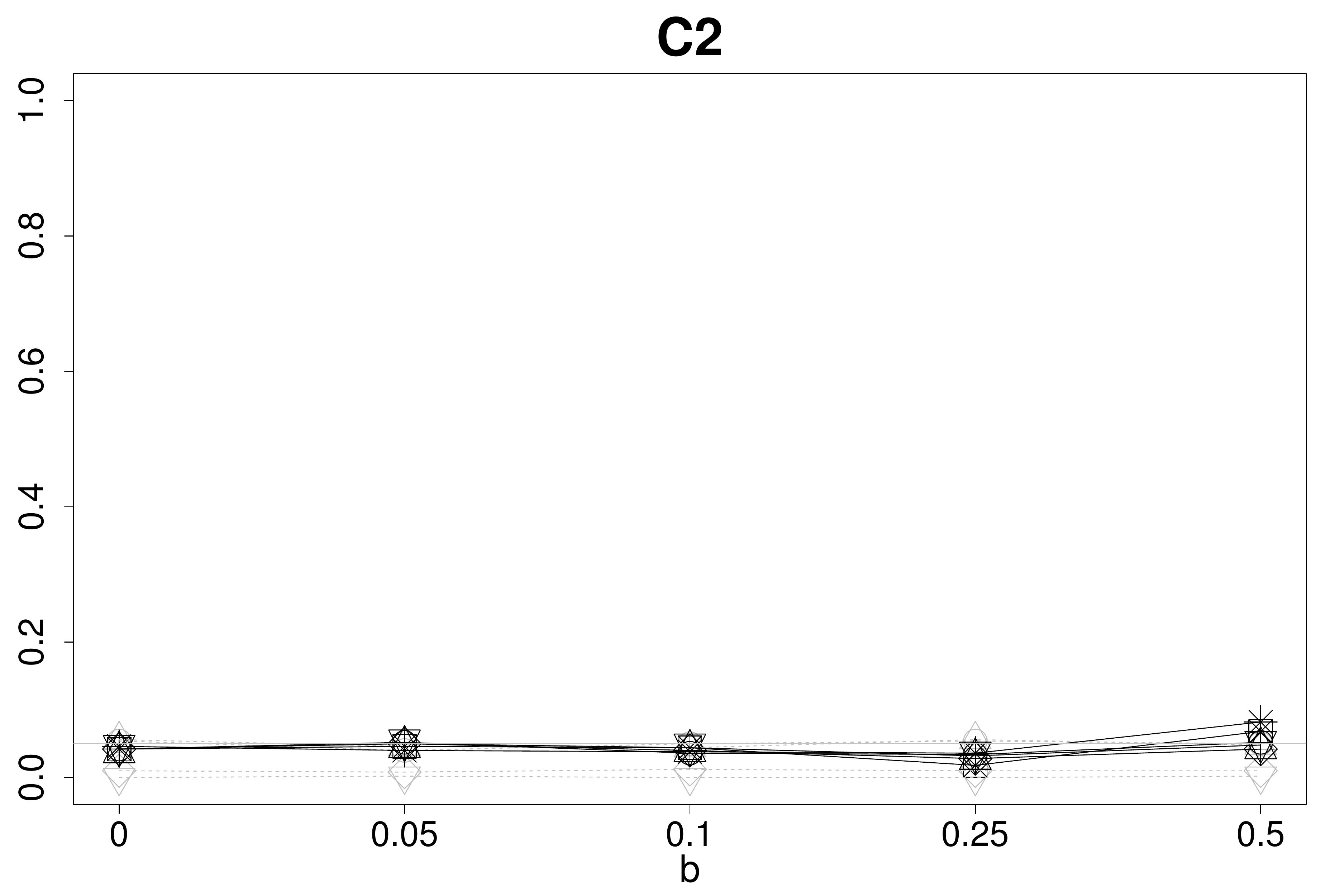}&\includegraphics[width=0.5\textwidth]{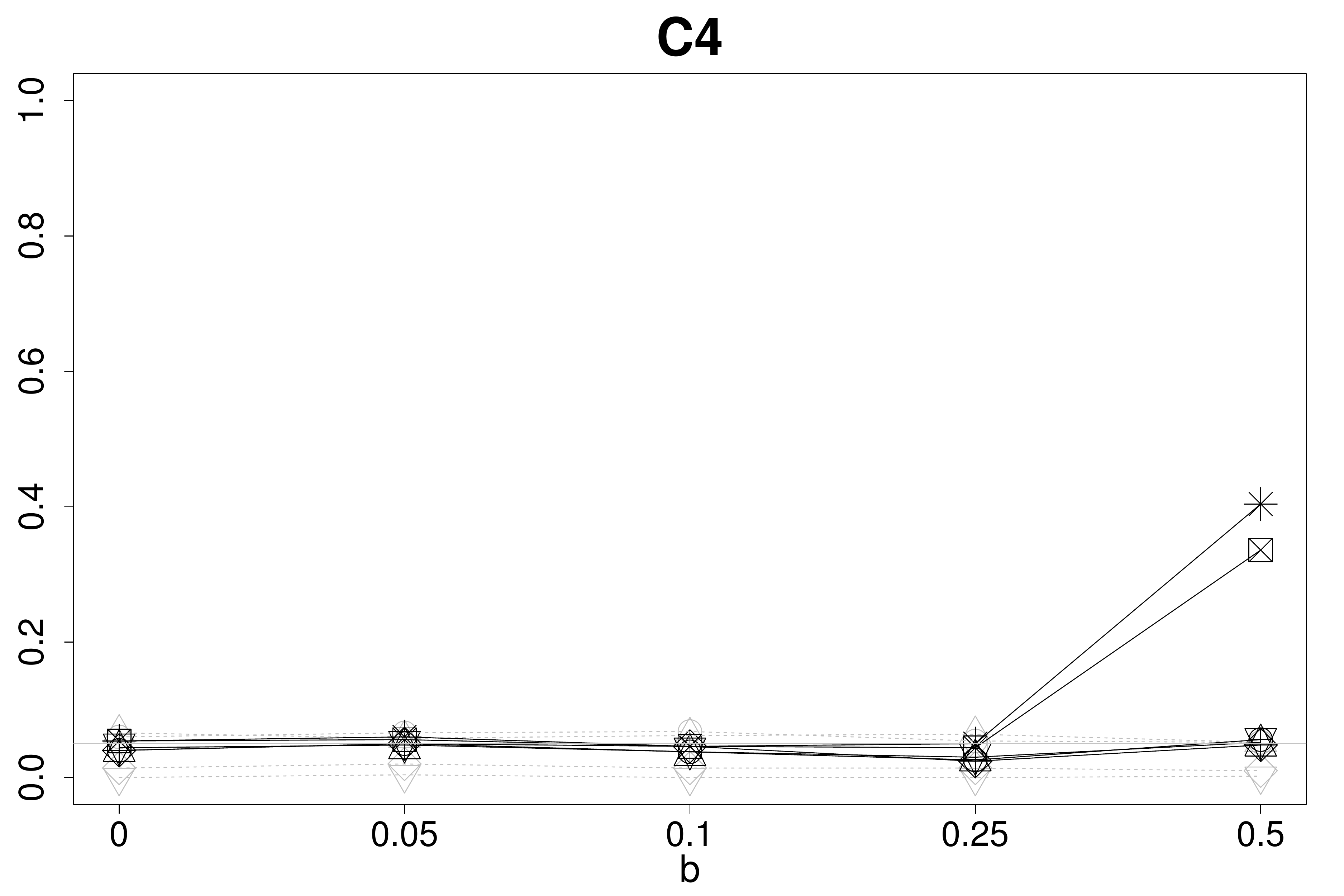}&\includegraphics[width=0.5\textwidth]{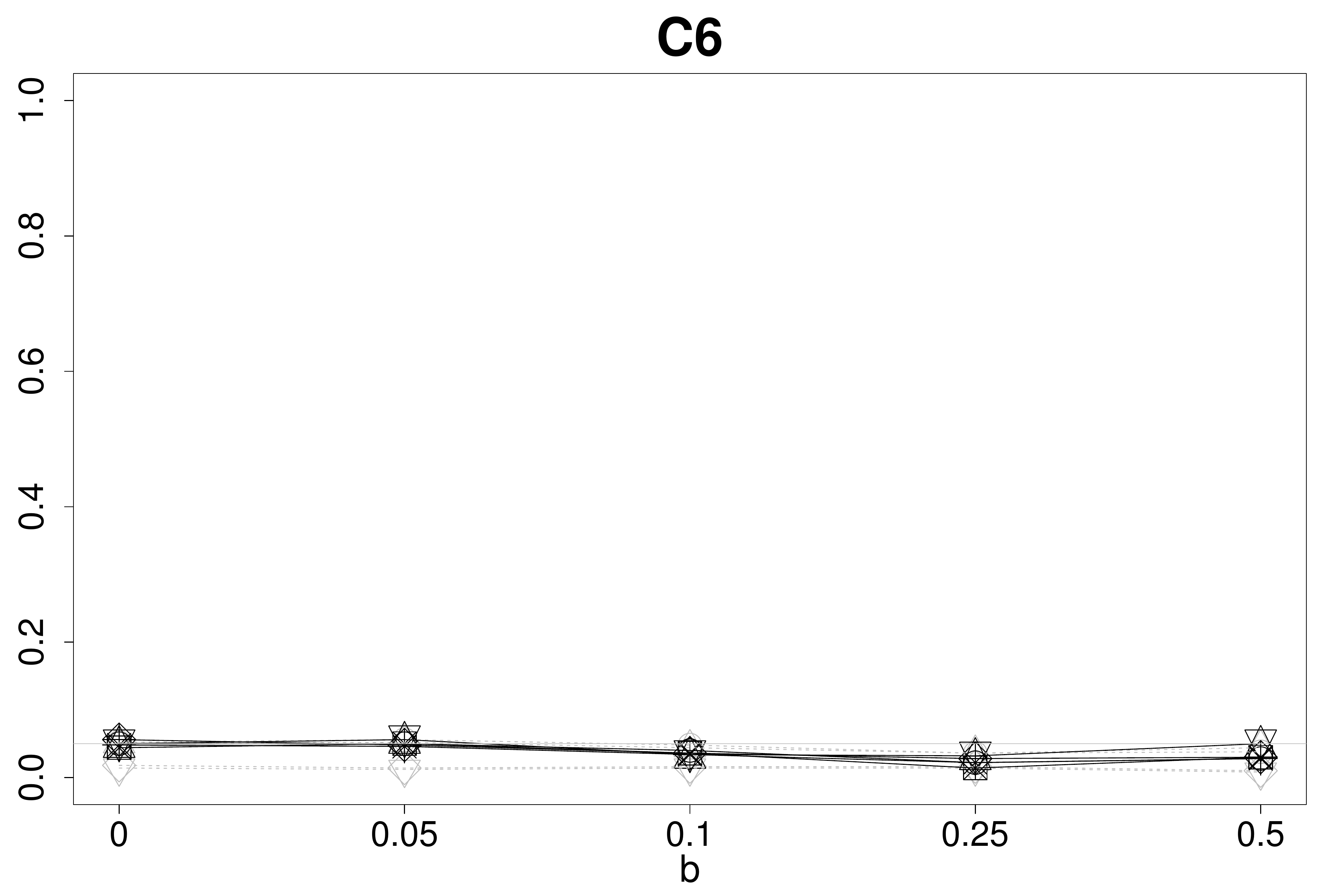}\\[3cm]
			\multicolumn{4}{c}{\huge $ \bm{a=0.25} $}\\[0.2cm]
				\multirow{2}{*}{\includegraphics[width=0.5\textwidth]{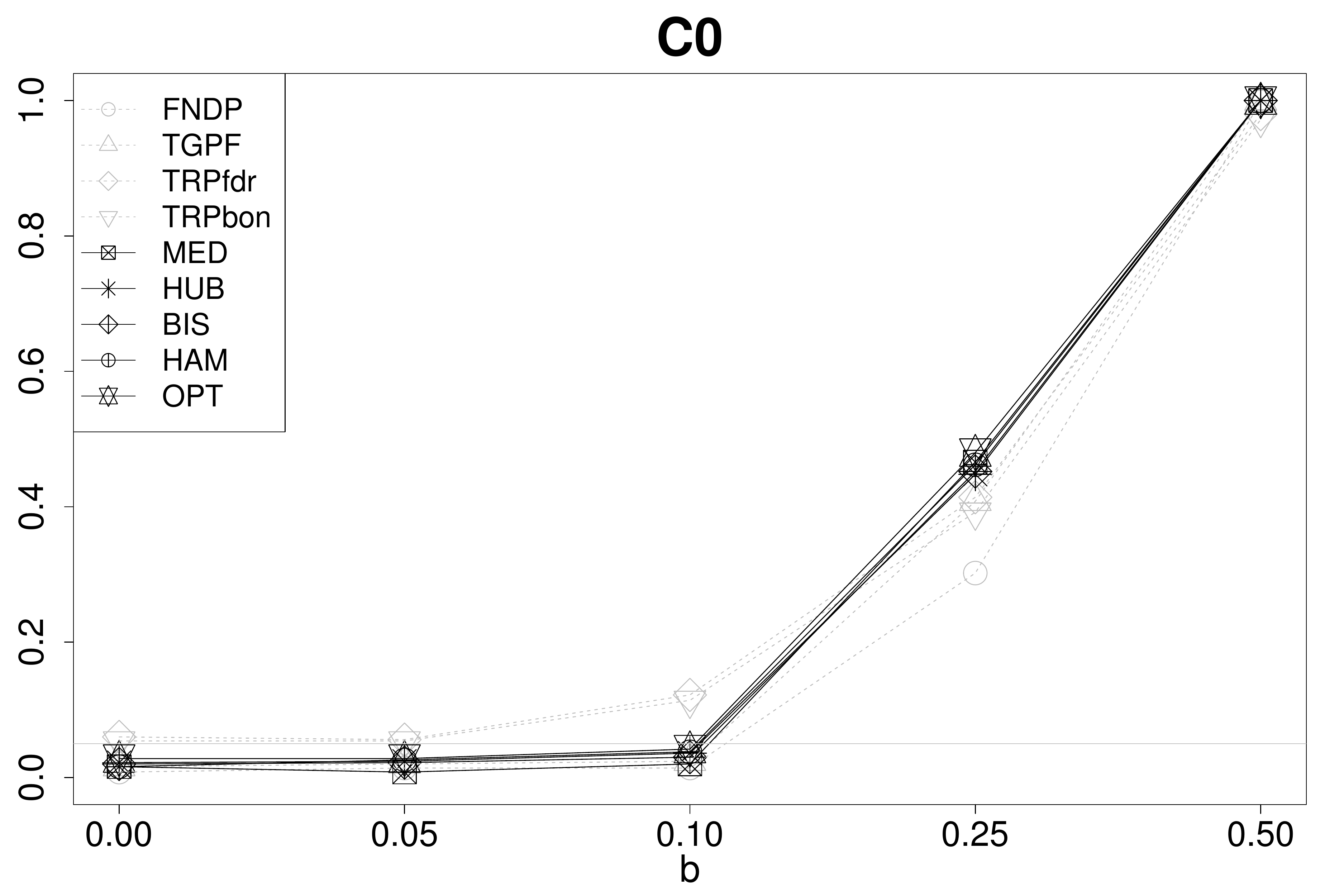}}&\includegraphics[width=0.5\textwidth]{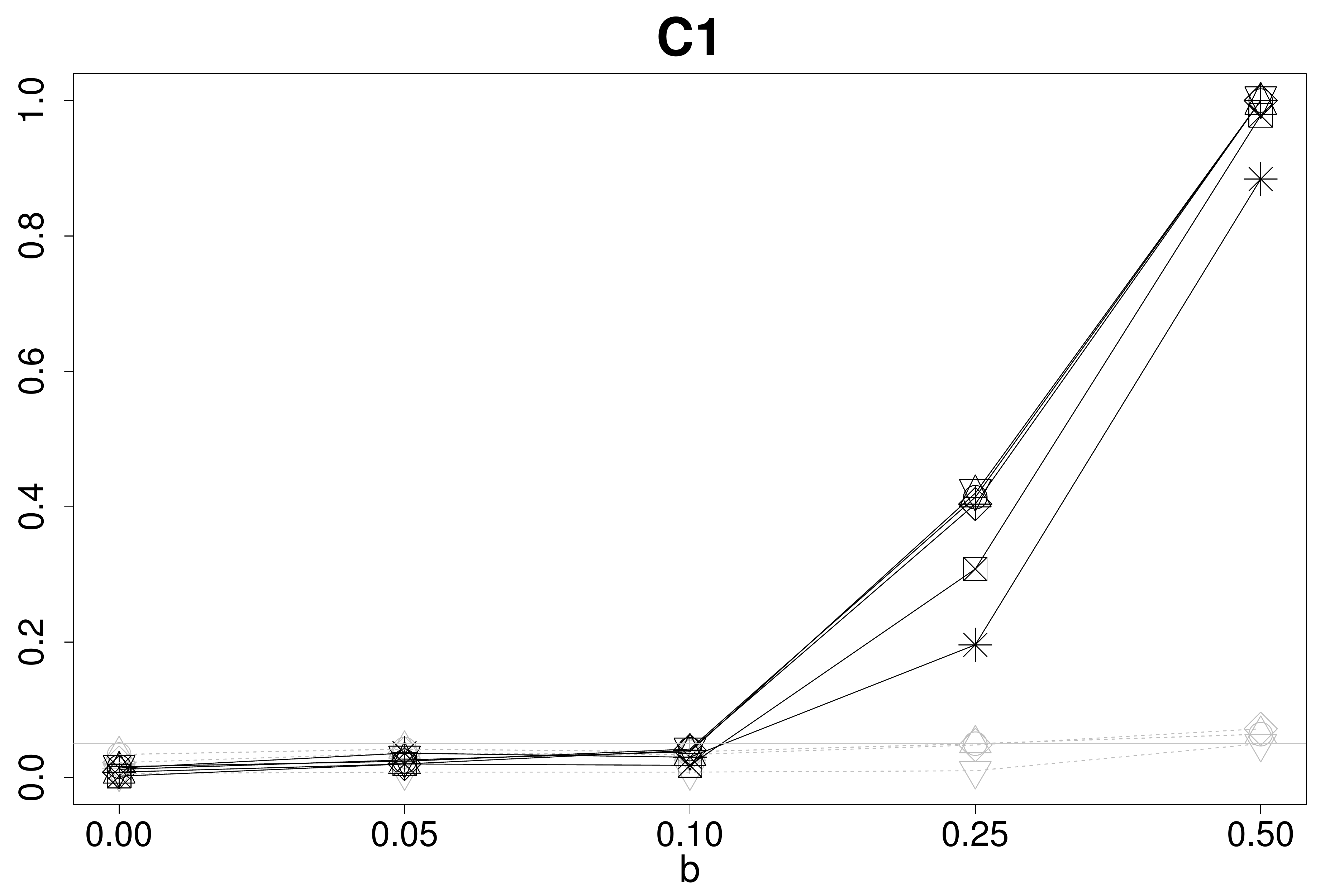}&\includegraphics[width=0.5\textwidth]{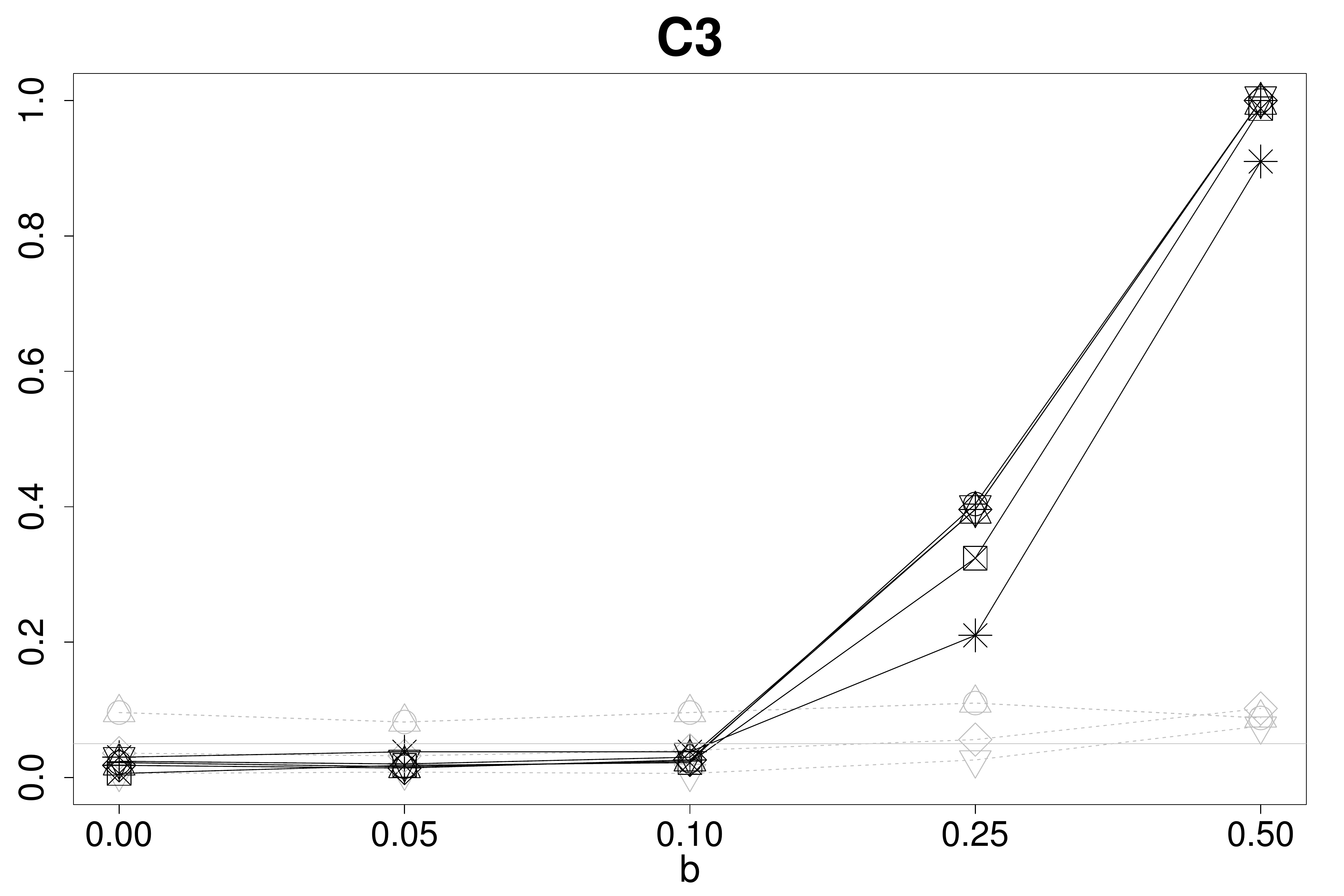}&\includegraphics[width=0.5\textwidth]{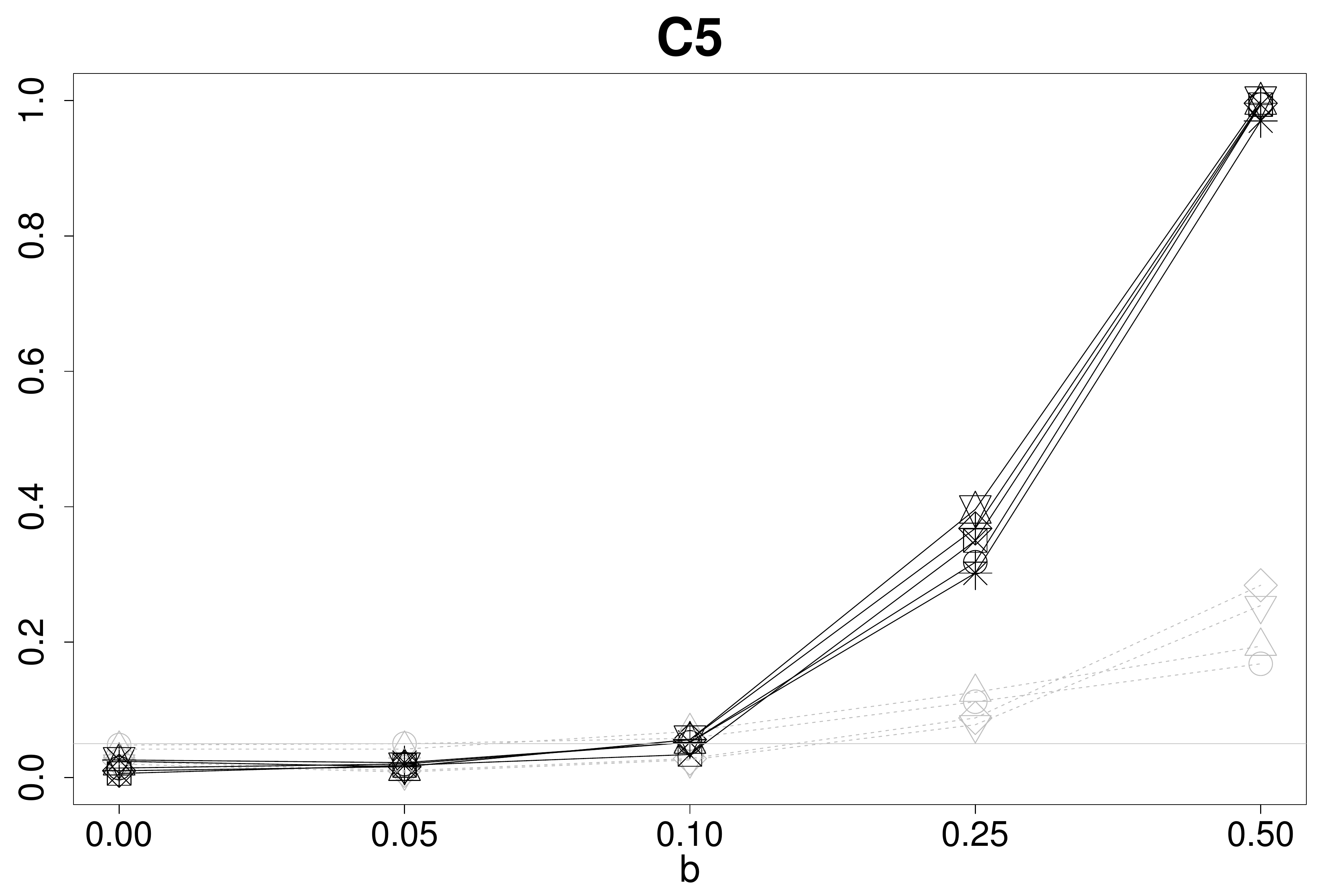}\\
			&\includegraphics[width=0.5\textwidth]{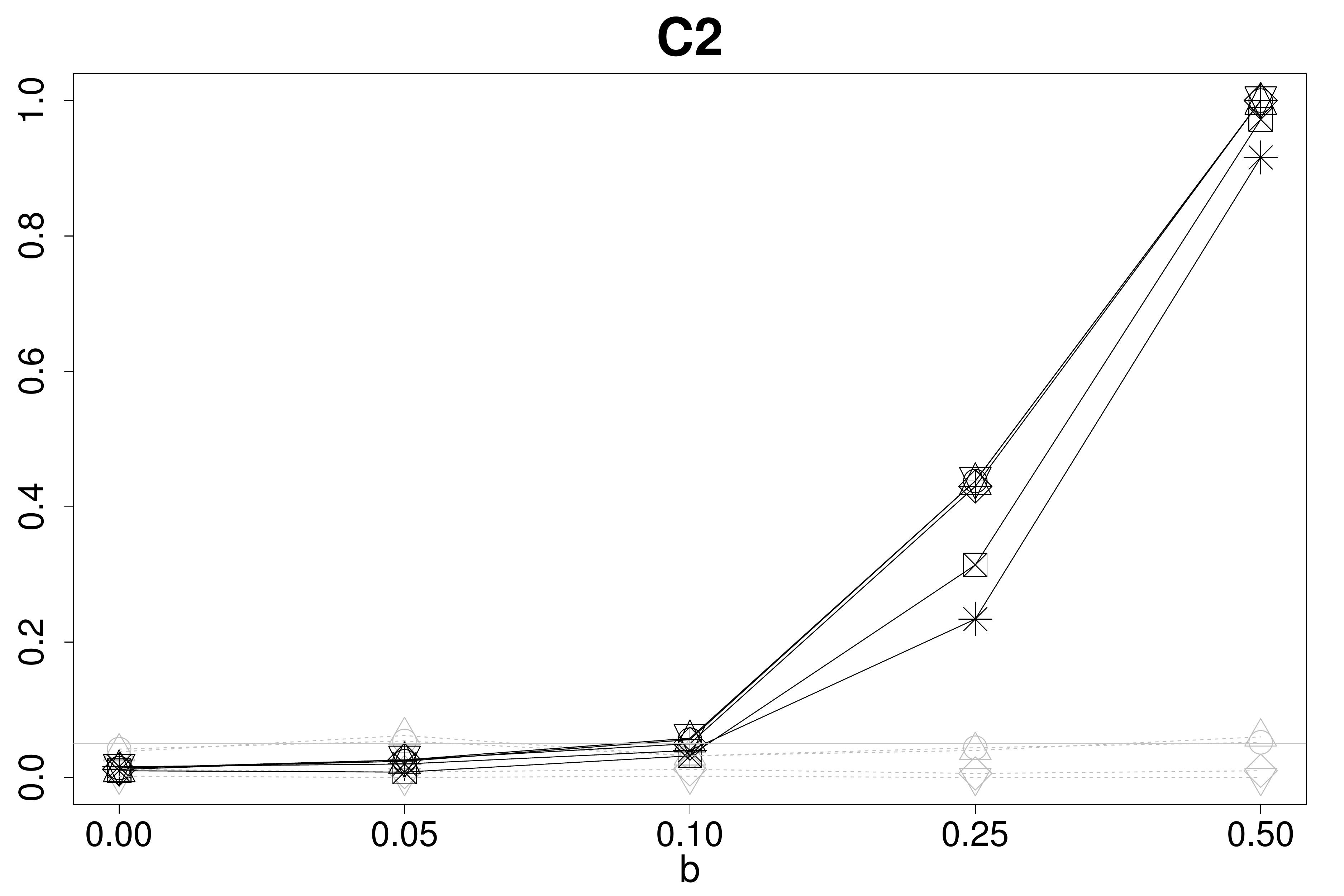}&\includegraphics[width=0.5\textwidth]{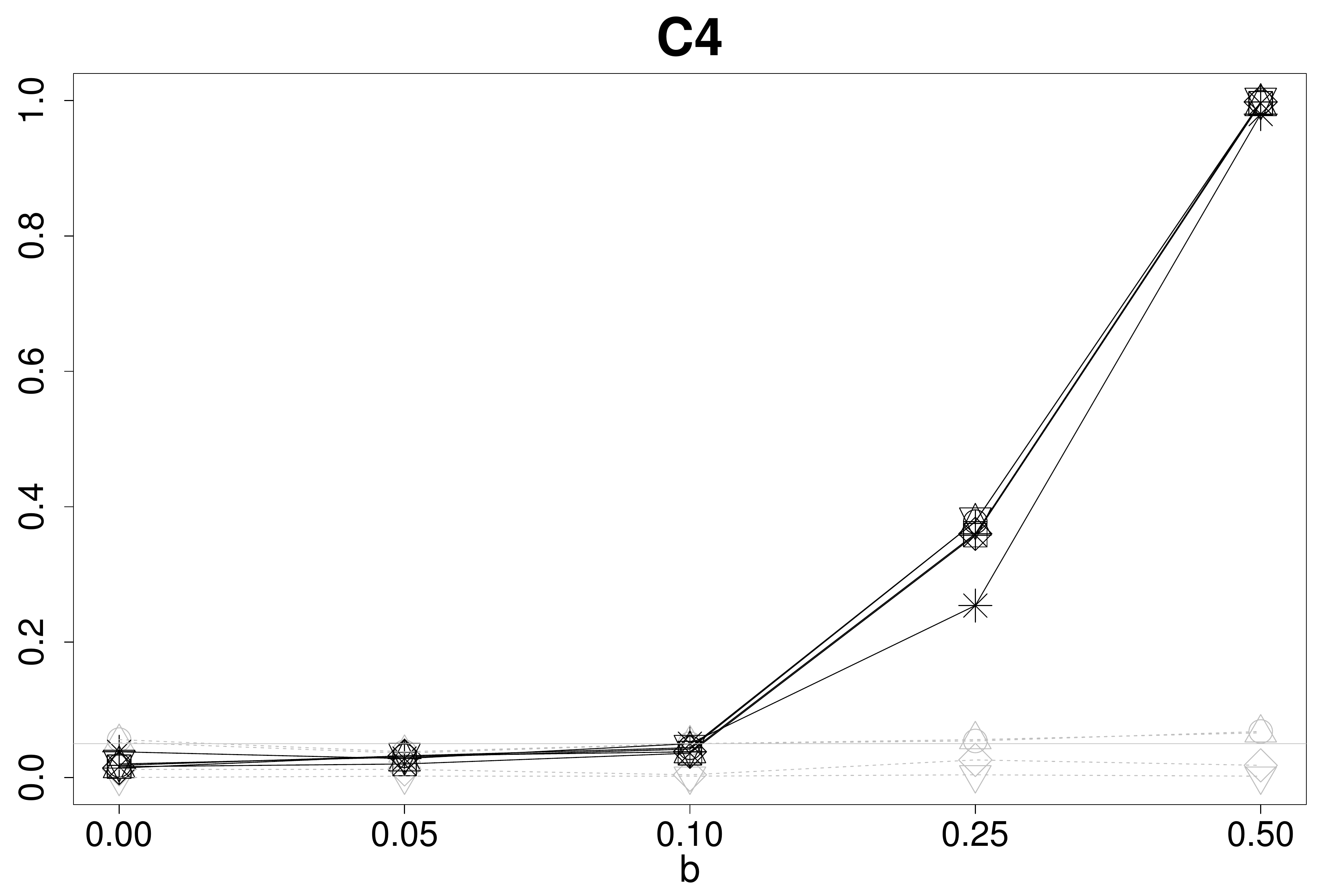}&\includegraphics[width=0.5\textwidth]{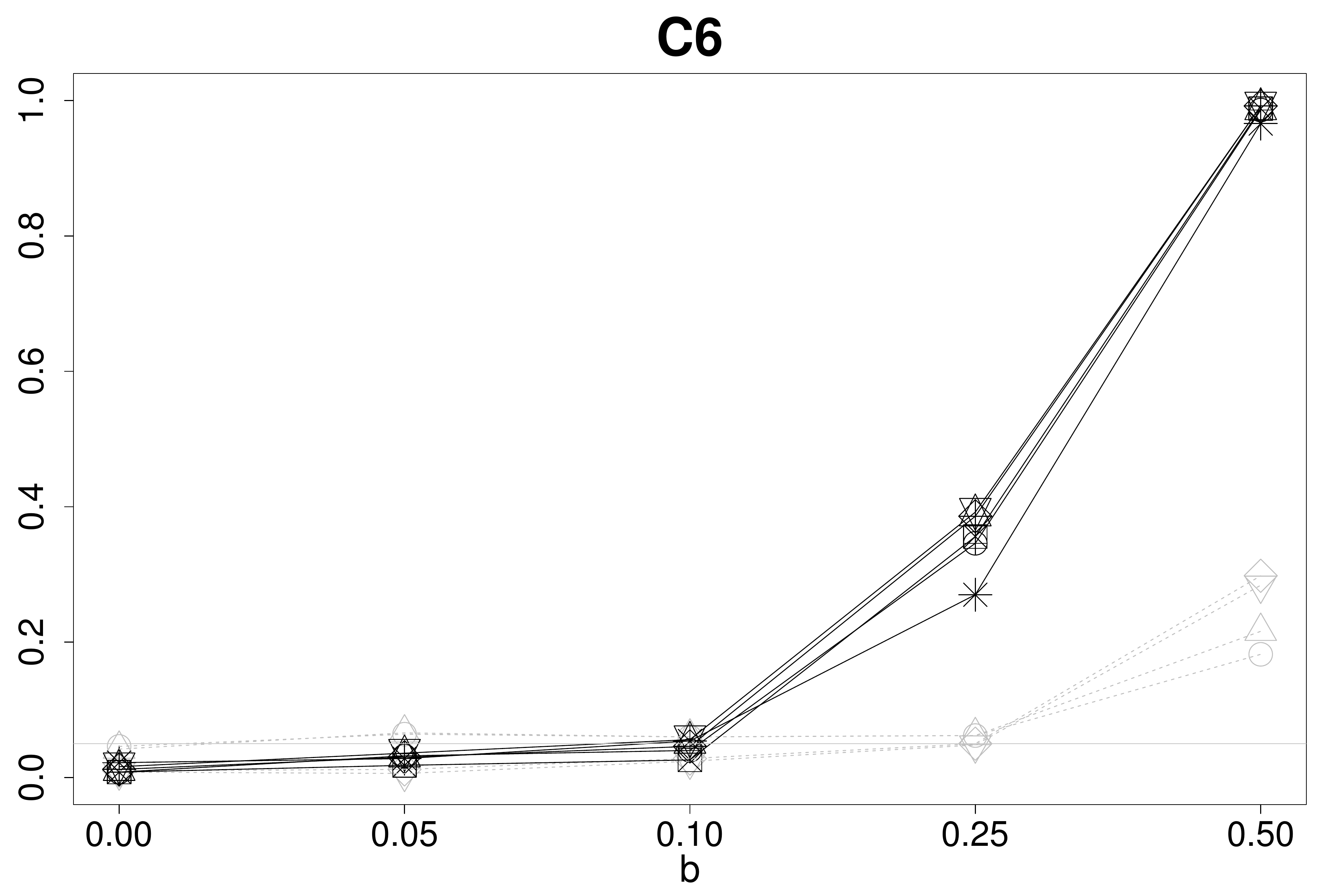}\\[3cm]
			\multicolumn{4}{c}{\huge $ \bm{a=0.50} $}\\[0.2cm]
			\multirow{2}{*}{\includegraphics[width=0.5\textwidth]{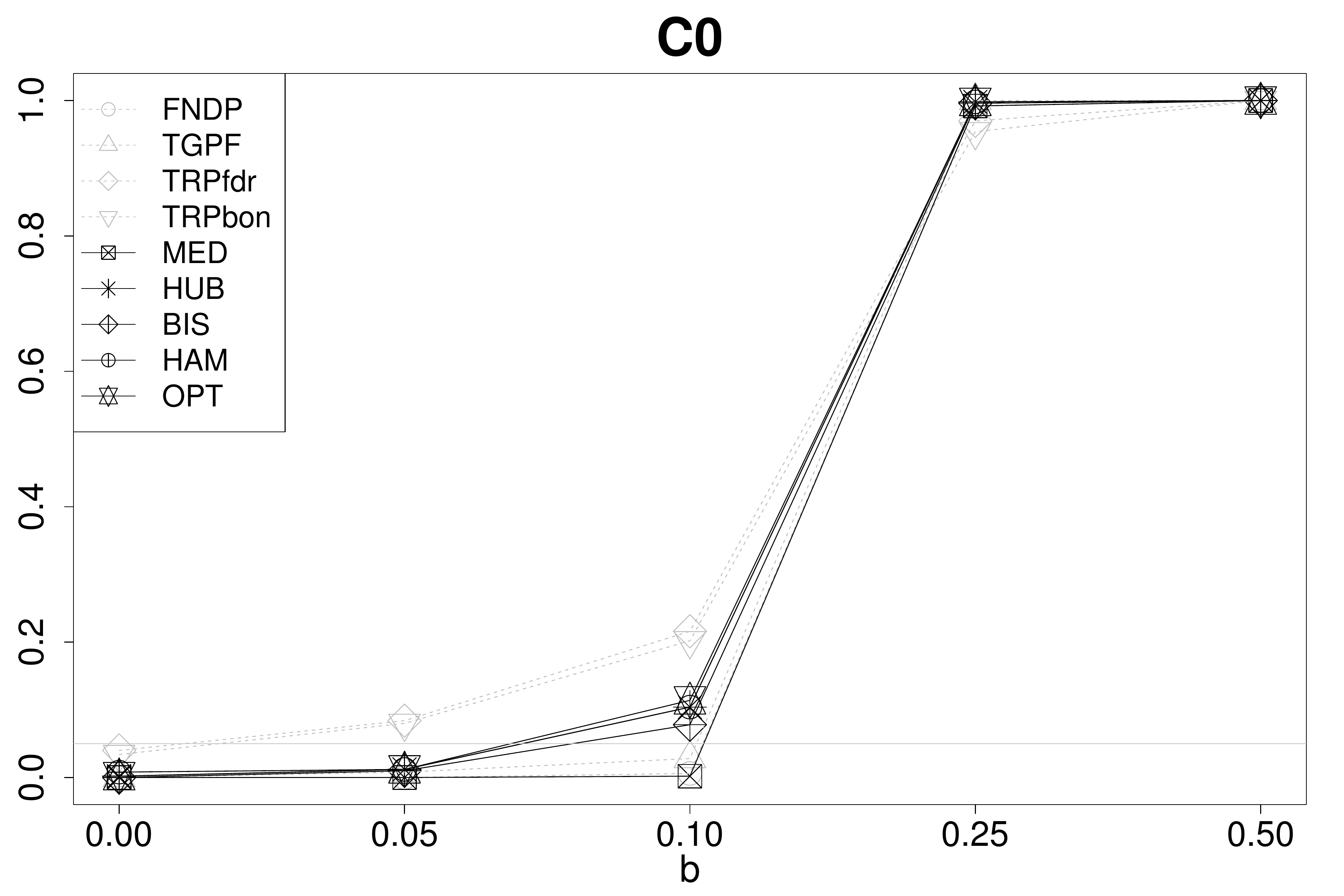}}&\includegraphics[width=0.5\textwidth]{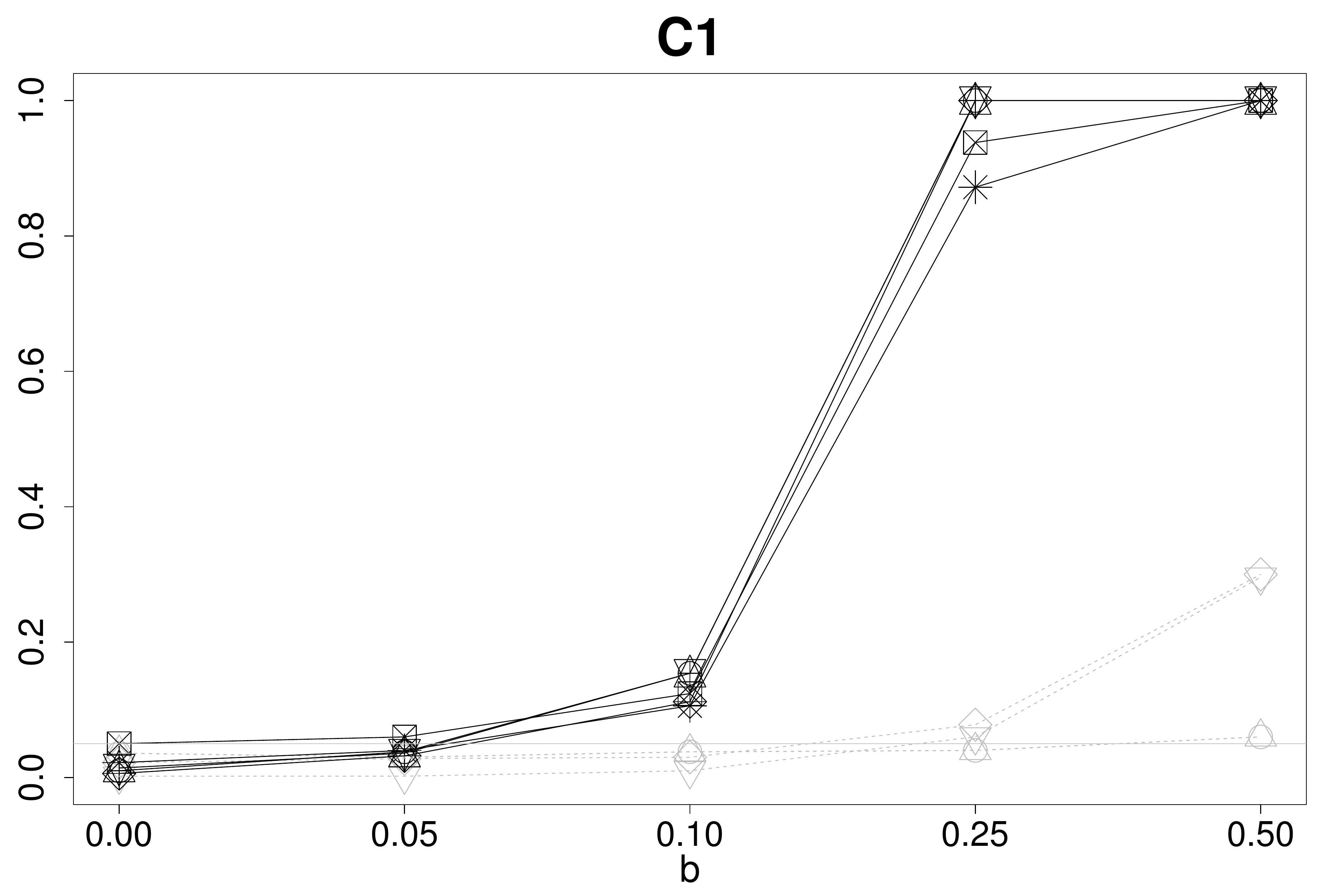}&\includegraphics[width=0.5\textwidth]{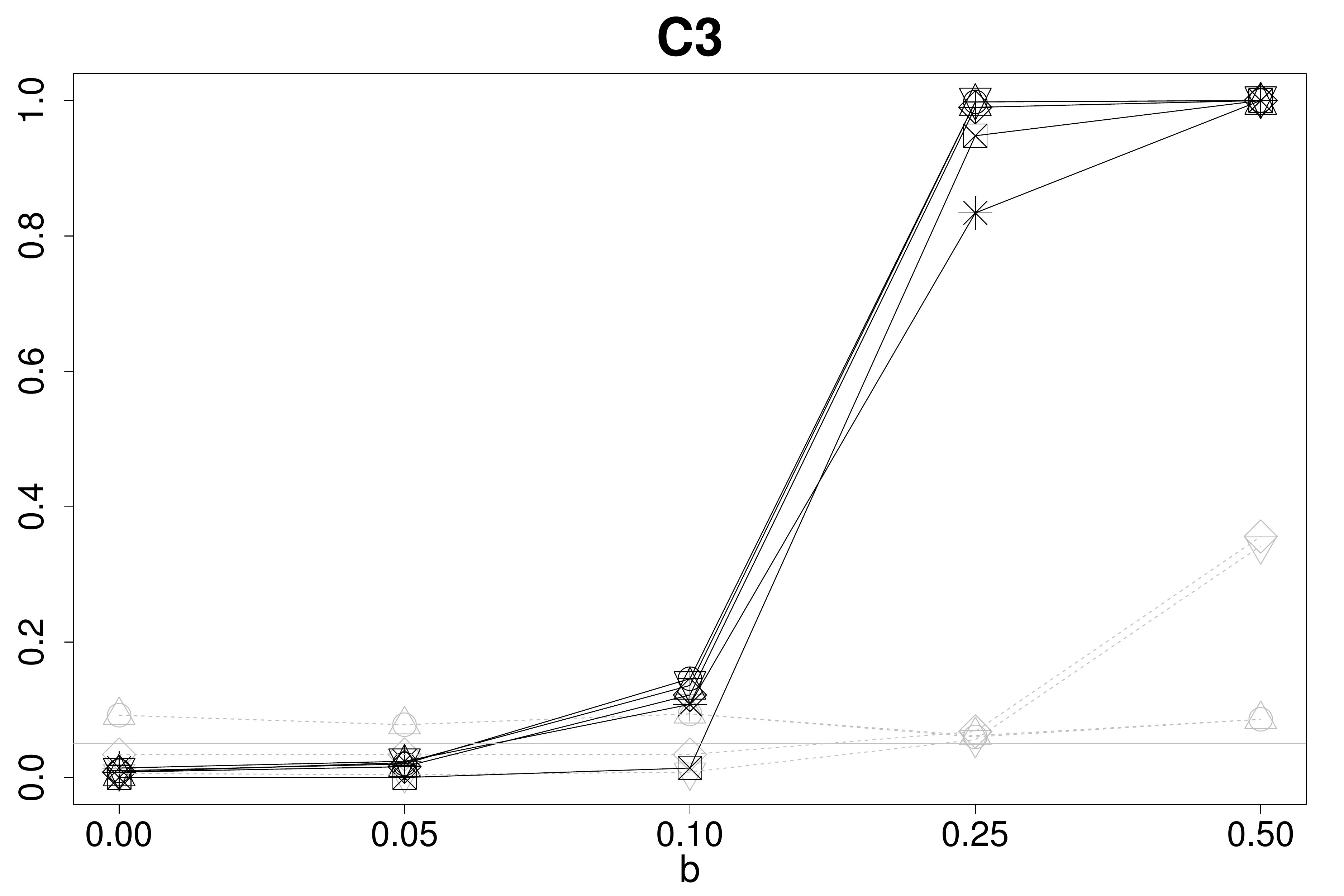}&\includegraphics[width=0.5\textwidth]{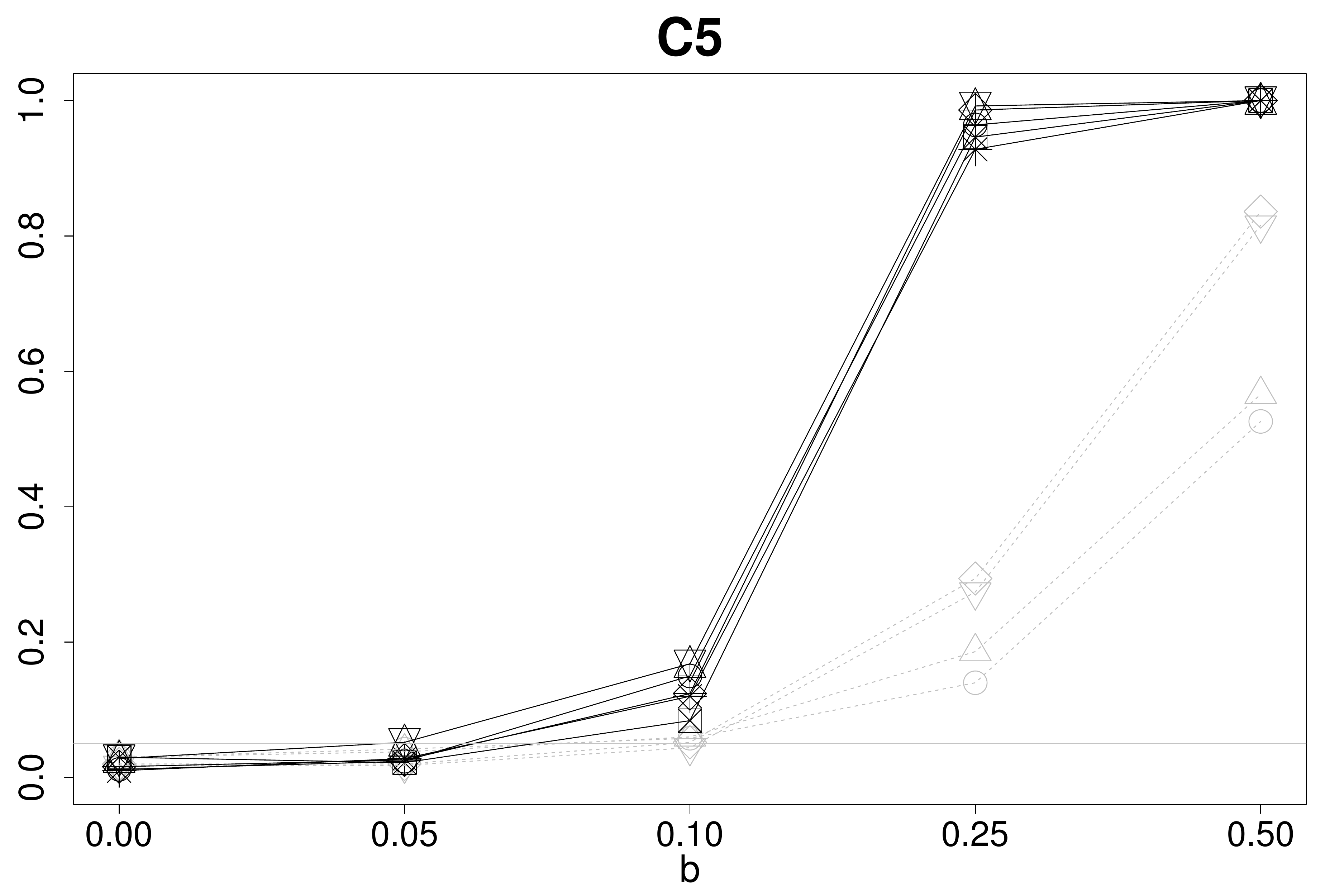}\\
			&\includegraphics[width=0.5\textwidth]{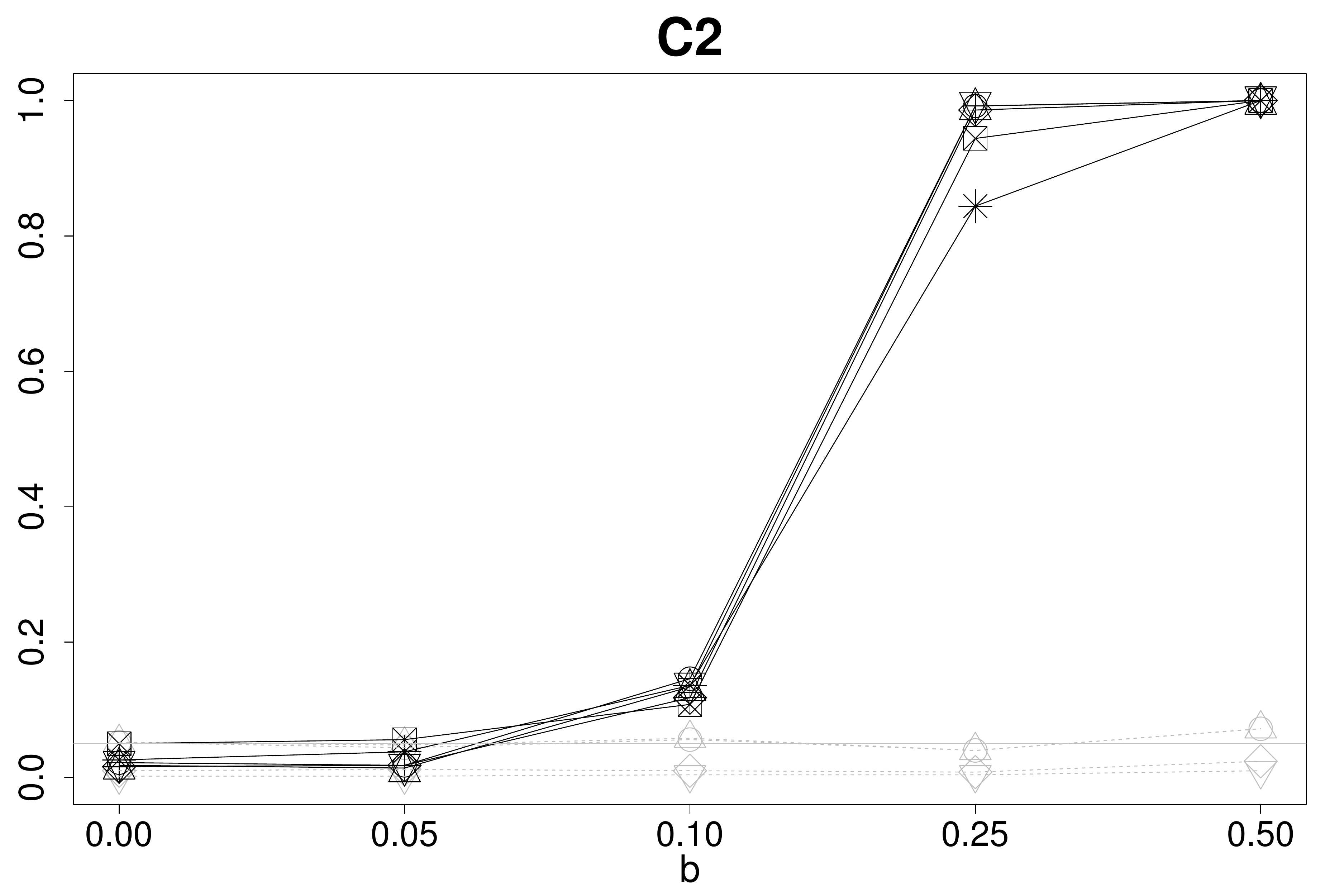}&\includegraphics[width=0.5\textwidth]{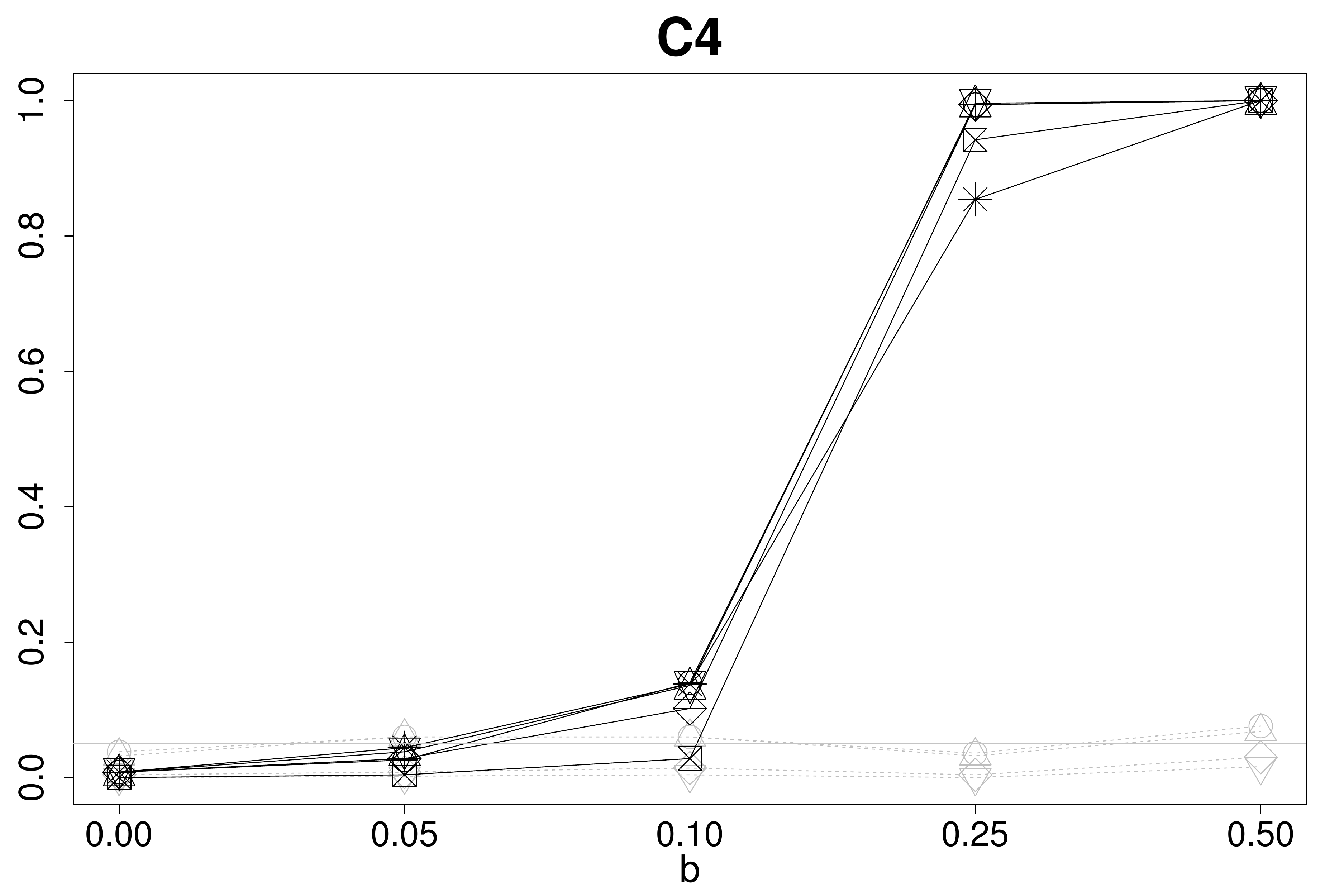}&\includegraphics[width=0.5\textwidth]{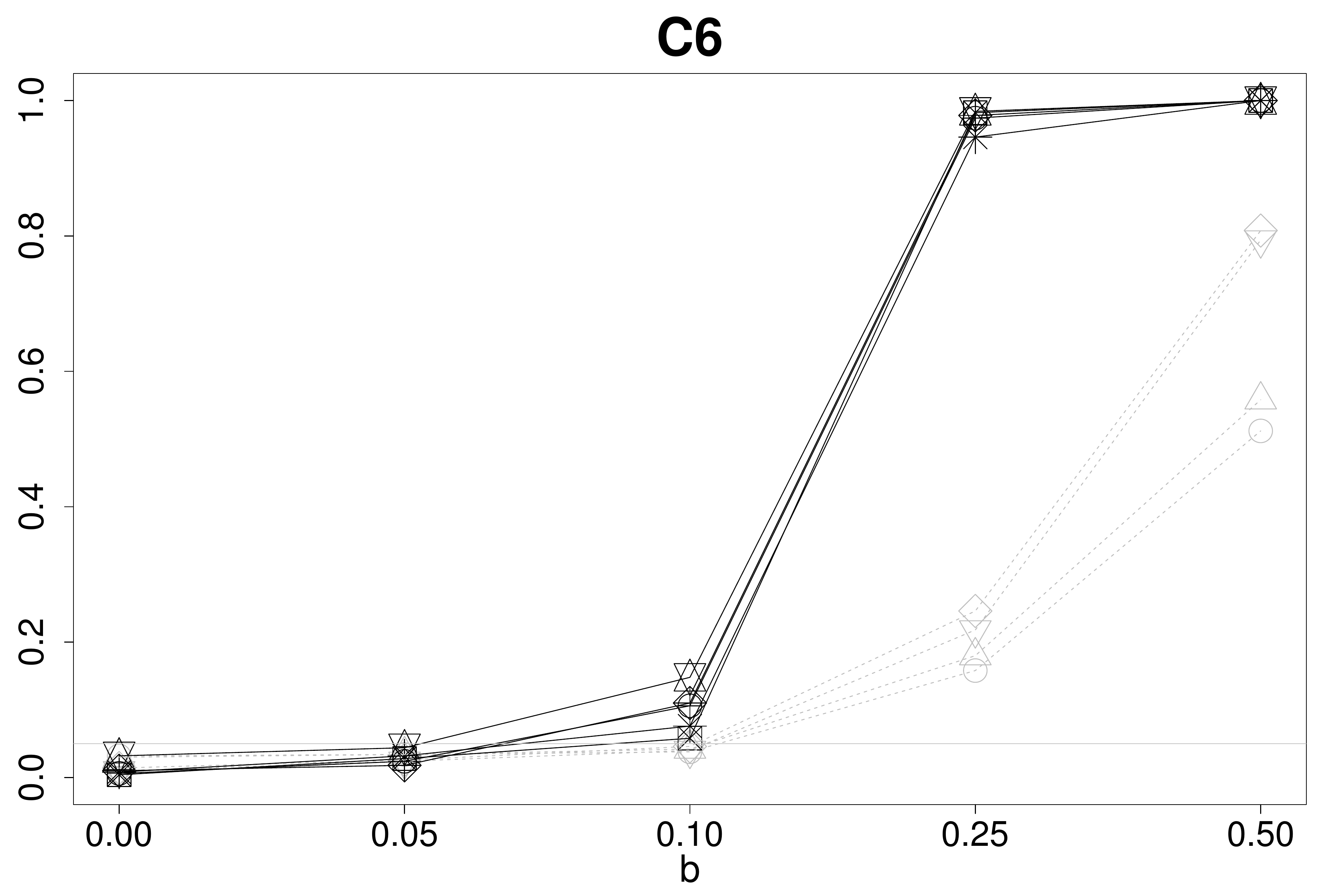}\\
	\end{tabular}}
	
	\caption{\label{fig_twoint}Empirical size ($ a\leq 0.10 $ and $ b=0 $) and  power  ($ a= 0.25,0.50 $ and $ b\neq0 $) of all tests for $ H_{0,AB} $ against  $ H_{1,AB} $ (at level $ \alpha=0.05 $) as a function of $ b $, for different contamination models (C0-6) in Scenario 2. The proposed and competing tests are displayed as  black and grey lines, respectively.}
\end{figure}

Fig. \ref{fig_twoint} shows the empirical size ($ a\leq 0.10 $ and $ b=0 $) and  empirical power  ($ a= 0.25,0.50 $ and $ b\neq0 $) of all tests for $ H_{0,AB} $ against  $ H_{1,AB} $ (at level $ \alpha=0.05 $) as a function of $ b $, for different contamination models (C0-6).
In terms of empirical size ($ a\leq 0.10 $ and $ b=0 $), all the tests are able to approximately control the level $ \alpha $, except for the FNDP and TGPF tests for model C3 and for RoFANOVA-MED and RoFANOVA-HUB  for model C4 at $ b=0.50 $.
For $ a= 0.25,0.50 $ and $ b\neq0 $, the empirical power of the proposed tests is much larger than that of the competitors, for all the contamination model C1-6. Moreover, in case of no contamination (C0), the power of the RoFANOVA tests is comparable to that of the competitors for $ a= 0.25,0.50 $.
Also in this case, among the RoFANOVA tests, the RoFANOVA-BIS, RoFANOVA-HAM and RoFANOVA-OPT  are the best ones.

\section{Additional details about the real case-study}
Table \ref{TABLE A} presents the process parameters and corresponding energy density levels for the production of the specimens in the real case-study. 
Table \ref{TABLE B} shows the locations of the six analysed layers along the specimen build direction (distance from the baseplate) and orientation of the laser scan direction relative to the shielding gas flow in each layer.

\begin{table}
	\caption{\label{TABLE A} Process parameters and corresponding energy density levels.}
	\centering
    \scriptsize
    % \hspace{-50cm}
    	\resizebox{1\textwidth}{!}{
	\begin{tabular}{c|c|c|c|c|c|c}
	\toprule
	\multirow{5}{*}{\shortstack{	Energy\\ density  level }} &\multirow{5}{*}{\shortstack{Laser exposure\\ time\\ $t$ ($\mu$s)} }&\multirow{5}{*}{\shortstack{Distance between\\ exposed points\\ along laser scan\\ track\\ $dp$ ($\mu$m)}}&\multirow{5}{*}{\shortstack{Distance between\\ parallell laser\\ scan tracks \\$dh$ ($\mu$m)}}&\multirow{5}{*}{\shortstack{Laser power\\ $P$ (W)}} &\multirow{5}{*}{\shortstack{Powder bed \\thickness\\  $z$ ($\mu$m )}}&\multirow{5}{*}{\shortstack{Energy density\\  $F$ (kJ/cm\textsuperscript{3})}}\\
	&&&&&&\\
	&&&&&&\\
	&&&&&&\\
	&&&&&&\\\midrule
	1&39&65&80&200&50&30\\\midrule
	2&85&85&80&200&50&50\\\midrule
	3&104&65&80&200&50&80\\\midrule
	4&125&62.5&80&200&50&100\\\midrule
	5&115&50&80&200&50&115\\\midrule
	6&104&40&80&200&50&130\\
	\bottomrule
	\end{tabular}
}
\end{table}

\begin{sidewaysfigure} 
	\centering
	
	\footnotesize
	\begin{tabular}{M{0.14\textwidth}|M{0.090\textwidth}|M{0.090\textwidth}M{0.090\textwidth}M{0.090\textwidth}M{0.090\textwidth}M{0.090\textwidth}M{0.090\textwidth}}
	\diagbox[height=2cm,width=1.4\textwidth,innerwidth=2.9cm]{ Fluency level}{ Scan strategy}	& - &1&2&3&4&5&6 \\\hline&&&&&&&\\[-0.3cm]
-&\includegraphics[width=0.1\textwidth]{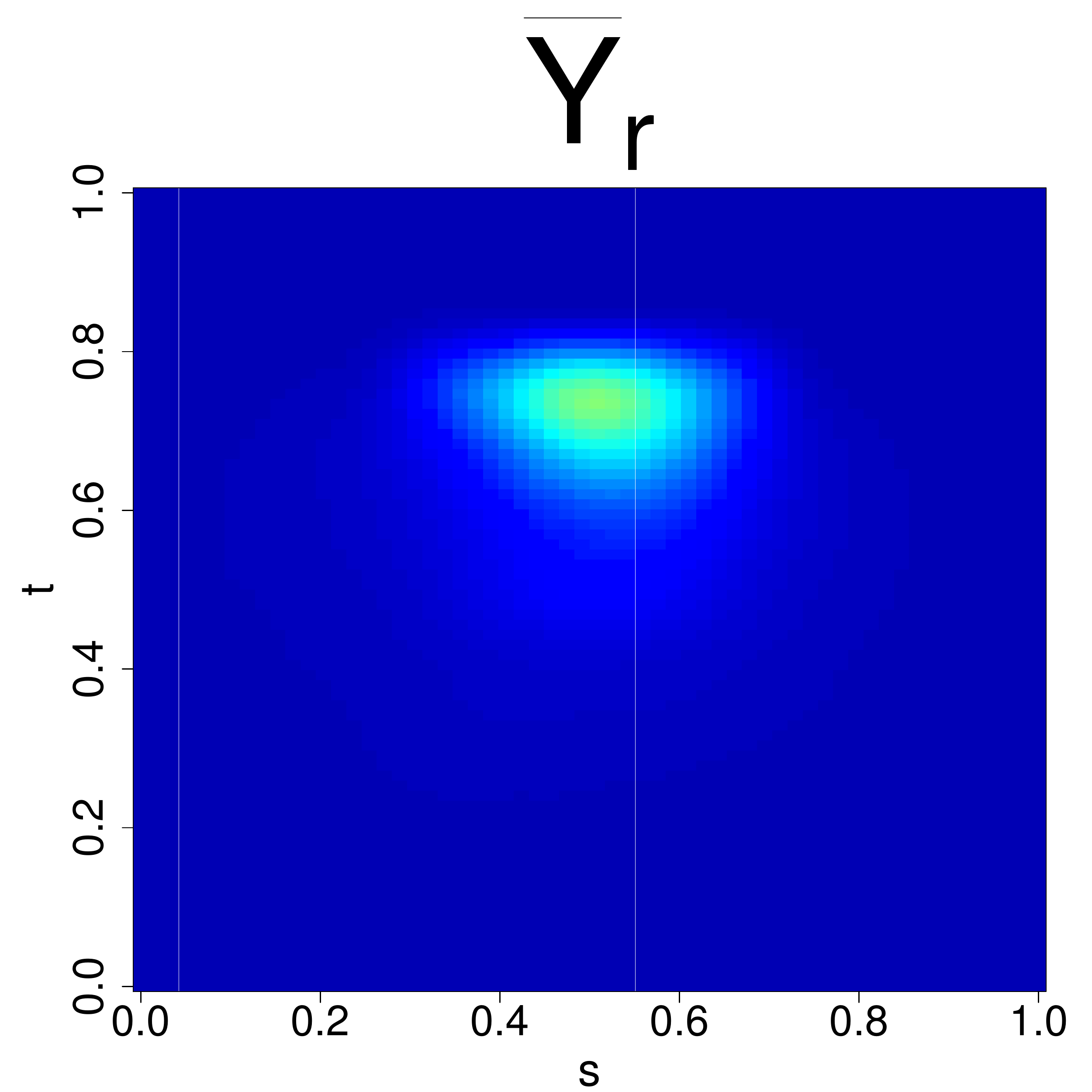}&\includegraphics[width=0.1\textwidth]{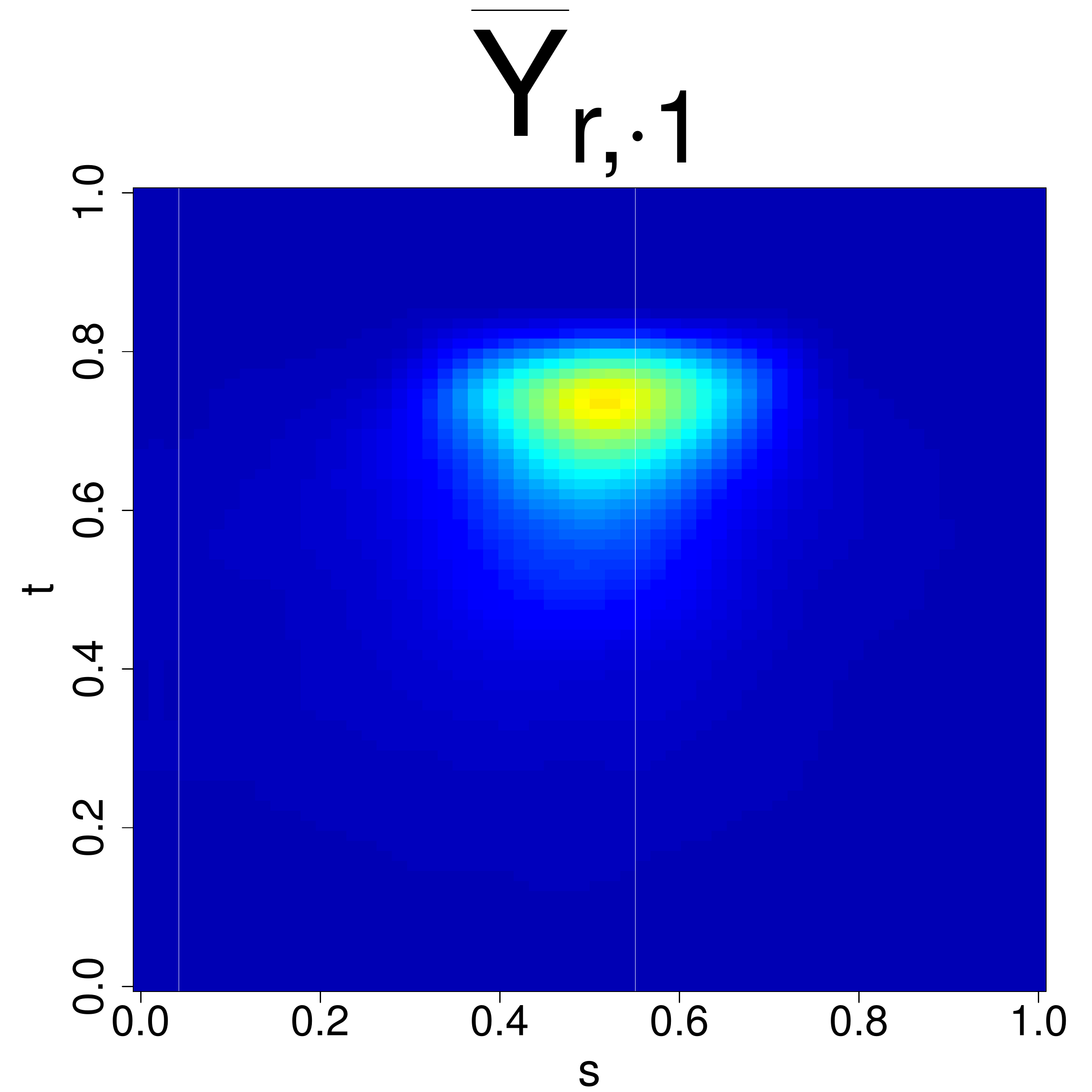}&\includegraphics[width=0.1\textwidth]{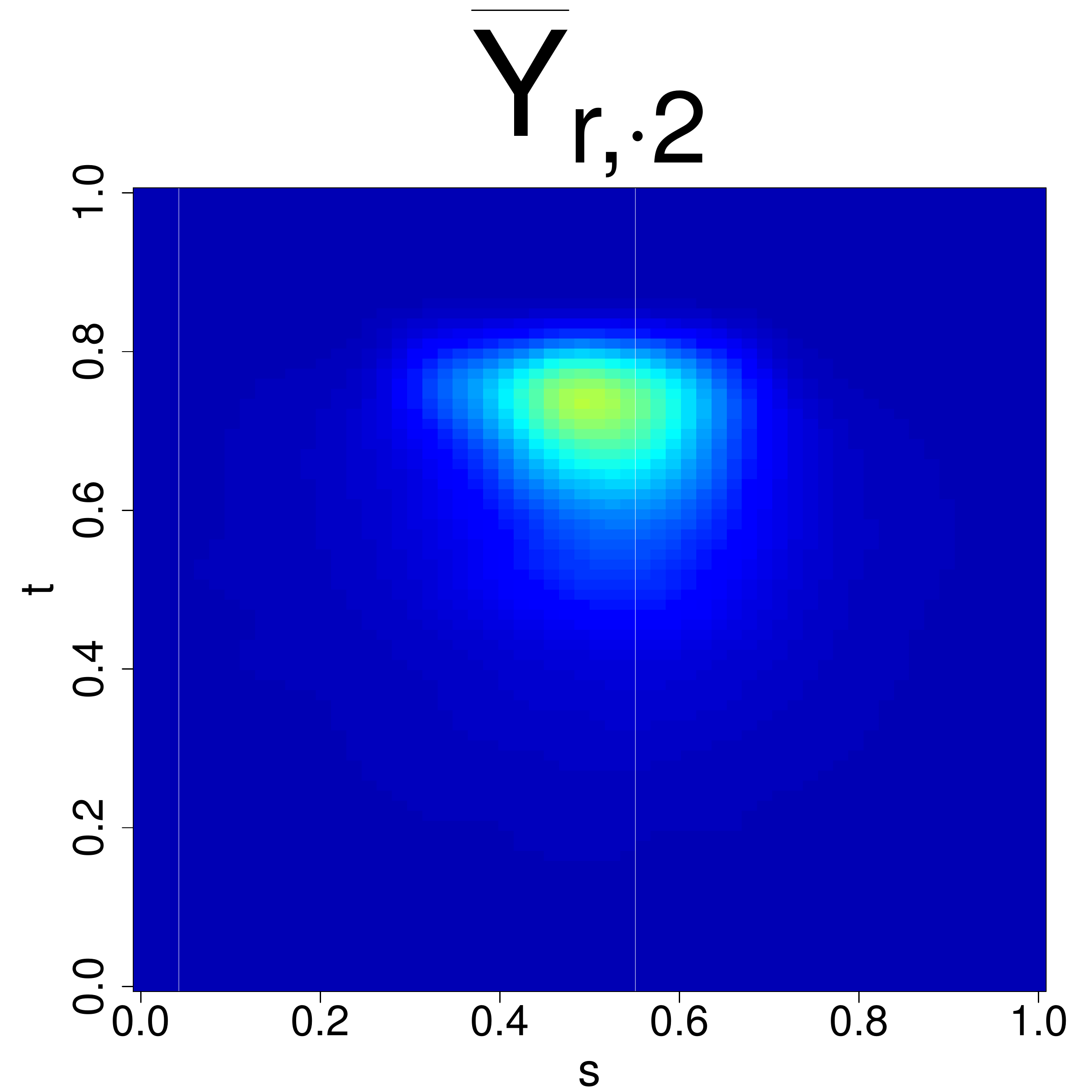}&\includegraphics[width=0.1\textwidth]{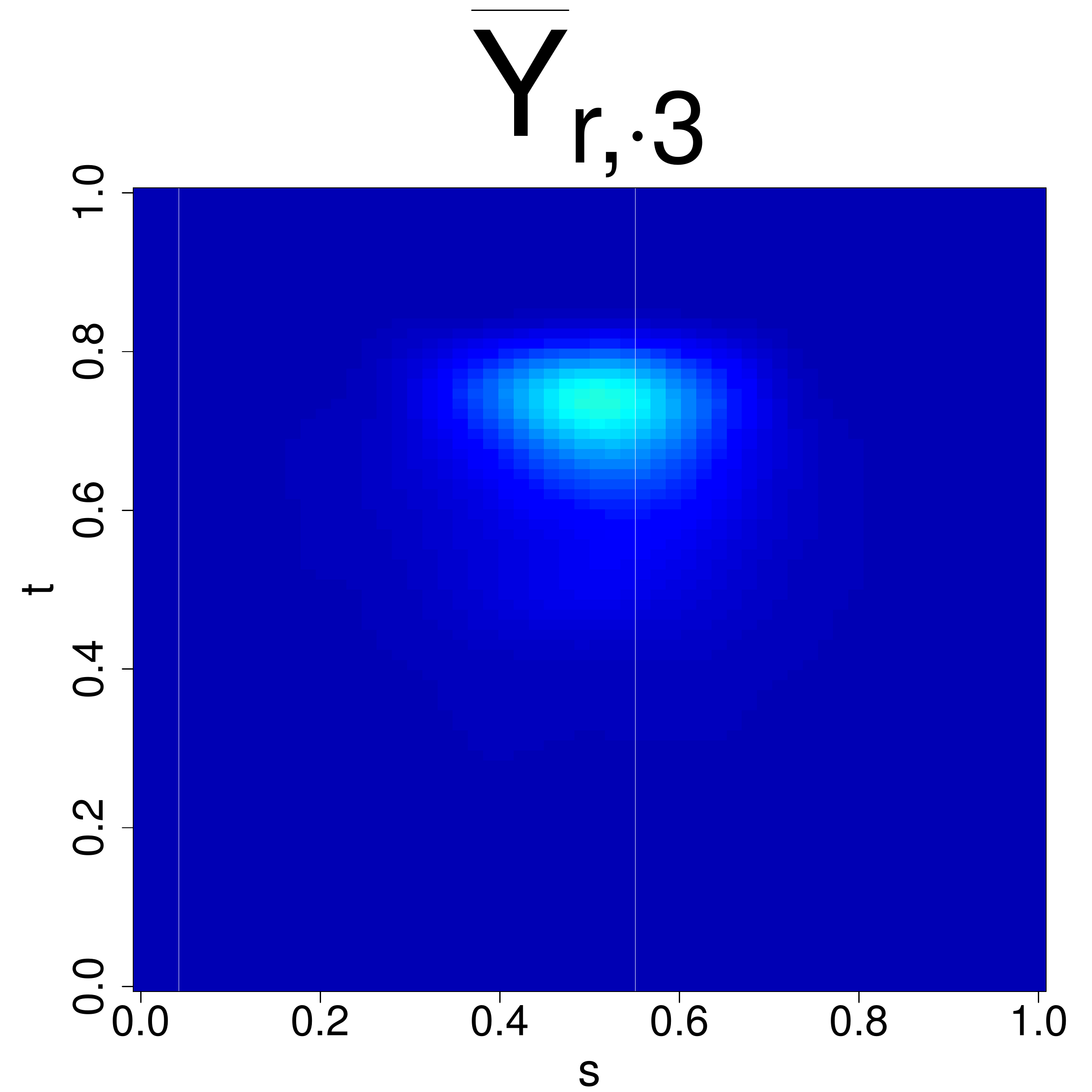}&\includegraphics[width=0.1\textwidth]{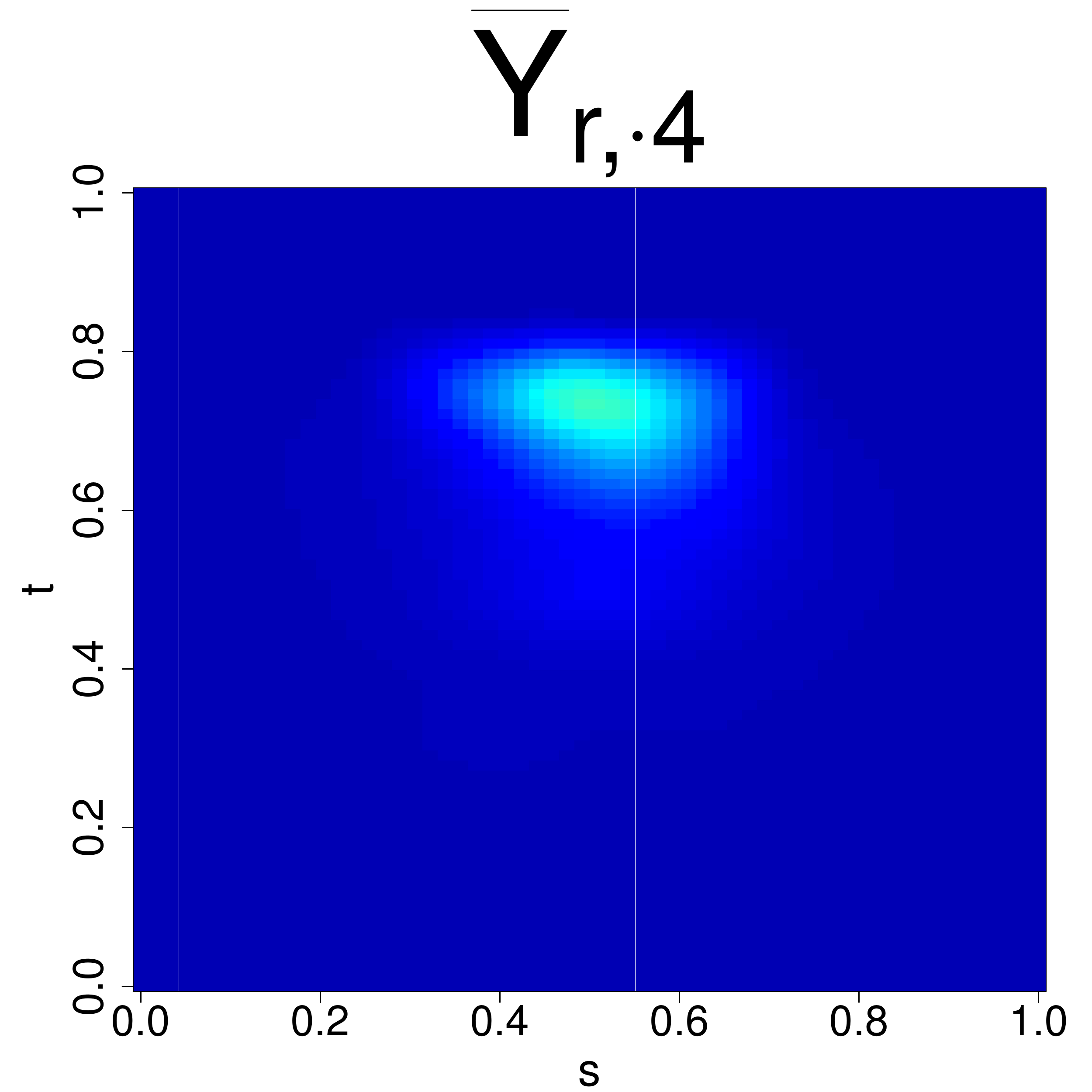}&\includegraphics[width=0.1\textwidth]{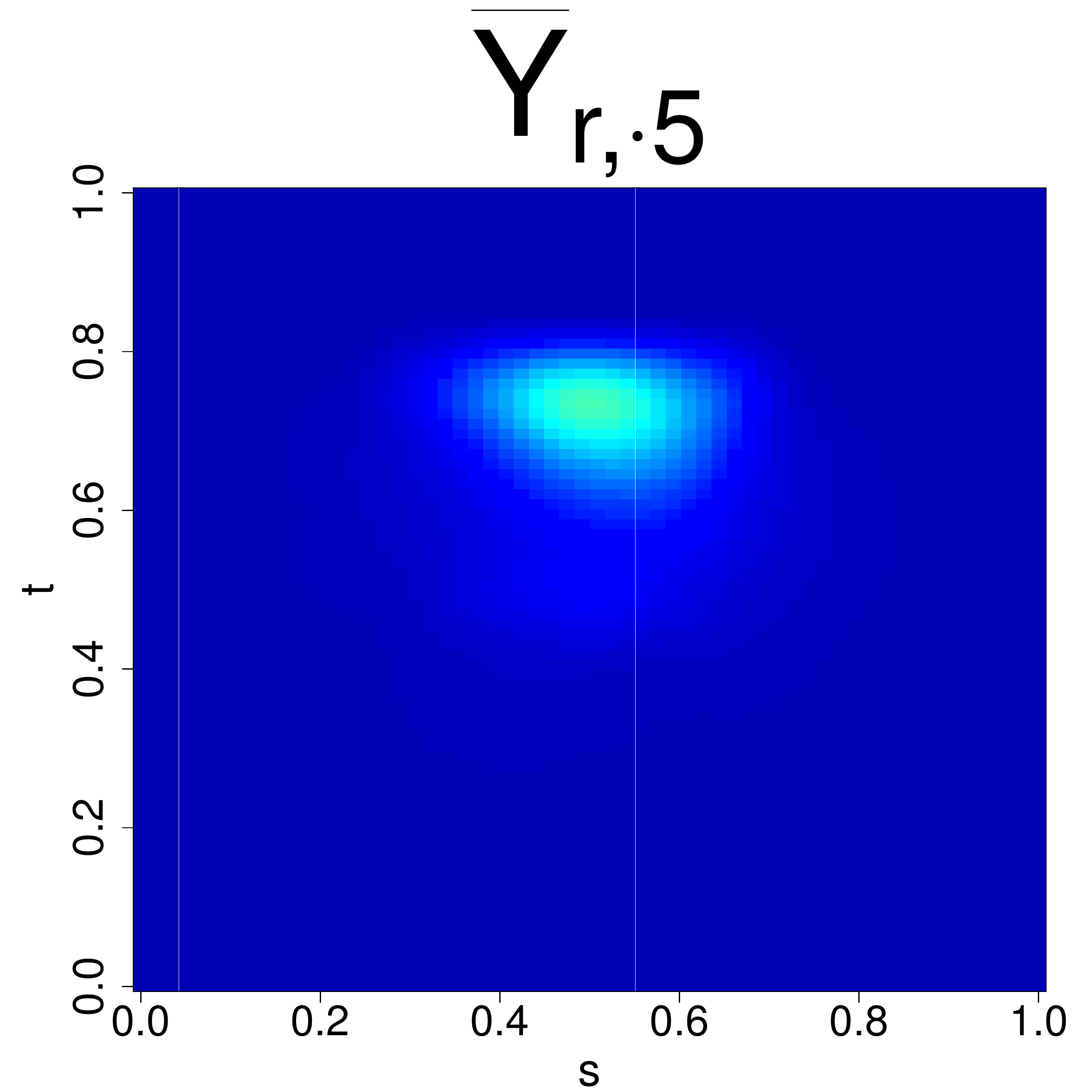}&\includegraphics[width=0.1\textwidth]{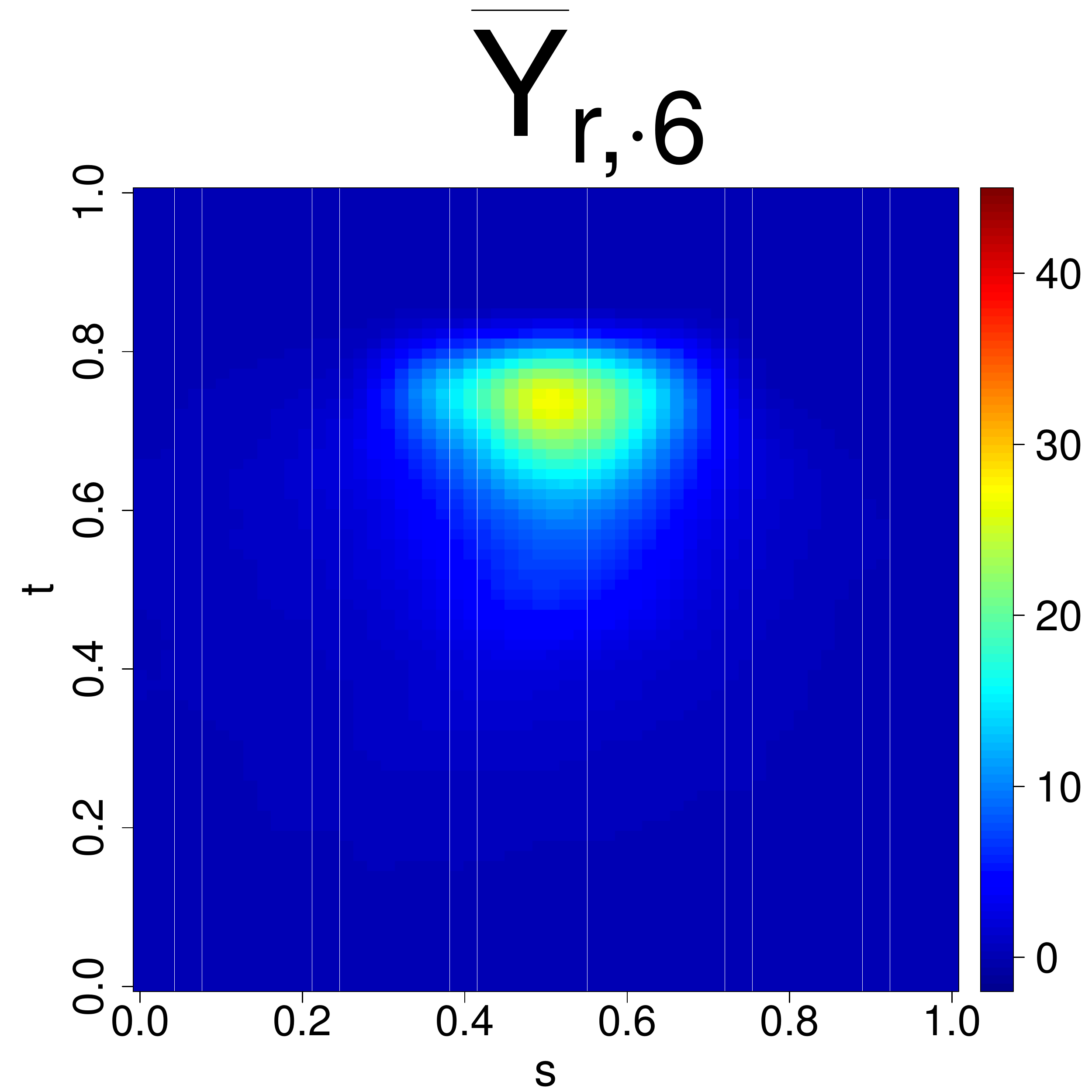}\\\hline&&&&&&&\\[-0.3cm]
	1&\includegraphics[width=0.1\textwidth]{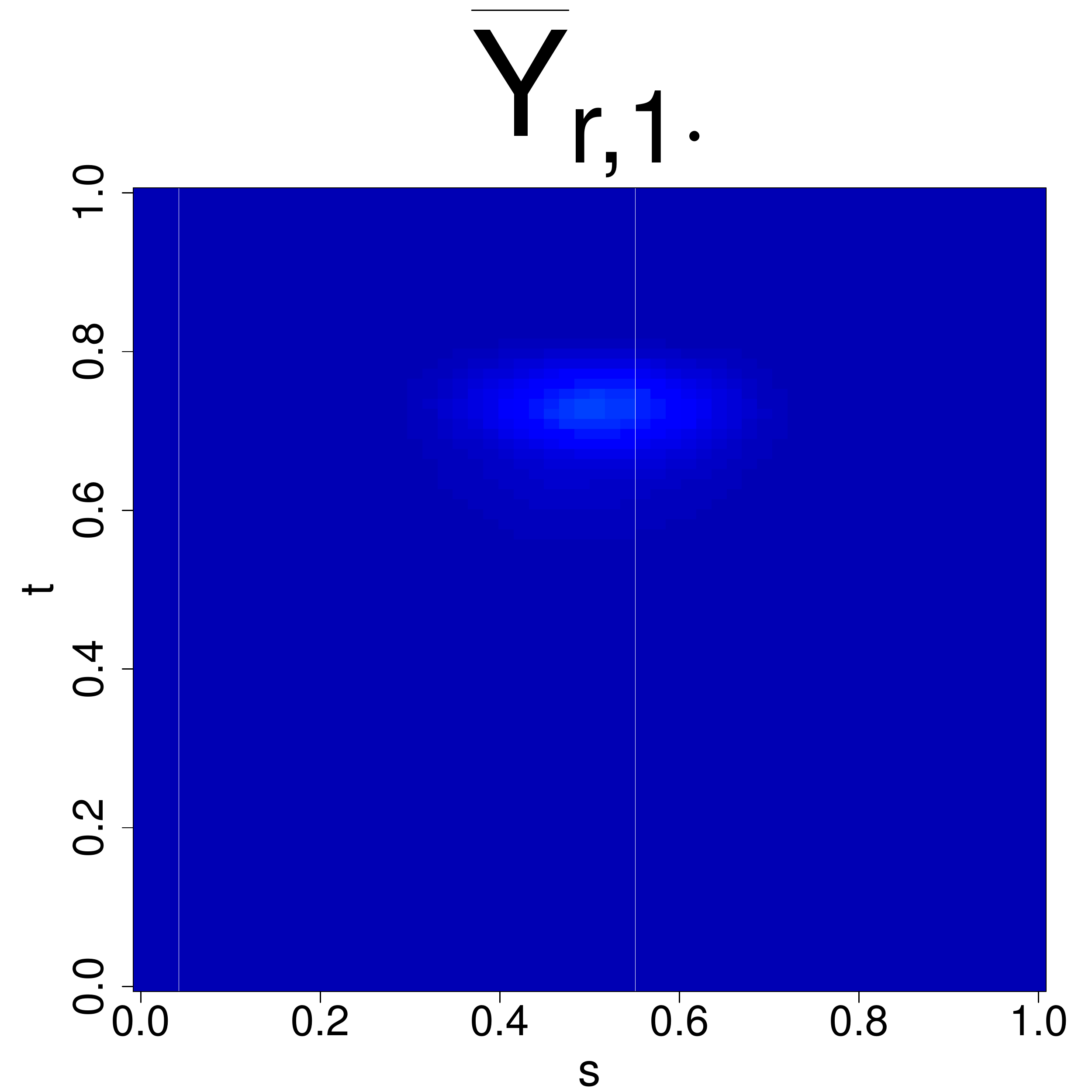}&\includegraphics[width=0.1\textwidth]{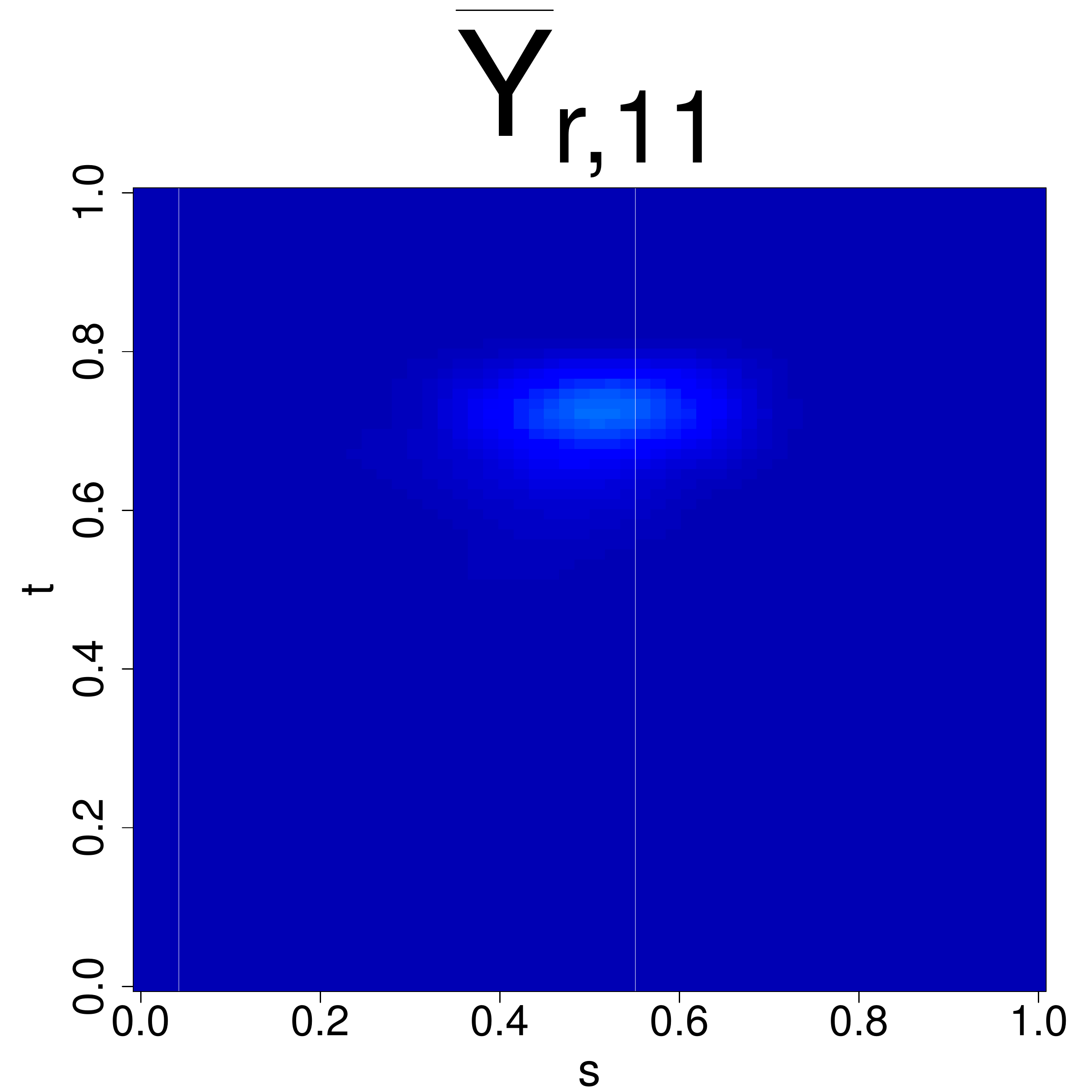}&\includegraphics[width=0.1\textwidth]{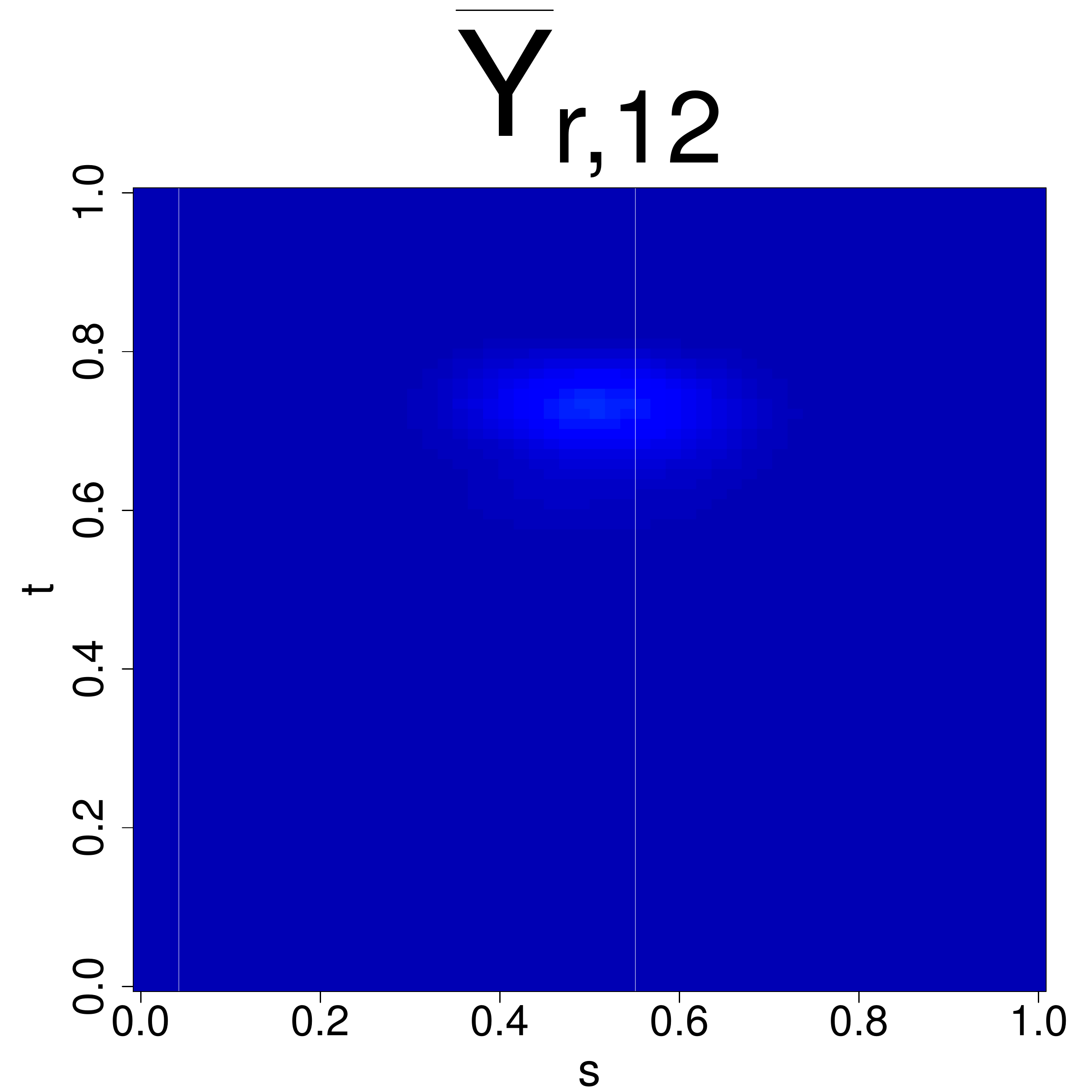}&\includegraphics[width=0.1\textwidth]{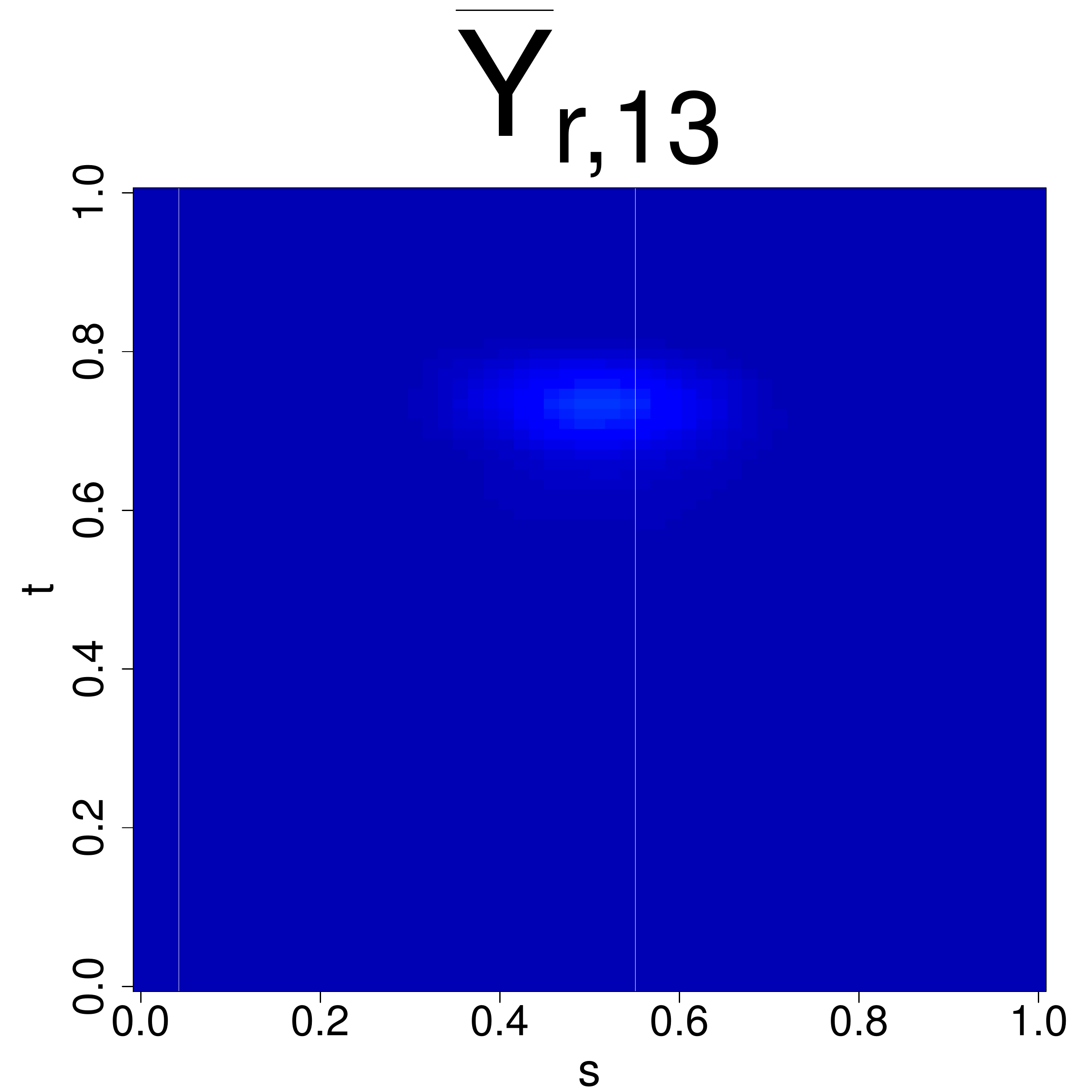}&\includegraphics[width=0.1\textwidth]{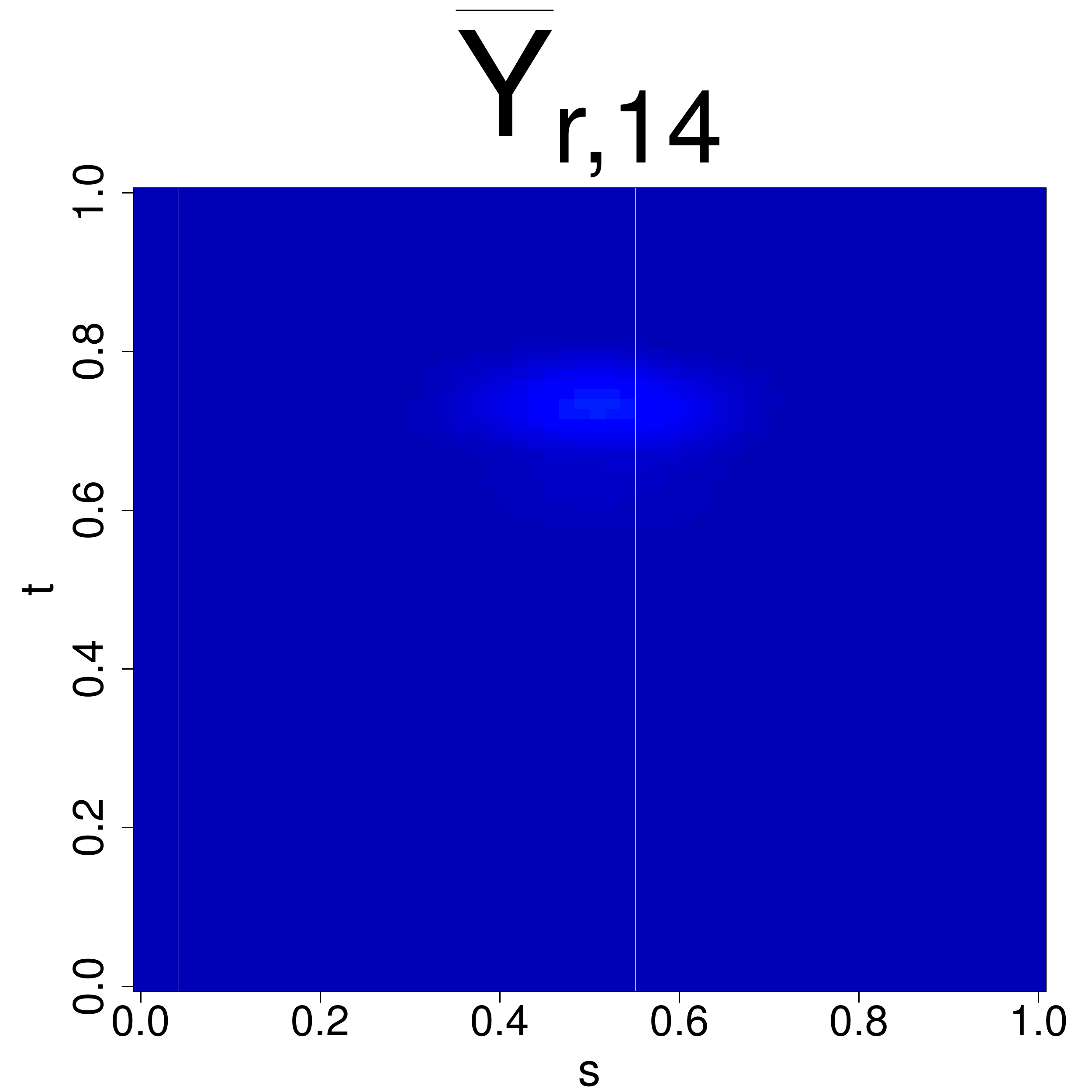}&\includegraphics[width=0.1\textwidth]{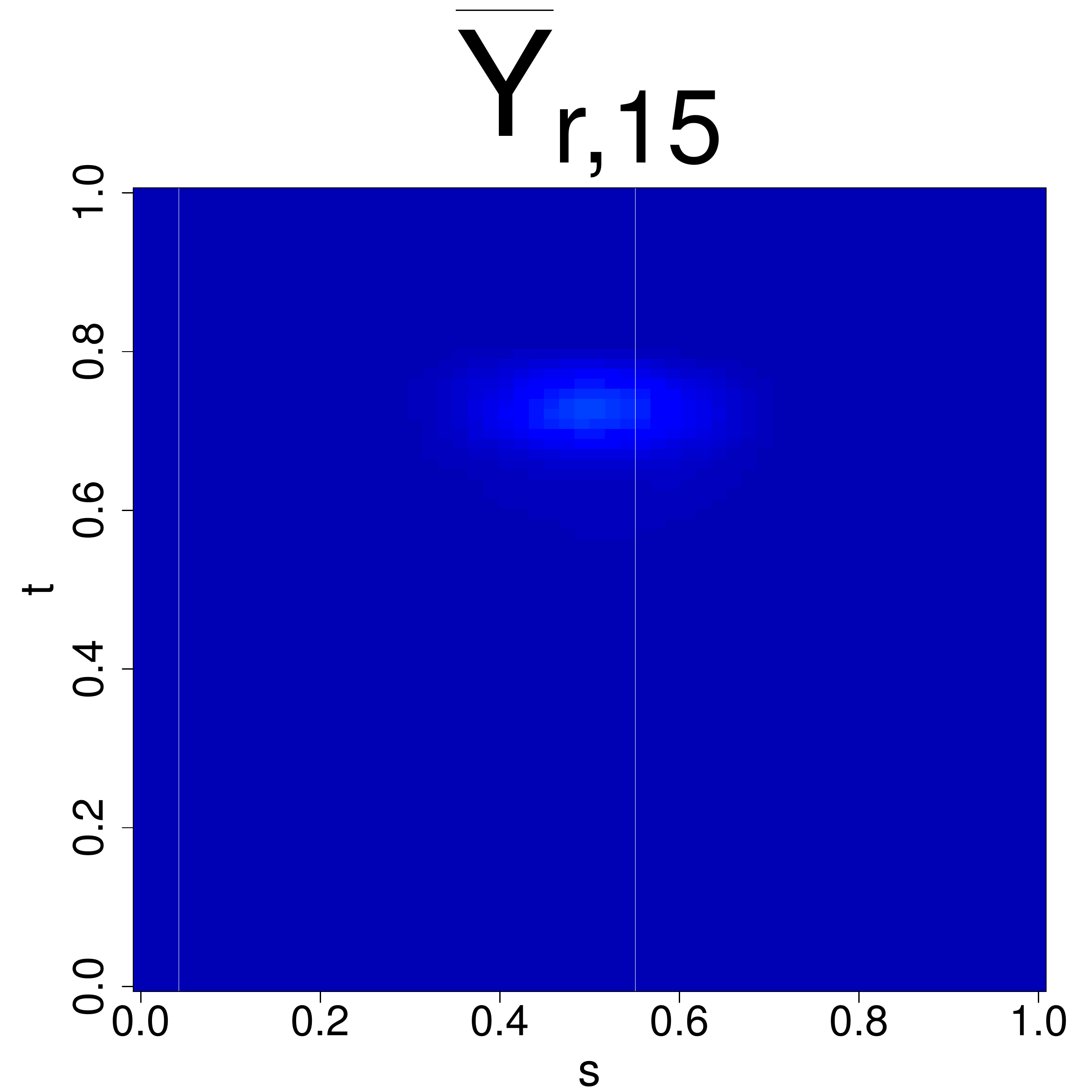}&\includegraphics[width=0.1\textwidth]{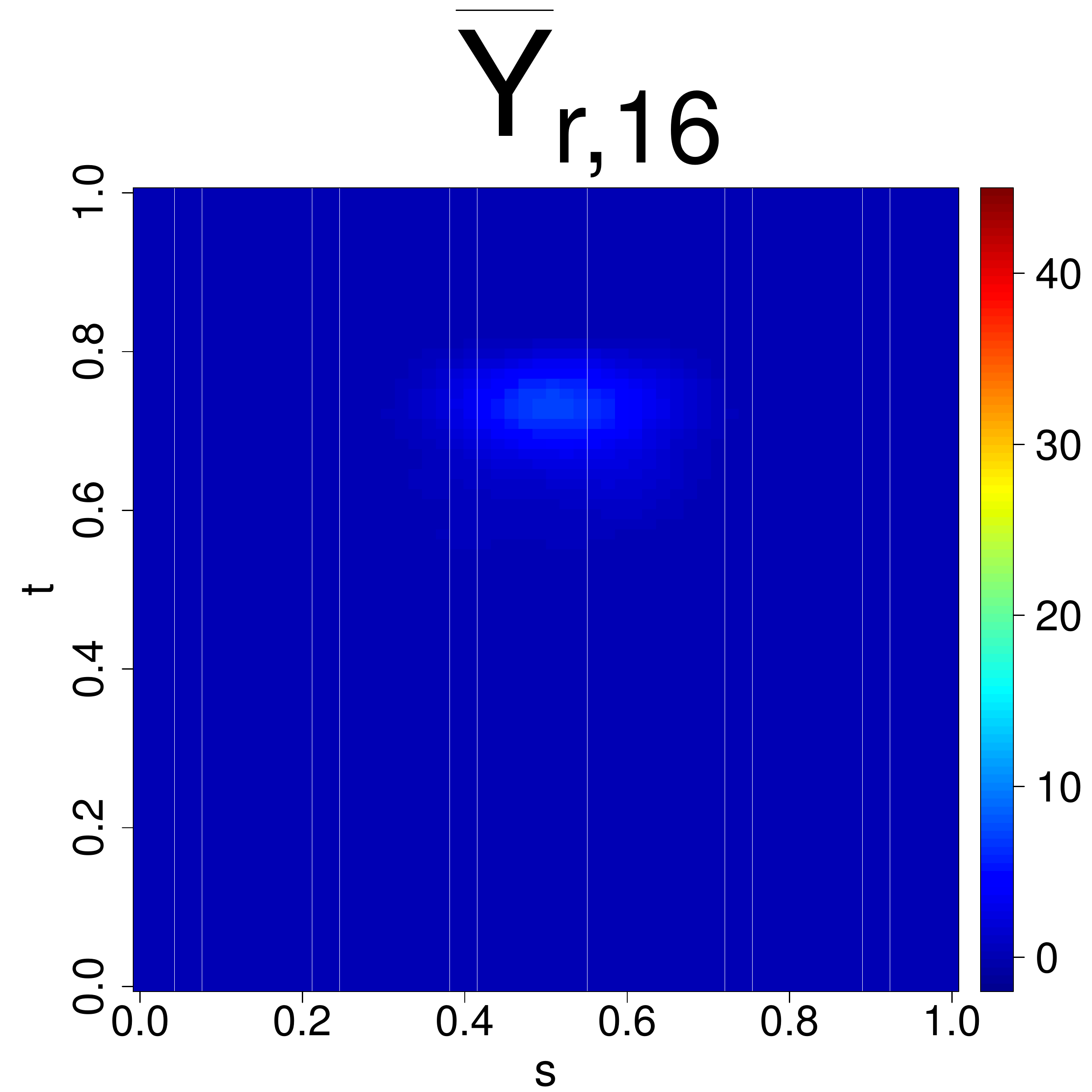}\\
	2&\includegraphics[width=0.1\textwidth]{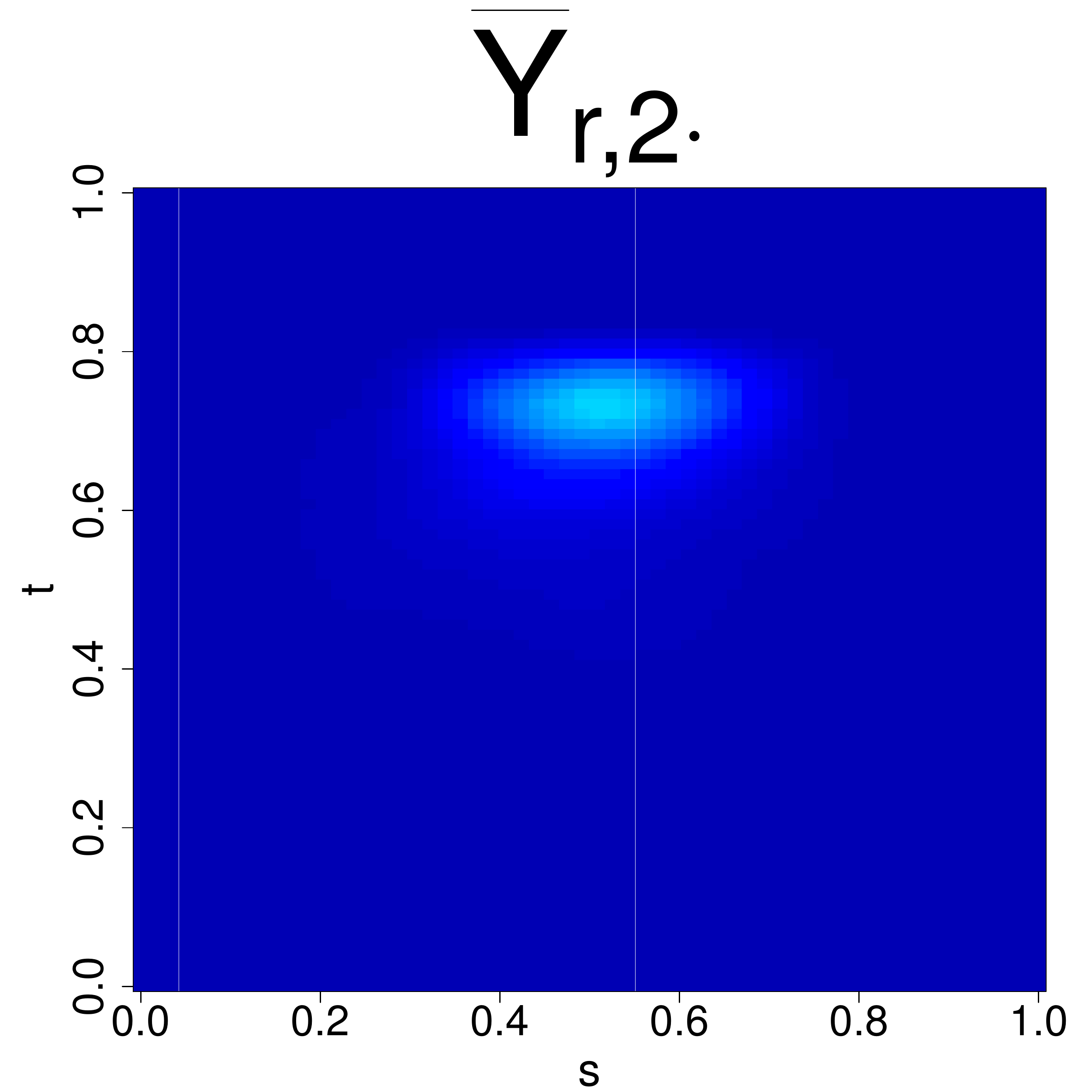}&\includegraphics[width=0.1\textwidth]{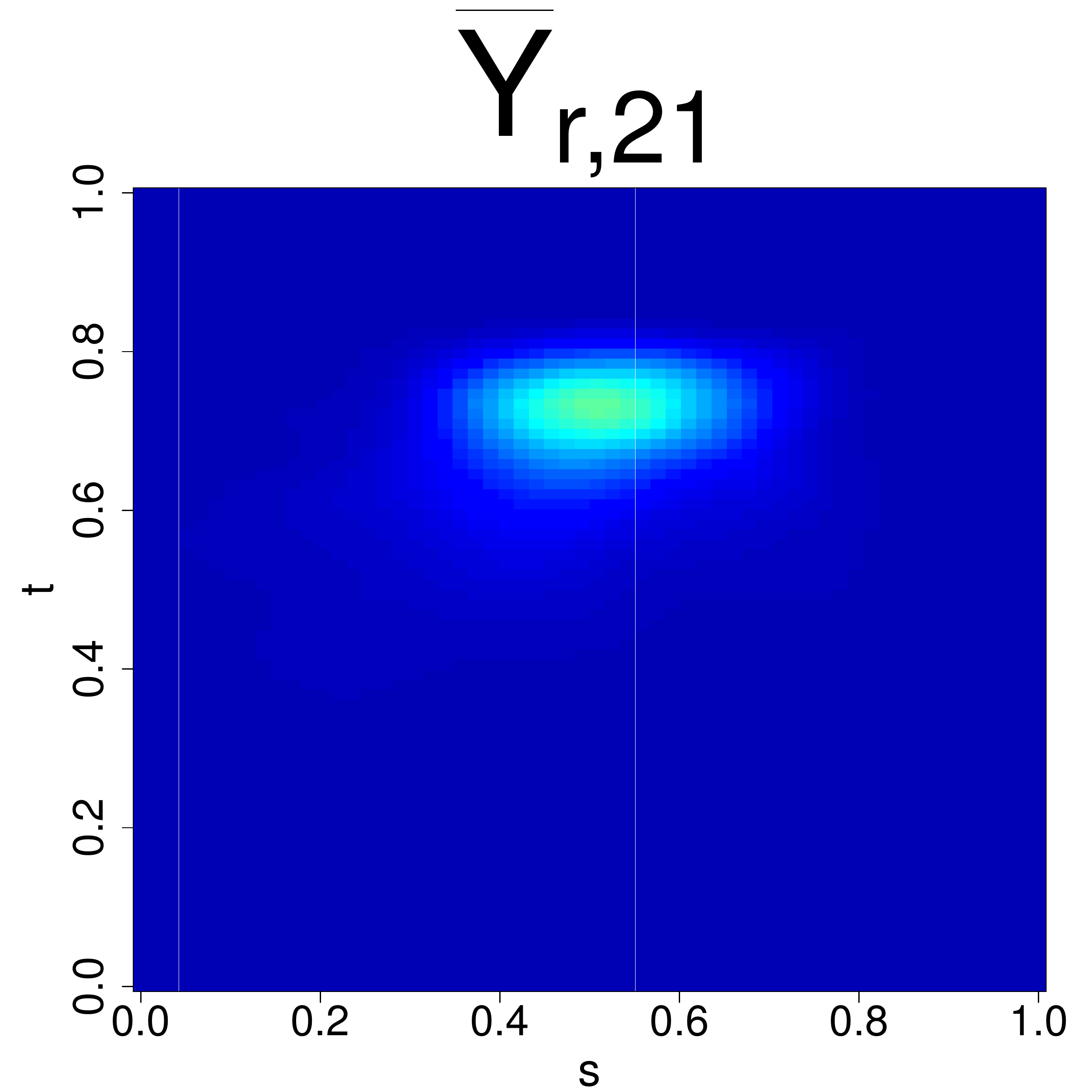}&\includegraphics[width=0.1\textwidth]{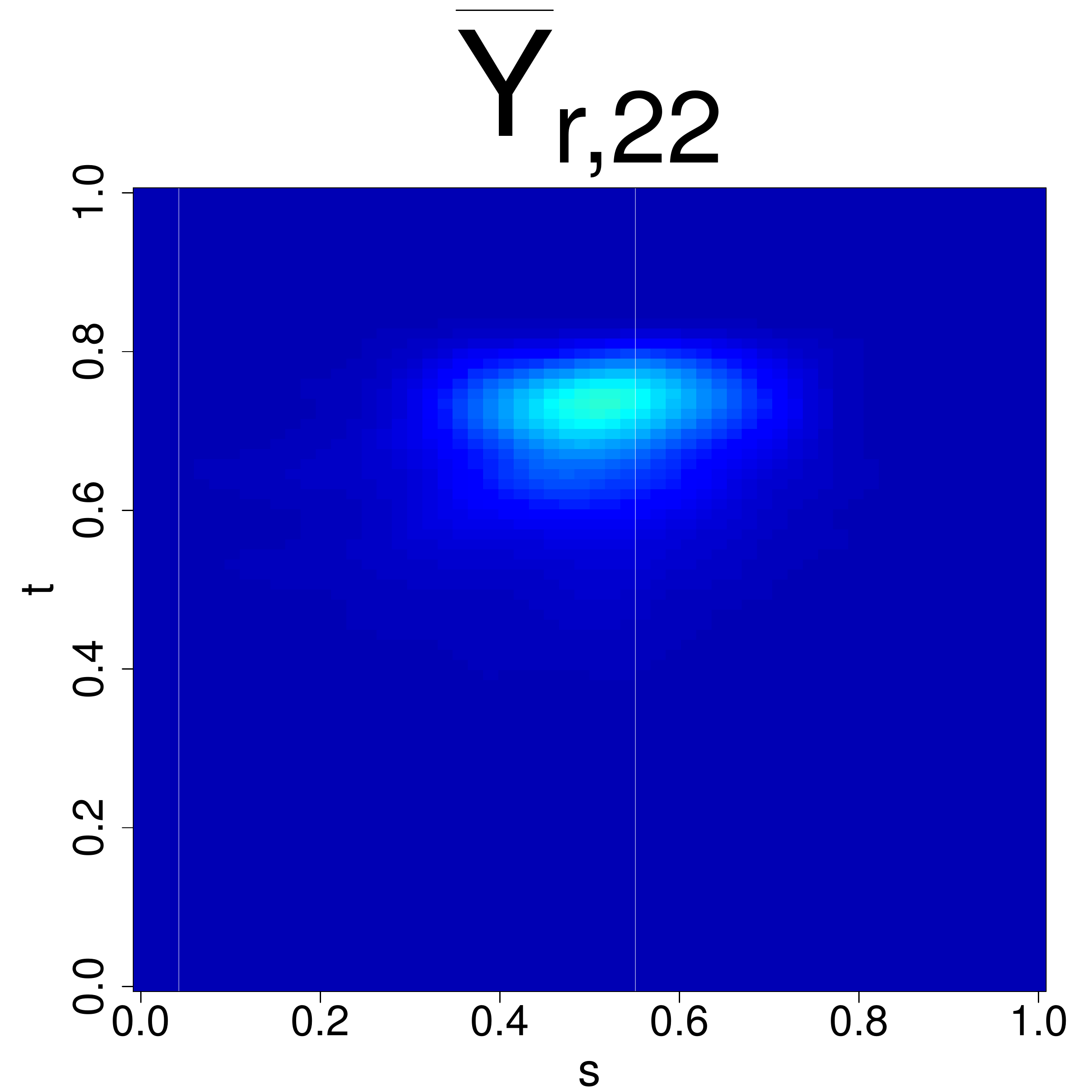}&\includegraphics[width=0.1\textwidth]{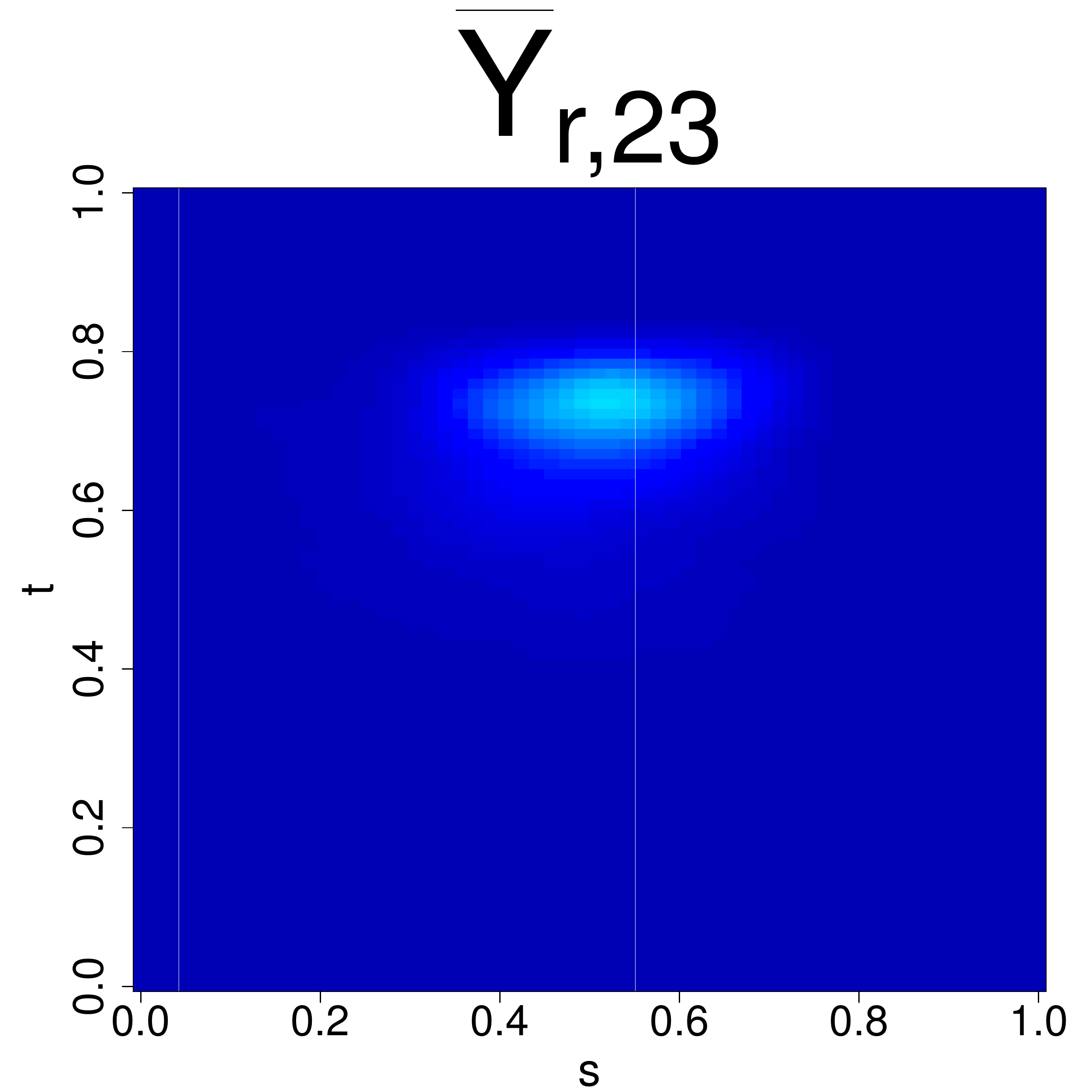}&\includegraphics[width=0.1\textwidth]{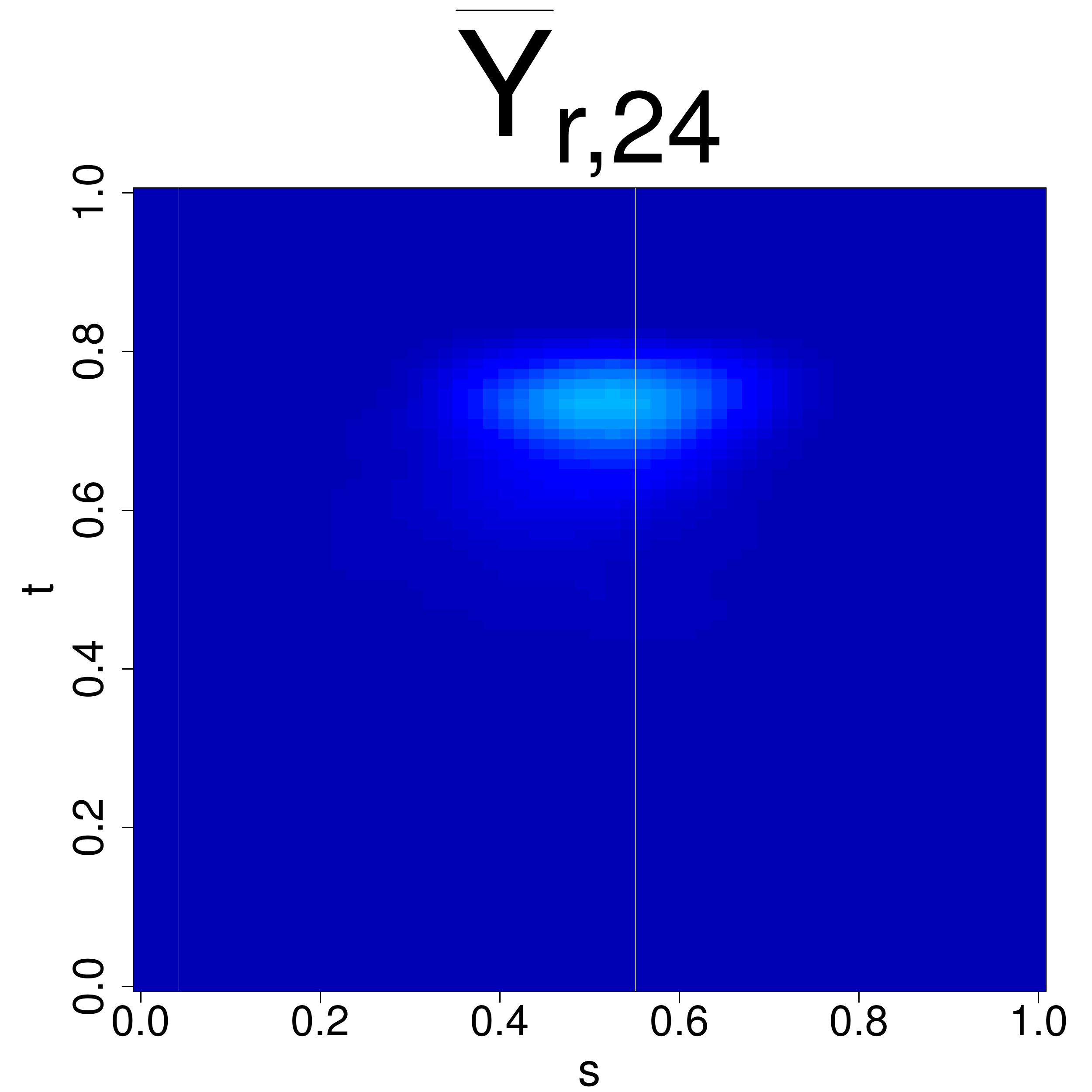}&\includegraphics[width=0.1\textwidth]{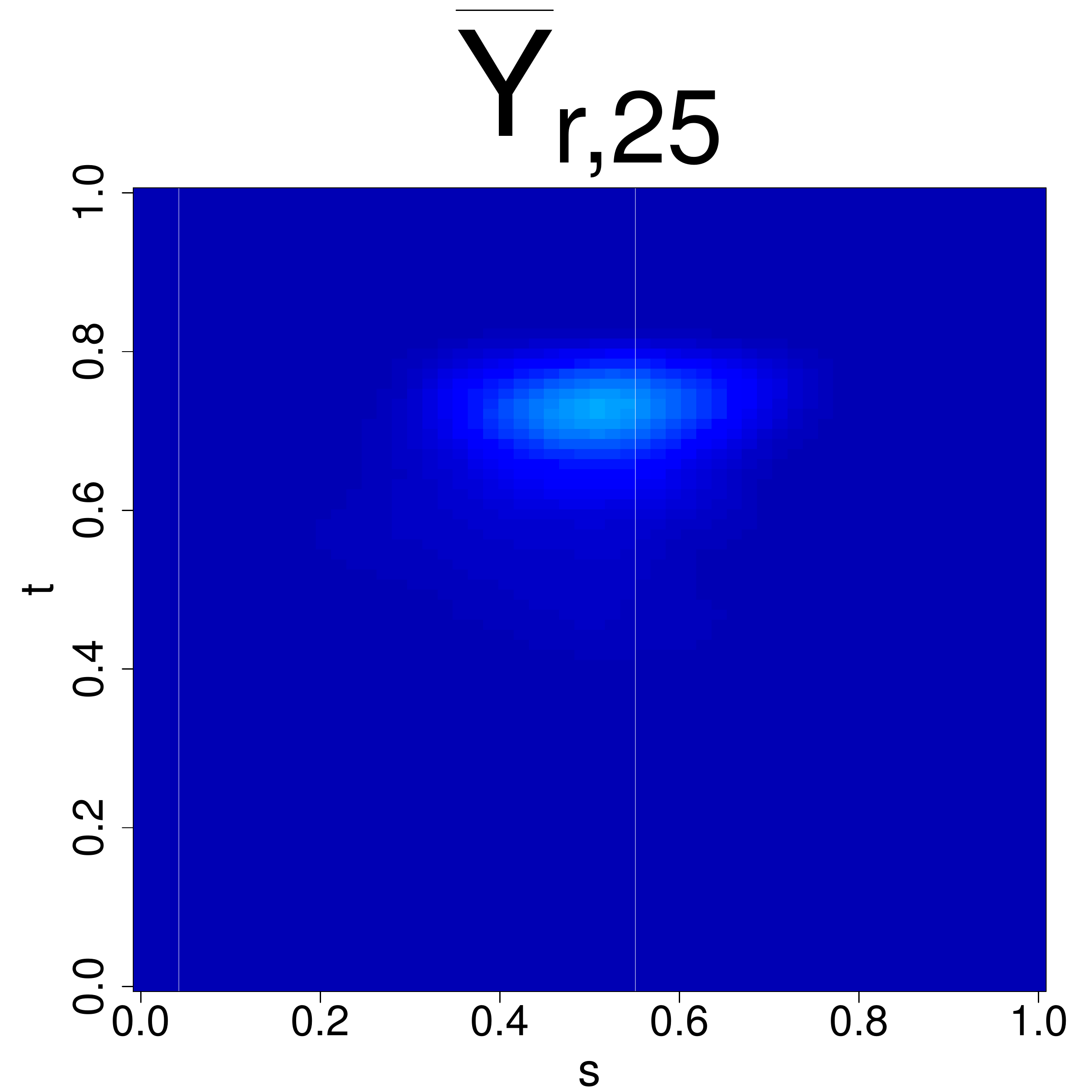}&\includegraphics[width=0.1\textwidth]{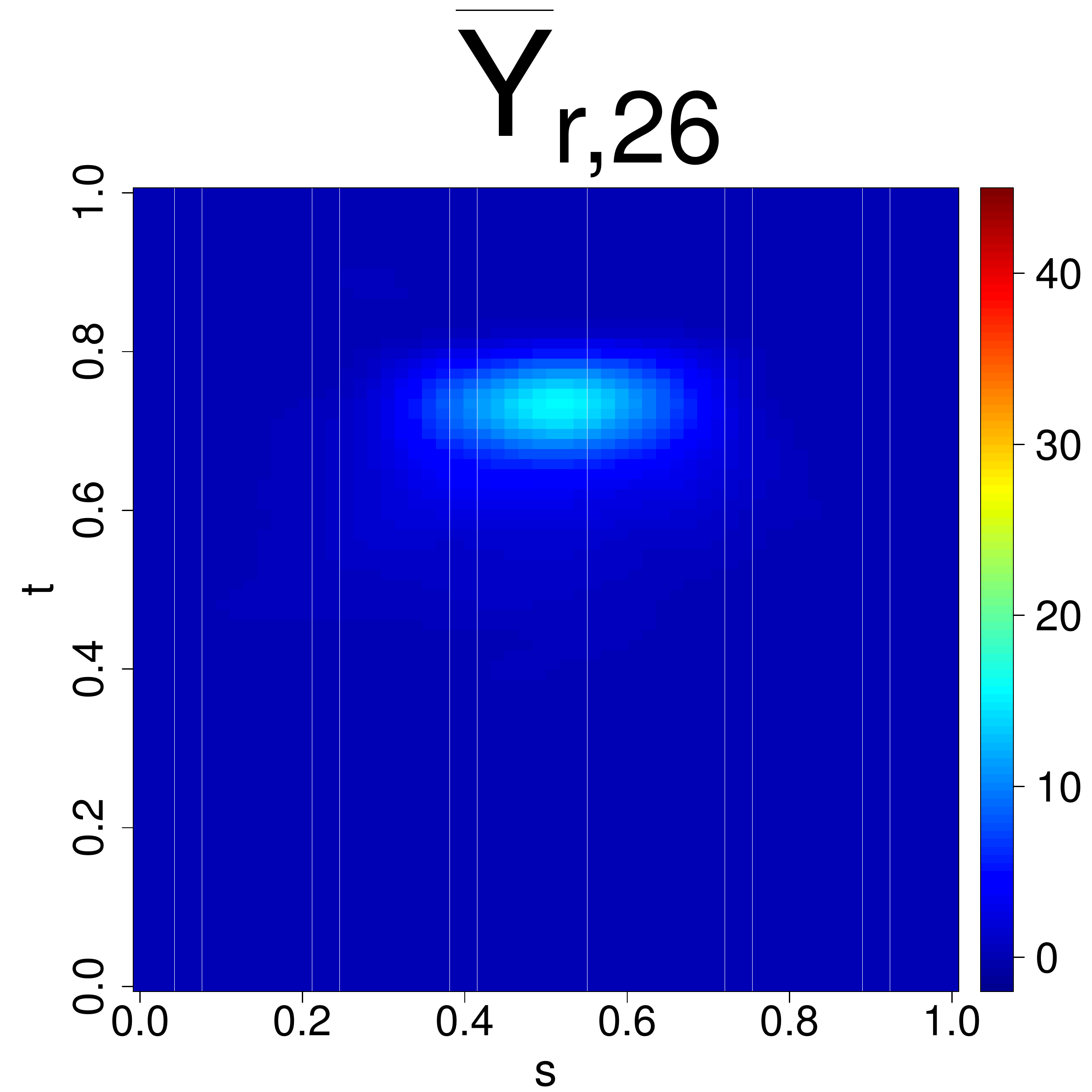}\\
	3&\includegraphics[width=0.1\textwidth]{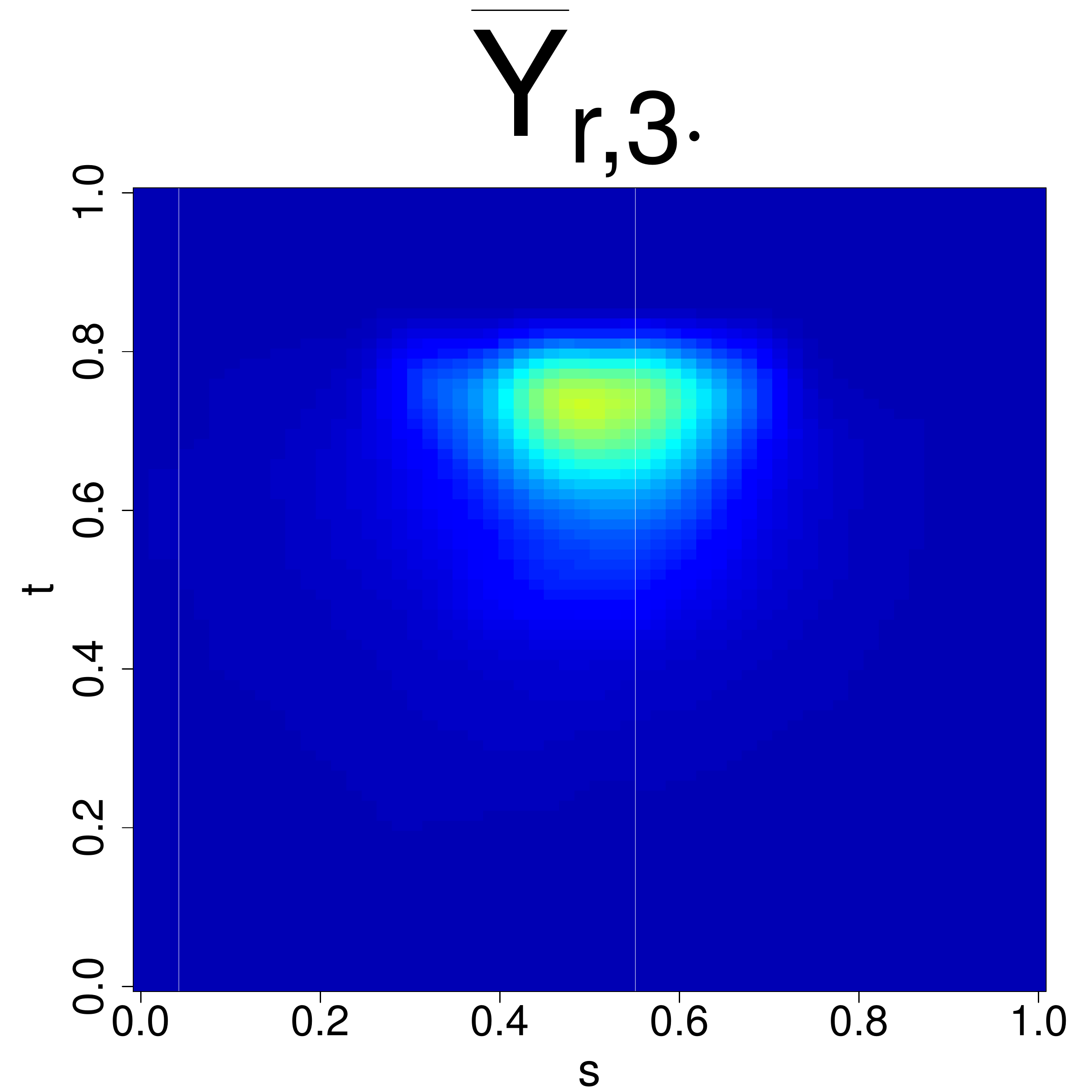}&\includegraphics[width=0.1\textwidth]{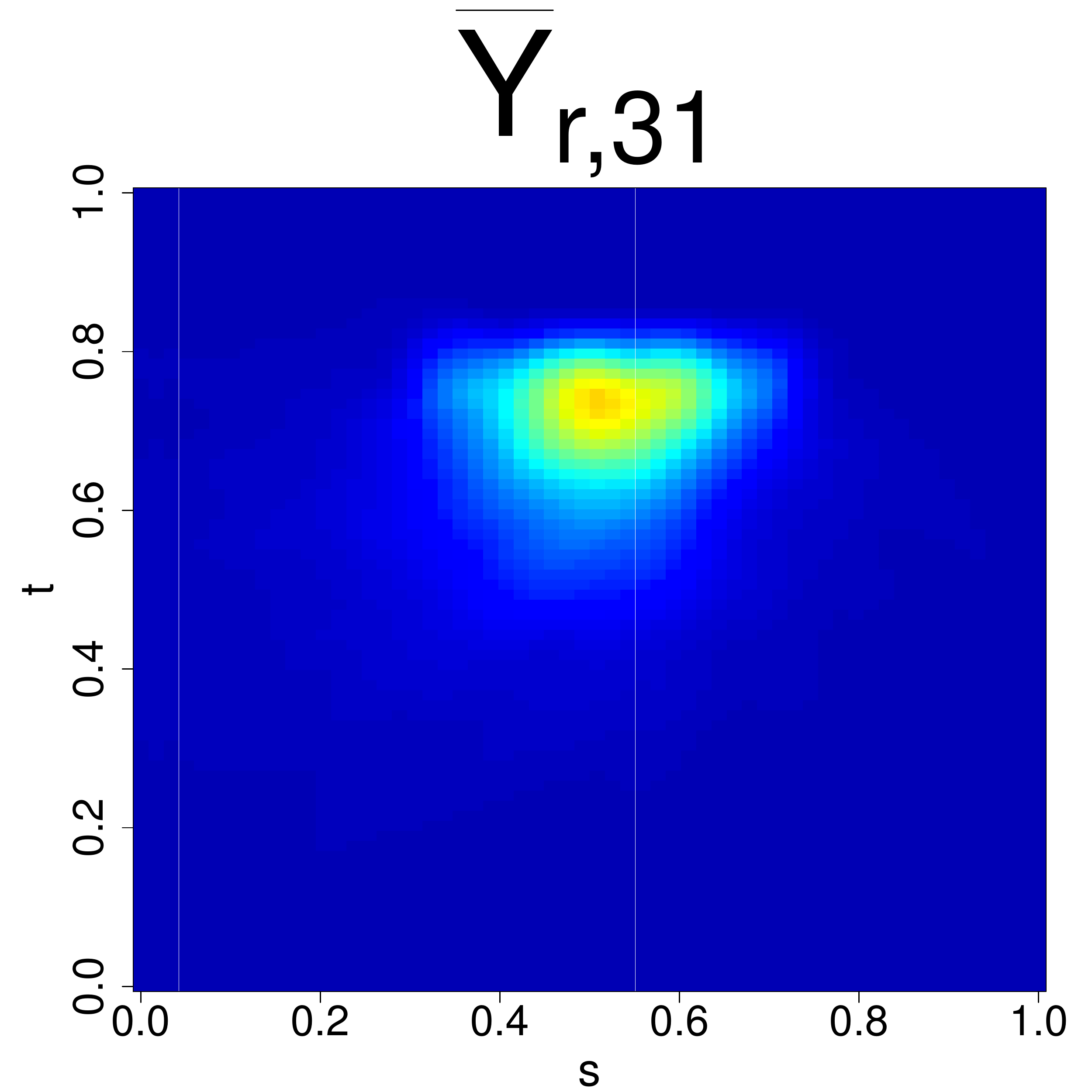}&\includegraphics[width=0.1\textwidth]{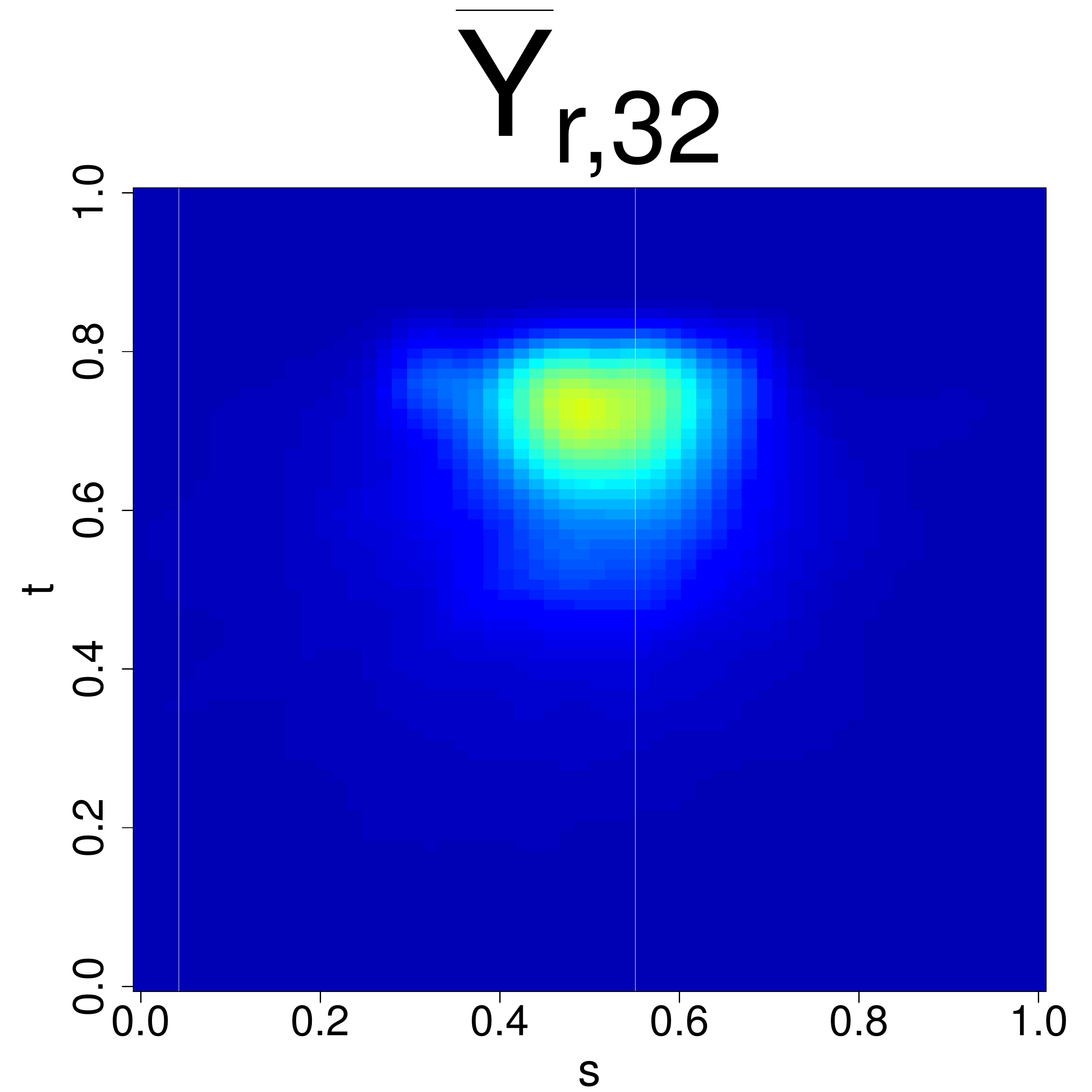}&\includegraphics[width=0.1\textwidth]{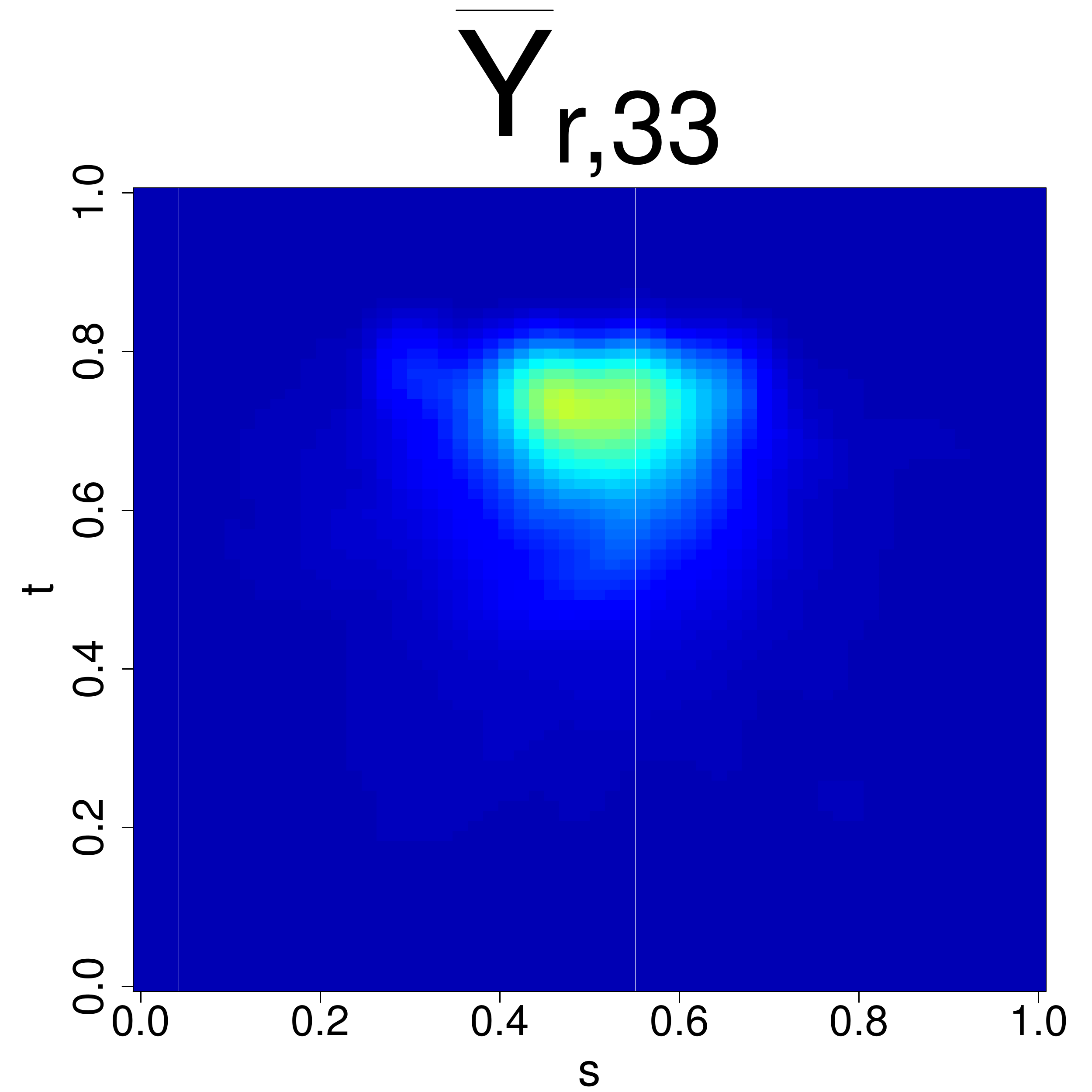}&\includegraphics[width=0.1\textwidth]{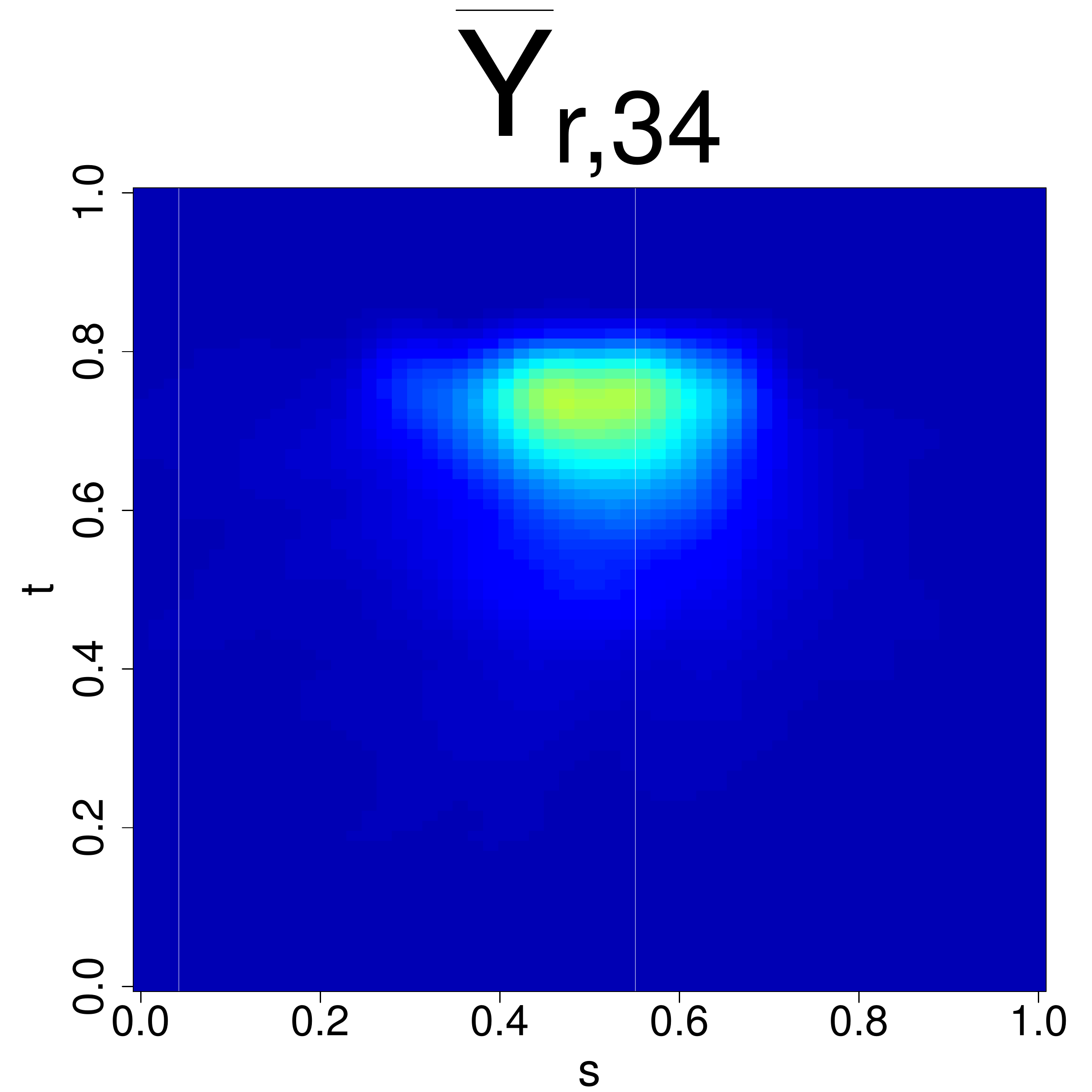}&\includegraphics[width=0.1\textwidth]{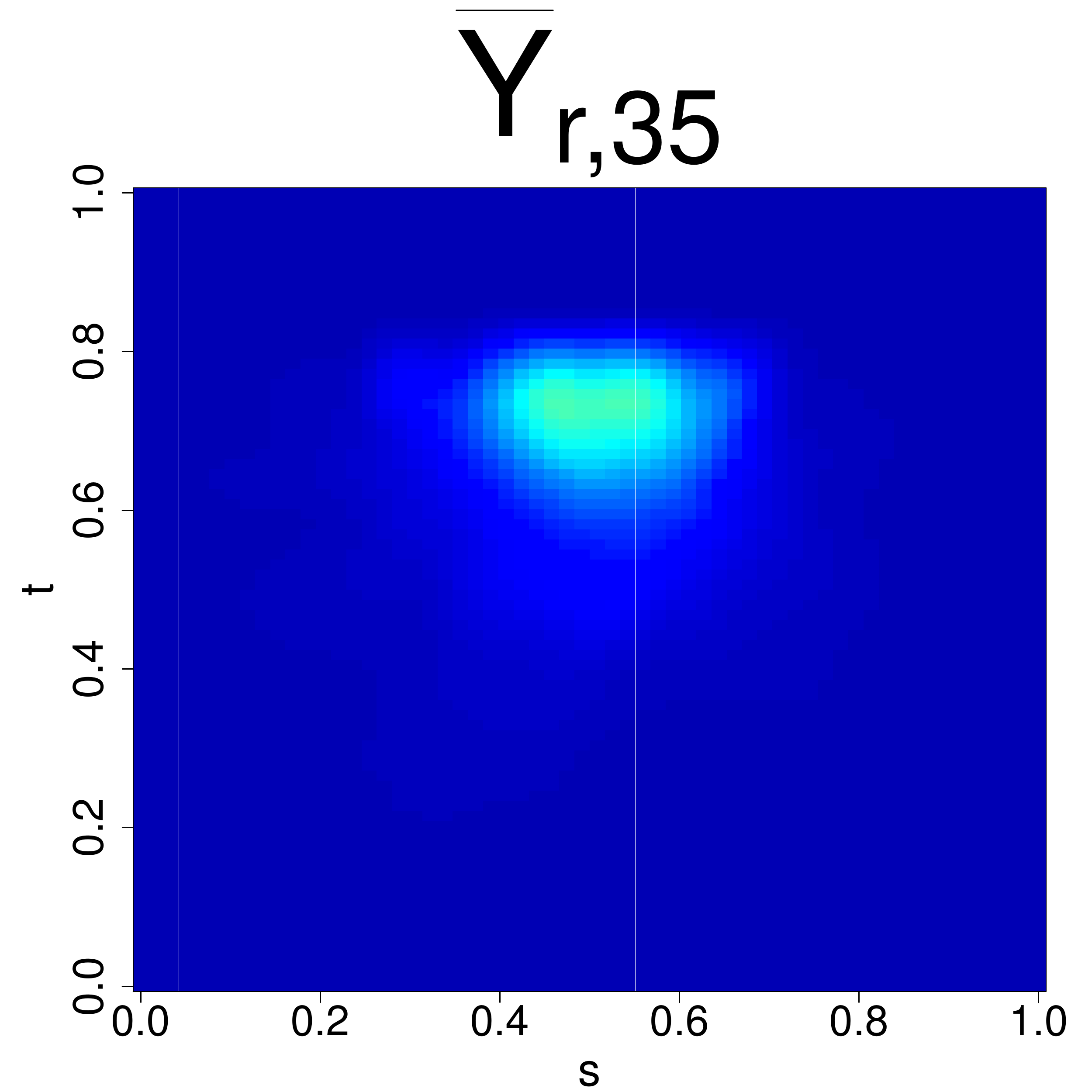}&\includegraphics[width=0.1\textwidth]{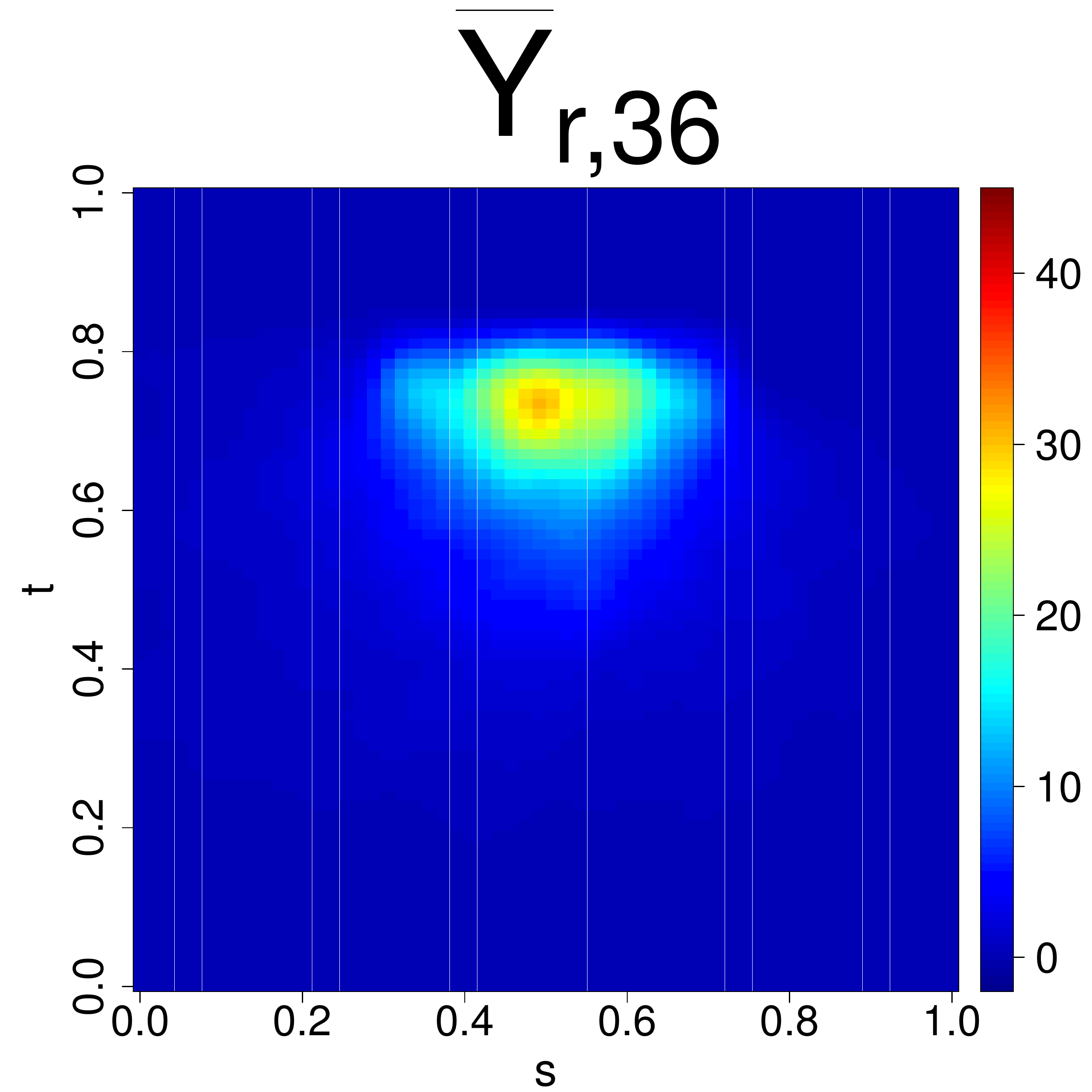}\\
4	&\includegraphics[width=0.1\textwidth]{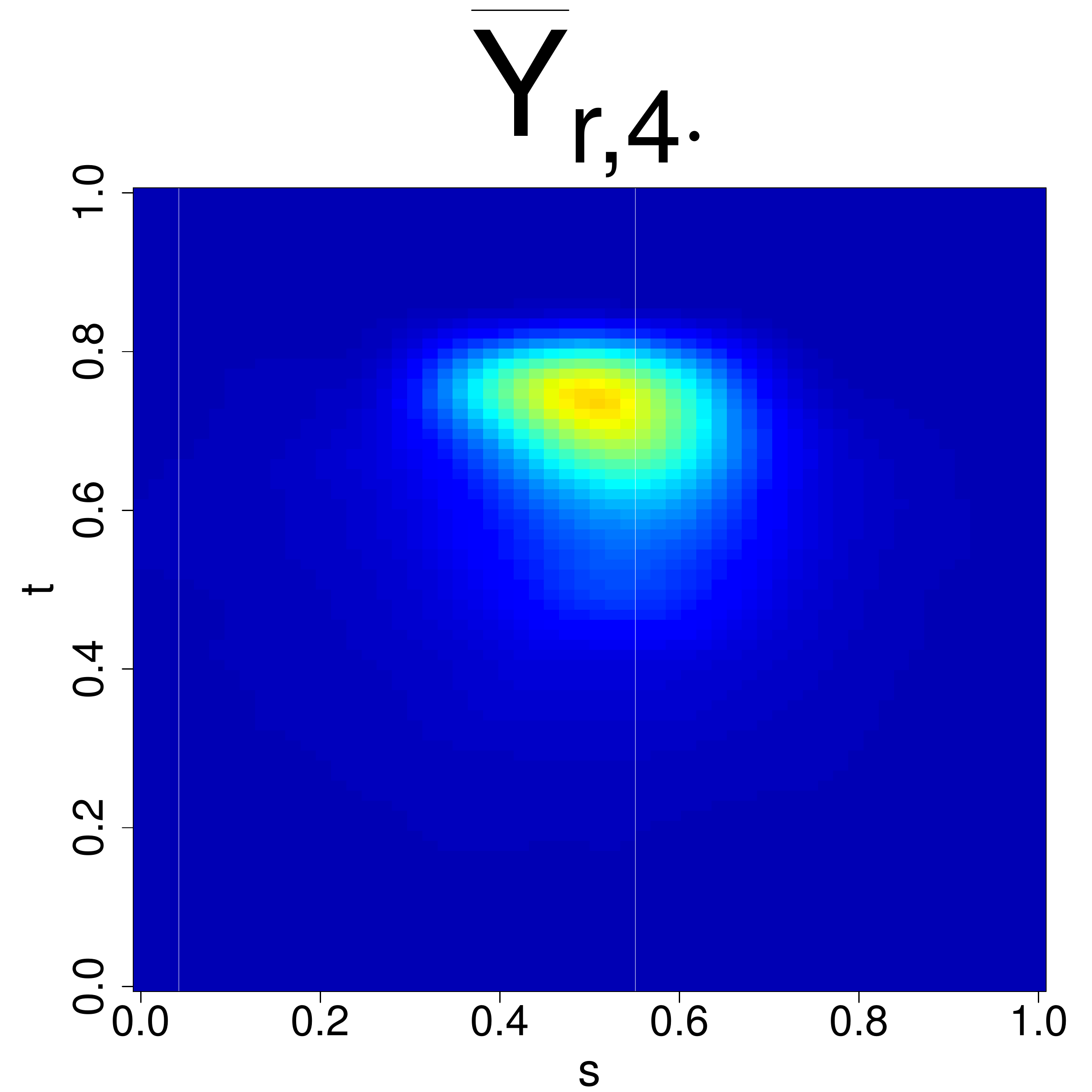}&\includegraphics[width=0.1\textwidth]{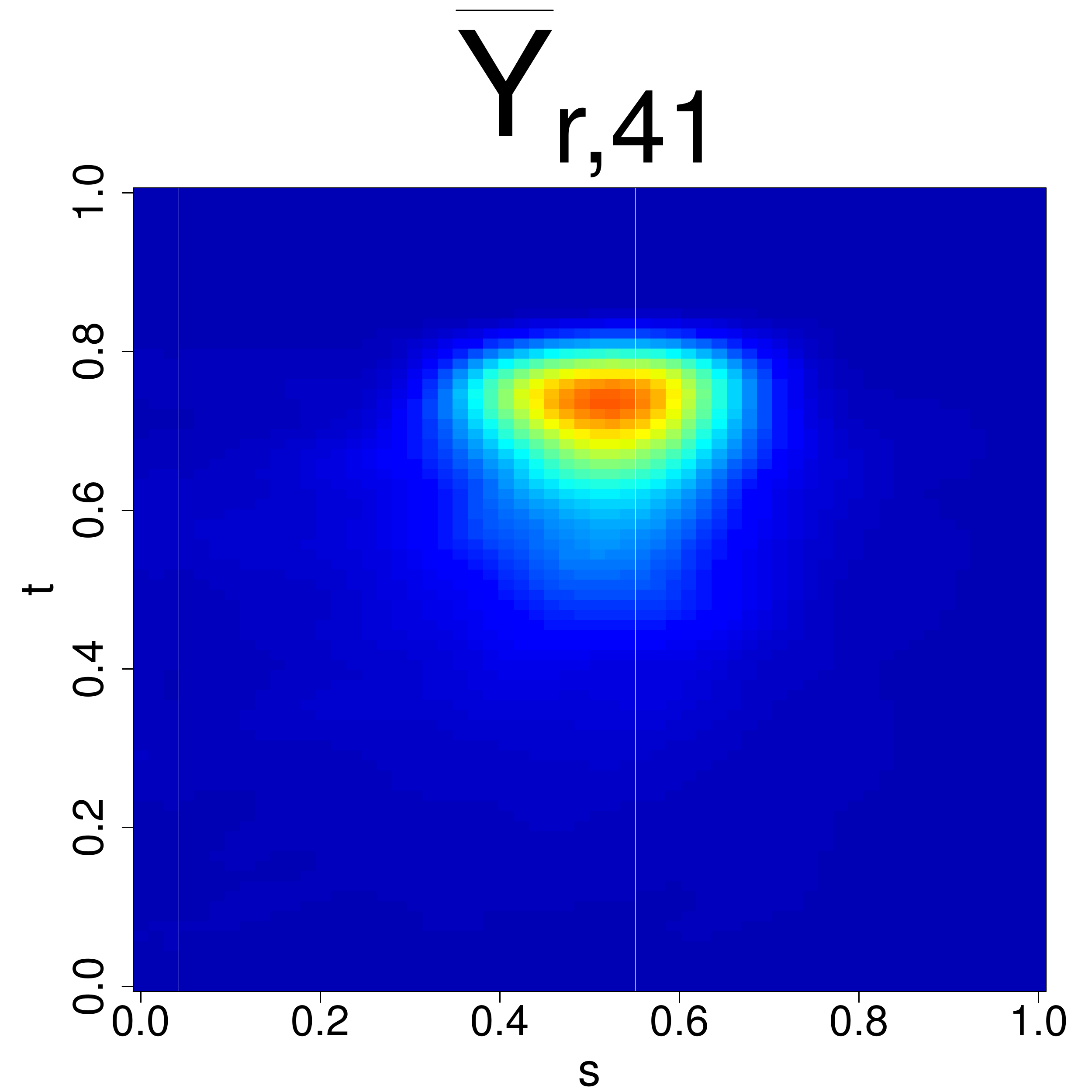}&\includegraphics[width=0.1\textwidth]{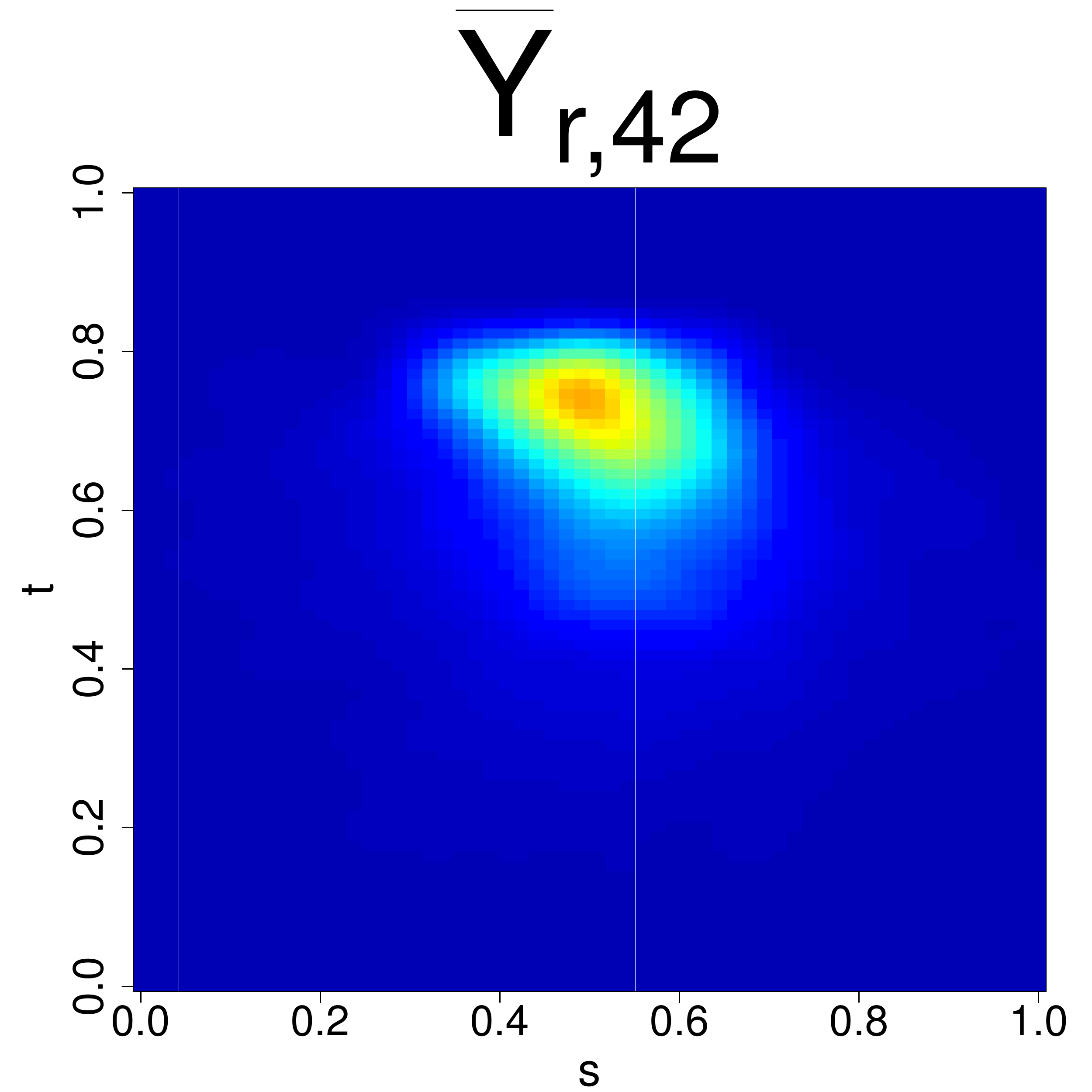}&\includegraphics[width=0.1\textwidth]{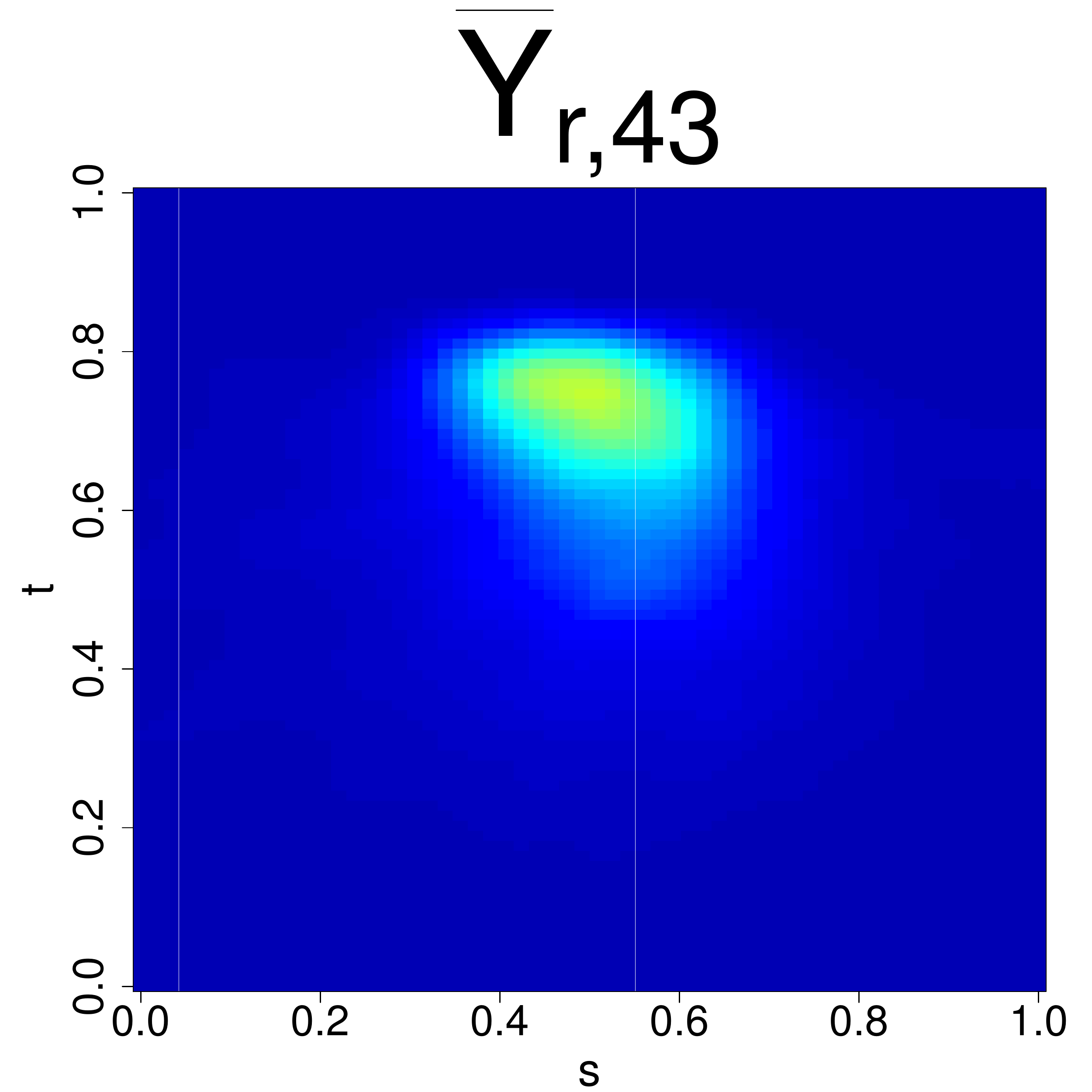}&\includegraphics[width=0.1\textwidth]{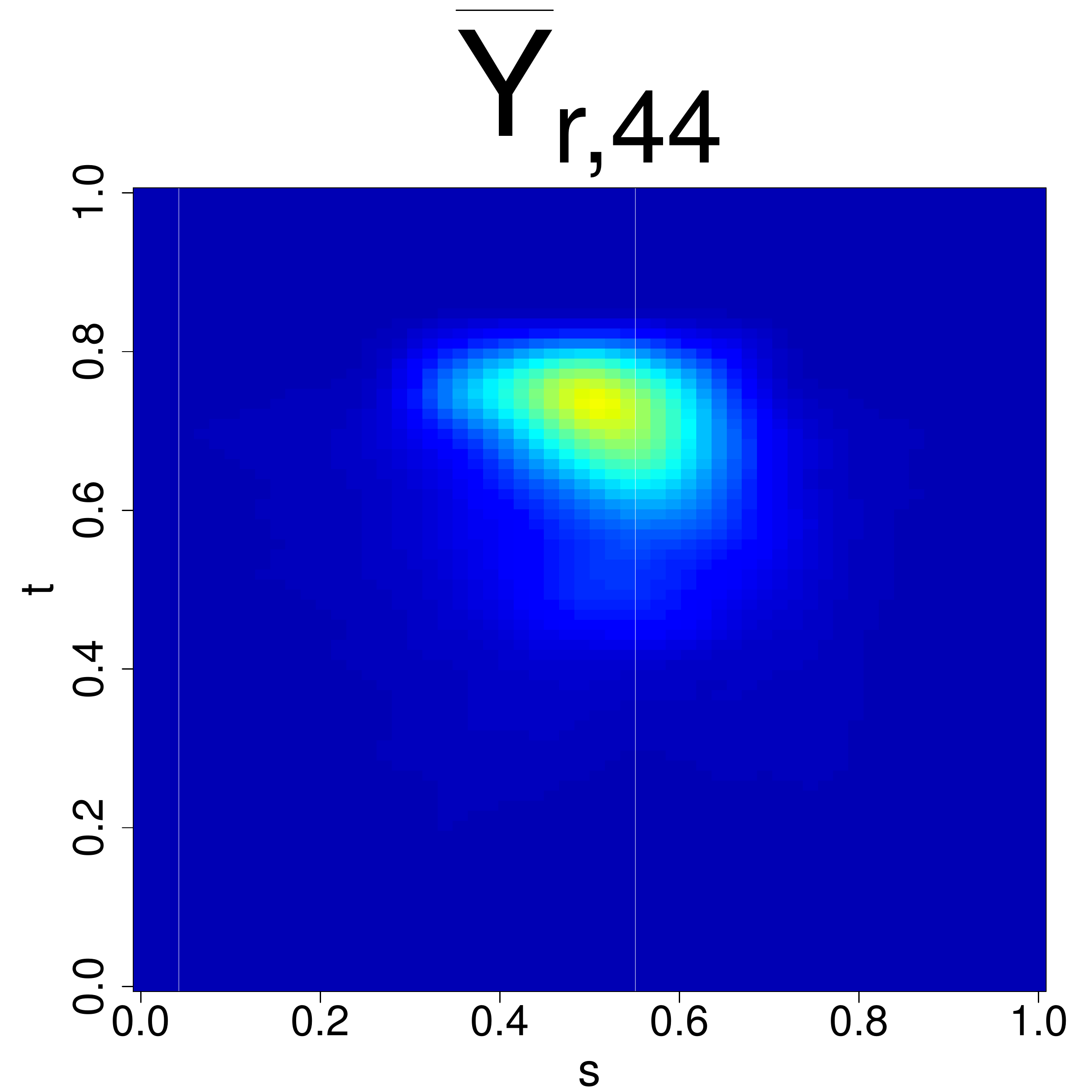}&\includegraphics[width=0.1\textwidth]{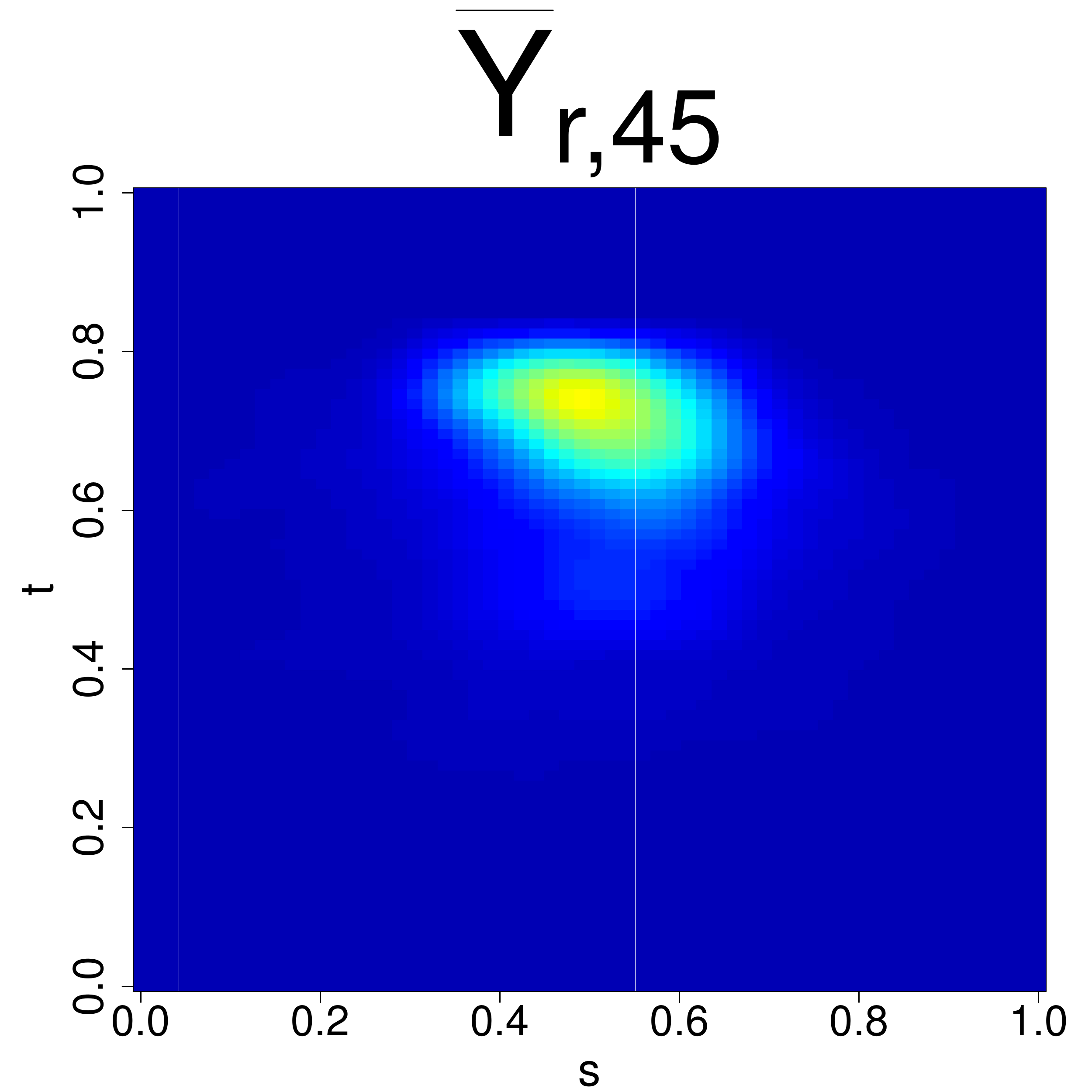}&\includegraphics[width=0.1\textwidth]{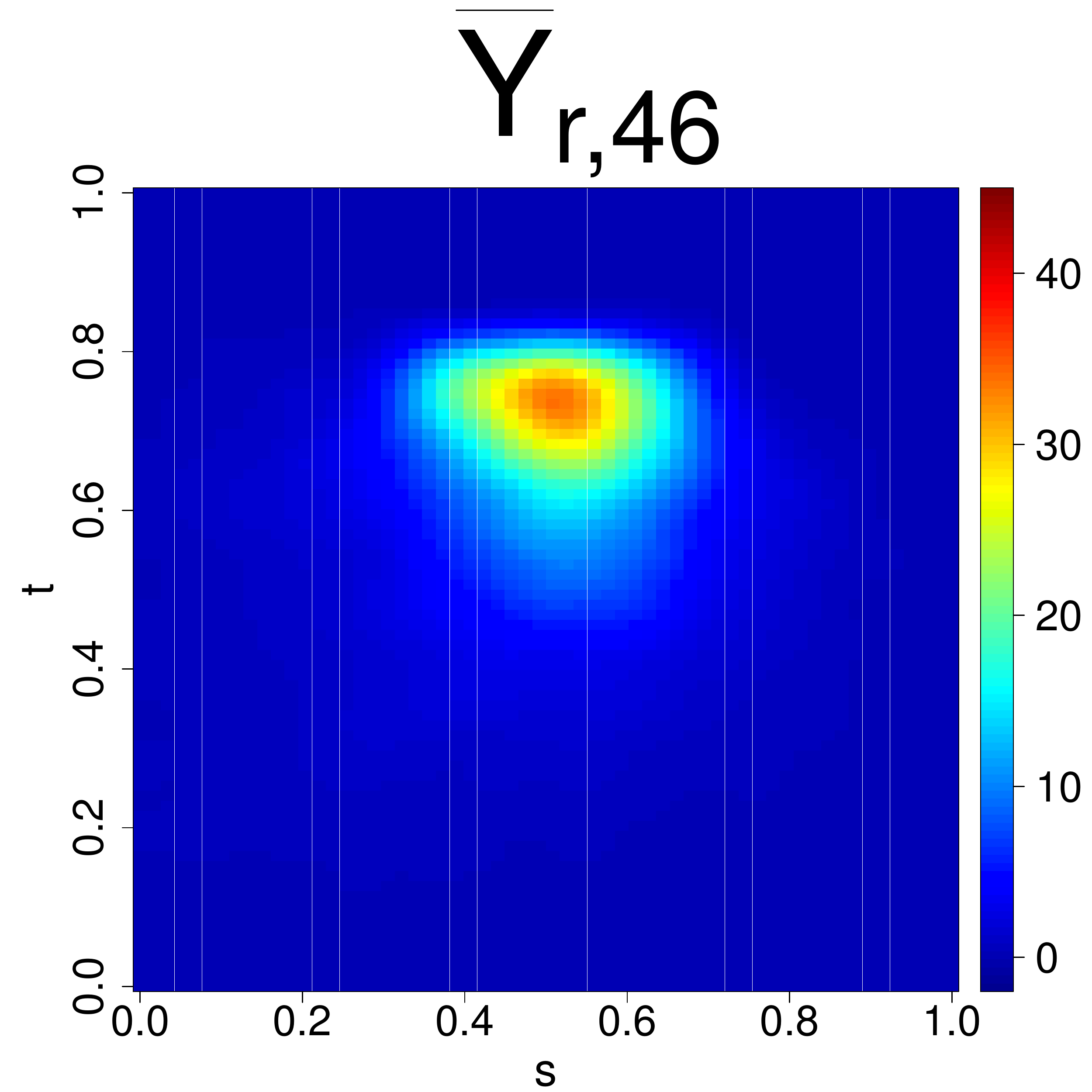}\\
	5&\includegraphics[width=0.1\textwidth]{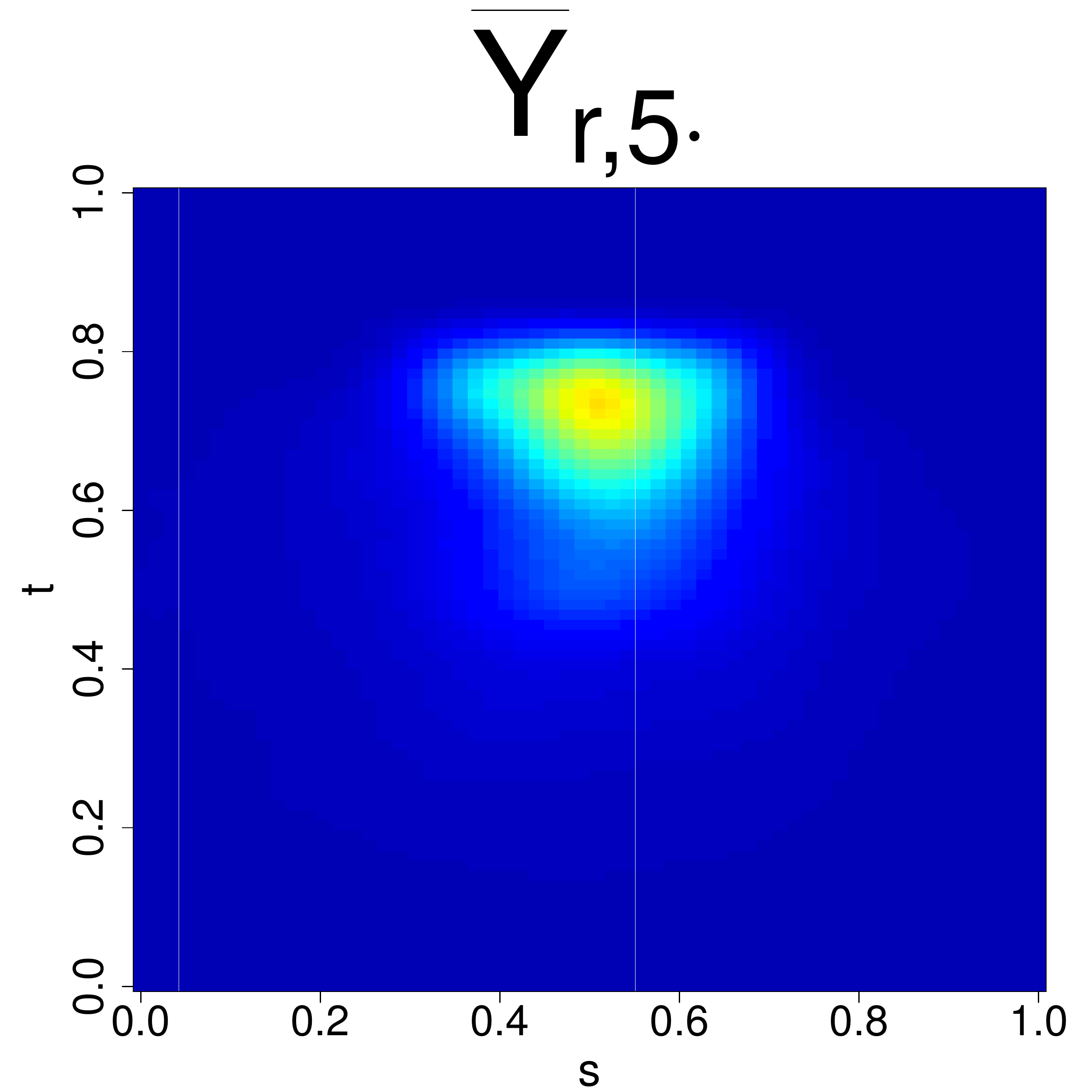}&\includegraphics[width=0.1\textwidth]{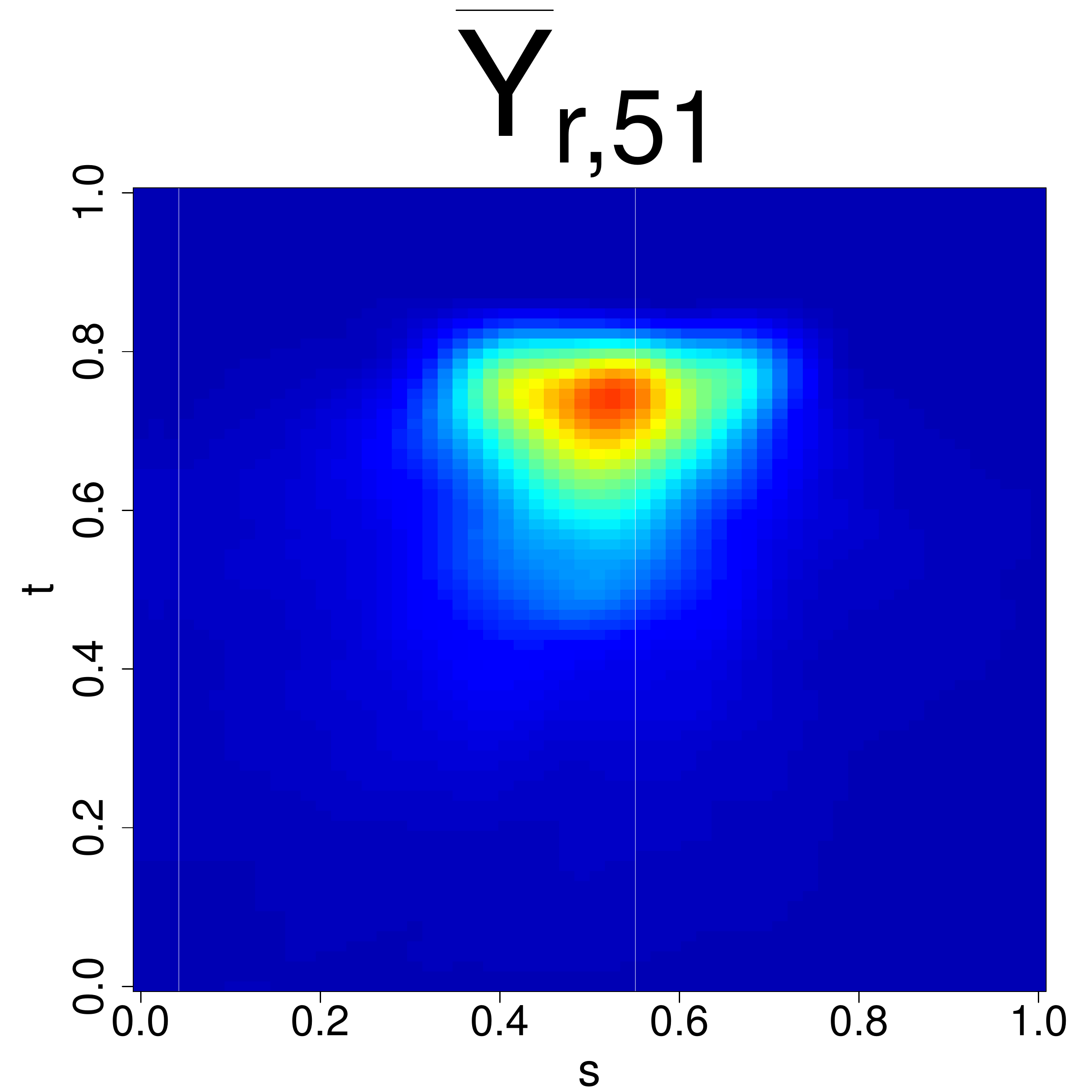}&\includegraphics[width=0.1\textwidth]{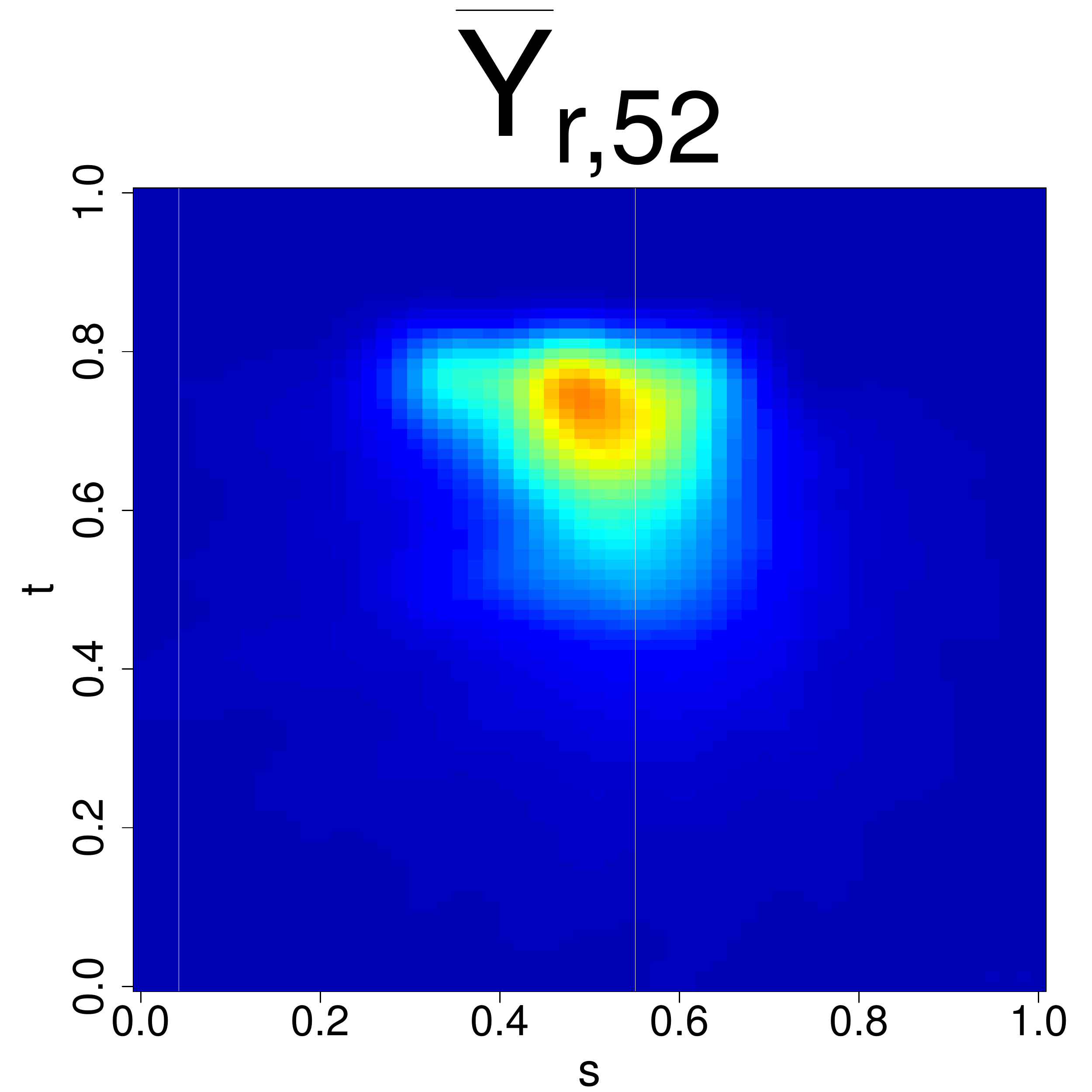}&\includegraphics[width=0.1\textwidth]{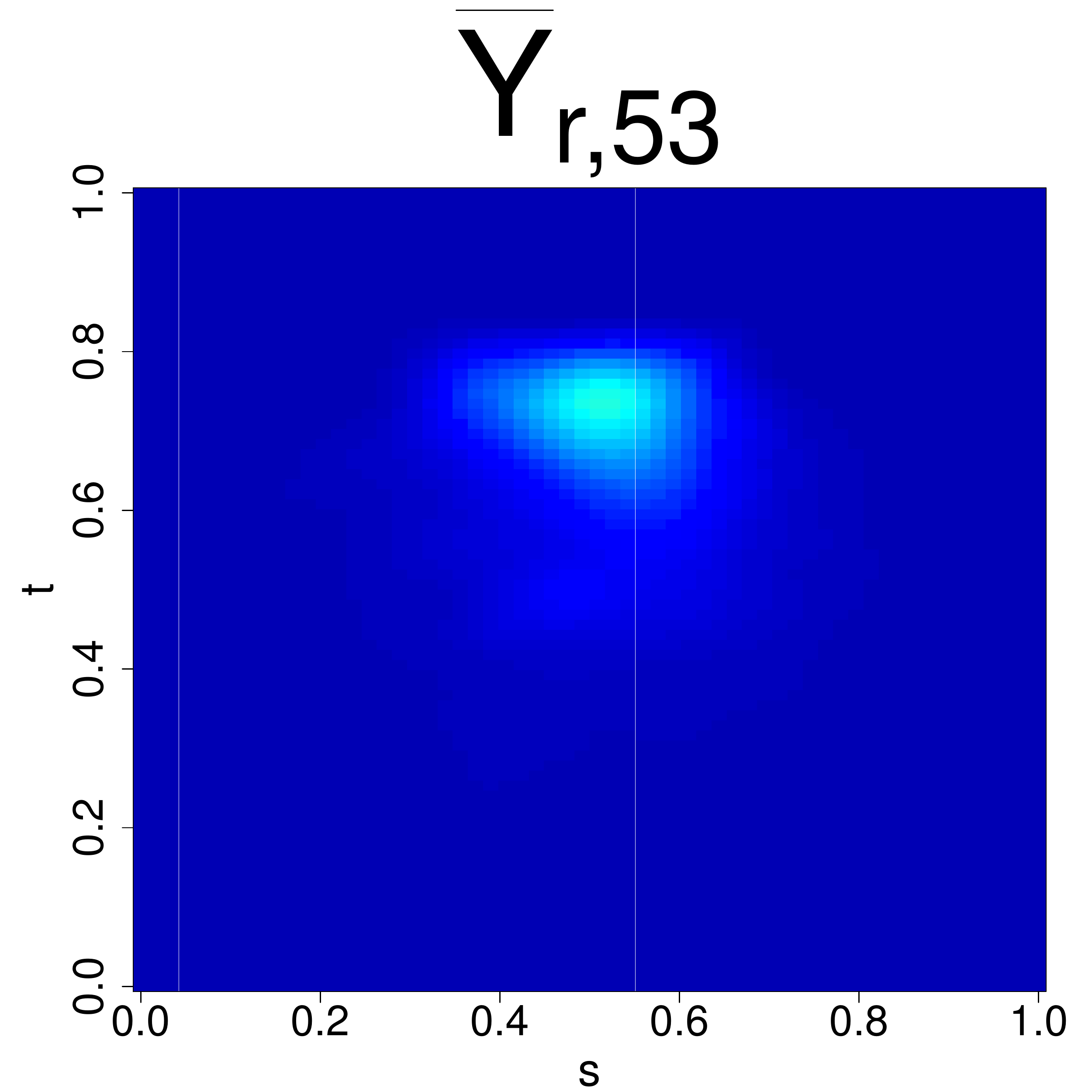}&\includegraphics[width=0.1\textwidth]{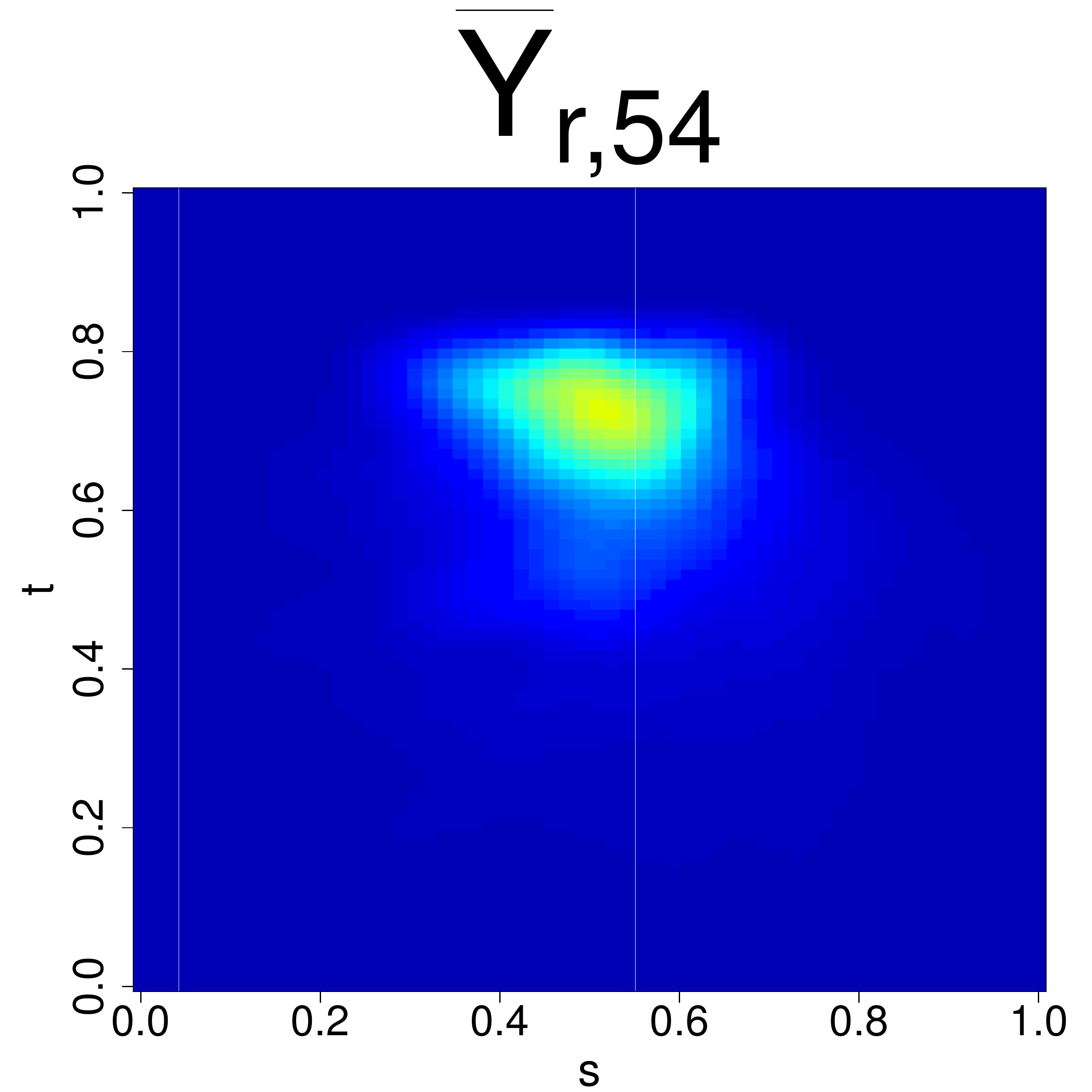}&\includegraphics[width=0.1\textwidth]{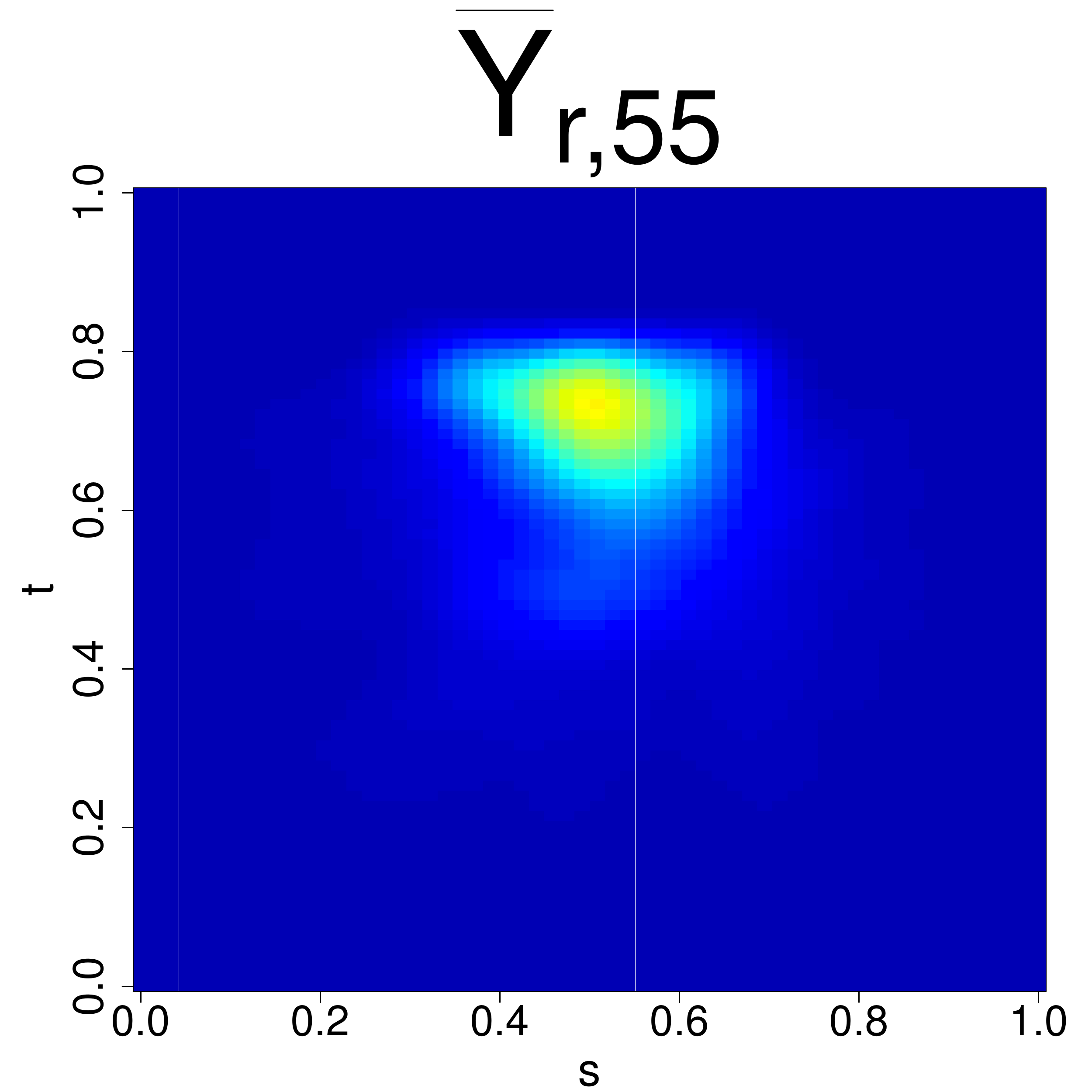}&\includegraphics[width=0.1\textwidth]{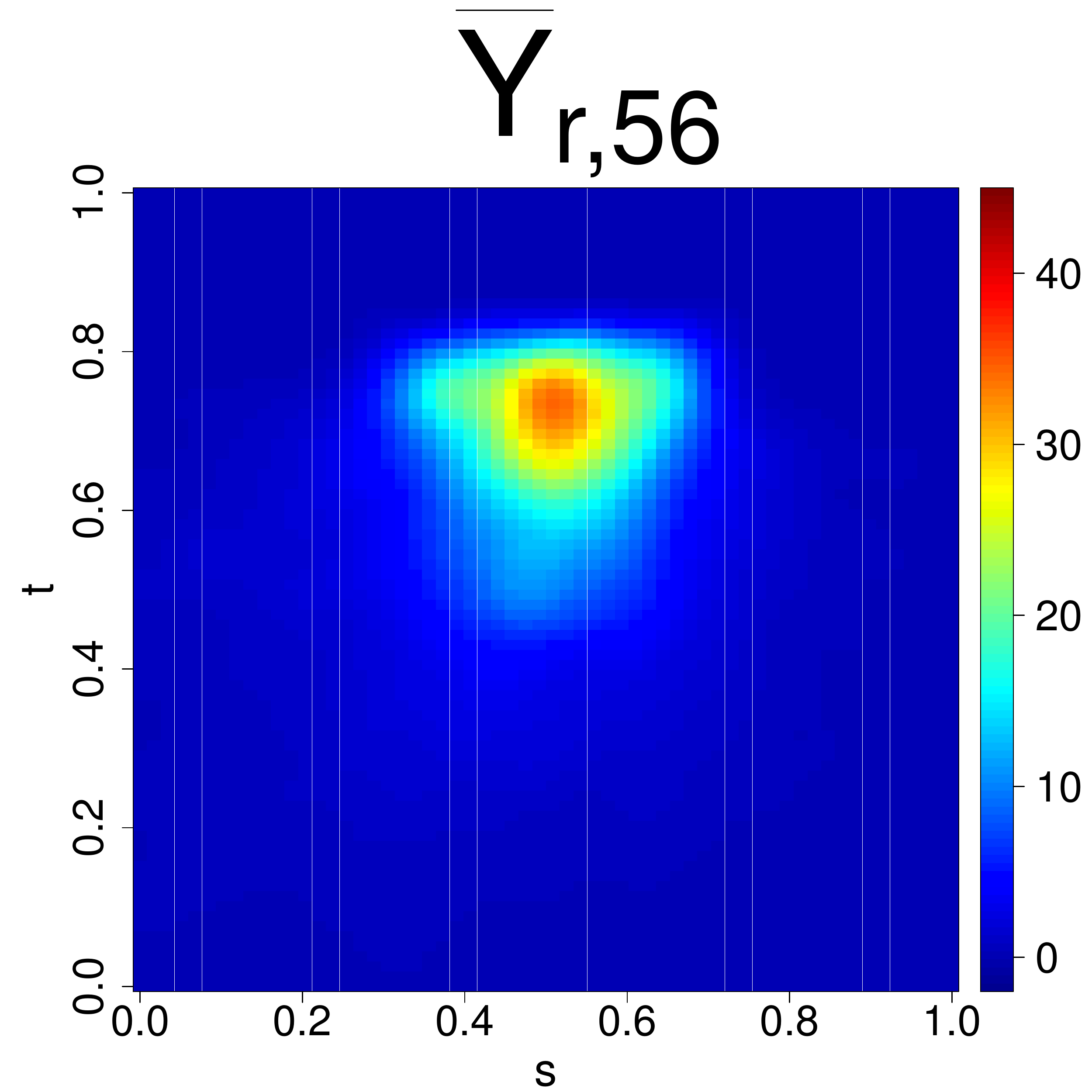}\\
6&\includegraphics[width=0.1\textwidth]{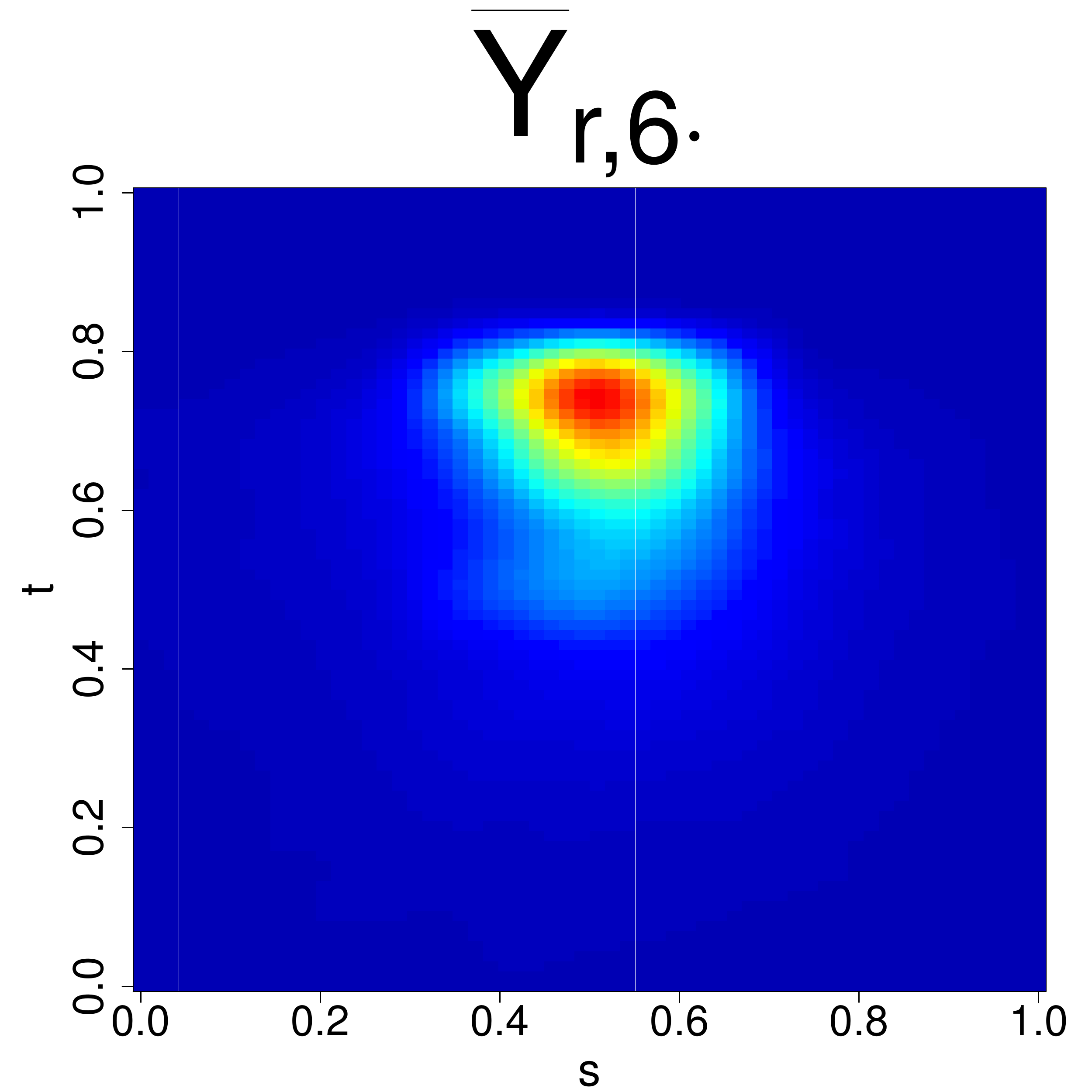}&\includegraphics[width=0.1\textwidth]{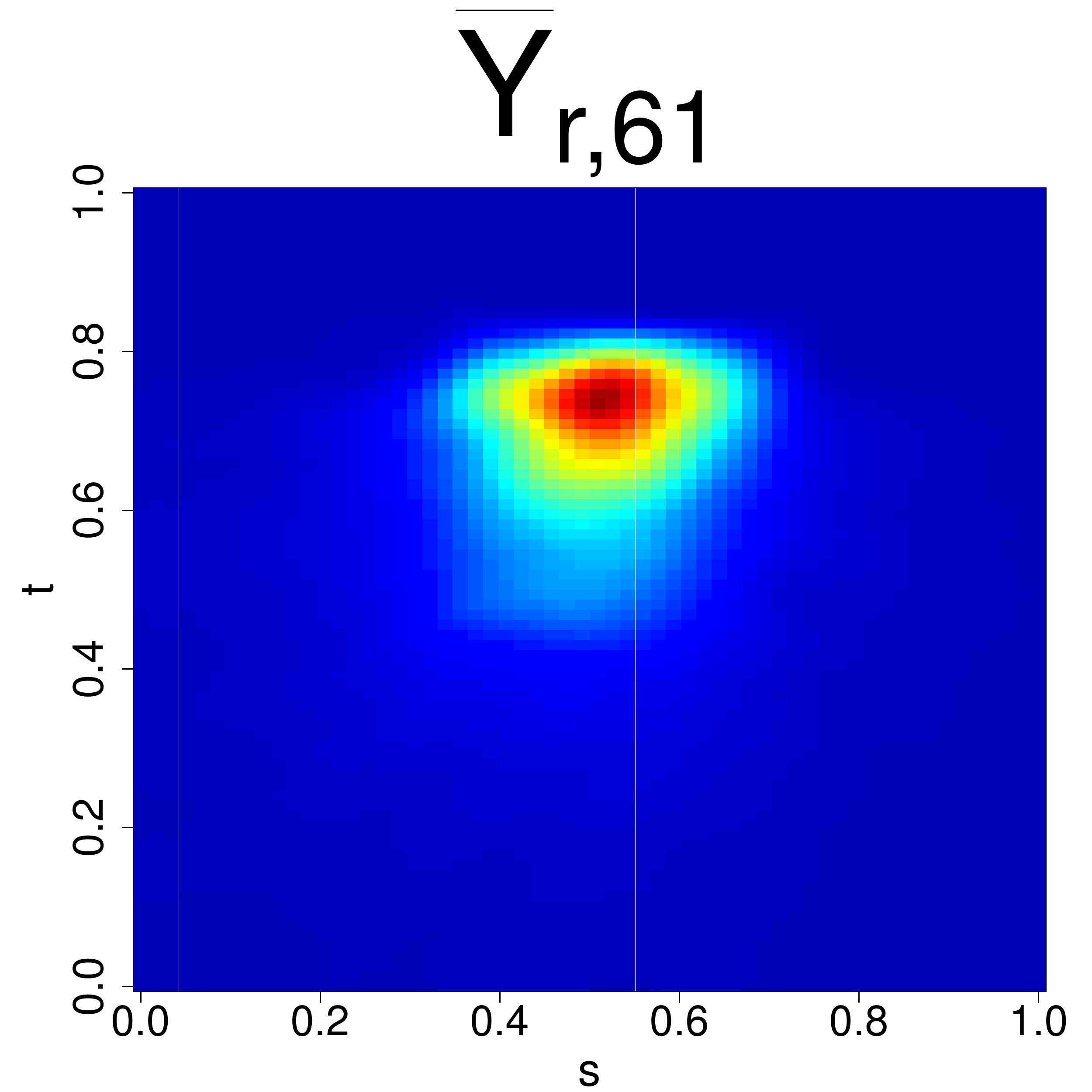}&\includegraphics[width=0.1\textwidth]{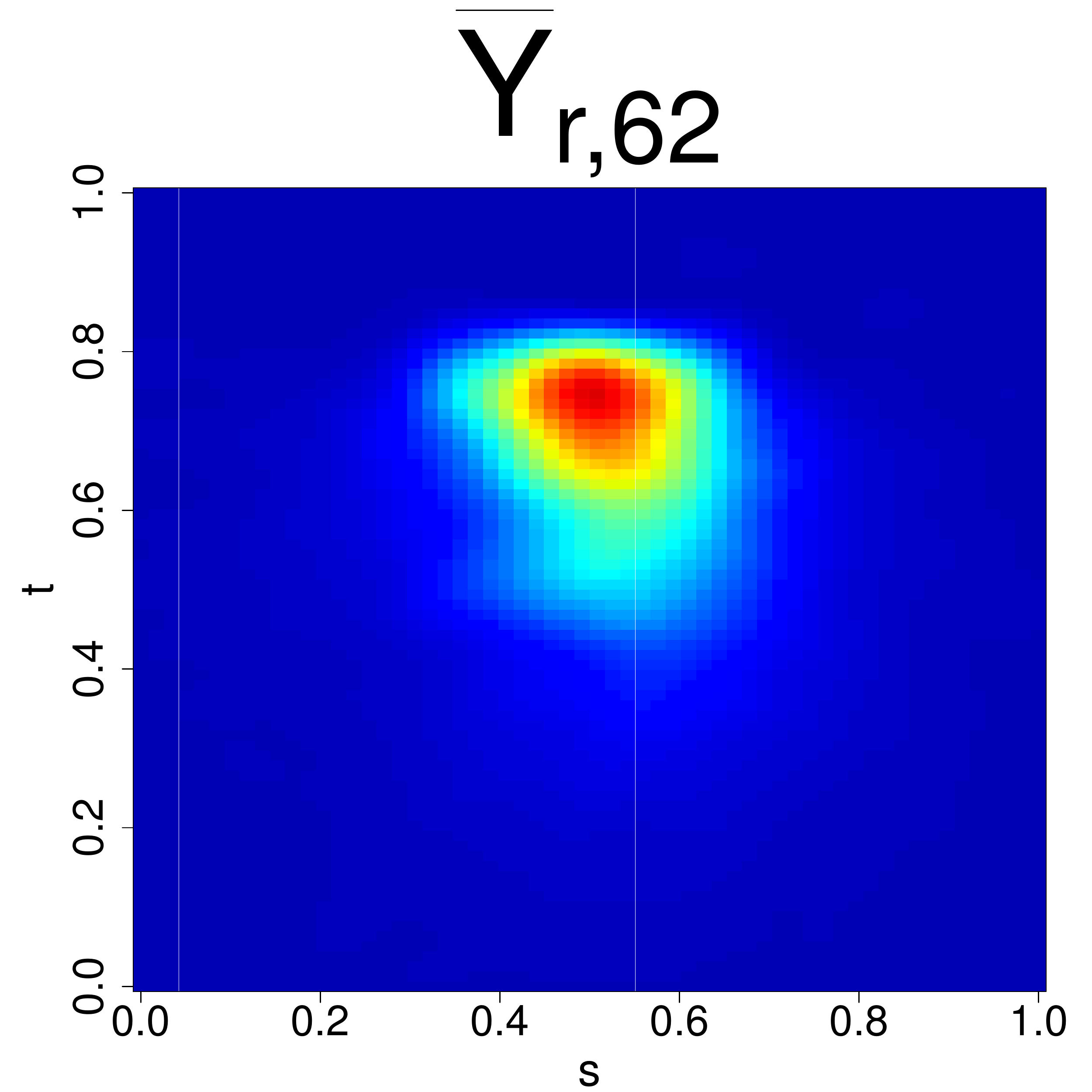}&\includegraphics[width=0.1\textwidth]{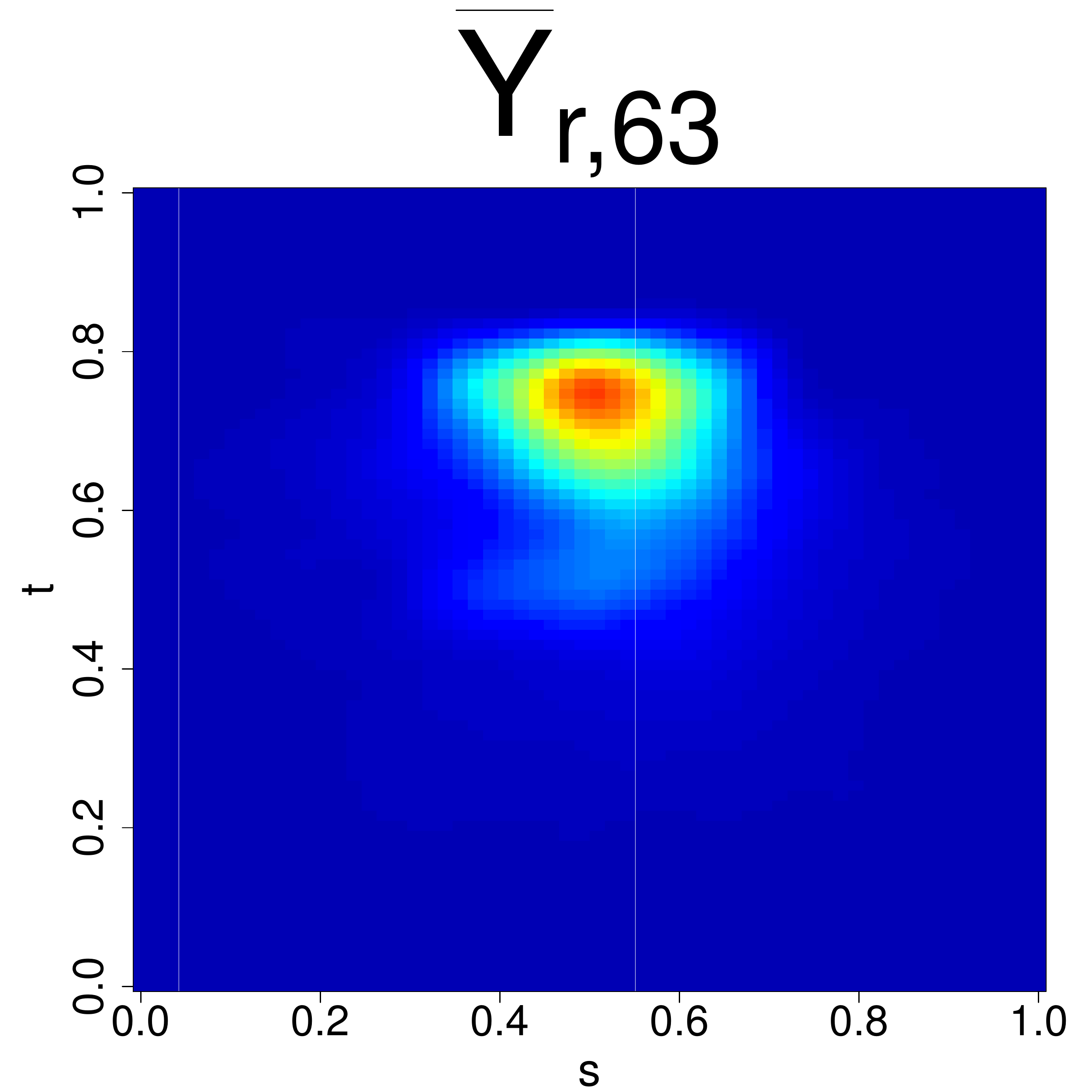}&\includegraphics[width=0.1\textwidth]{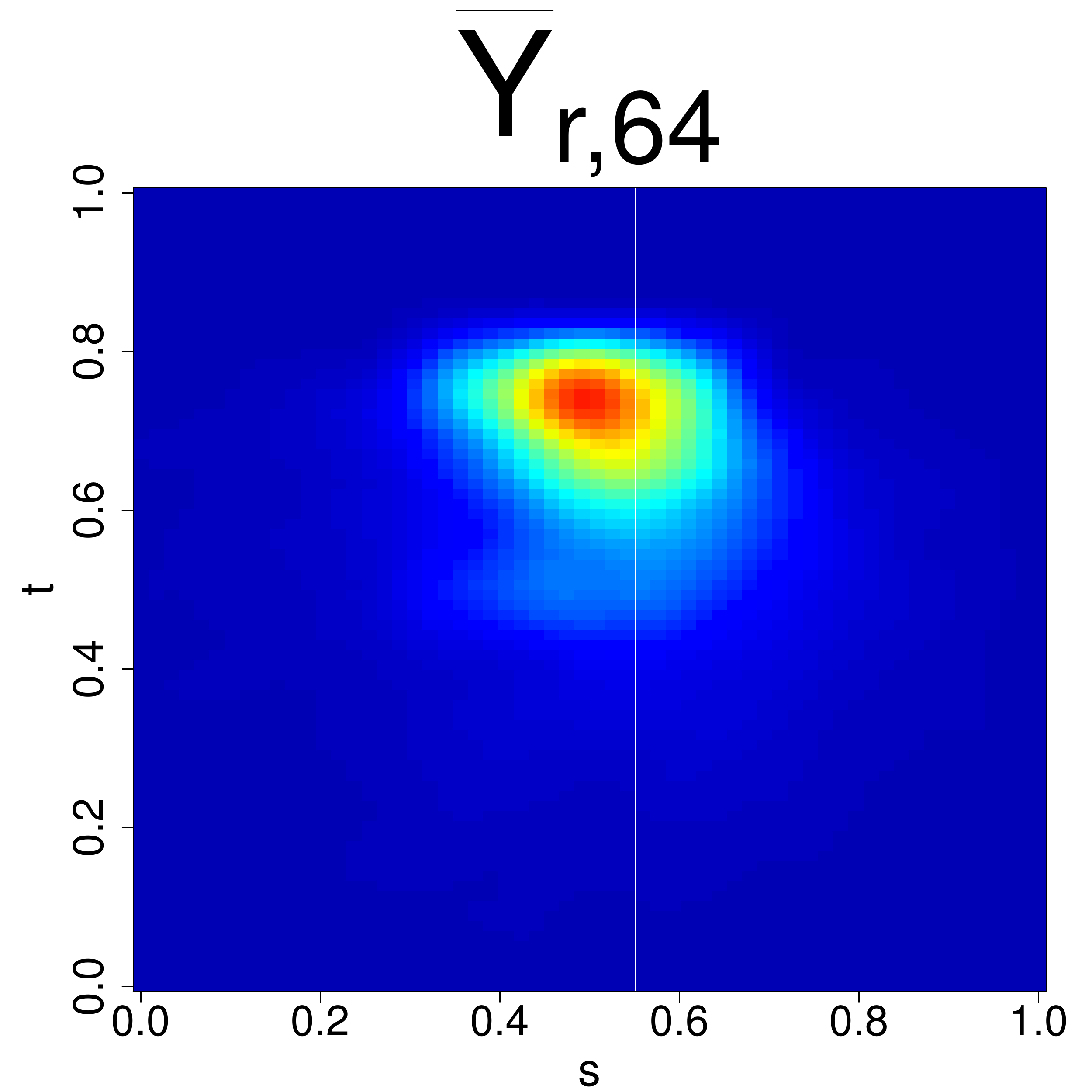}&\includegraphics[width=0.1\textwidth]{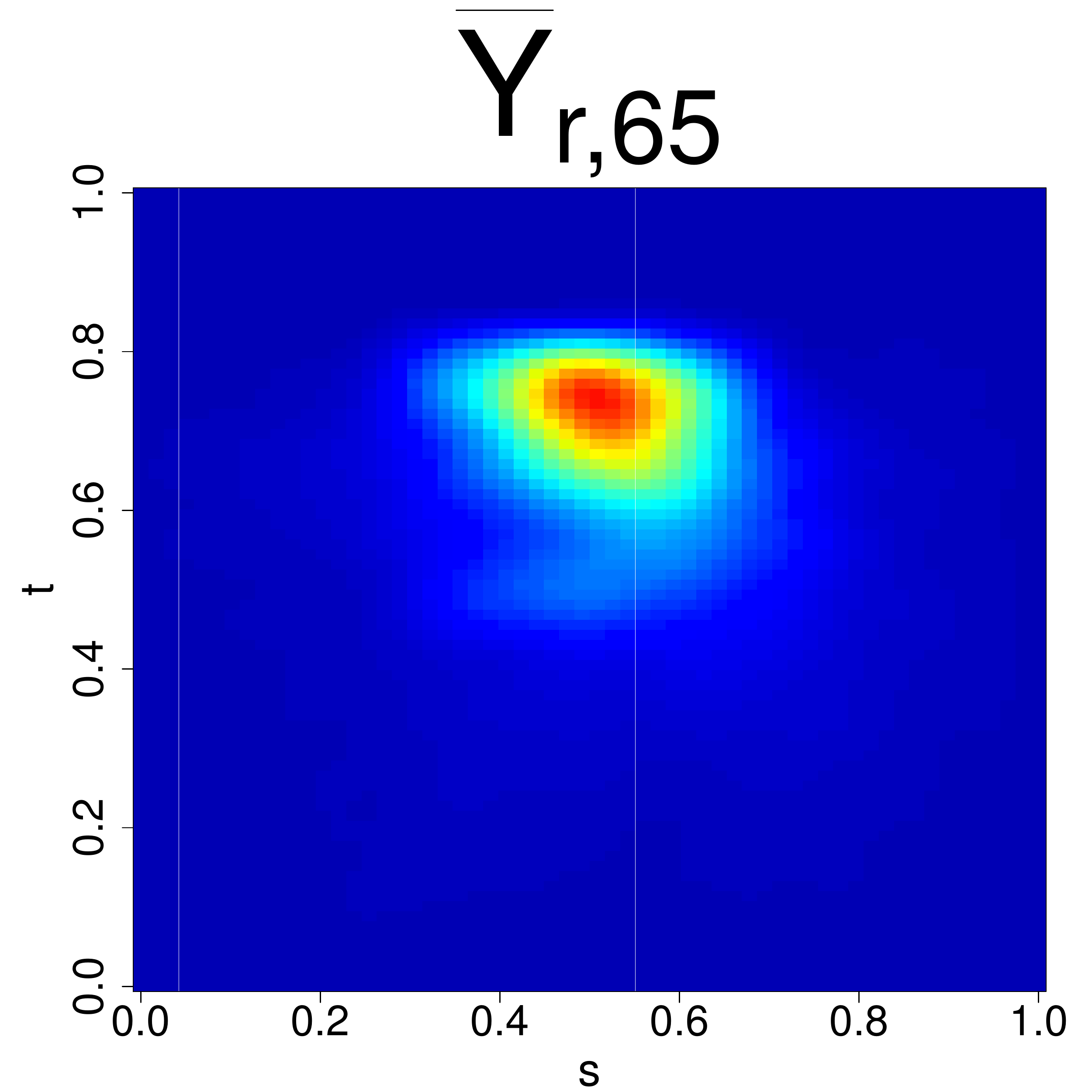}&\includegraphics[width=0.1\textwidth]{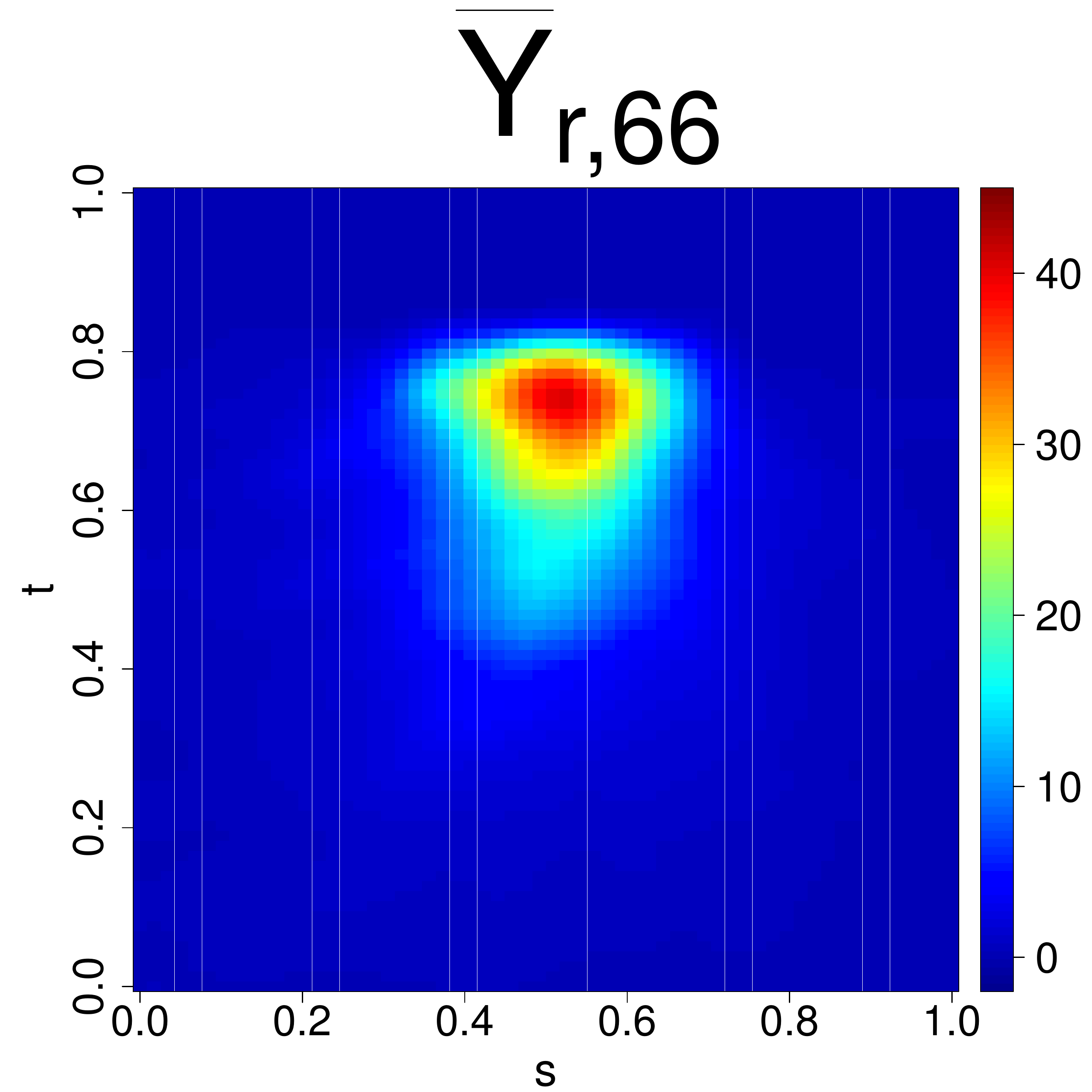}\\
\end{tabular}
	\caption{\label{fig_mat} Equivariant functional M-estimators $ \bar{Y}_{r} $, $ \bar{Y}_{r,i\cdot} $ (first column), $ \bar{Y}_{r,\cdot j} $ (first row) and $ \bar{Y}_{r,ij} $  of the functional grand mean, and of the group means  of $ \lbrace Y_{ijk} \rbrace_{k=1,\dots n_{ij},i=1,\dots 6 } $,  $ \lbrace Y_{ijk} \rbrace_{k=1,\dots n_{ij},j=1,\dots 6 } $, and  $ \lbrace Y_{ijk} \rbrace_{k=1,\dots n_{ij} } $, respectively. }
\end{sidewaysfigure}

\begin{table}
	\caption{\label{TABLE B} Location of analysed layers along the specimen build direction (distance from the baseplate)
	and orientation of the laser scan direction relative to the shielding gas flow in each layer.}
% 	\centering
	\scriptsize
    % \hspace{-58cm}
% 	\resizebox{0.6\textwidth}{!}{%
	\resizebox{0.6\textwidth}{!}{
	\begin{tabular}{c|c|c}
	\toprule
	\multirow{4}{*}{\shortstack{Analysed layer
 }} &\multirow{4}{*}{\shortstack{Layer height\\ along the build \\direction\\(mm)} }&\multirow{4}{*}{\shortstack{Laser scan angle\\ relative to the\\ shielding gas flow
}}\\
	&&\\
		&&\\
			&&\\\midrule
				1&31&$10^{\circ}$\\\midrule
				2&56&$40^{\circ}$\\\midrule
				3&83&$80^{\circ}$\\\midrule
				4&110&$85^{\circ}$\\\midrule
				5&137&$90^{\circ}$\\\midrule
				6&163&$30^{\circ}$ \\
				\bottomrule
	\end{tabular}
}
% 	}
\end{table}
To visually explore the effects of the energy density and the layer on the spatter intensity, Fig. \ref{fig_mat}   shows the equivariant functional M-estimators (Section \ref{subsec_mestimator})  $ \bar{Y}_{r} $, $ \bar{Y}_{r,i\cdot} $, $ \bar{Y}_{r,\cdot j} $ and $ \bar{Y}_{r,ij} $  of the functional grand mean, and of the group means  of $ \lbrace Y_{ijk} \rbrace_{k=1,\dots n_{ij},i=1,\dots 6 } $,  $ \lbrace Y_{ijk} \rbrace_{k=1,\dots n_{ij},j=1,\dots 6 } $, and  $ \lbrace Y_{ijk} \rbrace_{k=1,\dots n_{ij} } $, respectively. 

\bibliographystyle{chicago}
\bibliography{References}